\documentclass[useAMS,usenatbib,usegraphicx]{mn2e}
\usepackage{color}
\newcounter{subfigure}

\title[IFU spectroscopy of 10 ETG nuclei: III]
  {IFU spectroscopy of 10 early-type galactic nuclei - III. Properties of the circumnuclear gas emission}
\author[Ricci et al.]
  {T.V.~Ricci,$^1$\thanks{tvricci@iag.usp.br}
  J.E.~Steiner,$^1$ R.B.~Menezes$^1$ \\
  $^1$Instituto de Astronomia, Geof\'isica e Ci\^encias Atmosf\'ericas, Universidade de S\~ao Paulo, 05508-900, S\~ao Paulo, Brasil }
\date{Released 2015}

\pagerange{\pageref{firstpage}--\pageref{lastpage}} \pubyear{2015}

\def\LaTeX{L\kern-.36em\raise.3ex\hbox{a}\kern-.15em
    T\kern-.1667em\lower.7ex\hbox{E}\kern-.125emX}

\begin{document}

\label{firstpage}

\maketitle

\begin{abstract}

Many Early-type galaxies have ionized gas emission in their centres that extends to scales of $\sim$ 1kpc. The majority of such objects are classified as low-ionization nuclear emission regions (LINERs), but the nature of their ionizing source is still not clear. The kinematics associated with these gaseous structures usually shows deviations from a pure rotational motion due to non-gravitational effects (e.g. outflows) or to non-axisymmetric potentials (e.g. bars or tri-axial systems).  This is the third of a series of papers that describes a sample of 10 nearby (d $<$ 30 Mpc) and massive ($\sigma$ $>$ 200 km s$^{-1}$) early-type galaxies observed with the Gemini Multi-Object Spectrograph in Integral Field mode installed on the Gemini-South telescope. In paper II, we performed spectral synthesis to subtract the stellar components from the data cubes of the sample galaxies in order to study their nuclear spectra. In this work, we analyse the circumnuclear gas emission (scales of $\sim$ 100 pc) of the sample galaxies and we compare the results with those obtained with PCA Tomography in paper I. We detected circumnuclear gas emission in seven galaxies of the sample, all of them classified as LINERs. Pure gaseous discs are found in three galaxies. In two objects, gaseous discs are probably present, but their kinematics are affected by non-Keplerian motions. In one galaxy (IC 5181), we detected a spiral structure of gas that may be caused either by a non-axisymmetric potential or by an outflow together with a gaseous disc. In NGC 3136, an ionization bicone is present in addition to five compact structures with LINER-like emission. In galaxies with a gaseous disc, we found that ionizing photons emitted by an AGN are not enough to explain the observed H$\alpha$ flux along this structure. On the other hand, the H$\alpha$ flux distribution and equivalent width along the direction perpendicular the gaseous disc suggest the presence of low-velocity ionized gas emission which seem to be related to the nuclear activity. We propose a scenario for LINER-like circumnuclear regions where a low-velocity ionization cone is formed by a collimating agent aligned with the gaseous disc.

\end{abstract}


\section{Introduction} \label{sec:intro}

The presence of ionized gas in the central regions of early-type galaxies (ETGs) has been studied by several works in the literature \citep{1986AJ.....91.1062P,1989ApJ...346..653K,2006MNRAS.366.1151S,2010MNRAS.402.2187S,2012ApJ...747...61Y}. However, the source of ionization is still not clear. Since most of this emission is characterized as low-ionization nuclear emission regions (LINERs) \citep{1980A&A....87..152H,2008ARA&A..46..475H}, several mechanisms may be responsible for ionizing the gas: active galactic nuclei (AGNs) - \citet{1983ApJ...264..105F,1983ApJ...269L..37H}; shocks - \citet{1980A&A....87..152H,2003adu..book.....D}; post-asymptotic giant branch star populations (pAGBs) - \citet{1994A&A...292...13B,2008MNRAS.391L..29S,2011MNRAS.413.1687C,2012ApJ...747...61Y}. The study of unresolved nuclear sources is very useful in detecting the presence of an AGN. Broad components in permitted lines and the presence of compact radio and X-ray cores are typical signatures of nuclear activity in LINERs \citep{2008ARA&A..46..475H}. In addition, circumnuclear (scales of $\sim$ 100 pc) and extended regions (scales of $\sim$ 1 kpc) may also help to constrain the source of ionization in the central regions of ETGs.

Hubble Space Telescope (HST) observations of circumnuclear regions of LINERs have shown that their kinematics may be more complex than pure rotational motion typical of gaseous discs \citep{2000ApJ...532..323P,2008AJ....136.1677W}. Moreover, only one LINER of the sample studied by \citet{2000ApJ...532..323P} seems to have the same structure found in Seyfert galaxies, i.e. ionization cones and linear structures. The exception is NGC 1052, which clearly shows an ionization cone structure aligned with a radio jet. With regard to photoionization, these authors suggest that a central UV source is responsible for the emission of the circumnuclear region, although they do not speculate on the nature of this source (it could be an AGN or hot stars). \citet{2008AJ....136.1677W}, using spectra obtained with the Space Telescope Imaging Spectrograph (STIS), detected significant electron-density gradient in four LINERs and showed that, in most galaxies, the gas velocity dispersion peaks in the region dominated by the black hole and decreases in the outer regions. They argue that, in some galaxies, outflows may increase the gas velocity dispersion and that the presence of virial motions close to the central supermassive black hole (SMBH) may either increase or decrease gas velocity dispersion in the nuclear region, depending on the emission line surface brightness or even on the inclination of the disc. \citet{2007ApJ...654..125S}, comparing HST observations with data from the Palomar Survey \citep{1985ApJS...57..503F,1997ApJS..112..315H}, showed that the gas emission is spatially distributed, with an extension of $\sim$ 100 pc, and that accretion-powered objects reveal a concentration of the emission in the nuclear region.

In more extended regions, \citet{2006MNRAS.366.1151S} showed that the gas kinematics does not behave as a pure disc in a large fraction of the ETGs sample taken from SAURON integral field unit (IFU) data, although they proposed that the disturbances are probably caused by non-axisymmetric potentials (e.g. bars or tri-axial systems) on kpc-scales. In the same sample, \citet{2010MNRAS.402.2187S} noticed that the contribution of AGNs to the photoionization of LINERs is restricted to the nuclear region and that pAGB stars are likely responsible for the extended gas emission. The same findings were reported by \citet{2013A&A...558A..43S}, using IFU data from the Calar Alto Legacy Integral Field Area (CALIFA) Survey. Both \citet{2010MNRAS.402.2187S} and \citet{2013A&A...558A..43S} highlight the fact that the association of LINER emission with nuclear activity may overestimate the population of AGNs in the Universe, since aperture effects may include regions dominated by extended photoionization sources (see e.g. \citealt{2012ApJ...747...61Y}). In fact, \citet{2011MNRAS.413.1687C} proposed that a comparison of the equivalent width (EW) of the H$\alpha$ emission line with the [N II]/H$\alpha$ ratio (the WHAN diagram) may be used to separate AGN from pAGB stars as the source of photoionization in LINER galaxies, even when aperture effects are present in the data (e.g. data from the Sloan Digital Sky Survey - SDSS). \citet{2013A&A...555L...1P}, also using data from the CALIFA survey, showed that LINER emission from ETGs may be divided in two types: one where AGNs and pAGB stars are a dominating photoionization source in the nuclear and extended region, respectively, and the other where the leakage of Lyman continuum photons is important, leading to reduced emission-line luminosities and EWs. This could explain why galaxies with strong AGN emission in radio continuum are weak in the optical region. Furthermore, they argued that the relative distribution of gas when compared to the stars may be an important issue as well. Since in tri-axial systems the volume of the emitting gas is limited when compared to the stellar component, the nuclear EW is lowered in these cases, leading to an observational bias against detection of AGN activity in gas-poor galaxy spheroids. Another possible scenario for weak optical emission and strong radio sources is that most of the energy released by the AGN may be mechanical and not radiative.

This is the third of a series of papers that aim to characterize the nuclear and circumnuclear regions of a sample of 10 ETGs observed with the IFU installed on the Gemini South Telescope. In this paper, we intend to characterize the circumnuclear region of these 10 ETGs.  In \citet{2014MNRAS.440.2419R} (hereafter paper I), we analysed this sample using Principal Component Analysis (PCA) Tomography \citep{1997ApJ...475..173H,2009MNRAS.395...64S}. In \citet{2014MNRAS.440.2442R} (hereafter paper II), we analysed the emission lines from the nuclear region after subtracting the stellar component from each spectrum of the data cubes. The results from both works showed that all 10 galaxies have typical LINER emission and, in seven of them, we detected bona fide AGNs. We proposed that eight galaxies have circumnuclear emission and in seven of them the kinematics may be interpreted as a gaseous disc, while an ionization cone is likely to be present in one galaxy. However, using only the PCA Tomography technique, it is hard to detect kinematic perturbations in a gaseous disc by means of, let's say, an outflow or even a non-axisymmetric potential. Thus, in this work, we will analyse the emission lines from the circumnuclear region after subtracting the stellar component from the data cubes. It is worth mentioning that the CALIFA and the SAURON surveys also performed such a study, but in more extended regions (scales of kpc). However, both projects have poor spatial resolution to resolve the circumnuclear region. In the CALIFA project, the fibres of the IFU instrument have a diameter of 2.7 arcsec \citep{2012A&A...538A...8S} and in the SAURON integral field spectrograph, the instrument allows a spatial sampling of 0.94 arcsec in the spatial low resolution mode \citep{2001MNRAS.326...23B}. Although HST observations allow for such information in a scale from 10 to 20 pc, it is restricted to long-slit data. In the Gemini Multi-Object Spectrograph (GMOS) in IFU mode, the fibres have a diameter of 0.2 arcsec and the spatial resolution is limited by the seeing of the observations. Although limited to a field of view (FOV) of 3.5 arcsec x 5 arcsec, it allows one to characterize the circumnuclear region of the ETGs using 3D spectroscopy with superb spatial resolution for ground-based telescopes. 

Section \ref{data_description} contains a brief description of the data. Section \ref{FOV_properties} describes the emission lines extracted from the circumnuclear regions. In Section \ref{diagdiagextendedemission}, we use BPT diagnostic diagrams to classify the emission type of the circumnuclear regions. In Section \ref{circumnuclear_maps}, we build maps of the radial velocity, velocity dispersion, H$\alpha$ and [N II] fluxes and EWs, [N II]/H$\alpha$ ratios and electron density for the whole FOV of the observations. In Section \ref{agn_flux_distribution}, we discuss if the output from AGNs are enough to explain the circumnuclear emission of ionized gas. In Section \ref{discs_or_cones}, we discuss the presence of discs or ionization cones in the sample galaxies. In Section \ref{hst_nuclear}, we show images taken with the HST of some galaxies of the sample. Finally, in Section \ref{sec:conclusions}, we present a brief discussion of the results and our main conclusions. In appendix \ref{comments_on_objects}, we provide comments for individual objects, summarizing the findings presented in papers I, II and III for each galaxy.

\section{Brief description of the data} \label{data_description}

The sample analysed in this paper was previously discussed in paper I. In short, it is composed by 10 massive ETGs ($\sigma_* >$ 200 km s$^{-1}$) with distances up to 30 Mpc. The galaxies were observed with the Gemini South Telescope using the GMOS-IFU in one-slit mode (programmes GS-2008A-Q51 and GS-2008B-Q21). The data were reduced with the standard {\sc IRAF} package for the Gemini telescopes. Bias, flat-fields and CuAr lamp images, in addition to spectrophotometric standards, were acquired to properly correct and calibrate the data. The final product is a data cube with a FOV of 3.5 arcsec x 5 arcsec, where each spaxel has a size of 0.05 arcsec. Besides the basic reduction process, we applied a Butterworth filter (low-pass) to each image of the data cubes with the aim of removing high frequency noise from the spatial dimension. A spectral low frequency instrumental fingerprint was identified and duly removed with the PCA Tomography technique (\citealt{2009MNRAS.395...64S}, paper I). The data cubes were also corrected for the differential atmospheric refraction (DAR) effect using an algorithm developed by us. Finally, we deconvolved each image of the data cubes with the Richardson-Lucy deconvolution process \citep{1972JOSA...62...55R,1974AJ.....79..745L} with six iterations and assuming a Gaussian point spread function (PSF), whose FWHM values are equal to the seeing of the observations shown in paper I, where a more detailed description of the data cube reduction process and treatment of the noise is presented.

We also performed a stellar population spectral synthesis in each spectrum of the data cubes. For this task, we used the {\sc starlight} software \citep{2005MNRAS.358..363C}. The stellar population basis in our synthesis was described by \citet{2009MNRAS.398L..44W}. This basis uses the MILES library \citep{2010MNRAS.404.1639V} with solar abundances, whose fluxes were recalibrated pixel by pixel for non-solar abundances using the theoretical models of simple stellar populations (SSPs) of \citet{2007MNRAS.382..498C}. The basis has 120 SSPs with a spectral resolution of 2.51 \AA\ \citep{2011A&A...532A..95F,2011A&A...531A.109B}, ages between 3 and 12 Gyr with steps of 1 Gyr, abundances [Fe/H] = -0.5, -0.25, 0.0 and 0.2 and [$\alpha$/Fe] = 0.0, 0.2 and 0.4. More details on the spectral synthesis performed on the data cubes of the sample galaxies, as well as the best-fitting stellar models for the spectra of the sample galaxies, are presented in paper II (Appendix A). Finally, we subtracted the stellar features from each spectrum of the data cubes, resulting in gas cubes, i.e., data cubes with only the spectral lines related to the gaseous component. In paper II, we analysed only the spectra extracted from the nuclear region of the galaxies. In the present work, the entire FOV of the gas cubes is studied to extract information from the circumnuclear regions of the sample galaxies.

\section{Circumnuclear emission line properties for each sample galaxies} \label{FOV_properties}

We extracted representative spectra from the circumnuclear regions of the sample galaxies. For each gas cube, we summed all spectra within a circular region. Since each gas cube has a distinct spatial dimension (due to DAR correction) and also because of the position of the galactic nucleus of each object in the FOV, the extracted radius varies for each galaxy and it is limited by the smaller spatial dimension of the cube. These radii are presented in Table \ref{tab_cin_NLR_ext}. The final circumnuclear spectrum of each galaxy is the result of the sum described above minus the respective nuclear spectrum that was extracted and analysed in paper II.

The line profiles were fitted by means of the Gauss-Newton algorithm. It fits non-linear functions using the least squares method. More details about the fitting procedure may be found in paper II. The profiles were fitted with the sum of two Gaussian functions for each emission line. We fitted first the [S II]$\lambda\lambda$6716, 6731 doublet. For the other emission lines, the only free parameters of the fitting procedure were the amplitudes of the Gaussian functions, once the gas velocity dispersion and the radial velocities of each Gaussian were constrained by the results found for the [S II] doublet (similar procedure was adopted by \citealt{1997ApJS..112..391H}). Although different lines may have different profiles, since they arise from different regions, the kinematic constraint is very helpful to describe the emission lines in spectral regions where the signal-to-noise ratio is quite low. Even with these constraints, we did not detect the H$\beta$ line of NGC 3136. The fitting procedure did indeed find a solution for H$\beta$ in this galaxy, but the amplitude of the sum of the two Gaussian functions is $\sim$ 1.3$\sigma$, where $\sigma$ is the standard deviation of the spectral region of this line. In ESO 208-G21, we did not detect the [O I]$\lambda$6300 line to a level of 1.2$\sigma$. We did not detect any emission lines in the circumnuclear region of NGC 1399, NGC 1404 and NGC 2663. 

\begin{figure*}
\begin{center}
\includegraphics[width=70mm,height=55mm]{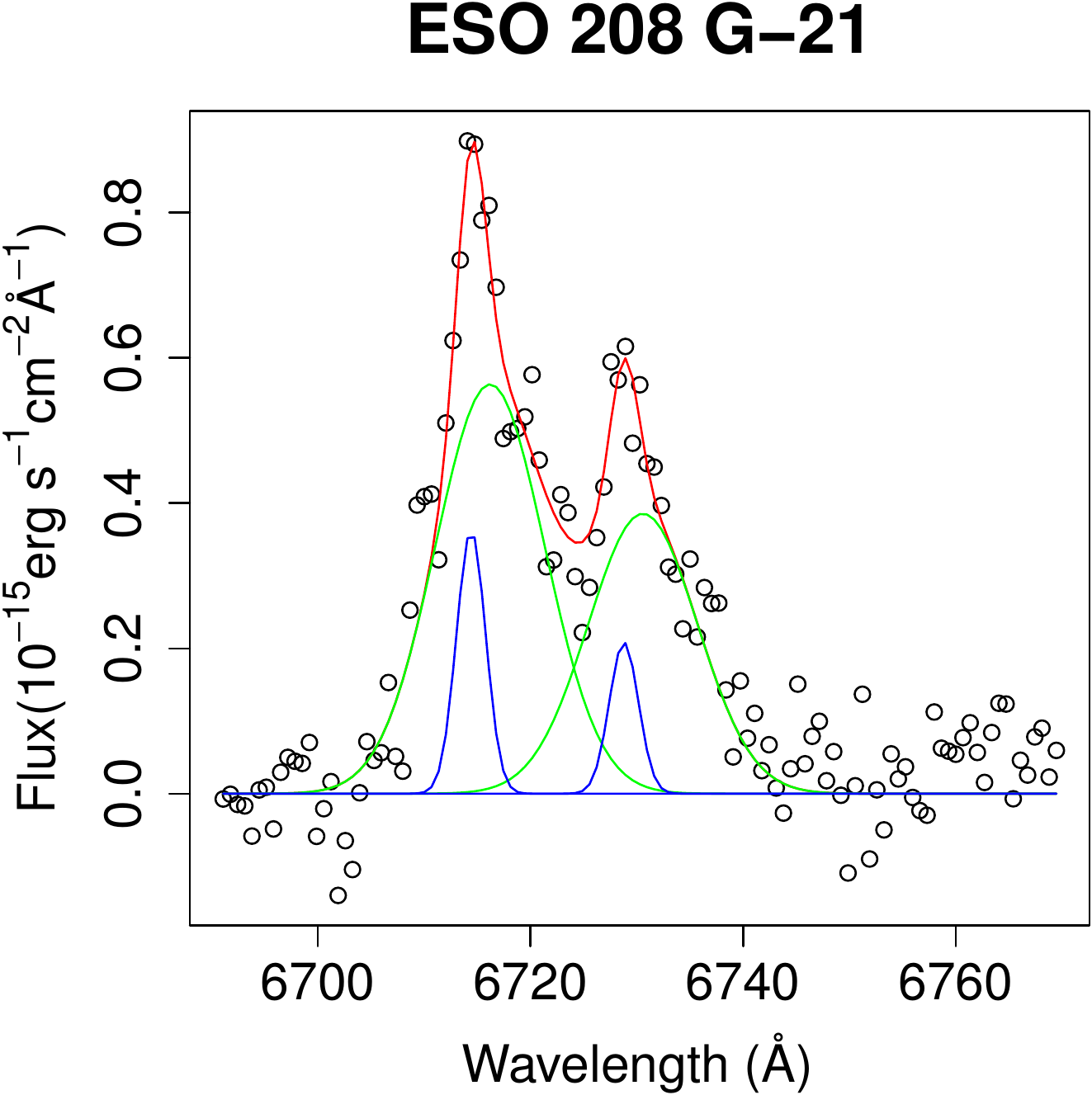}
\includegraphics[width=70mm,height=55mm]{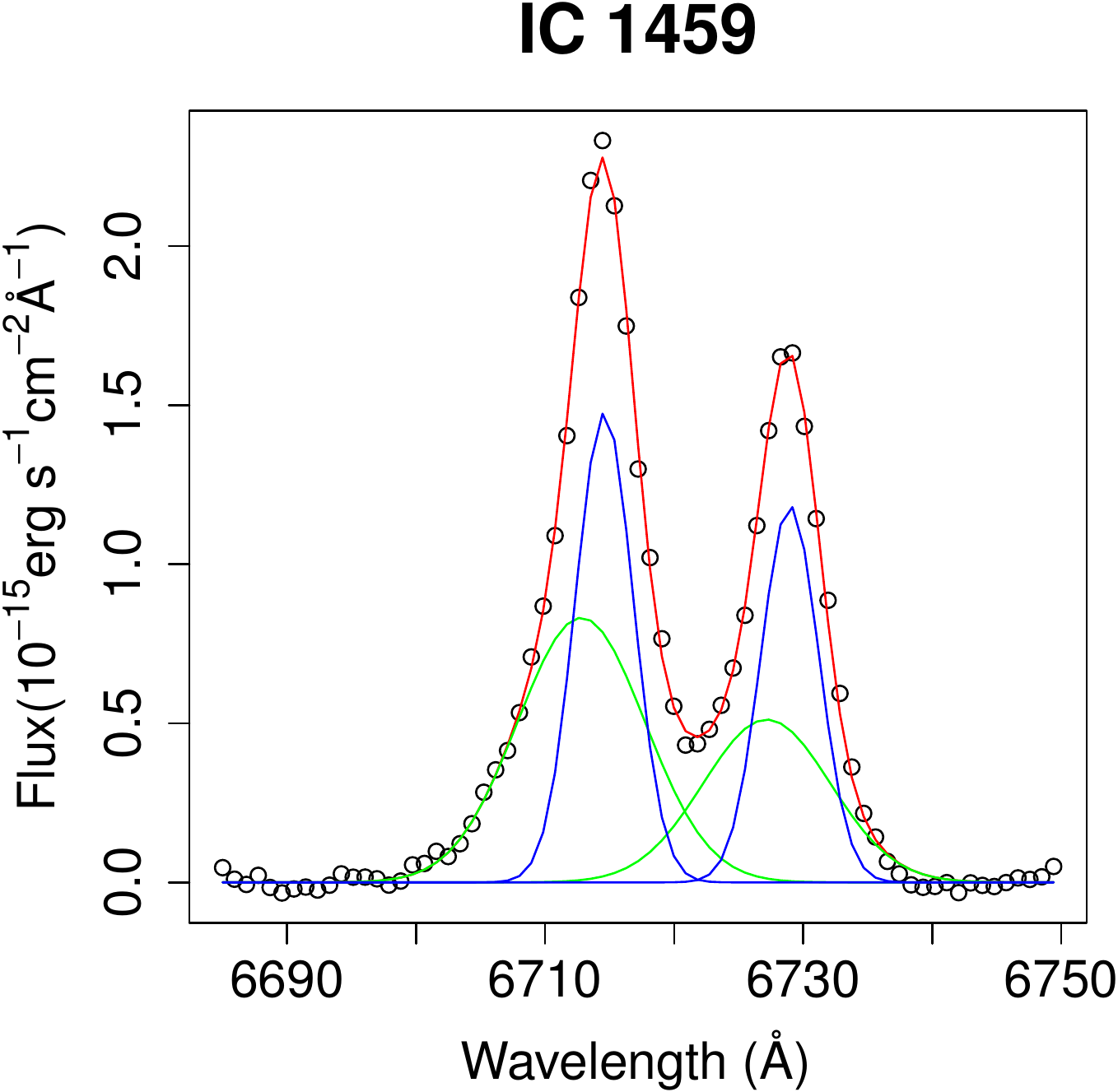}
\includegraphics[width=70mm,height=55mm]{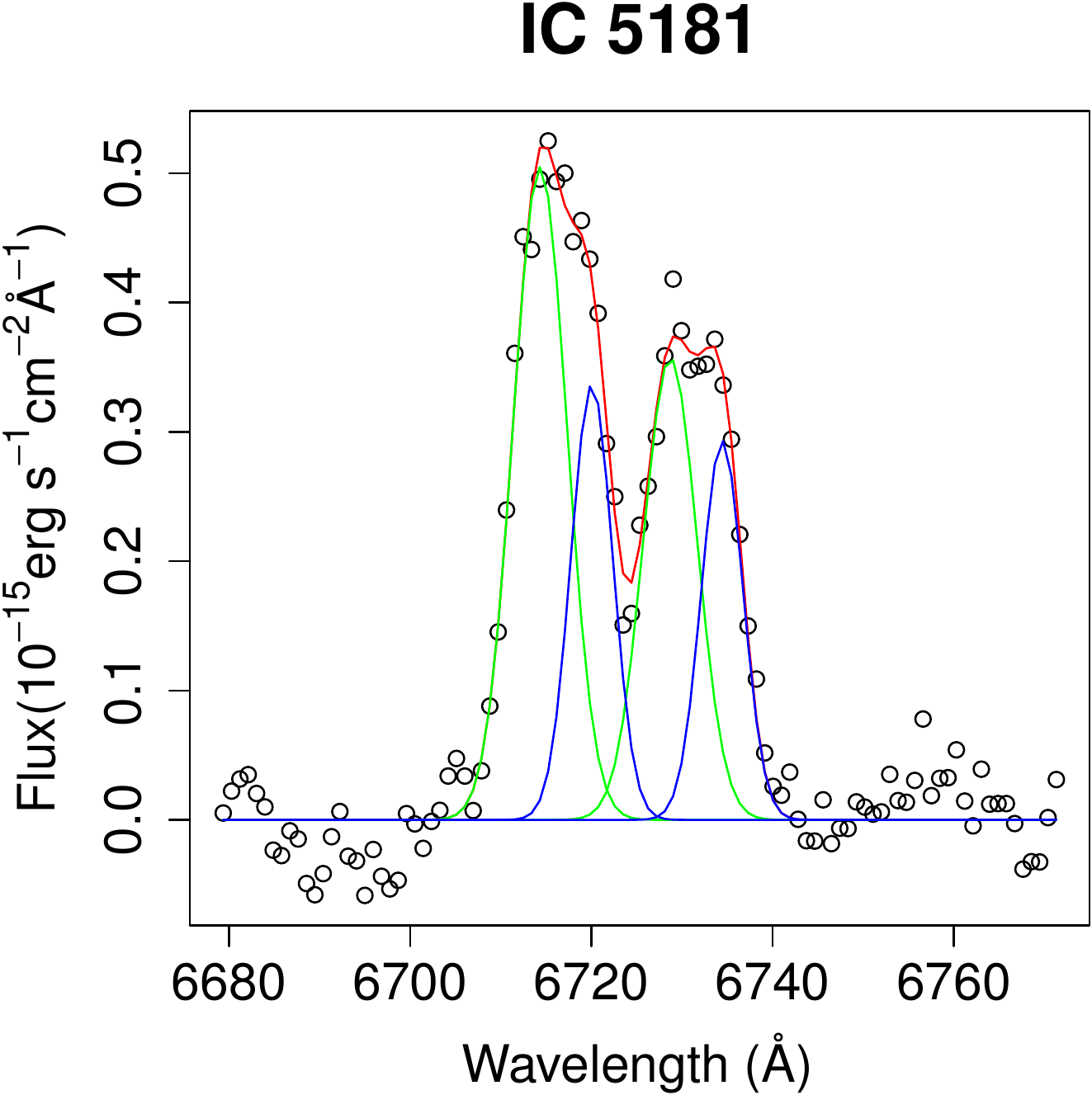}
\includegraphics[width=70mm,height=55mm]{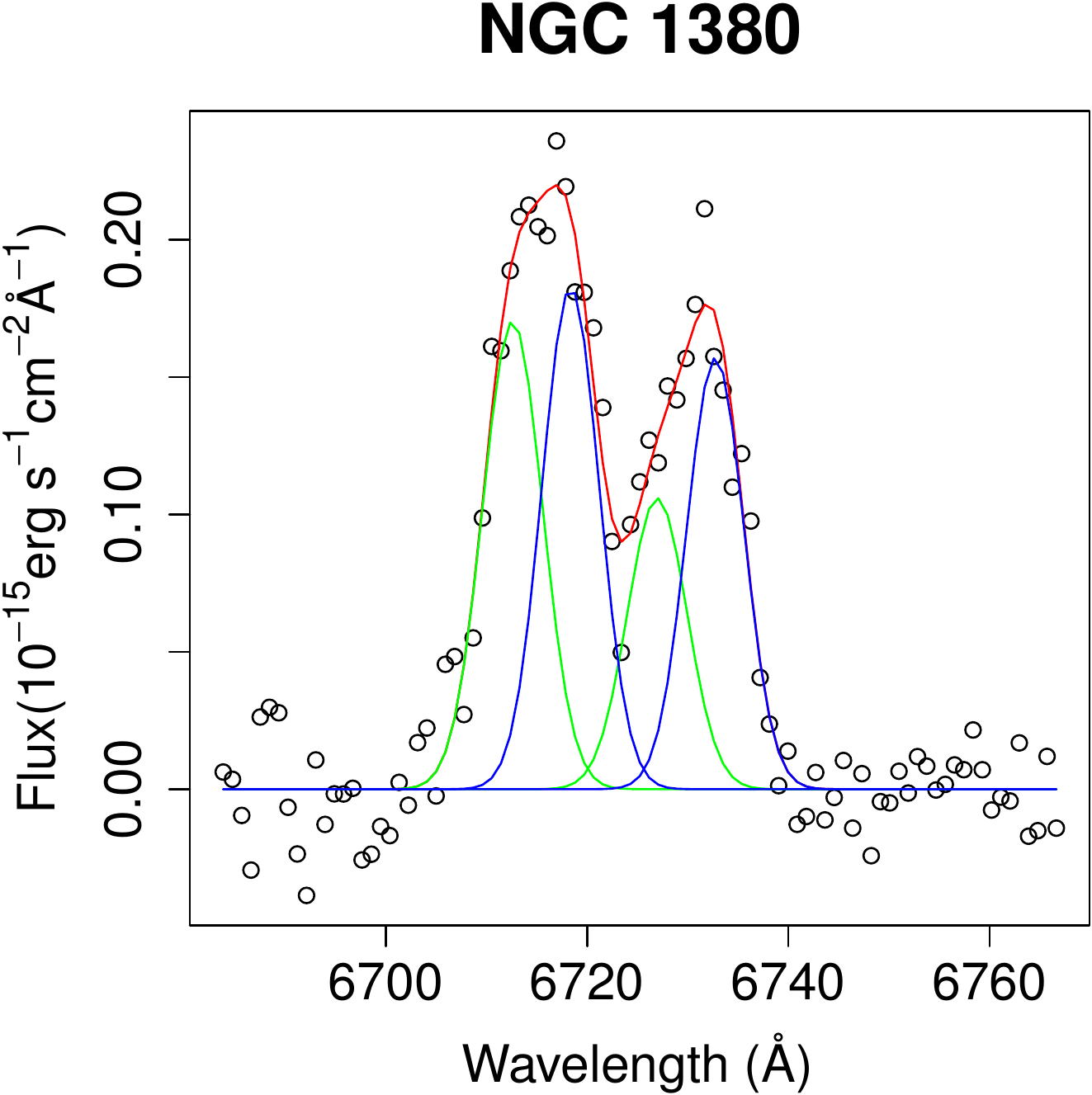}
\includegraphics[width=70mm,height=55mm]{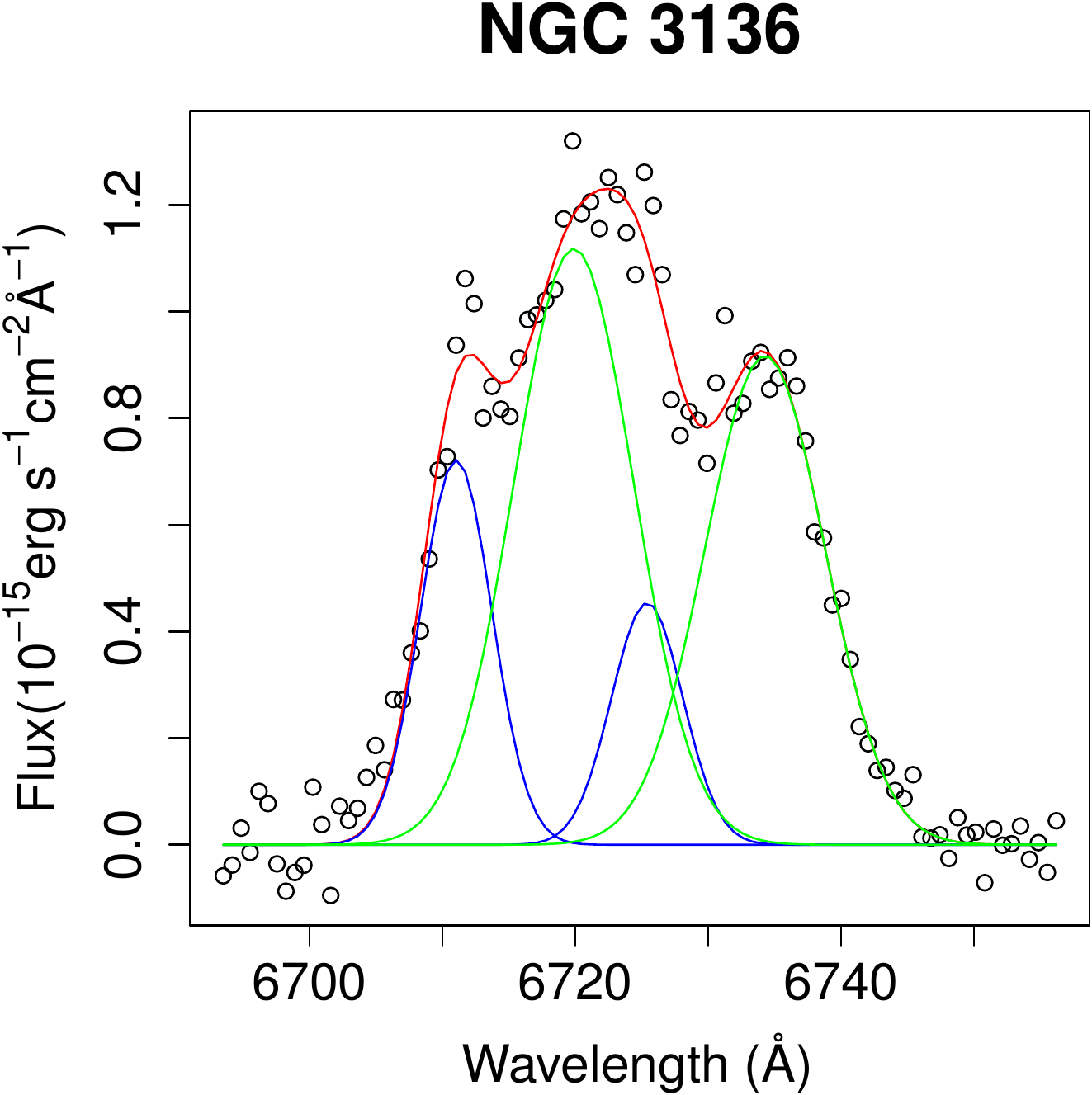}
\includegraphics[width=70mm,height=55mm]{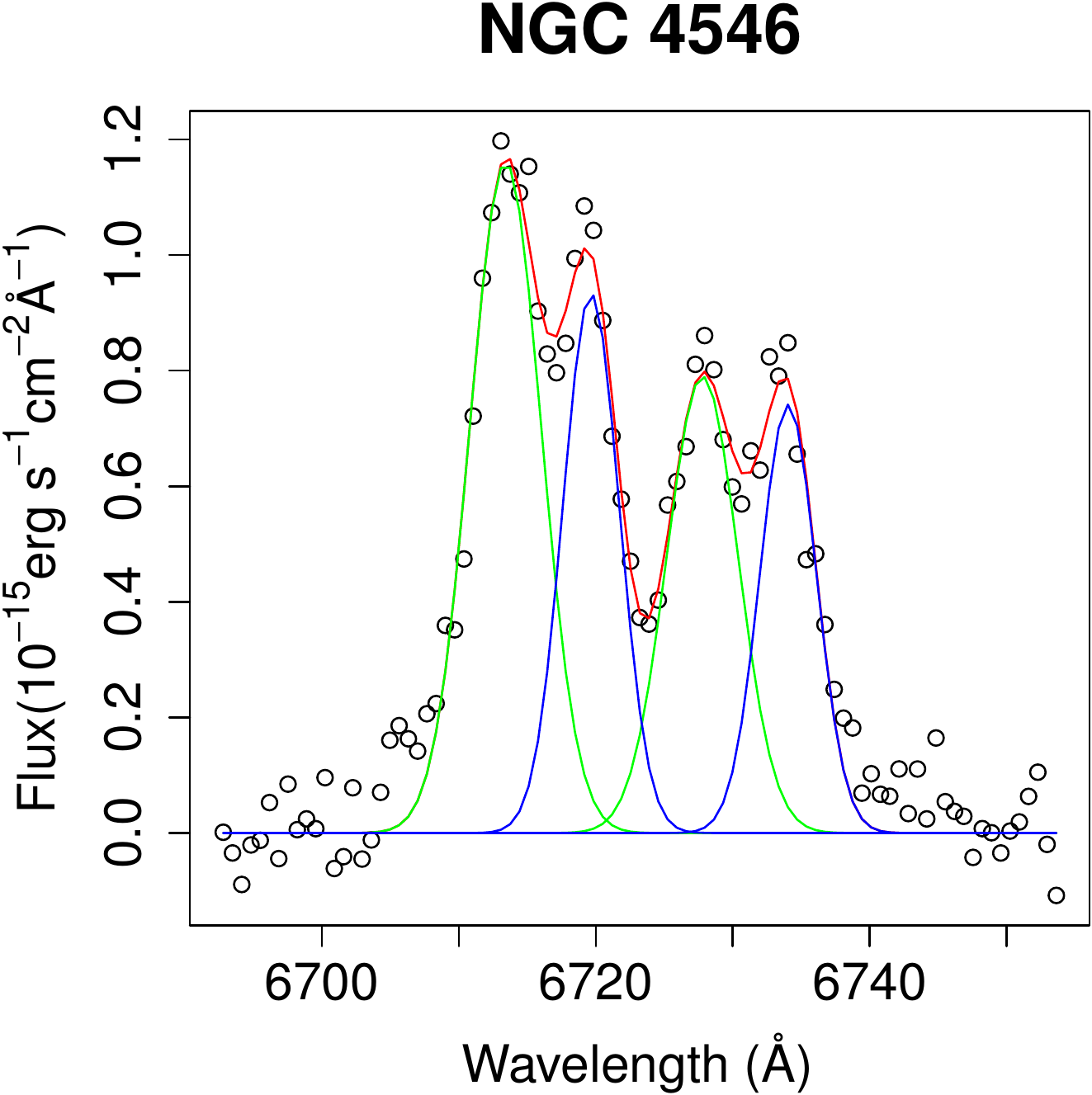}
\includegraphics[width=70mm,height=55mm]{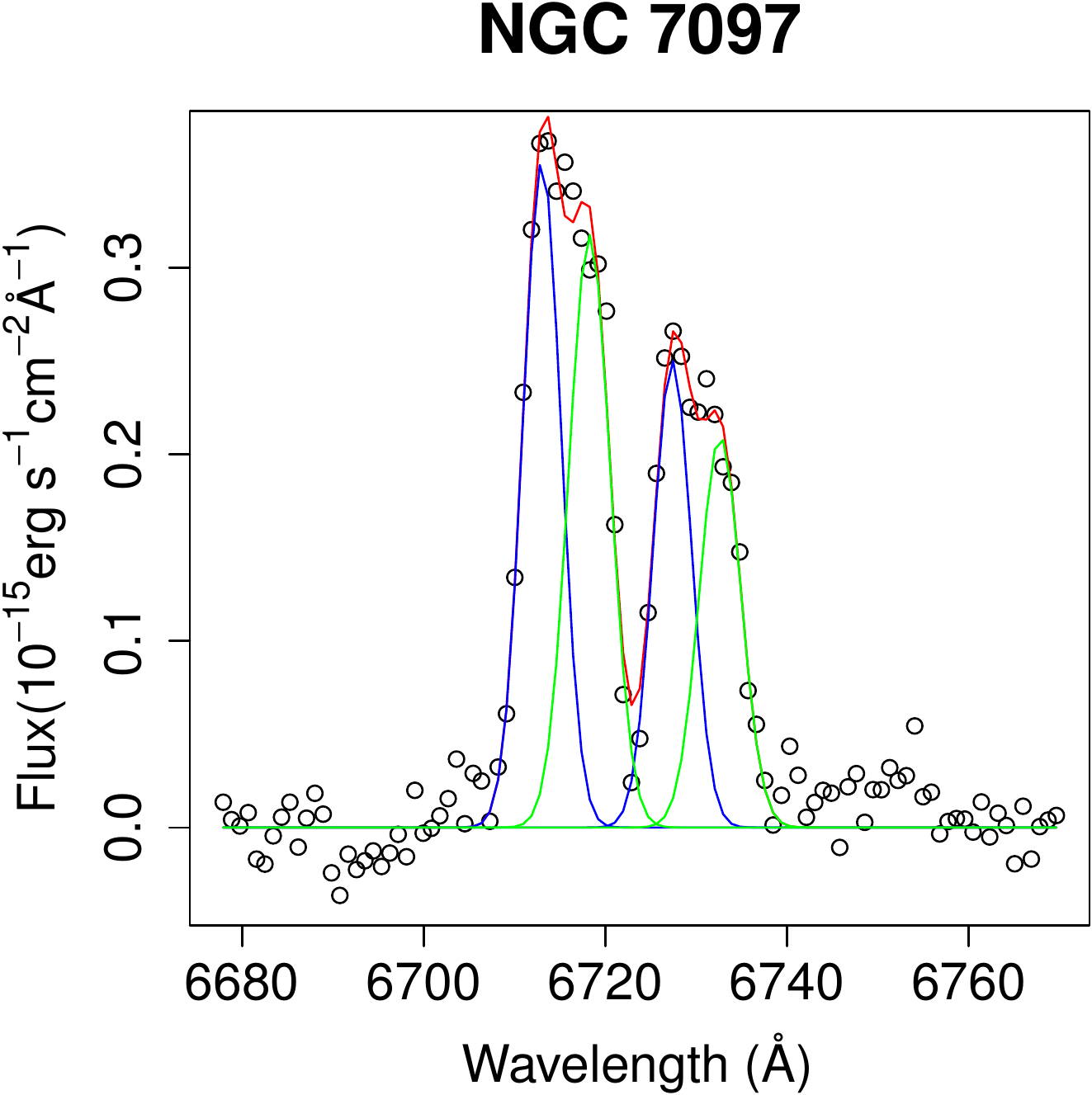}

\caption{[S II]$\lambda\lambda$6716, 6731 emission lines detected in the circumnuclear region of seven galaxies of the sample. The line profiles were fitted with a sum of two Gaussian functions, shown in blue and in green in the graphs above. For both [S II] lines, each Gaussian has the same kinematic parameters (radial velocity and velocity dispersion). The full fitted profile is shown in red. \label{perfil_SII_ext}
}
\end{center}
\end{figure*}

\begin{figure*}
\begin{center}
\includegraphics[width=70mm,height=55mm]{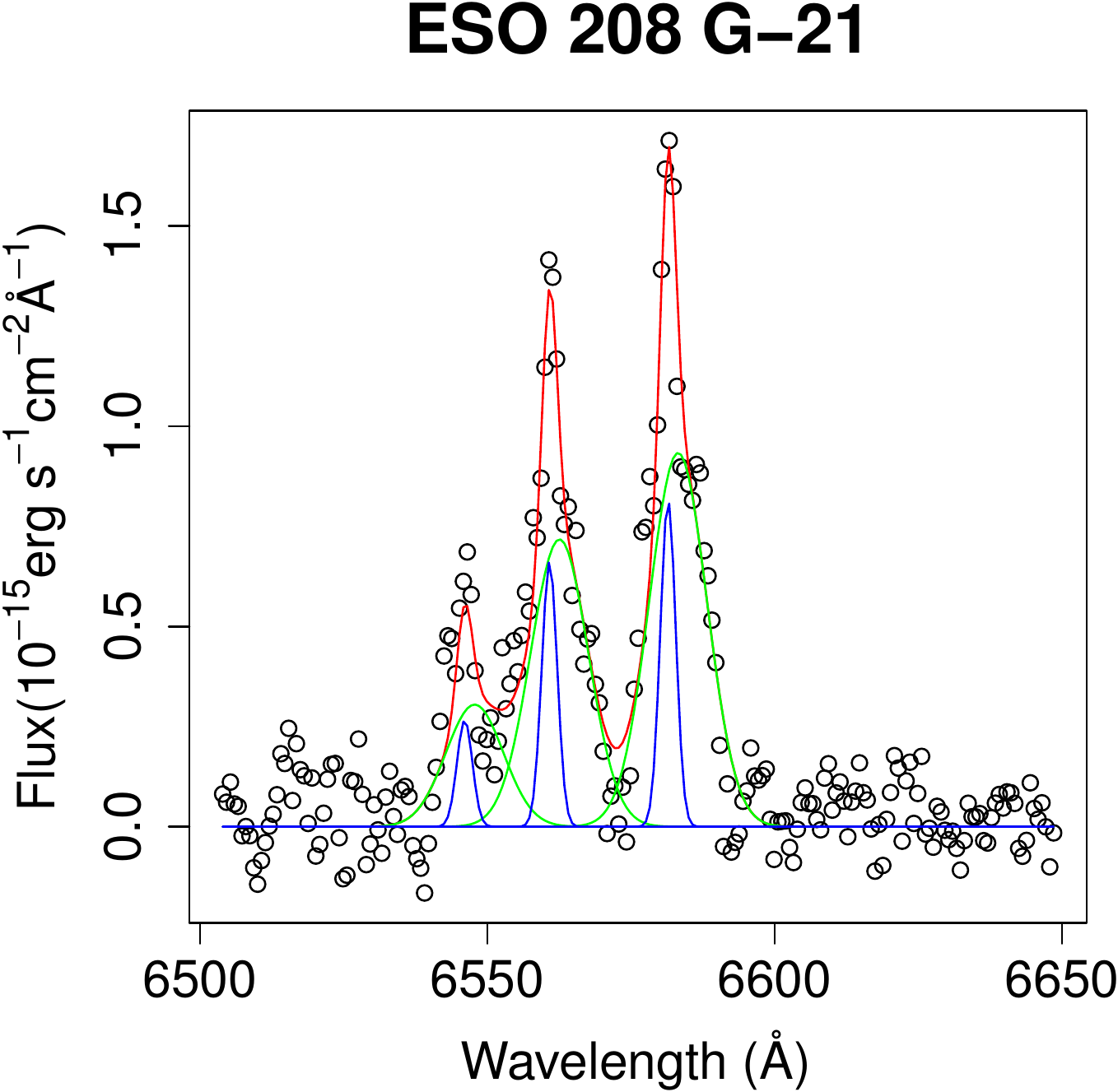}
\includegraphics[width=70mm,height=55mm]{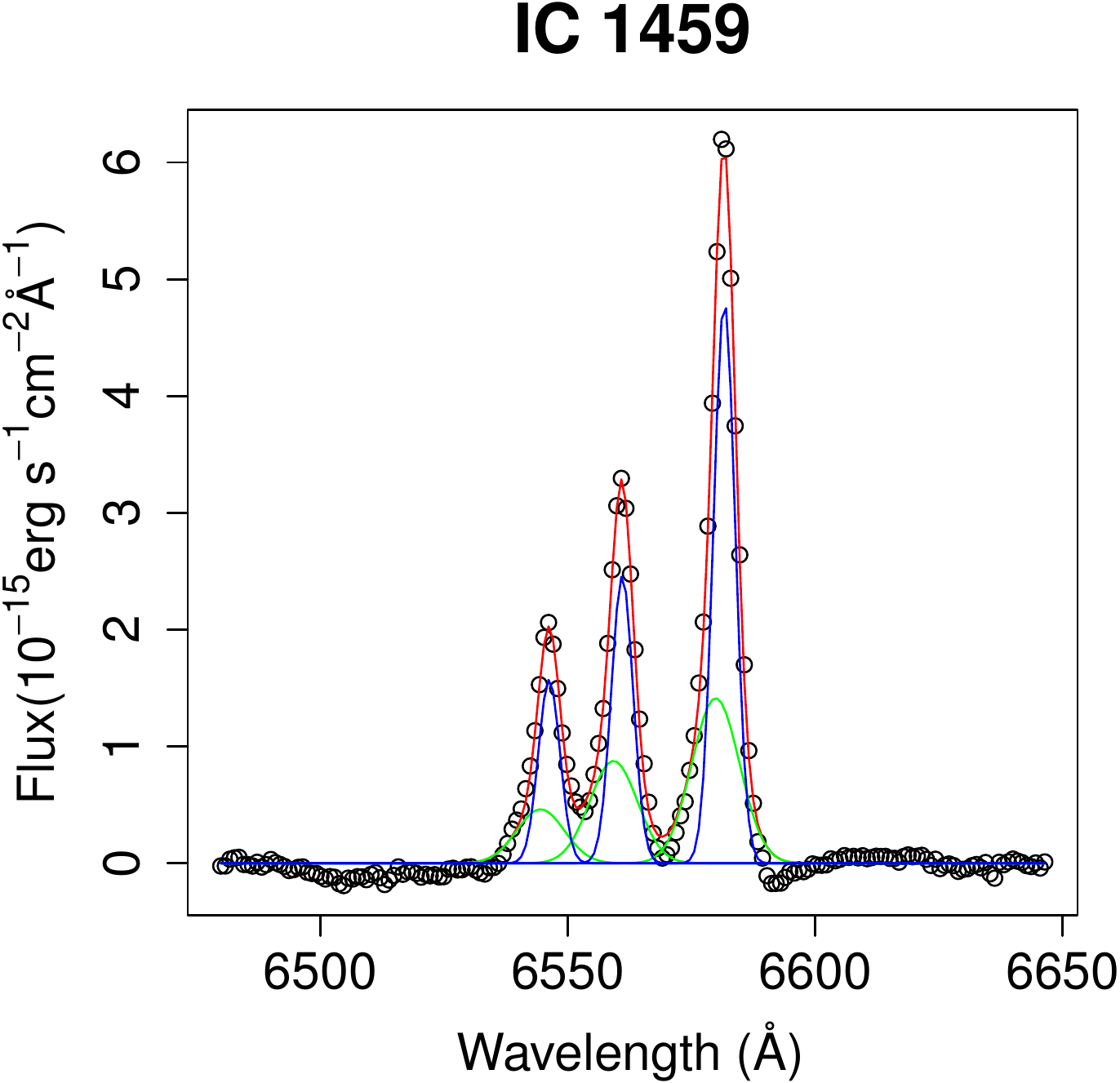}
\includegraphics[width=70mm,height=55mm]{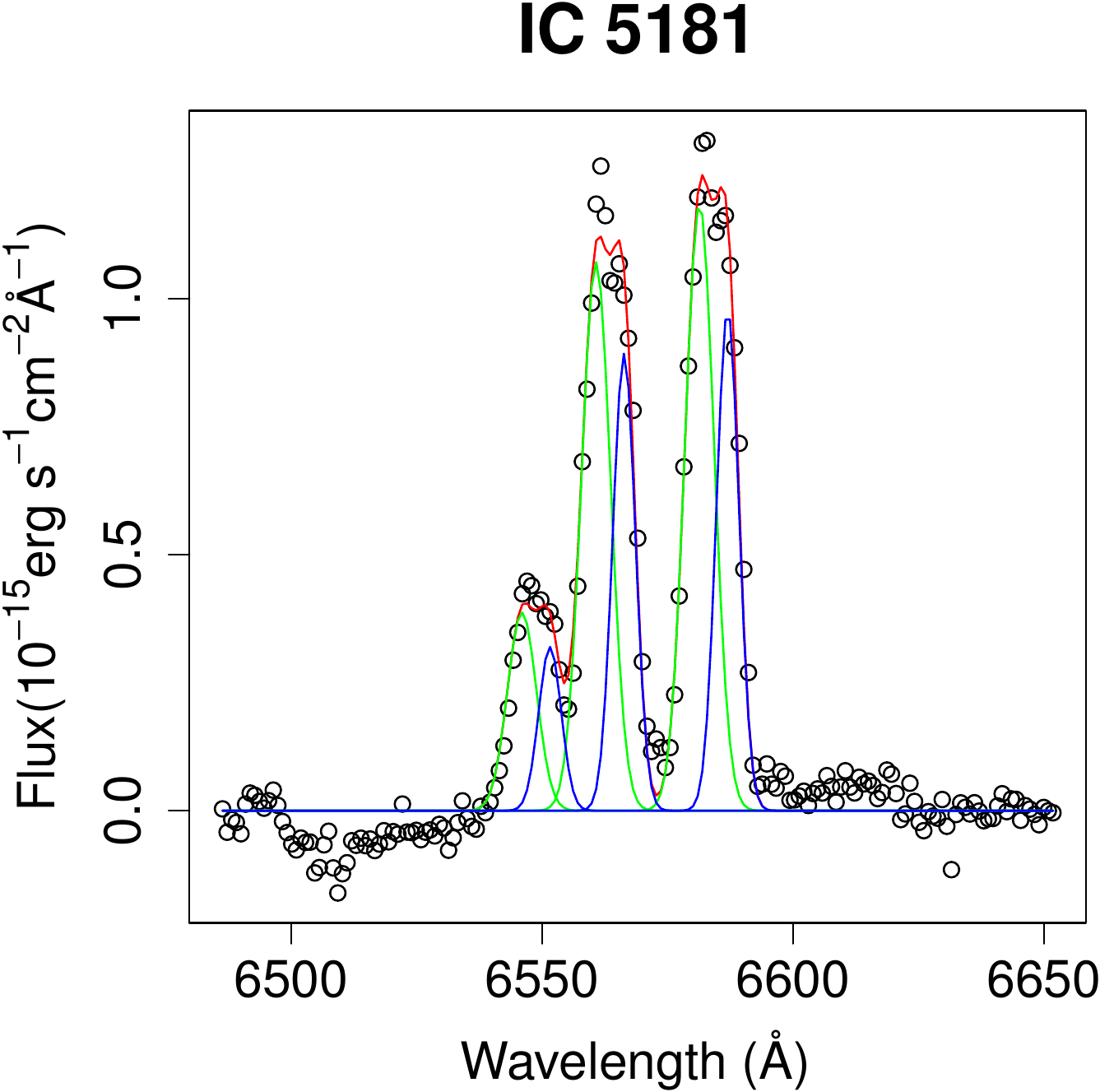}
\includegraphics[width=70mm,height=55mm]{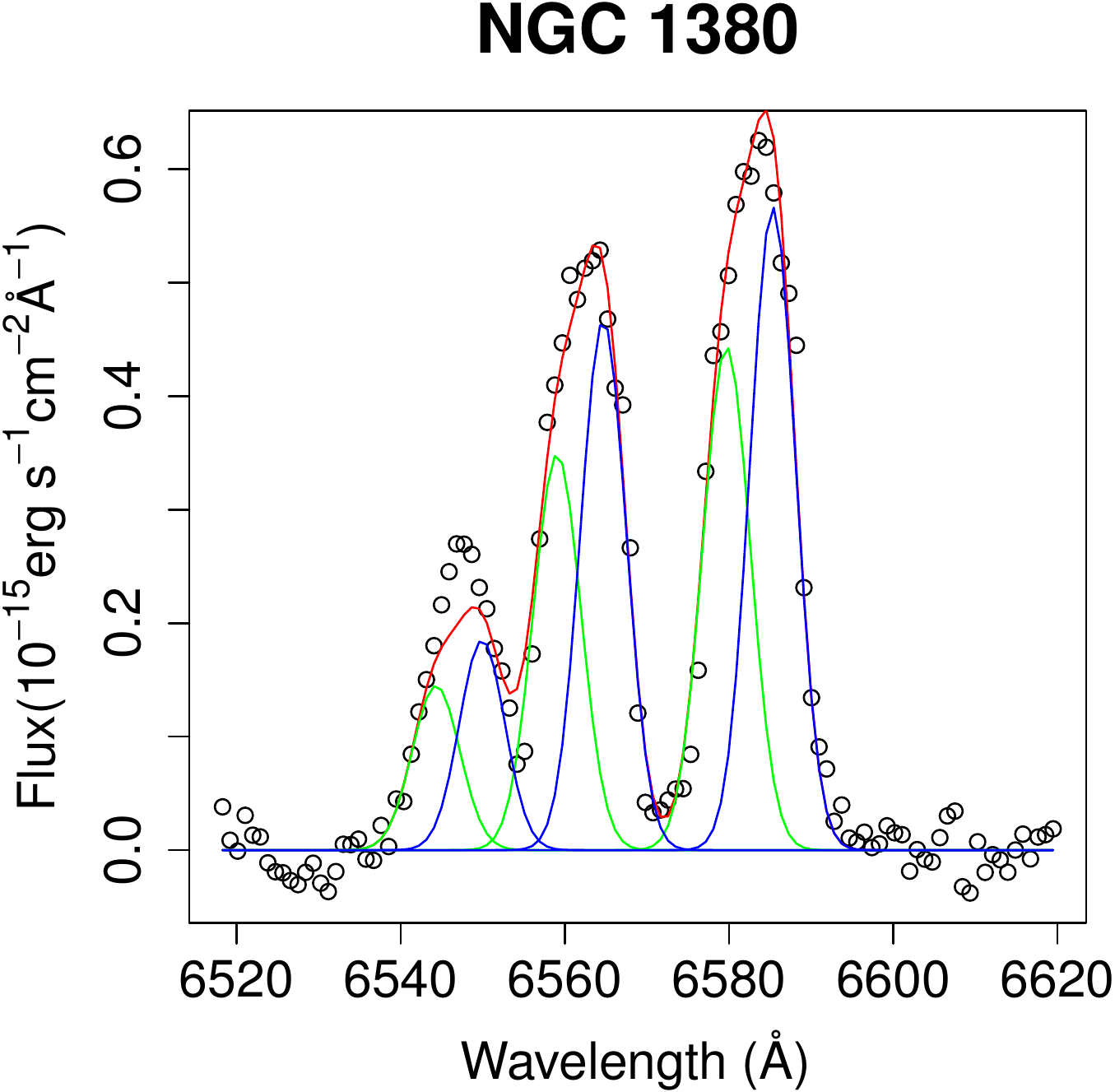}
\includegraphics[width=70mm,height=55mm]{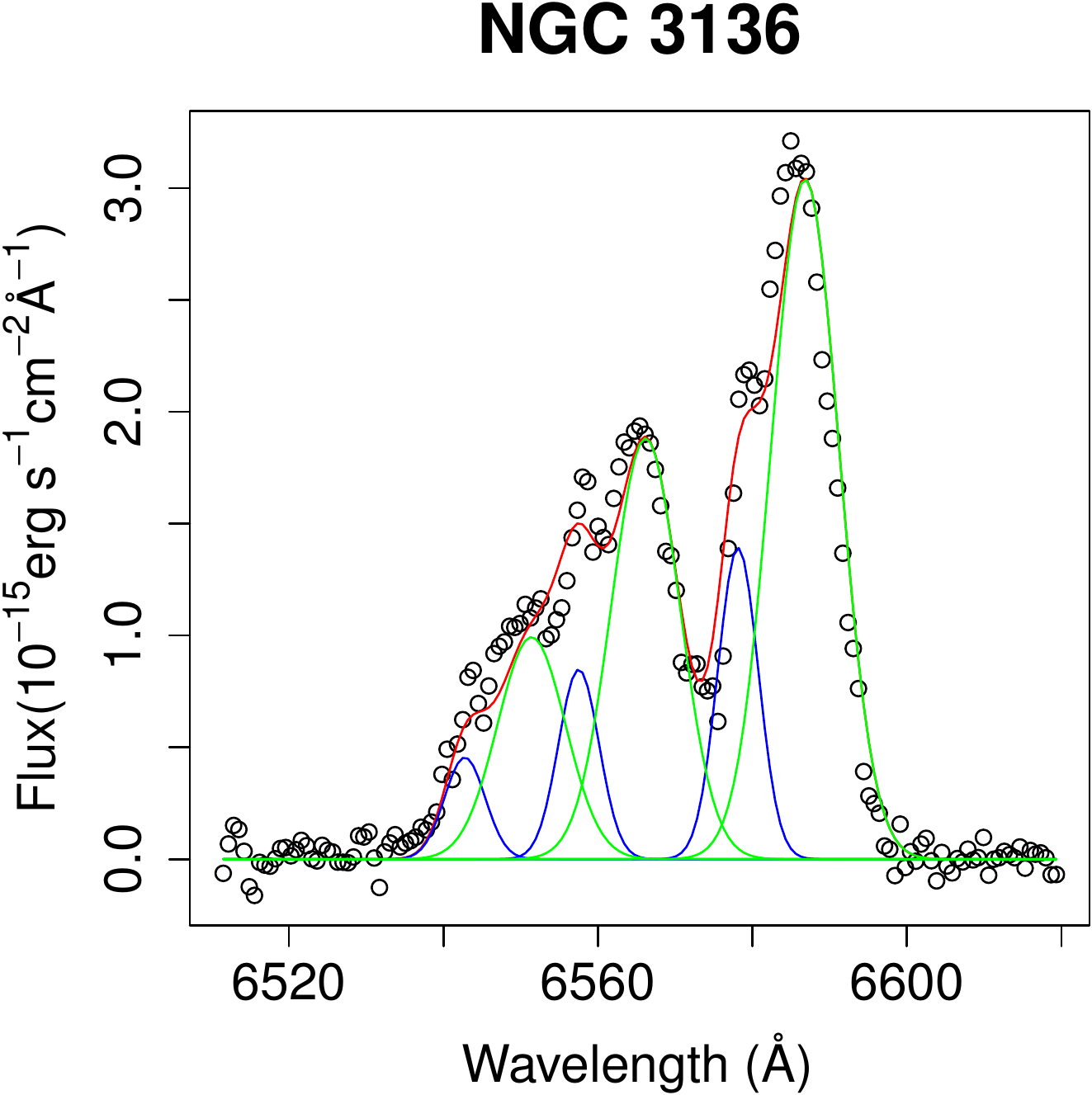}
\includegraphics[width=70mm,height=55mm]{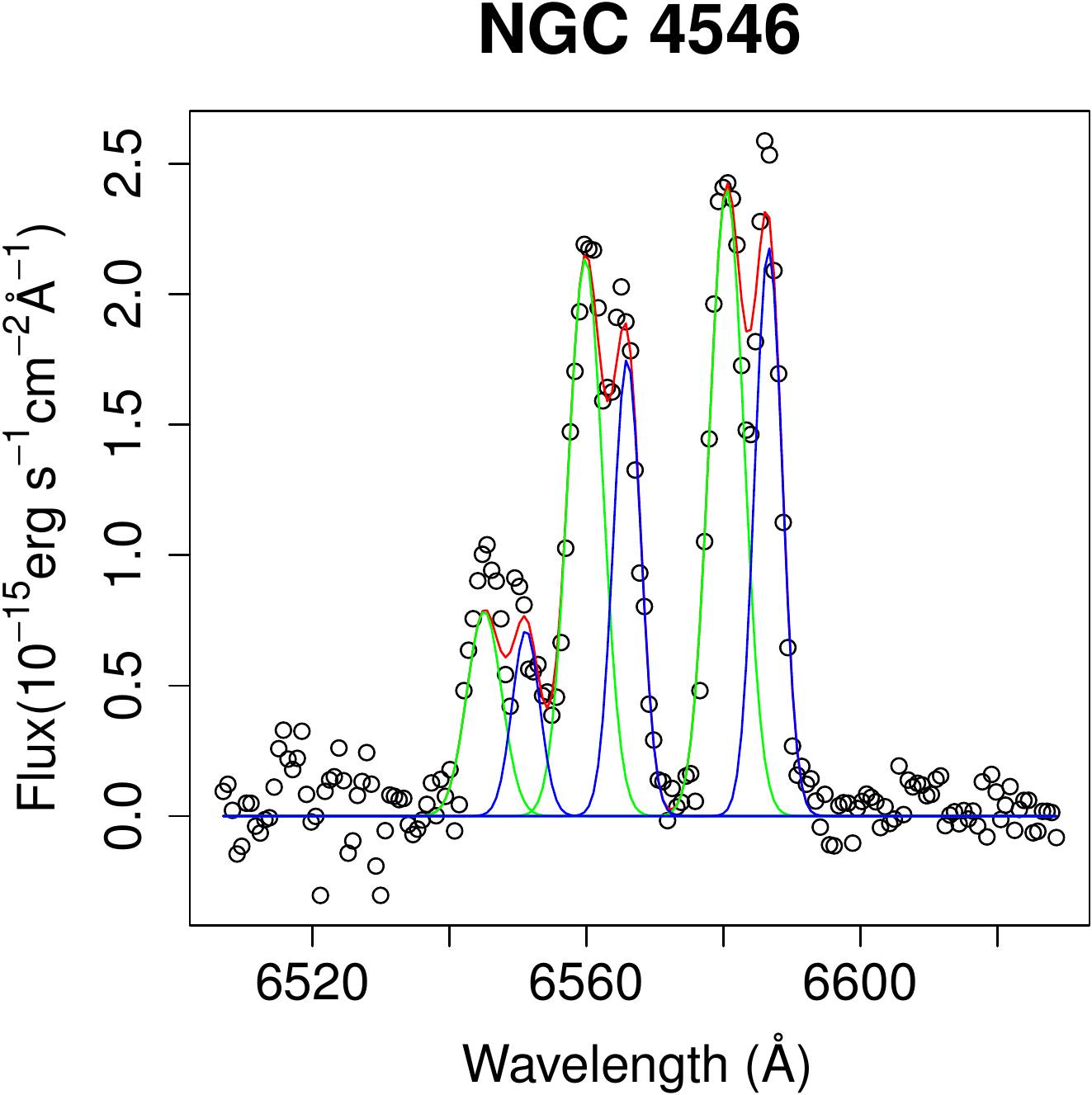}
\includegraphics[width=70mm,height=55mm]{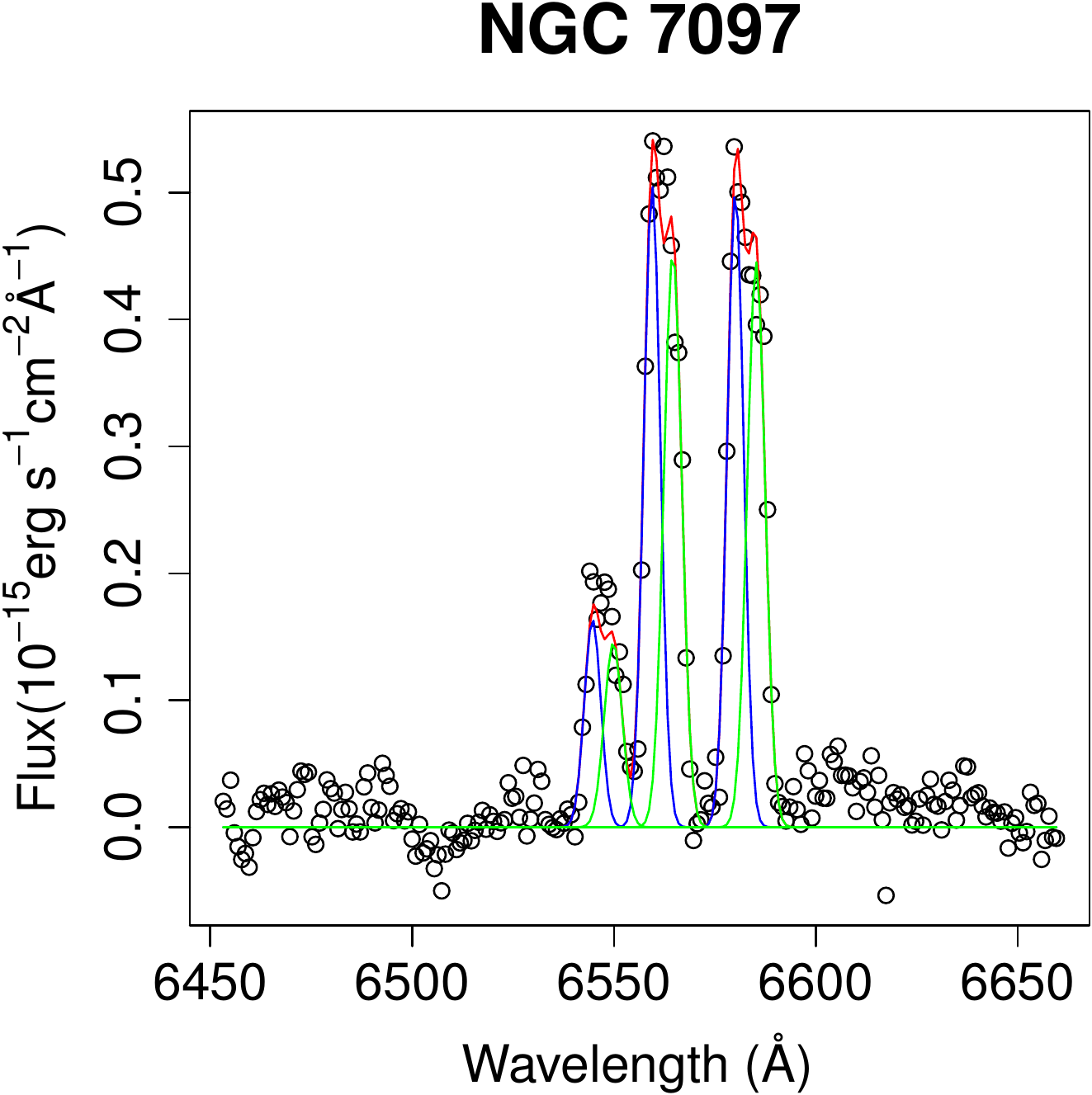}

\caption{H$\alpha$ and [N II]$\lambda \lambda$6548, 6583 emission lines detected in the circumnuclear region of seven galaxies of the sample. Both Gaussian functions were fitted with the same radial velocities and gas velocity dispersions as the ones fitted in the [S II] lines. \label{perfil_NII_Ha_ext}
}
\end{center}
\end{figure*}

\begin{figure*}
\begin{center}
\includegraphics[width=70mm,height=55mm]{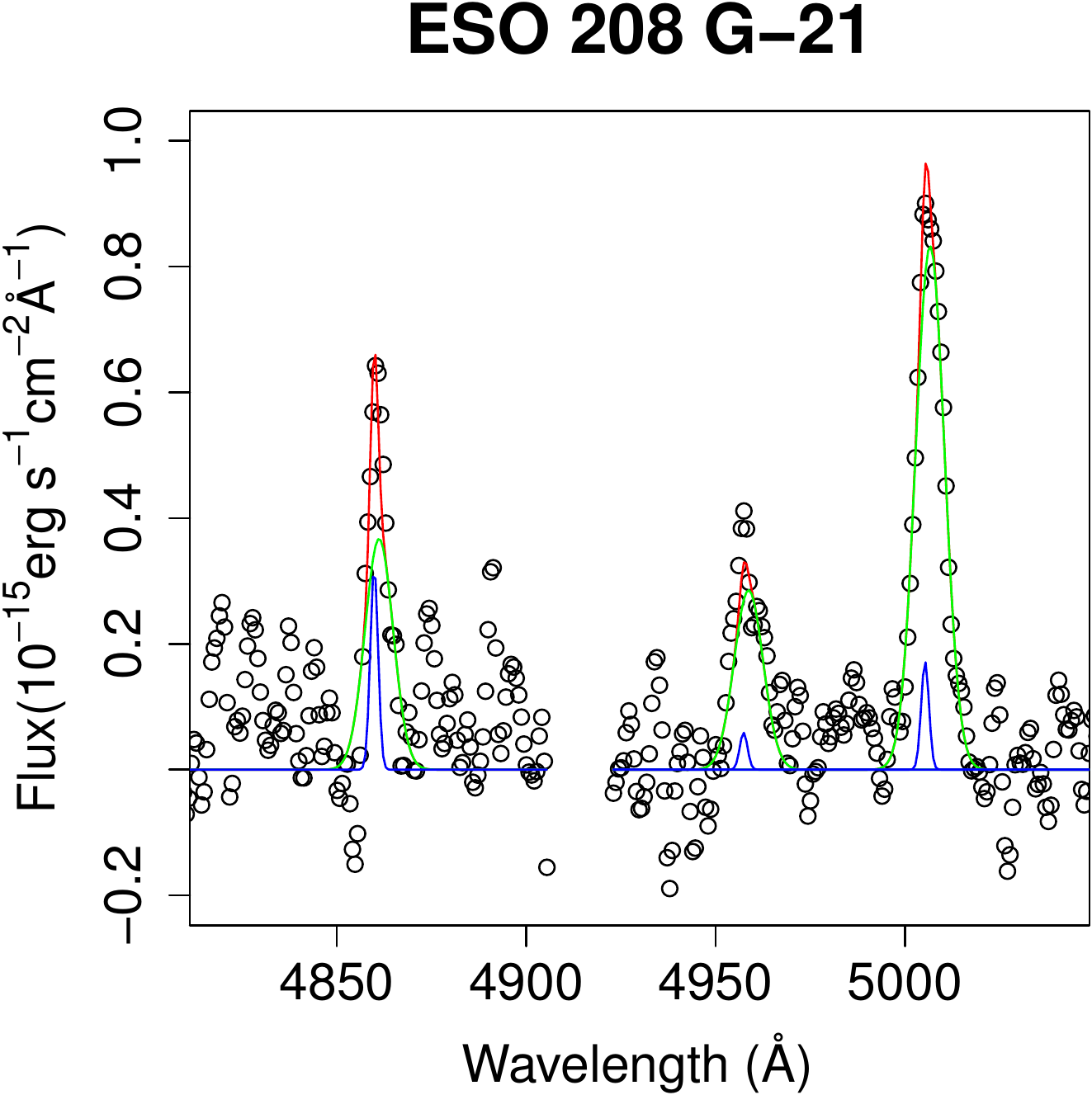}
\includegraphics[width=70mm,height=55mm]{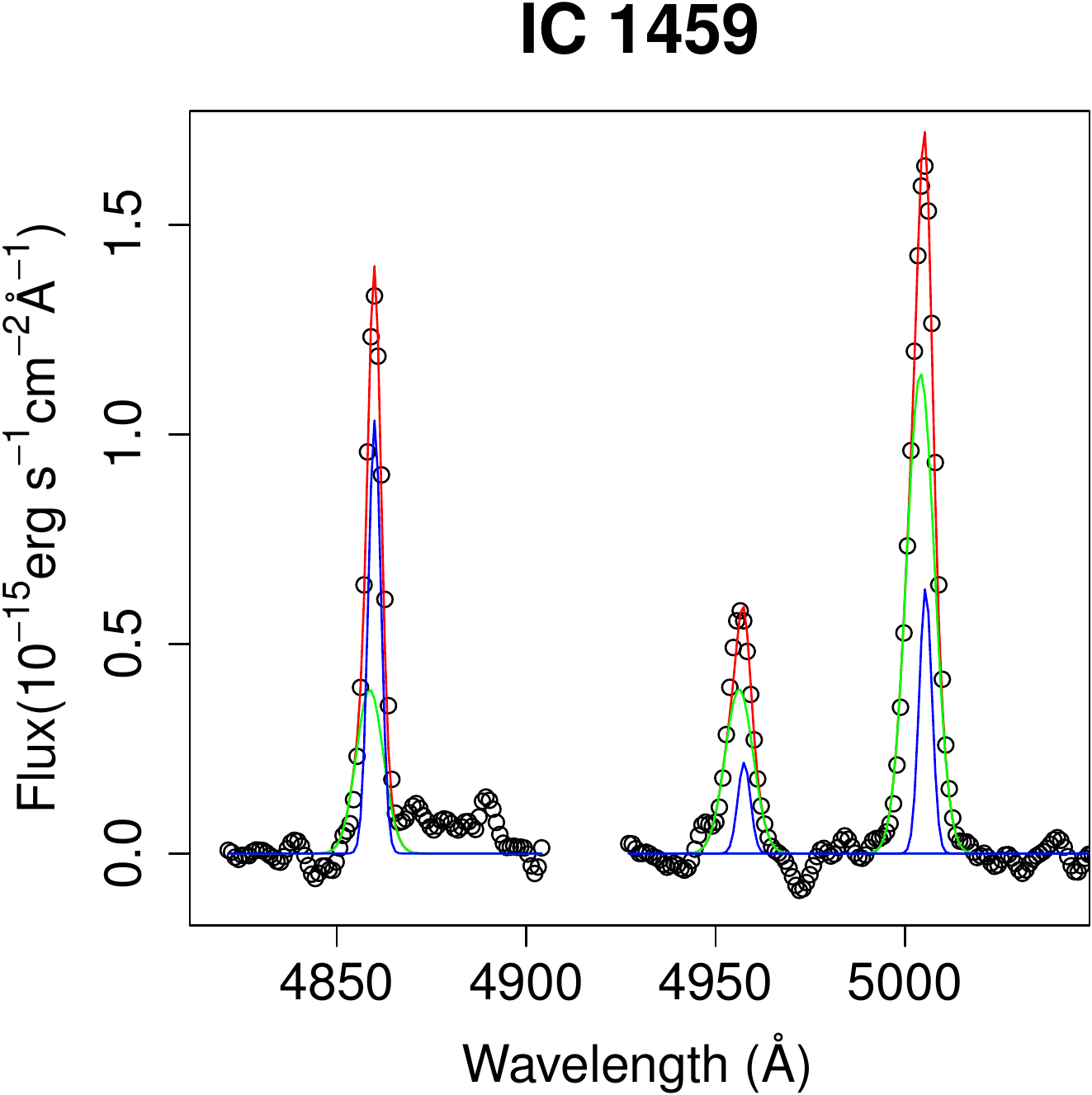}
\includegraphics[width=70mm,height=55mm]{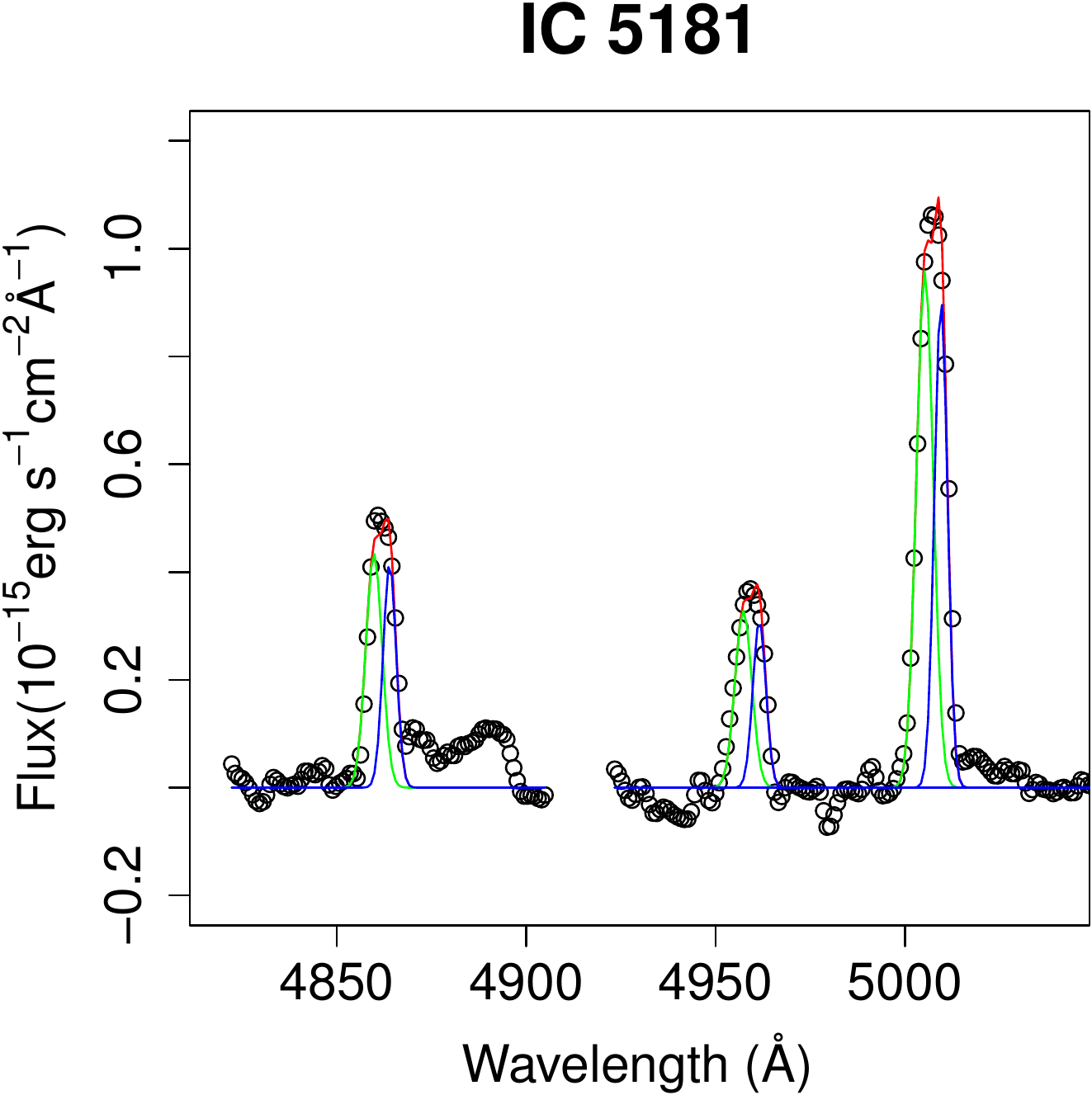}
\includegraphics[width=70mm,height=55mm]{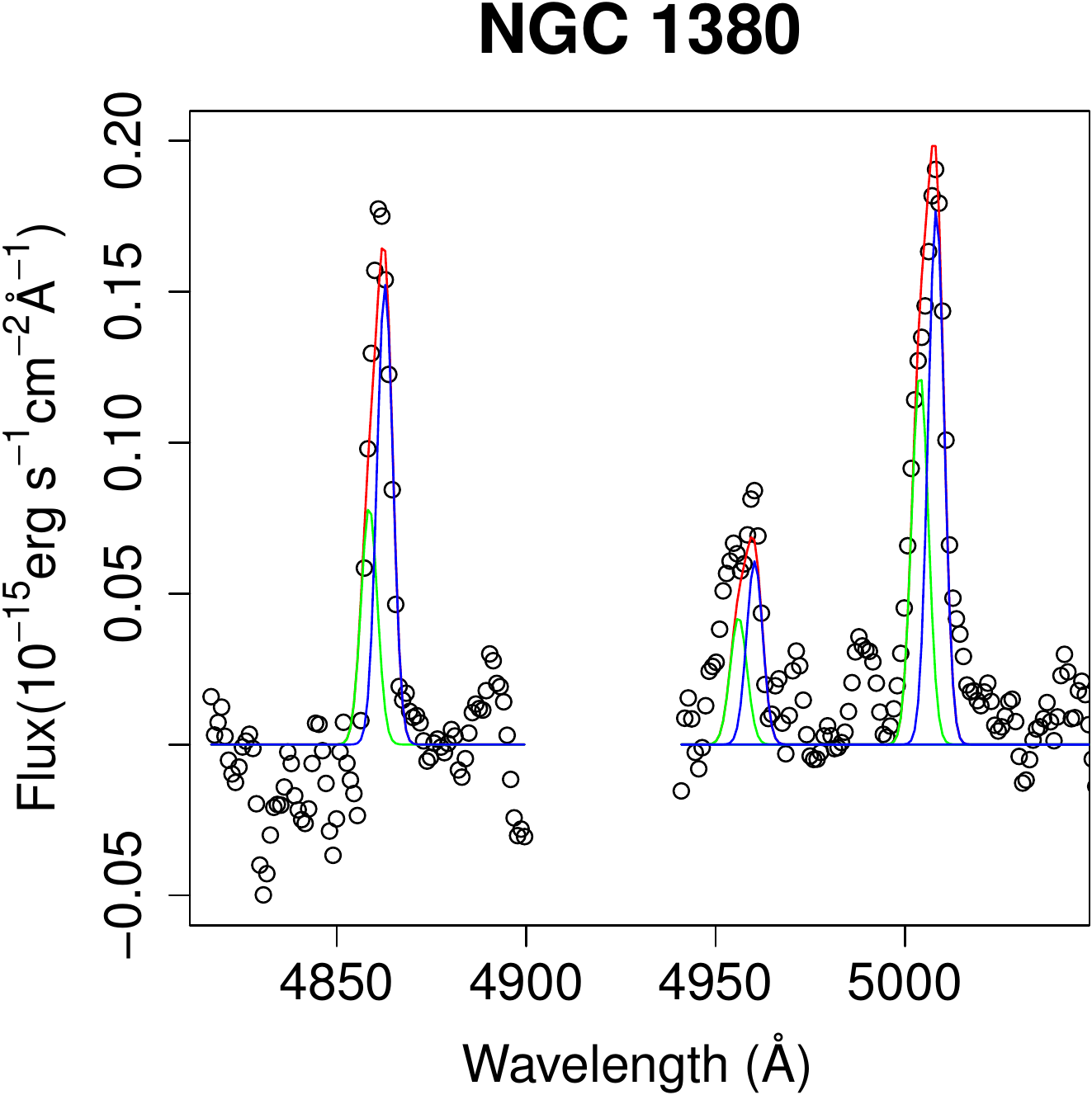}
\includegraphics[width=70mm,height=55mm]{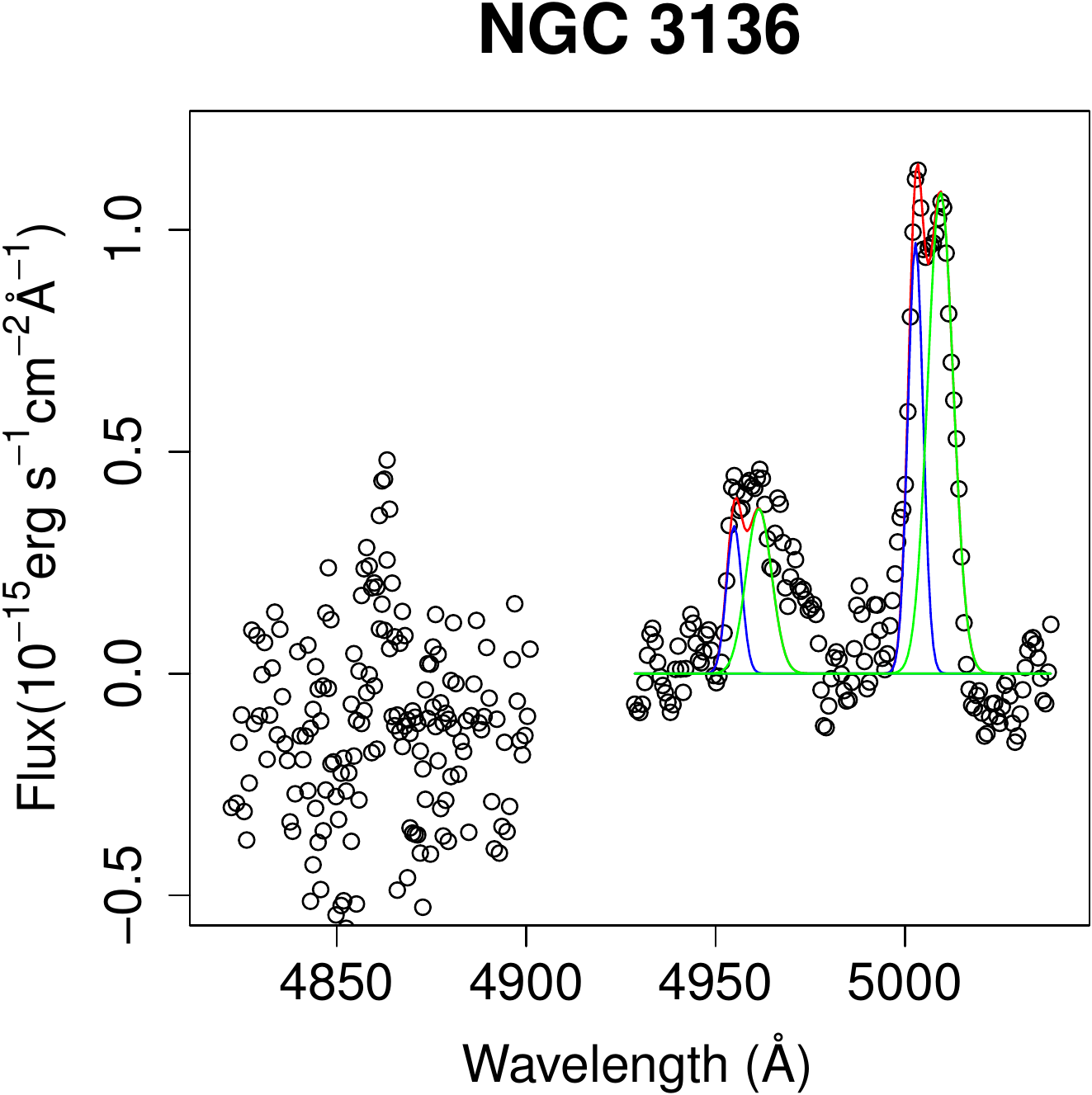}
\includegraphics[width=70mm,height=55mm]{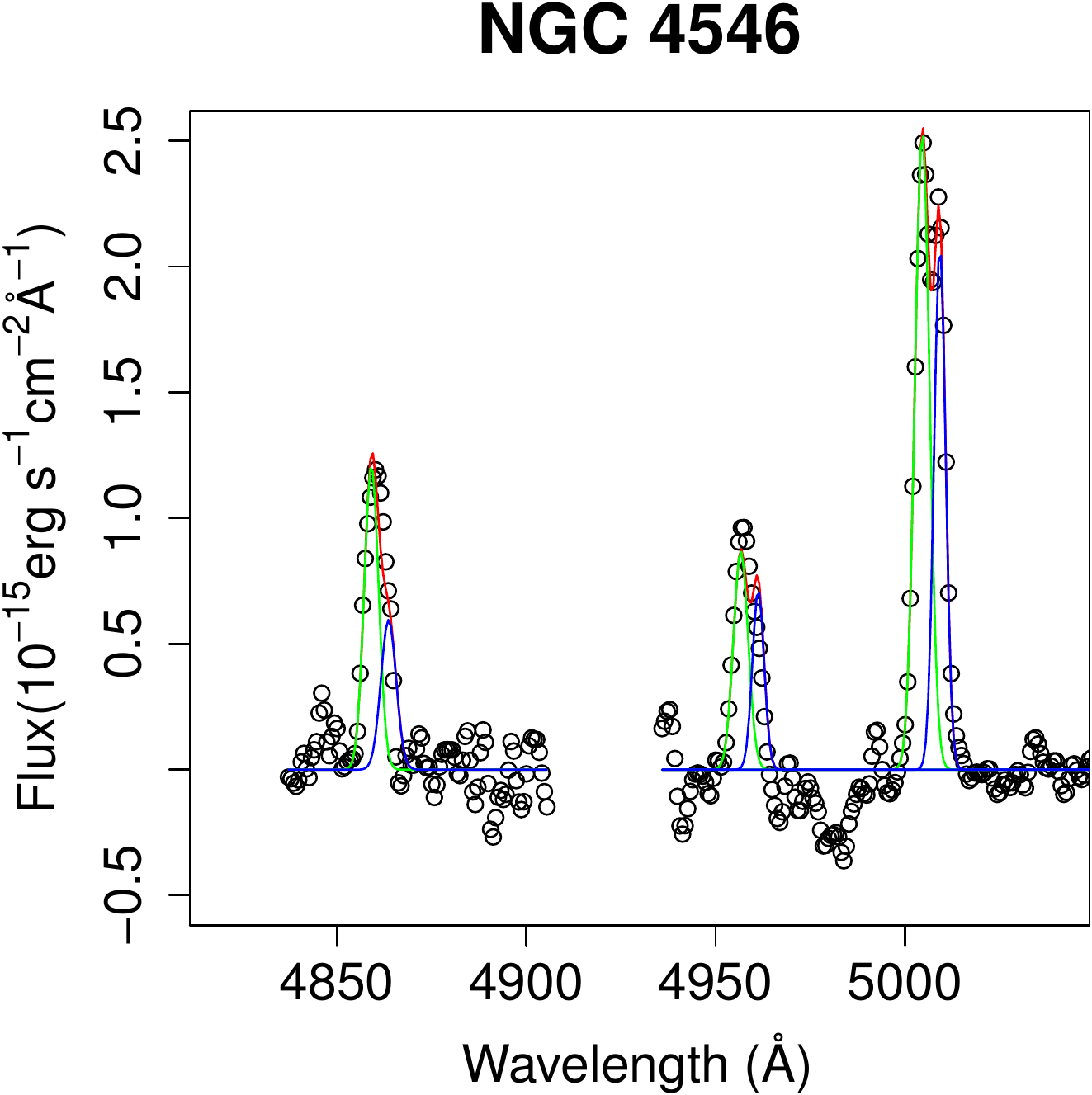}
\includegraphics[width=70mm,height=55mm]{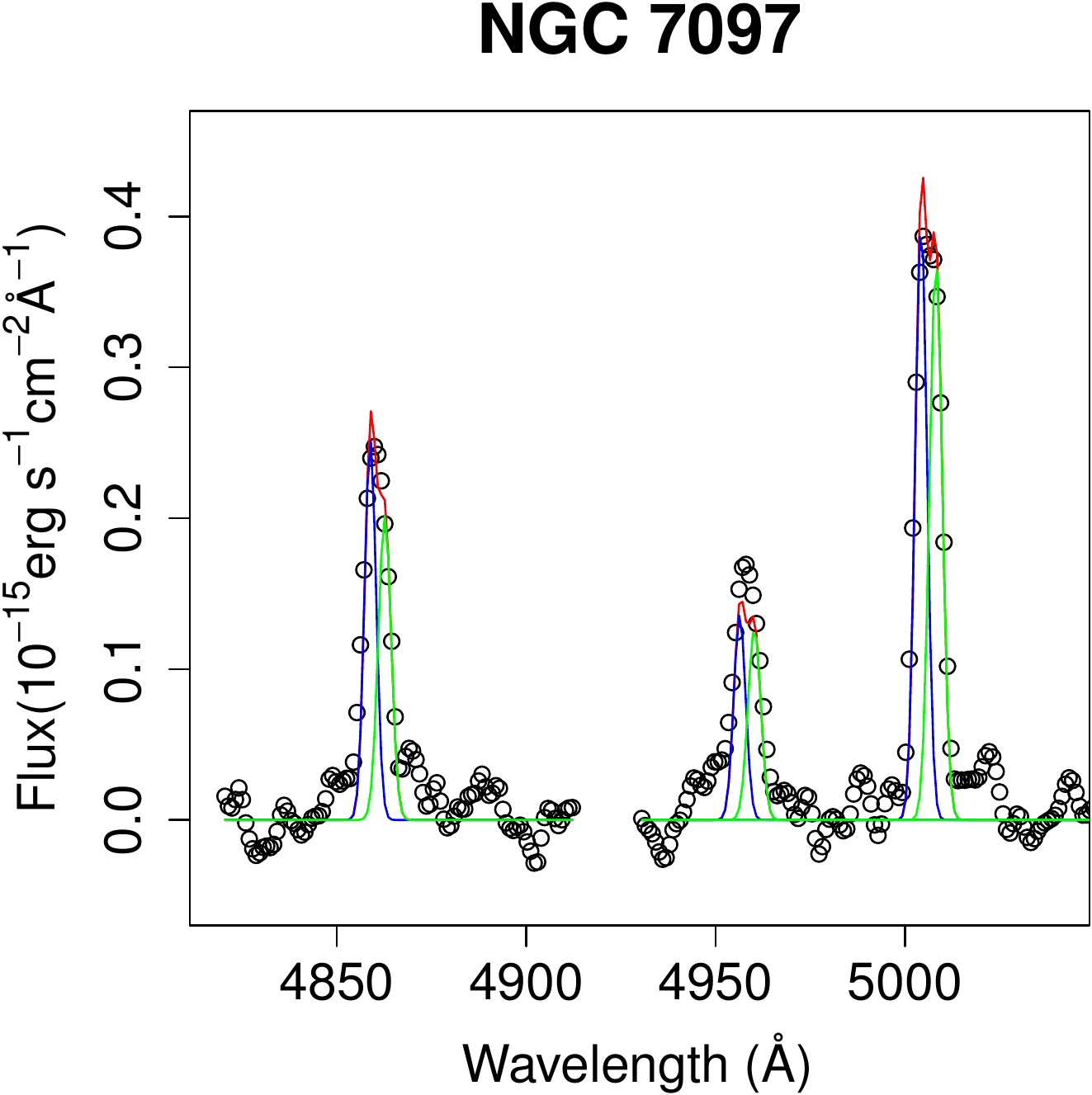}

\caption{H$\beta$ and [O III]$\lambda \lambda$4959, 5007 emission lines detected in the circumnuclear region of seven galaxies of the sample. Both Gaussian functions were fitted with the same radial velocities and gas velocity dispersions as the ones fitted in the [S II] lines. For a better visualization, we smoothed the observed spectra, although the fitting procedure was performed with the original data.\label{perfil_OIII_Hb_ext}
}
\end{center}
\end{figure*}

Figs \ref{perfil_SII_ext}, \ref{perfil_NII_Ha_ext} and \ref{perfil_OIII_Hb_ext} show the fitting results for the [S II], H$\alpha$+[N II] and H$\beta$+[O III] lines, respectively. The H$\alpha$/H$\beta$, [N II]/H$\alpha$, [S II]/H$\alpha$, [O I]/H$\alpha$ and [O III]/H$\beta$ line ratios, in addition to the flux calculated for the H$\alpha$ line, are shown in Table \ref{tab_f_NLR_ext}. The errors presented in this table are purely statistical and do not take into account systematic errors that may arise from an improper subtraction of the stellar component. However, it is worth mentioning that the absolute flux calibration is very uncertain for our data cubes, since the spectrophotometric standards and the galaxies were observed in different nights. Therefore, one should expect an error of about 30\% or even higher in the absolute flux (or luminosity) measurements. The colour excess E(B-V), also shown in Table \ref{tab_f_NLR_ext}, was calculated with the observed H$\alpha$/H$\beta$ ratio using the extinction curve proposed by \citet{1989ApJ...345..245C} with $R_V$ = 3.1 and assuming an intrinsic H$\alpha$/H$\beta$ ratio for LINERs of 3.1 \citep{1983ApJ...264..105F, 1983ApJ...269L..37H}. The H$\alpha$ luminosities, corrected for reddening effects (except for NGC 3136, whose H$\alpha$ luminosity is presented without this correction), are shown in Table \ref{tab_cin_NLR_ext}. For the electron densities $n_e$ of the circumnuclear regions, we used the [S II]$\lambda$6716/[S II]$\lambda$6731 ratio. We calculated the values for $n_e$ with the NEBULAR package under the IRAF environment \citep{1995PASP..107..896S}, assuming $T_e$ = 10000K and using the calculations proposed by \citet{1987JRASC..81..195D}. Due to the uncertainties associated with the [S II] lines ratio, we were able to assert only upper limits for the density of the circumnuclear regions of the sample galaxies. As a consequence, we estimated lower limits for the ionized gas mass of the circumnuclear regions, since this parameter is inversely proportional to $n_e$ (see Eq. 1 from paper II).

The annular regions from which we extracted the spectra of the circumnuclear regions are spatially resolved. This allows us to calculate the filling factor of the gas within the volume of these cylindrical shells. The filling factor $f$ is given by

\begin{equation}
	f = \frac{L(H\alpha)}{\epsilon\ V\ n_e^2 },
	\label{eq_fillingfactor}
\end{equation}
where $V$ is the volume of the cylindrical shell and $\epsilon$ is the H$\alpha$ line emissivity. Assuming case B and $T_e$ = 10000 K, $\epsilon$ = 3.84$\times10^{-25}$ \citep{2006agna.book.....O}. For simplicity, we assumed that the height of the cylinder is equal to its radius. Since we have only upper limits for $n_e$, we calculated the lower limits for the filling factors of the circumnuclear regions of the sample galaxies. The results are shown in Table \ref{tab_cin_NLR_ext}.

\begin{table*}
 \scriptsize
 \begin{center}
  \caption{Flux of the emission lines detected in the spectra of the circumnuclear regions of seven galaxies of the sample. The H$\alpha$ flux is in units of 10$^{-15}$ erg s$^{-1}$ cm$^{-2}$. The colour excess E(B-V) parameter was calculated through the H$\alpha$/H$\beta$ ratio. \label{tab_f_NLR_ext}
}

 \begin{tabular}{@{}lccccccc}
  \hline
  Galaxy name & $f$(H$\alpha$)$_n$ & (H$\alpha$/H$\beta$)$_n$ & E(B-V) & [N II]/H$\alpha$ & [S II]/H$\alpha$ & [O I]/H$\alpha$ & [O III]/H$\beta$ \\
  \hline
  ESO 208 G-21 & 11$\pm$1 & 2.68$\pm$0.60 & -0.14$\pm$0.22 & 1.28$\pm$0.13 & 1.26$\pm$0.14 & $-$ & 2.01$\pm$0.45 \\
  
  IC 1459 & 24$\pm$1 & 3.09$\pm$0.19 & 0.00$\pm$0.08& 1.80$\pm$0.08&1.31$\pm$0.07&0.17$\pm$0.01&1.70$\pm$0.13\\
  
  IC 5181 & 13$\pm$1 & 3.02$\pm$0.33 & -0.03$\pm$0.10&1.10$\pm$0.10&0.78$\pm$0.07&0.07$\pm$0.01&2.17$\pm$0.23\\
  
  NGC 1380 & 6$\pm$1 & 4.52$\pm$0.85 & 0.36$\pm$0.18&1.23$\pm$0.21&0.77$\pm$0.14&0.15$\pm$0.03&1.27$\pm$0.25 \\
   
  
  NGC 3136 & 26$\pm$1 & $-$ & $-$ &1.62$\pm$0.10&1.17$\pm$0.07&0.19$\pm$0.01& $-$ \\
  
  NGC 4546 & 22$\pm$1 & 2.58$\pm$0.23& -0.18$\pm$0.09 & 1.17$\pm$0.08&0.95$\pm$0.06&0.11$\pm$0.01&2.33$\pm$0.21\\
  
  NGC 7097 & 5$\pm$1&2.74$\pm$0.27&-0.12$\pm$0.09&0.98$\pm$0.09&1.21$\pm$0.10&0.14$\pm$0.02&1.68$\pm$0.17\\
  
  \hline
 \end{tabular}
 \end{center}
\end{table*}

\begin{table*}
 \scriptsize
 \begin{center}
  \caption{Physical characteristics of the circumnuclear component of the sample galaxies. Column (1): H$\alpha$ luminosity in erg s$^{-1}$. Column (2): Electron density in cm$^{-3}$. Column (3): Radius, in arcsec, of the outer region of the cylindrical shell used to extract the spectra from the circumnuclear regions of the galaxies. Column (4): Filling factor of the gas inside the cylindrical shell. Column (5): Ionized gas mass, in solar mass. In NGC 3136, the H$\alpha$ luminosity is presented assuming E(B-V) = 0, since we did not detect the H$\beta$ emission line in this object.  \label{tab_cin_NLR_ext}
}

 \begin{tabular}{@{}lccccc}
  \hline
  Galaxy name & log $L$(H$\alpha$)& $n_e$ & radius & $f$ & $M_{ion}$ \\
   &(1) &(2) &(3) &(4) &(5)  \\
  \hline
  ESO 208 G-21&38.58$\pm$0.29&$<$153&1.45&$>$2.7$\times$10$^{-4}$&$>$1.8$\times$10$^3$\\
  
  IC 1459 & 39.32$\pm$0.14&$<$90& 1.35&$>$1.2$\times$10$^{-3}$&$>$3.5$\times$10$^4$\\
  
  IC 5181 & 38.99$\pm$0.40&$<$312&1.5&$>$5.8$\times$10$^{-6}$&$>$5.7$\times$10$^2$\\
  
  NGC 1380 & 38.75$\pm$0.24&$<$658 &1.3&$>$1.2$\times$10$^{-5}$ &$>$8.4$\times$10$^2$ \\
  
  
  NGC 3136 & 39.25$\pm$0.27*&$<$226&1.25&$>$1.9$\times$10$^{-4}$&$>$6.7$\times$10$^3$\\ 
  
  NGC 4546 & 38.88$\pm$0.25&$<$142&1.45&$>$3.4$\times$10$^{-4}$&$>$5.0$\times$10$^3$\\
  
  NGC 7097 & 38.76$\pm$0.14&$<$100&1.25&$>$9.6$\times$10$^{-4}$&$>$8.7$\times$10$^3$\\
  \hline
 \end{tabular}
 \end{center}

\end{table*}

The H$\alpha$ luminosities of the circumnuclear regions are quite similar to those found for the nuclear regions (paper II, see also erratum). In five galaxies, the colour excess is consistent with zero, i.e., there are no significant reddening effects in the circumnuclear regions of these objects. Only for NGC 1380 a considerable E(B-V) was detected in this region. The upper limits of the circumnuclear densities are systematically lower than the nuclear densities. This result is in agreement with the proposals of \citet{1988ApJ...324..134F}, \citet{1996ApJ...462..183H} and \citet{2008AJ....136.1677W}, of a negative radial gradient for the ionized gas densities of the central regions of galaxies. For the ionized gas mass, the lower limits vary between 10$^3$ and 10$^4$ M$_\odot$. 

\section{Diagnostic diagrams} \label{diagdiagextendedemission}

\begin{figure*}
\begin{center}
\includegraphics[scale=0.75]{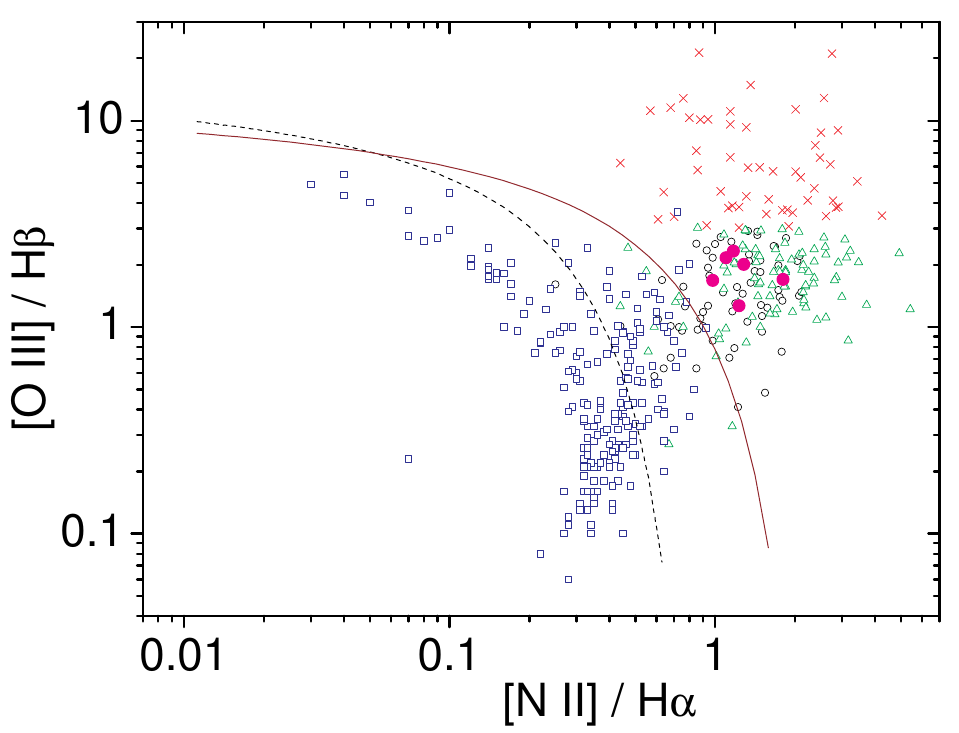}
\hspace{0.5cm}
\includegraphics[scale=0.75]{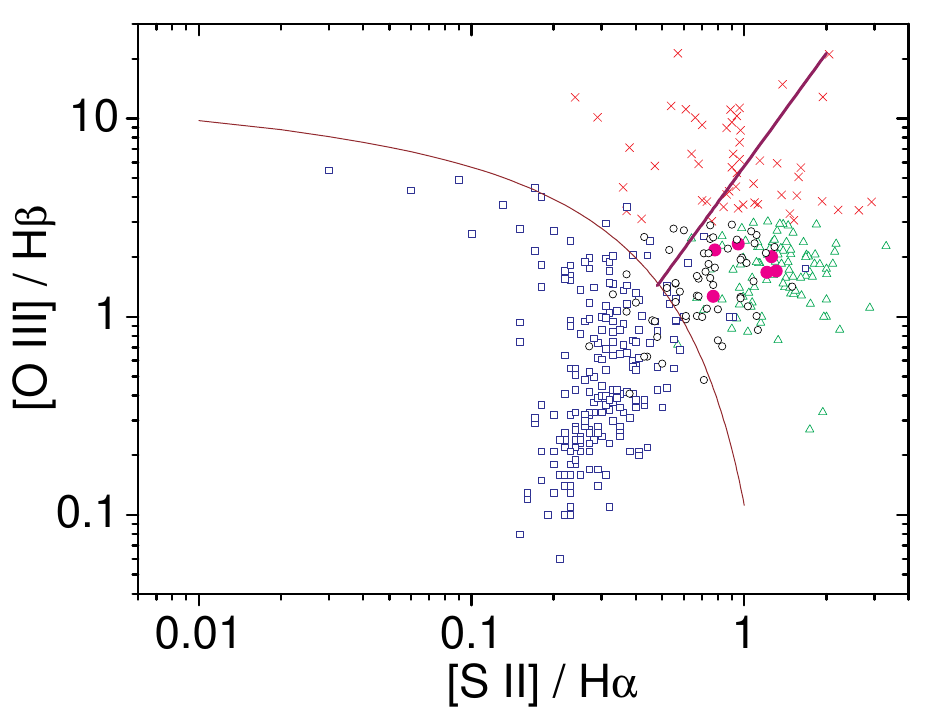}
\caption{BPT Diagram with the line ratios measured in the circumnuclear spectra of six galaxies of the sample shown as filled magenta circles. The line ratios are presented in Table \ref{tab_f_NLR_ext}. According to the classification proposed by \citet{1997ApJS..112..315H}, the red crosses are Seyferts, green triangles are LINERs, hollow black circles are TOs and blue squares are H II regions. The thin brown line is the maximum starburst line proposed by \citet{2001ApJ...556..121K}, the dashed black line is the empirical division between H II regions and AGNs \citep{2003MNRAS.346.1055K} and the thick purple line is the LINER-Seyfert division suggested by \citet{2006MNRAS.372..961K}. \label{BPTextended} 
 }
\end{center}
\end{figure*}

Diagnostic diagrams were used to classify the circumnuclear emission of six galaxies of the sample (NGC 1399, NGC 1404 and NGC 2663 have no circumnuclear emission and NGC 3136 has no detected H$\beta$). We used the BPT diagram \citep{1981PASP...93....5B} with the line ratios proposed by \citet{1987ApJS...63..295V}. The values shown in Table \ref{tab_f_NLR_ext} were inserted into the [N II]/H$\alpha$ versus [O III]/H$\beta$ and [S II]/H$\alpha$ versus [O III]/H$\beta$ diagrams, together with the line ratios of the galaxies from the Palomar Survey \citep{1997ApJS..112..315H}. We also added the maximum theoretical limit for H II regions proposed by  \citet{2001ApJ...556..121K}, the empirical separation between AGNs and starburst galaxies proposed by \citet{2003MNRAS.346.1055K} and the separation between LINERs and Seyferts proposed by \citet{2006MNRAS.372..961K}. According to \citet{2006MNRAS.372..961K}, the transition objects (TOs) are found in the region between the maximum theoretical limit for H II regions and the empirical separation between AGNs and starburst galaxies. The diagnostic diagrams are presented in Fig. \ref{BPTextended}. We found that the spectra of all six galaxies analysed with this tool have typical LINER line rations.

\section{Mapping the nuclear and circumnuclear properties of the narrow line regions} \label{circumnuclear_maps}

One of the advantages of using data cubes is that we are able to obtain information of a given spectral region in two spatial dimensions. In this section, we present maps containing information about the ionized gas of the gas cubes of the sample galaxies, such as line fluxes, kinematics and electron densities. All maps were drawn by fitting Gaussian functions to the emission lines of each spectrum of a given gas cube. Before the fitting procedure, we subtracted the broad components of the H$\alpha$ lines, related to the broad line regions (BLRs), from each spectrum of the gas cubes of the galaxies ESO 208 G-21, IC 1459, IC 5181, NGC 2663, NGC 4546 and NGC 7097. We performed this subtraction because we are interested in analysing only the properties of narrow line regions (NLRs) of the galaxies, even on the position of the nucleus of each galaxy. To do so, we built data cubes whose spectra contained only the broad component of the H$\alpha$ emission line. This was achieved by creating a normalized Gaussian image (i.e., the sum of all spaxels of this image equals one), where the FWHM is related to the PSF of the data cubes (papers I and II). Then, each spaxel was multiplied by a spectrum containing only the broad component of the H$\alpha$ line, which is also represented by a Gaussian function. The radial velocity, flux and FWHM of the Gaussian related to the BLR were measured in the nuclear spectra of the galaxies and are shown in paper II. The result is a data cube that contains only the BLR of a given object, which has the same spatial and spectral dimensions as the respective gas cube. The subtraction of the broad component from the spectra implies that the information taken from the H$\alpha$ and from the [N II] emission lines are associated only with their narrow component (in fact, the [N II] doublet would be affected by the broad component of the H$\alpha$ emission line), even in the nuclear region of these six galaxies. 

For each spectrum of the gas cubes, the fitting procedure is quite similar to that described in Section \ref{FOV_properties}. Nevertheless, in regions of the FOV far from the nucleus, the signal-to-noise ratio is not enough for the procedure to fit more than one Gaussian function per emission line. Thus, each emission line was described by only one Gaussian function. Moreover, the maps of NGC 1380 and NGC 2663 do not cover the whole spatial dimension of the gas cubes, since no emission lines exist in some regions in the FOV of both objects. In the case of NGC 1380, there is no gas emission in the upper region of the FOV. In NGC 2663, we were only able to map the emission lines of the region closest to the AGN. This is in accordance with the fact that we did not detect emission lines in the circumnuclear region of NGC 2663. In NGC 1399 and NGC 1404, the emission lines are too weak and, therefore, no maps were constructed for either object.

For the seven galaxies where we detected circumnuclear emission, and NGC 2663, we built maps of the gas kinematics, flux and EW of the H$\alpha$ and [N II]$\lambda$6583 emission lines. We also drew maps of the electron densities of the gas. In keeping with procedures in Section \ref{FOV_properties}, we also used the [S II]$\lambda\lambda$6716, 6731 emission lines to calculate $n_e$. Each map is discussed in the sub-sections below. 

\subsection{Gas kinematics maps} \label{kinematics_extended_emission}

For the kinematic properties, we fitted the [N II]$\lambda\lambda$6548, 6583 and H$\alpha$ emission lines of each spectrum of the gas cubes. The theoretical ratio [N II]$\lambda6583$/[N II]$\lambda6548$ = 3.06 \citep{2006agna.book.....O} was fixed for the fitting procedure. We extracted the radial velocity and the velocity dispersion from each spectrum of the gas cubes of the sample galaxies. All three Gaussians were assumed to have the same $\sigma$ and were separated in the H$\alpha$ and [NII] lines by their respective wavelength. In other words, there is only one value for the radial velocity and for the velocity dispersion per spectrum. The velocity dispersion, which is related to the $\sigma$ of the Gaussian function and not to the FWHM, was corrected for the instrumental broadening effect as $\sigma^2$ = $\sigma^2_{obs} - \sigma^2_{inst}$, where the $\sigma_{inst}$ values correspond to the spectral resolution of the data cubes (see paper I). The kinematic maps are shown in Figs. \ref{mapa_cin_gal_1}, \ref{mapa_cin_gal_2}, \ref{mapa_cin_gal_3} and \ref{mapa_cin_gal_4} for eight galaxies of the sample.

\renewcommand{\thefigure}{\arabic{figure}\alph{subfigure}}
\setcounter{subfigure}{1}

\begin{figure*}
\hspace{0.0cm}
\includegraphics[scale=0.32]{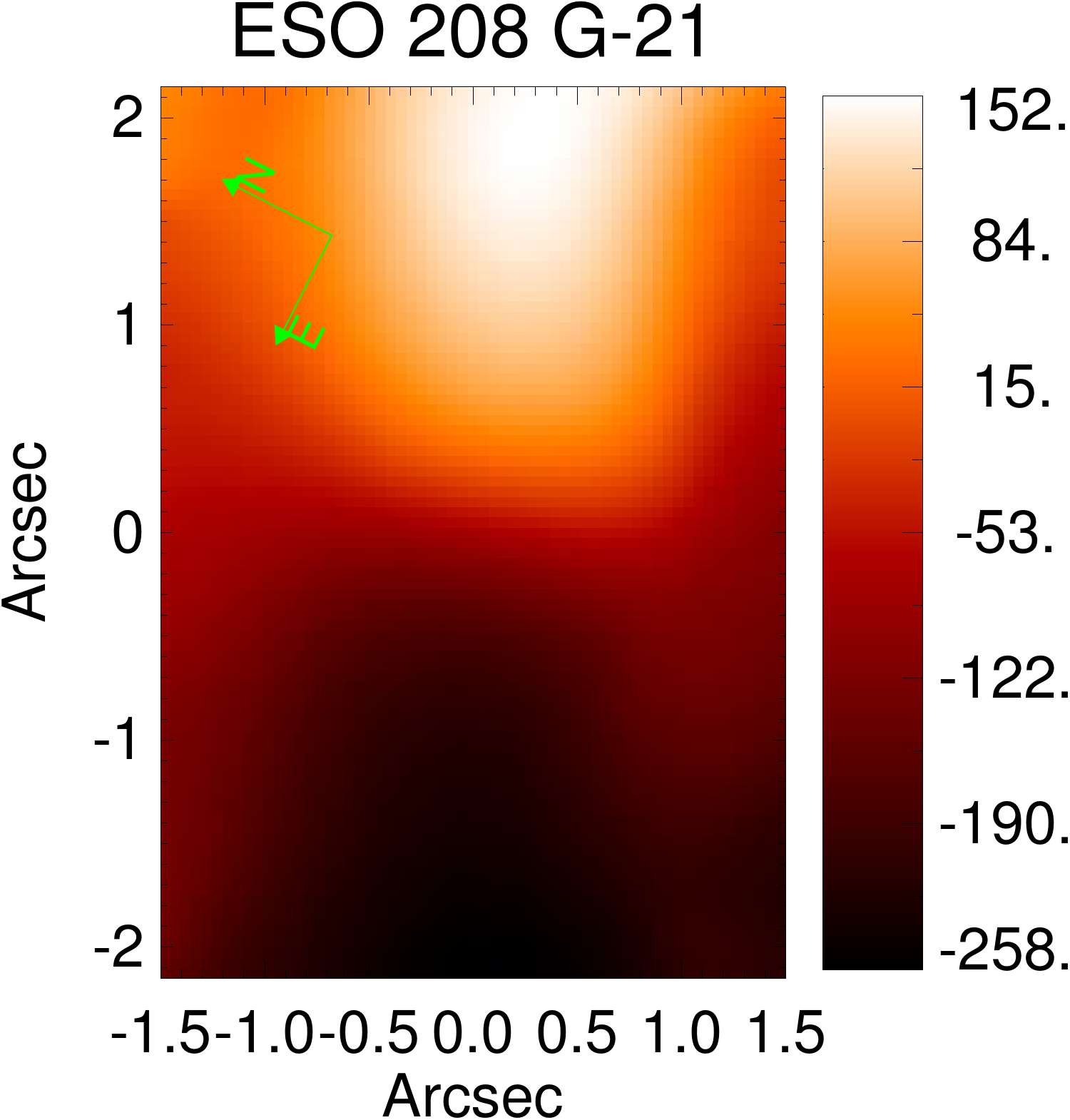}
\hspace{0.5cm}
\vspace{0.7cm}
\includegraphics[scale=0.32]{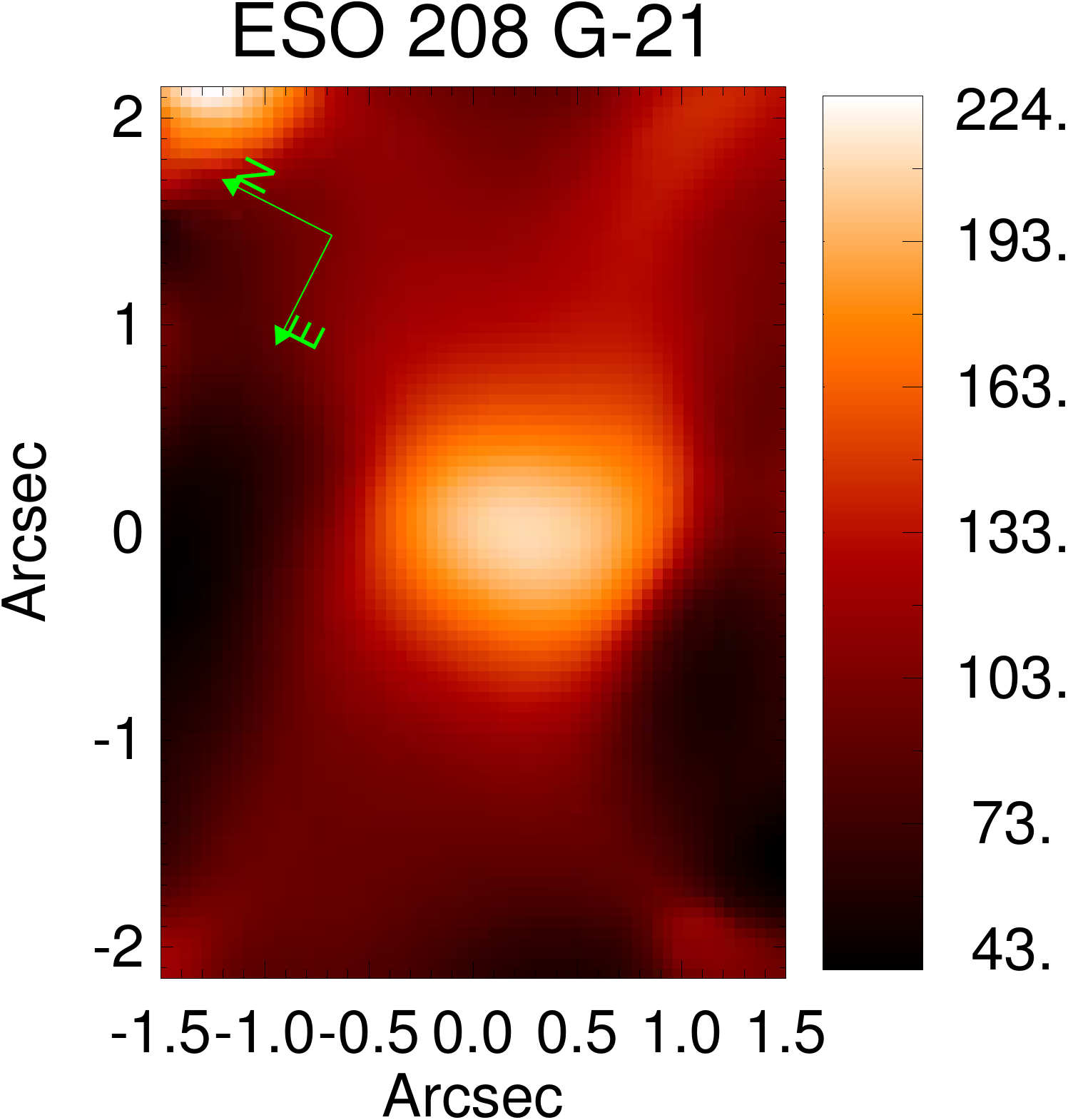}
\hspace{1.0cm}

\includegraphics[scale=0.36]{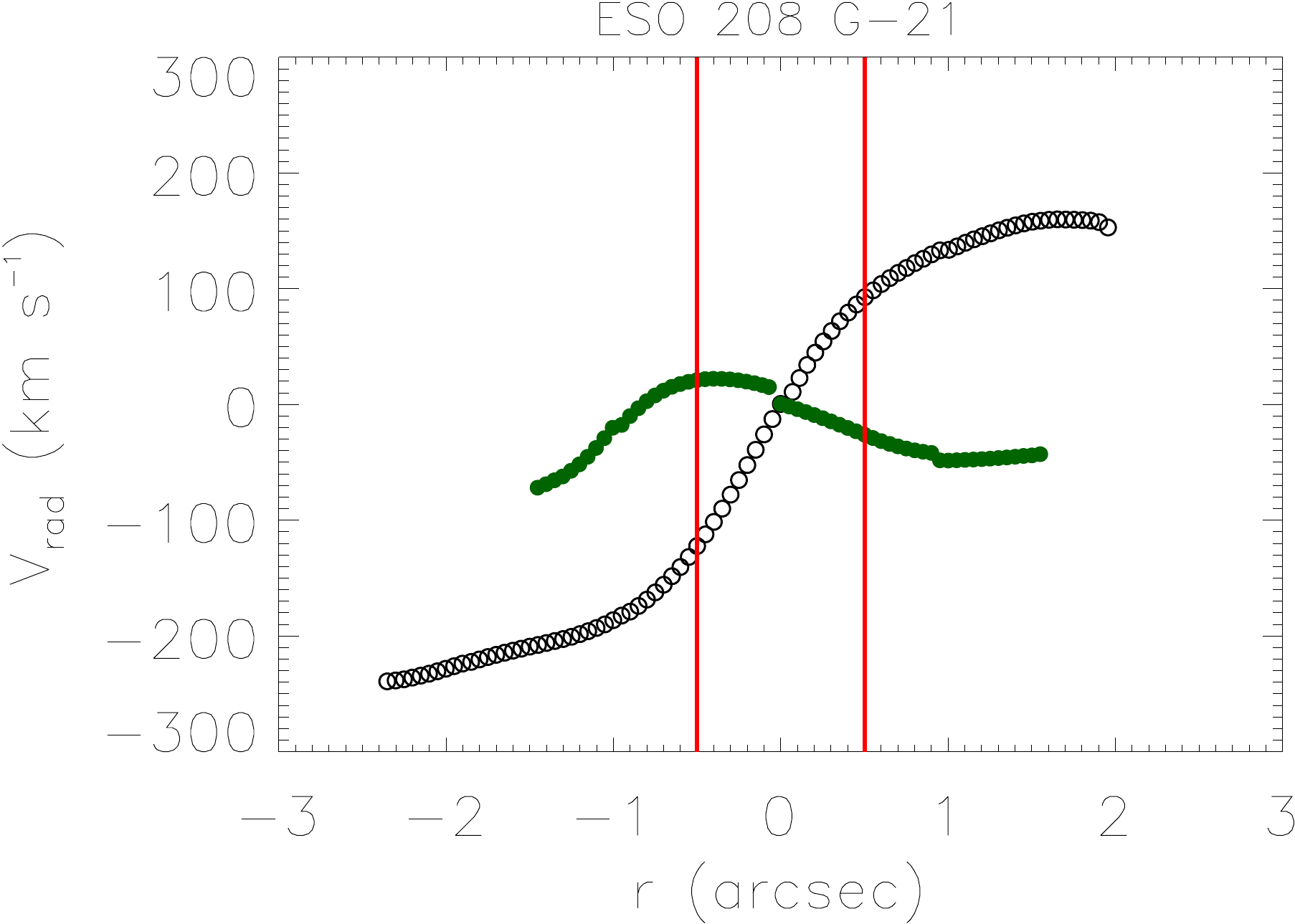}
\vspace{1.0cm}
\includegraphics[scale=0.36]{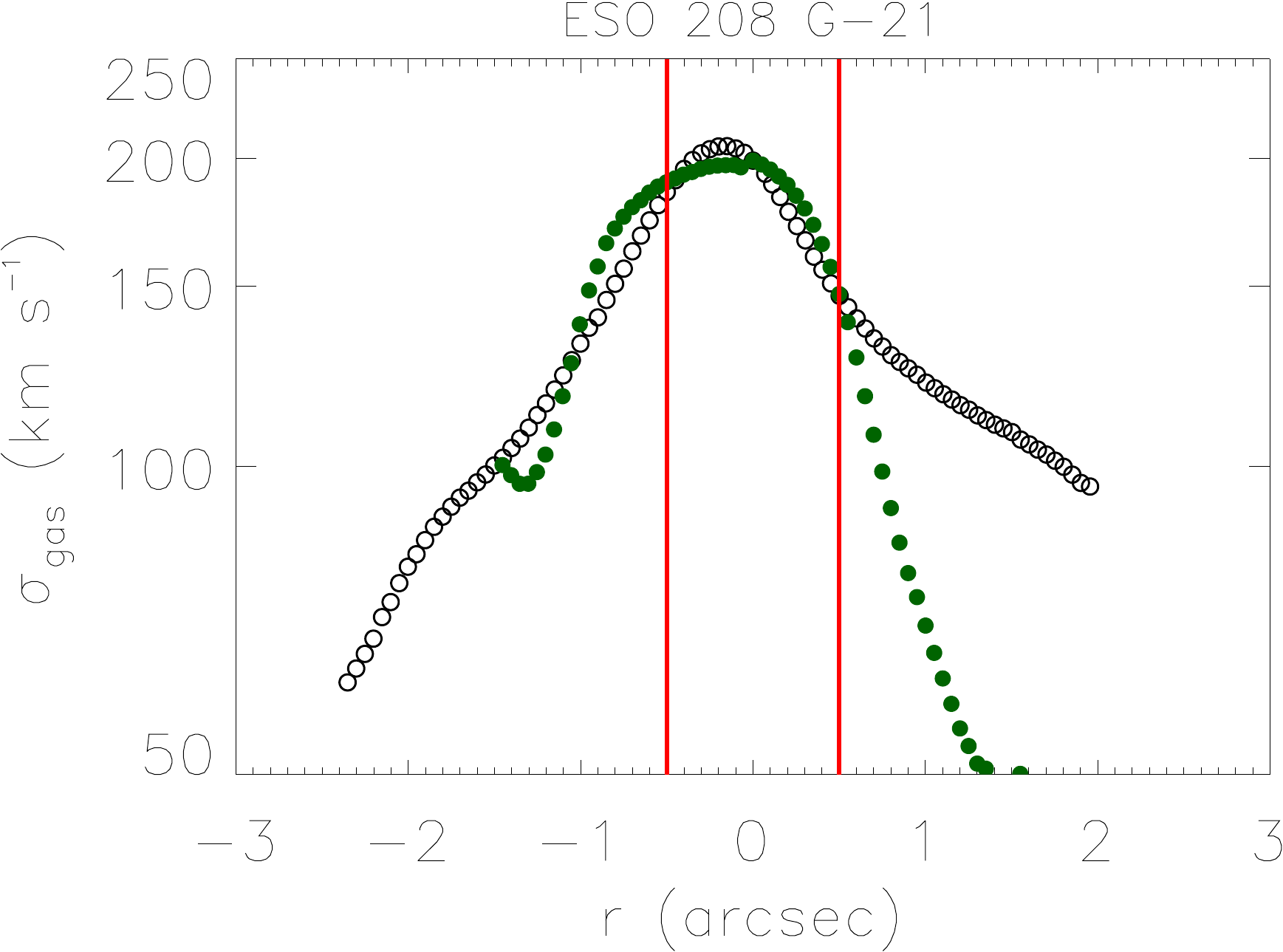}

\hspace{0.0cm}
\includegraphics[scale=0.32]{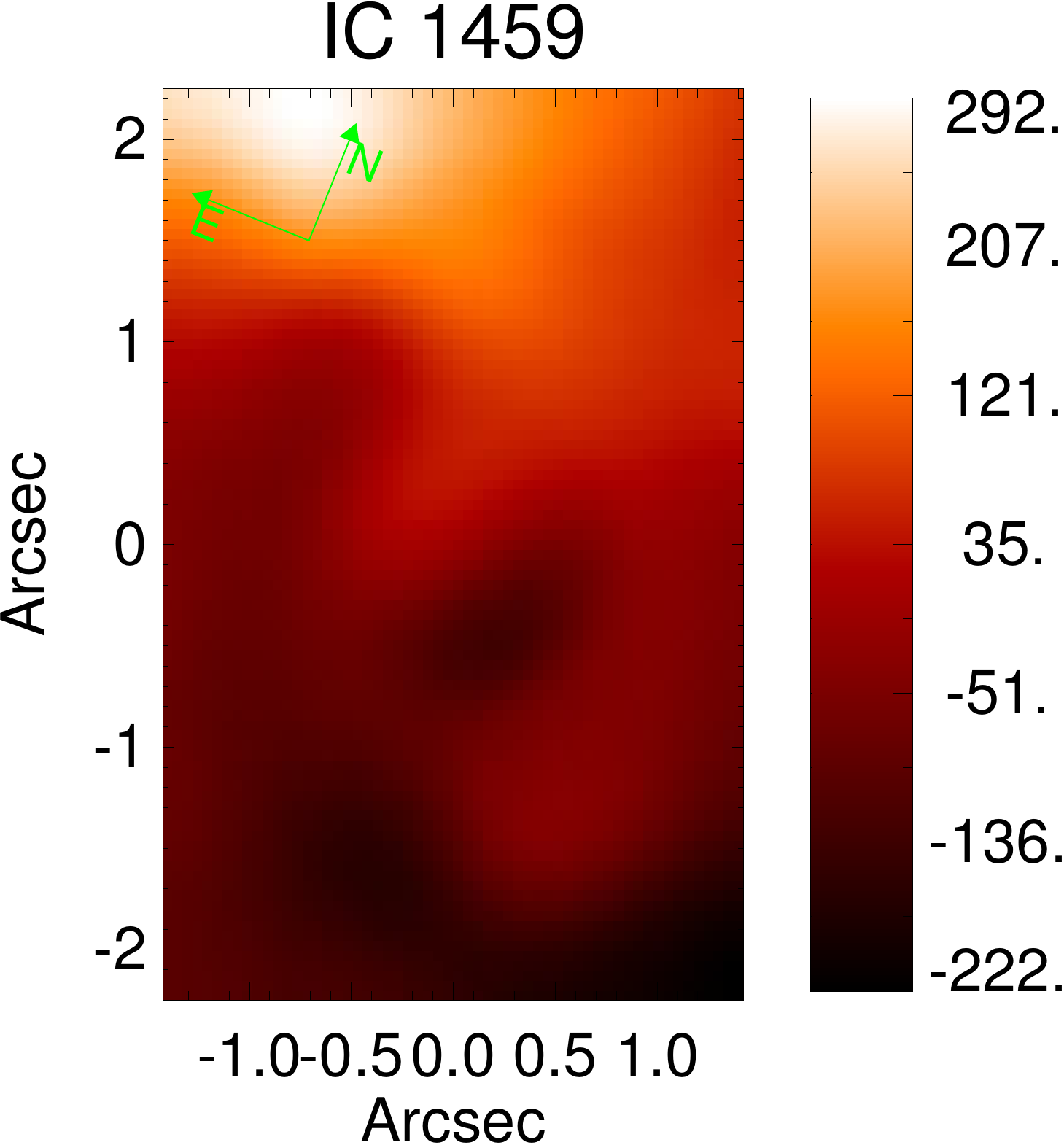}
\hspace{0.5cm}
\vspace{0.7cm}
\includegraphics[scale=0.32]{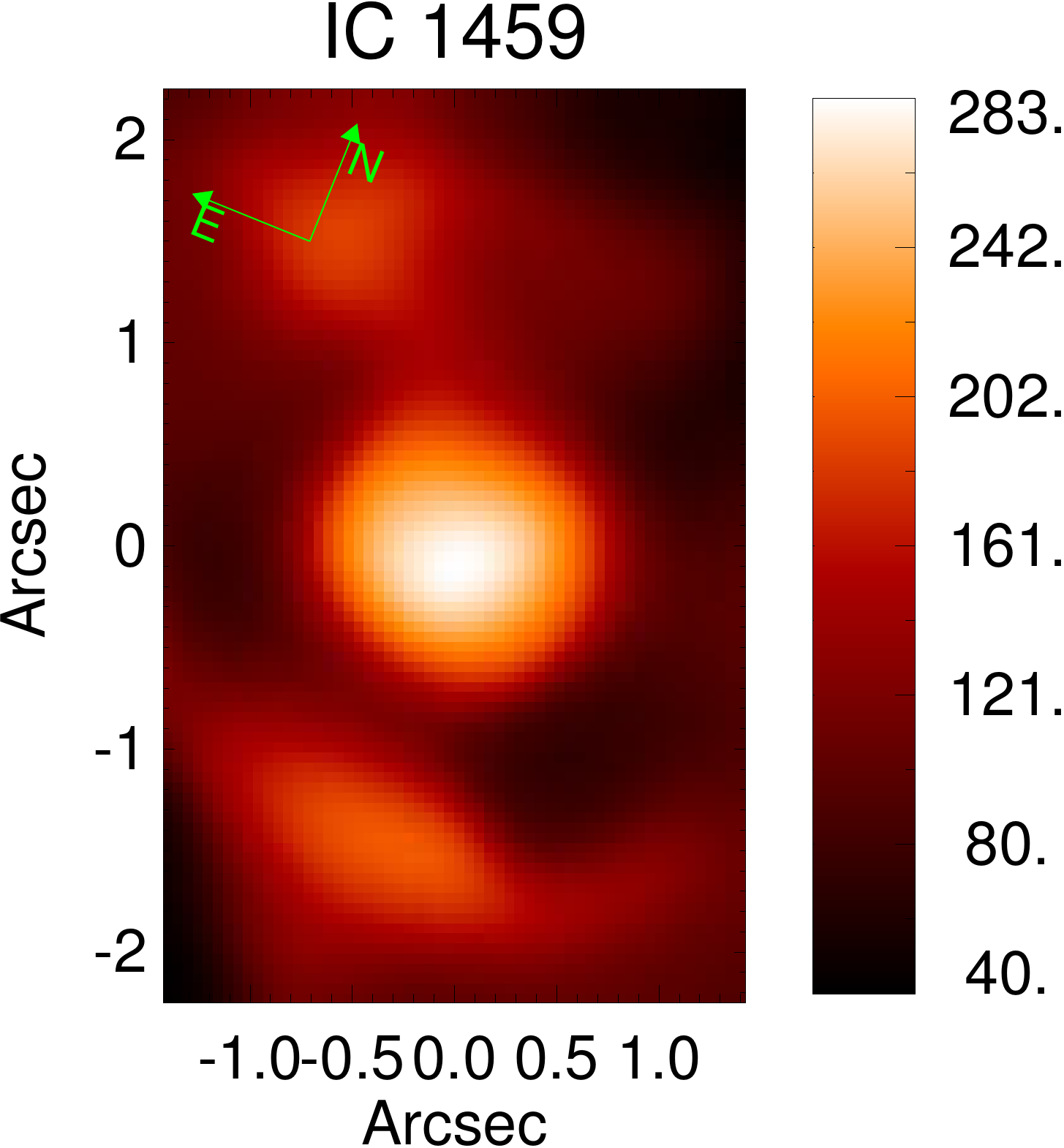}
\hspace{1.0cm}

\includegraphics[scale=0.36]{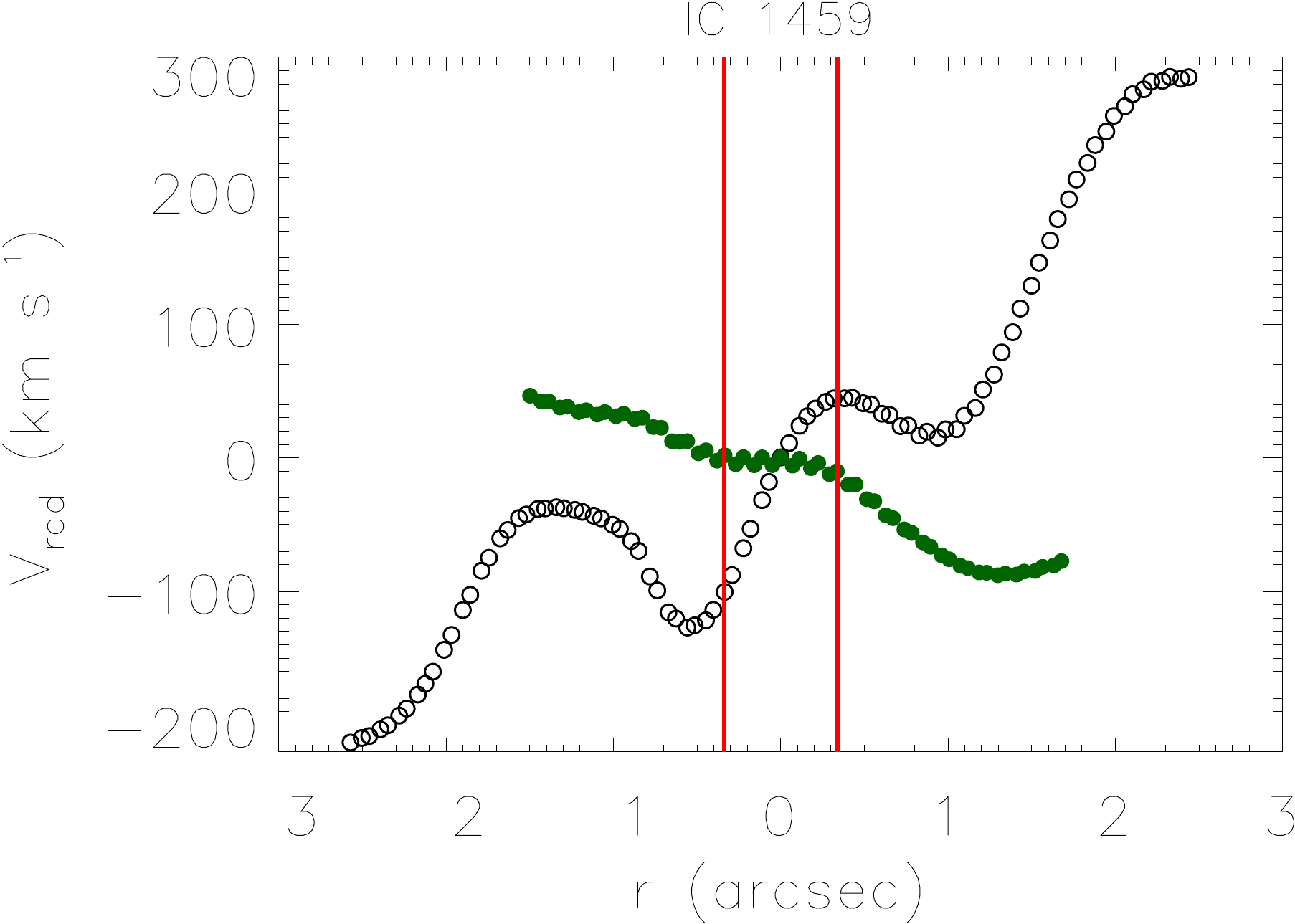}
\includegraphics[scale=0.36]{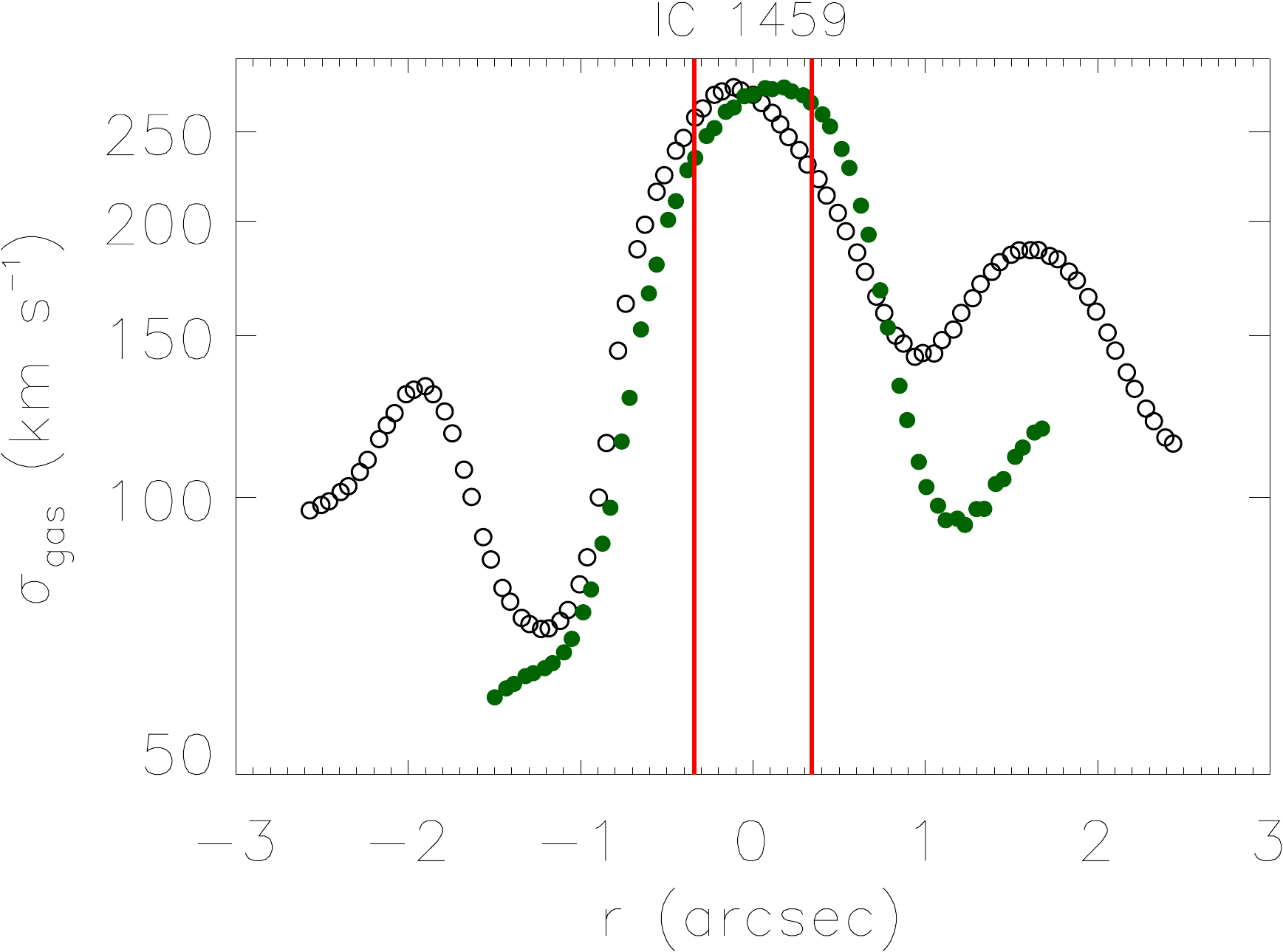}

\caption{Left: Gas radial velocity map, in km s$^{-1}$, taken from the [N II]+H$\alpha$ emission lines. Right: Gas velocity dispersion map, in km s$^{-1}$. In the profiles, the hollow black circles were extracted in the direction parallel to the P.A. of the bipolar gas structures, while the filled green circles are related to the direction perpendicular to the P.A. of the bipolar structures. The red vertical lines delimit the FWHM of the PSFs of the data cubes. All the velocities shown above are relative to the nuclei of the galaxies. The radial velocities originally measured for the nuclear regions of the galaxies are shown in Table \ref{tab_PA}. \label{mapa_cin_gal_1}}
\end{figure*}

\addtocounter{figure}{-1}
\addtocounter{subfigure}{1}

\begin{figure*}

\hspace{0.0cm}
\includegraphics[scale=0.32]{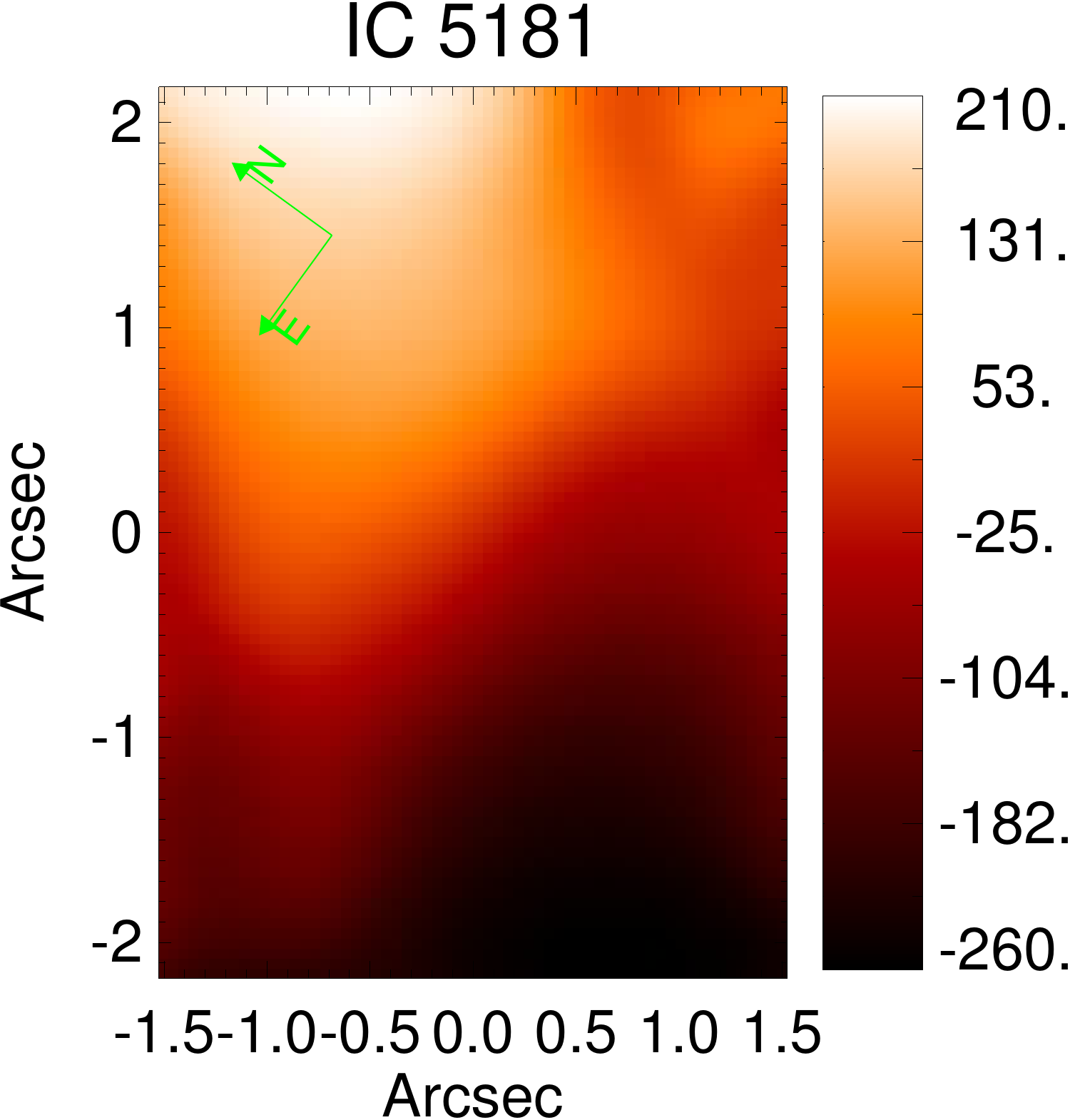}
\hspace{0.5cm}
\vspace{0.7cm}
\includegraphics[scale=0.32]{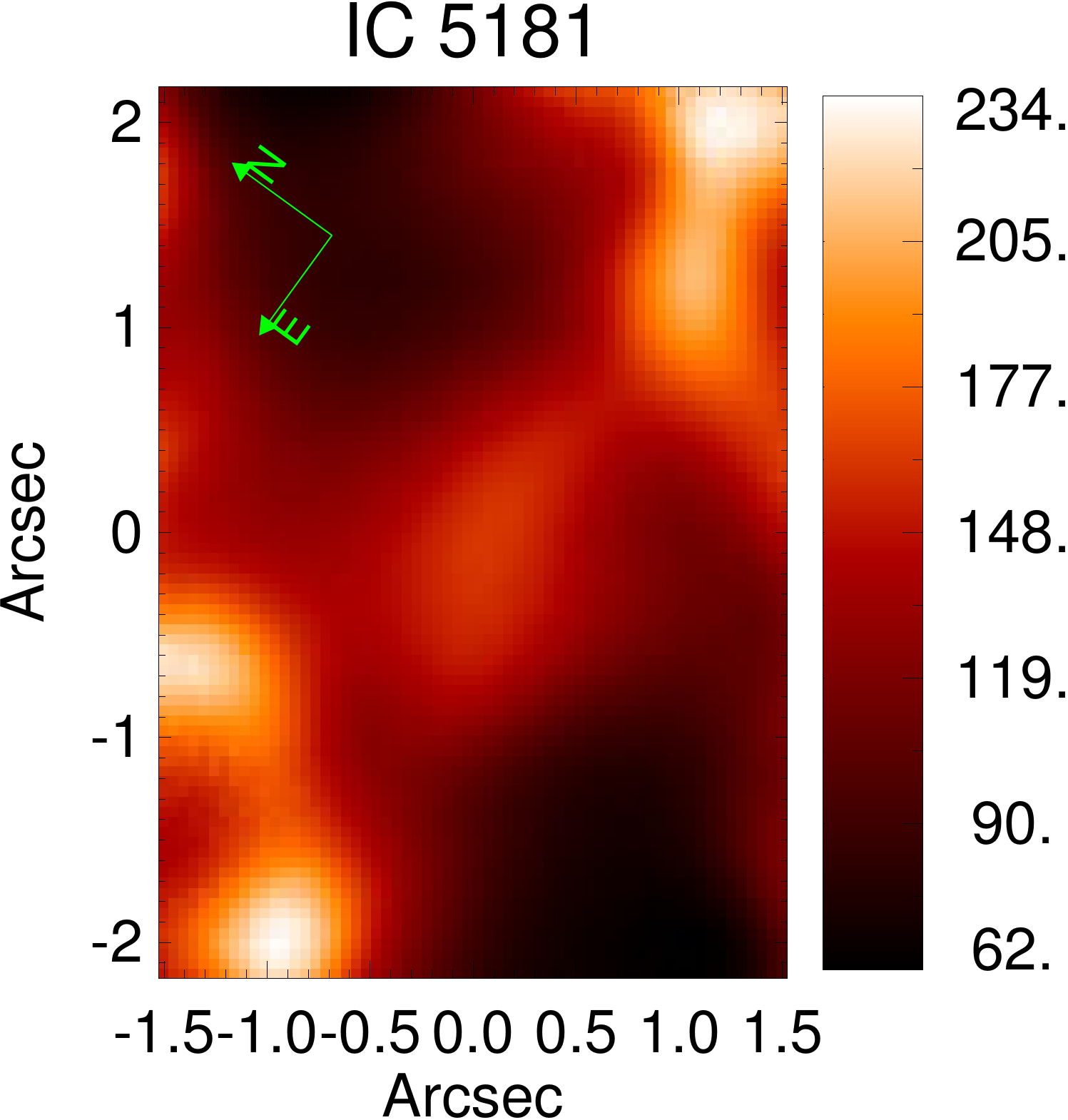}
\hspace{1.0cm}

\includegraphics[scale=0.36]{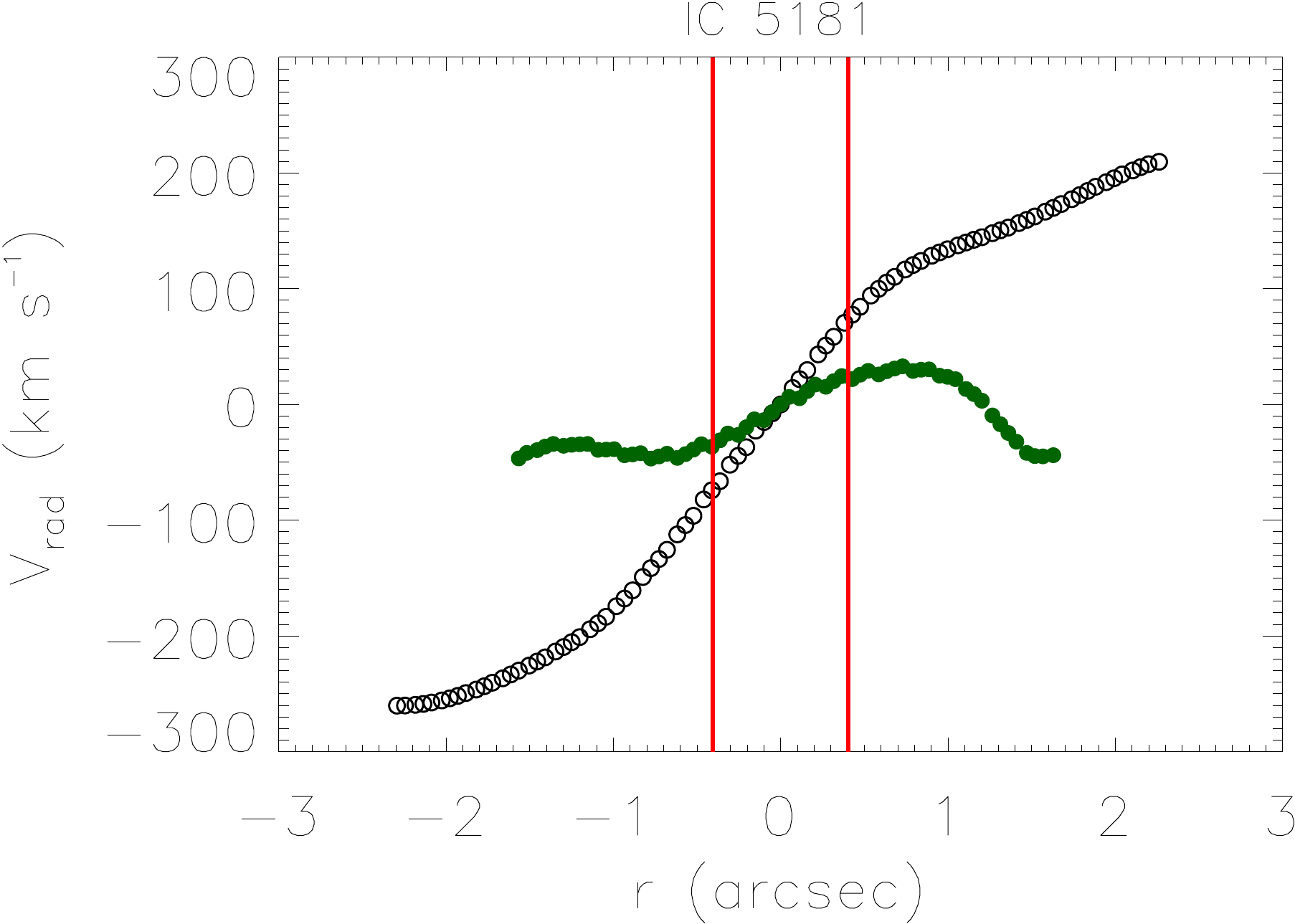}
\vspace{1.0cm}
\includegraphics[scale=0.36]{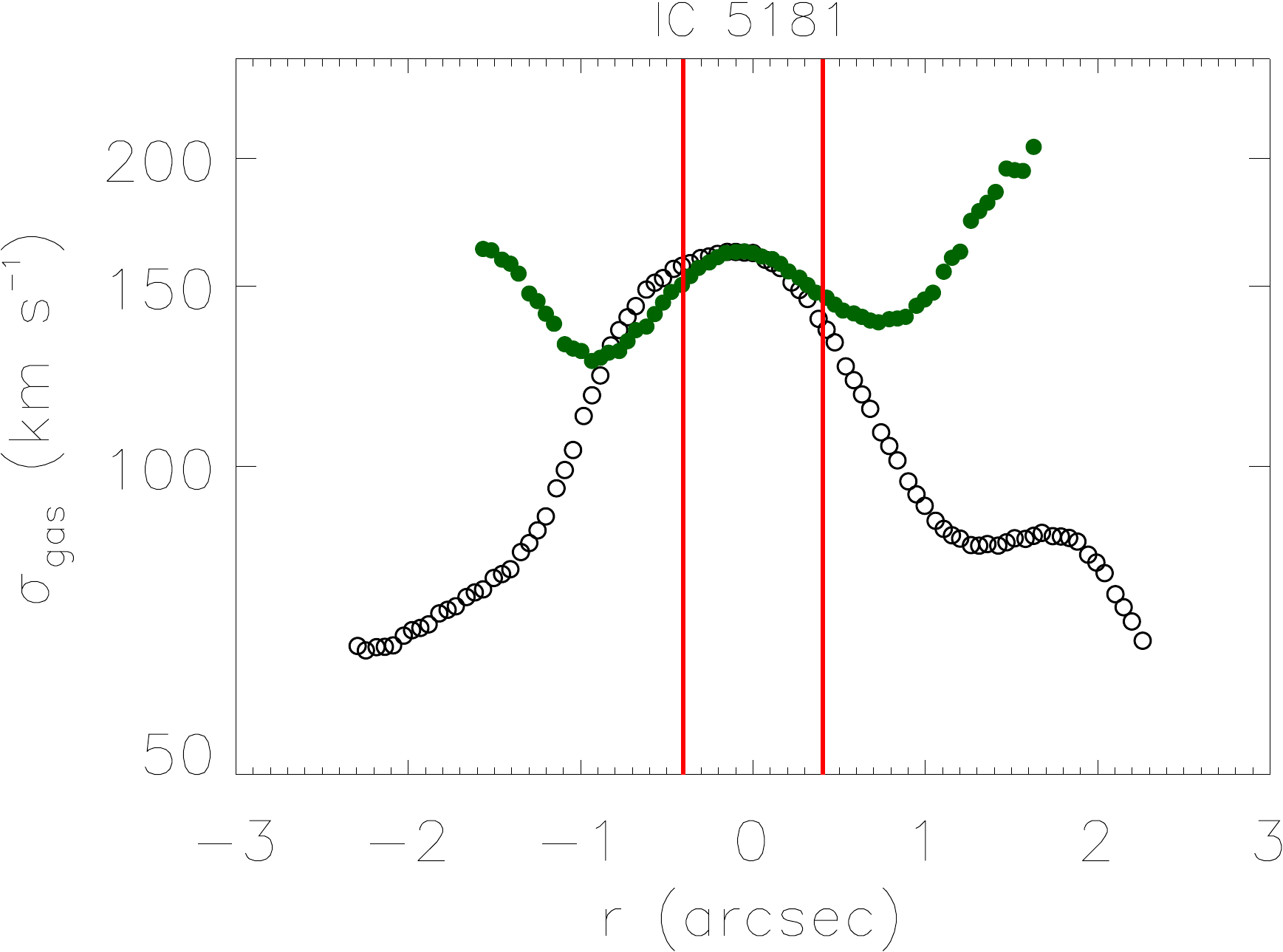}

\hspace{0.0cm}
\includegraphics[scale=0.32]{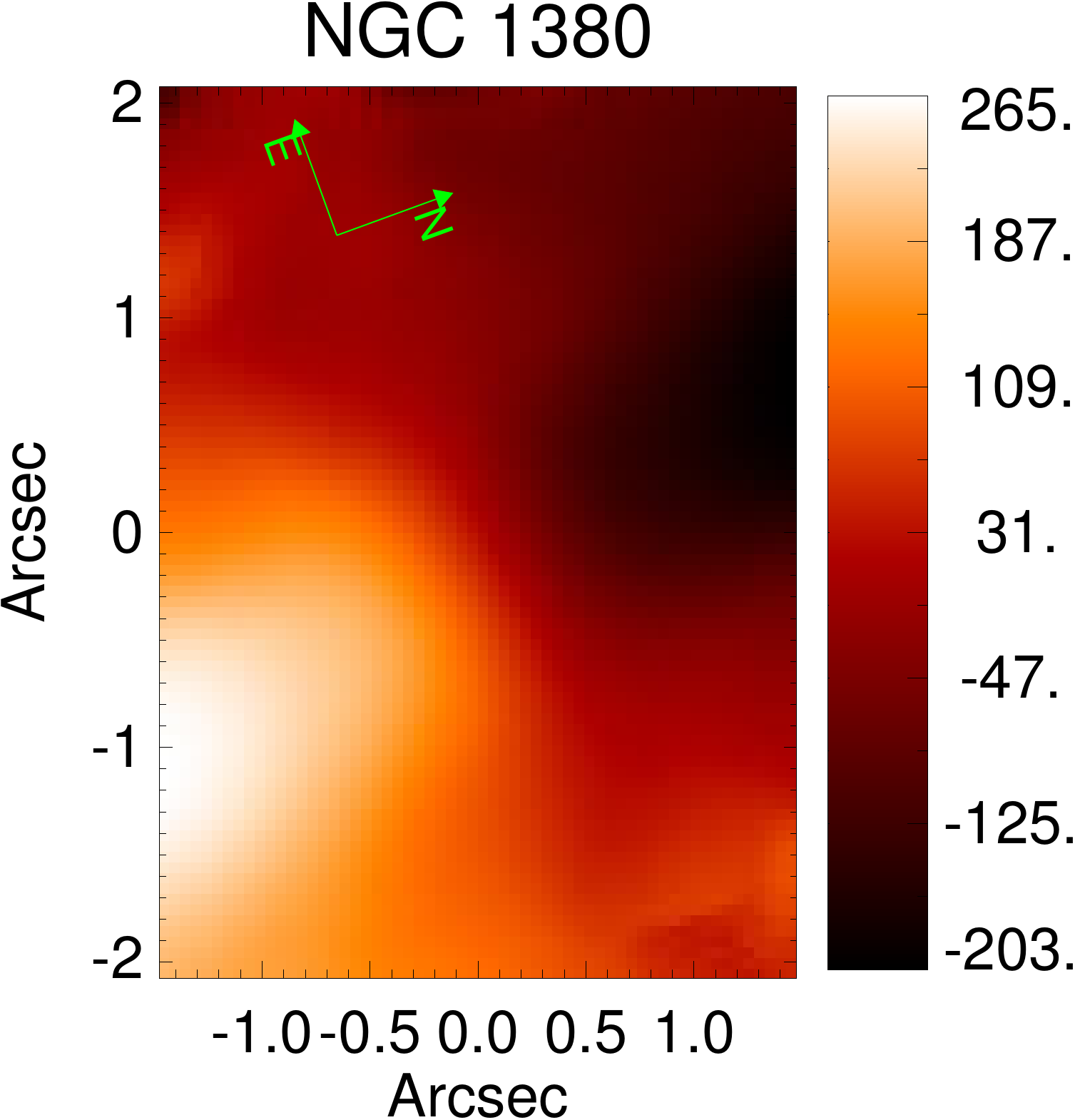}
\hspace{0.5cm}
\vspace{0.7cm}
\includegraphics[scale=0.32]{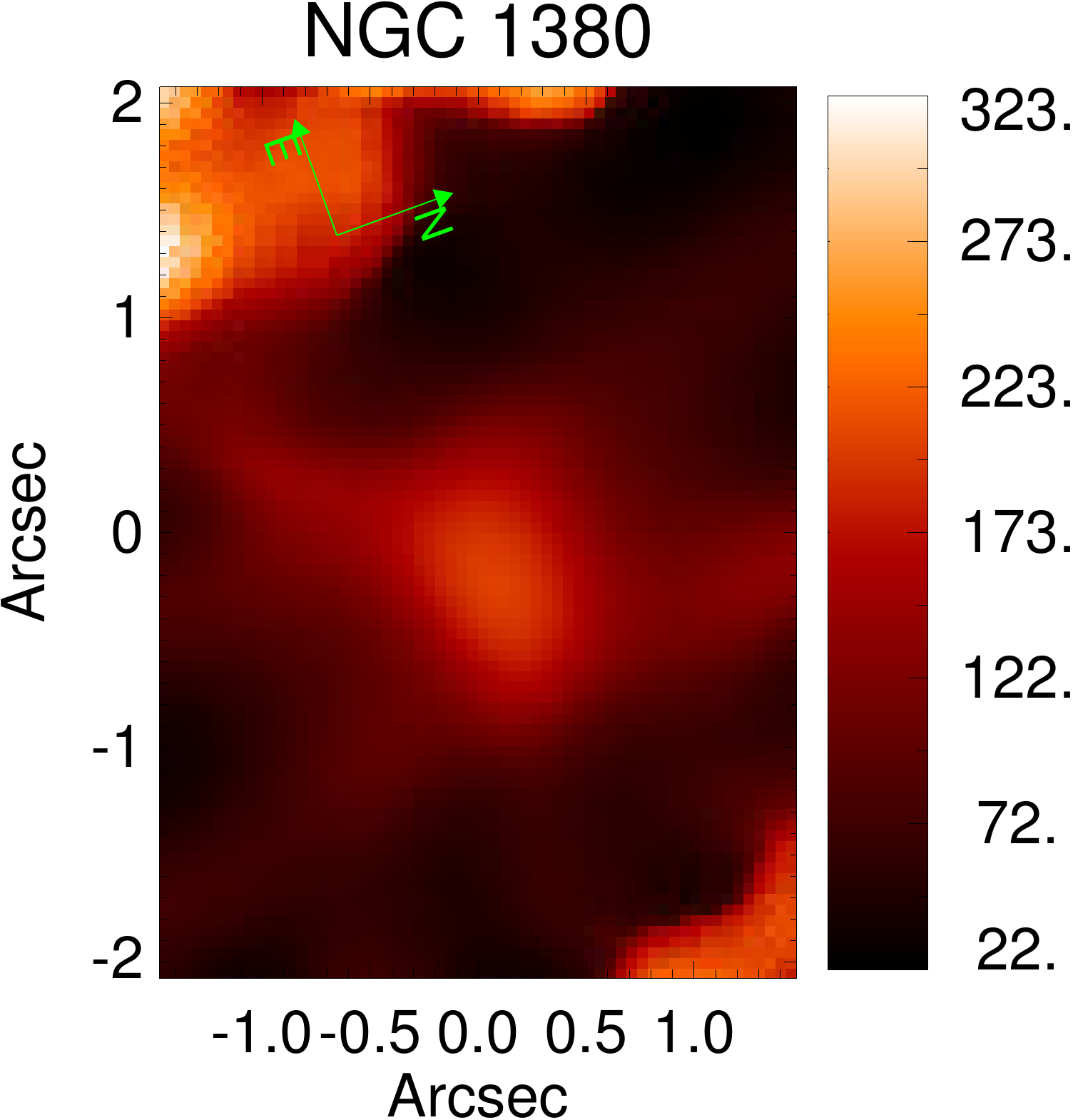}
\hspace{1.0cm}

\includegraphics[scale=0.36]{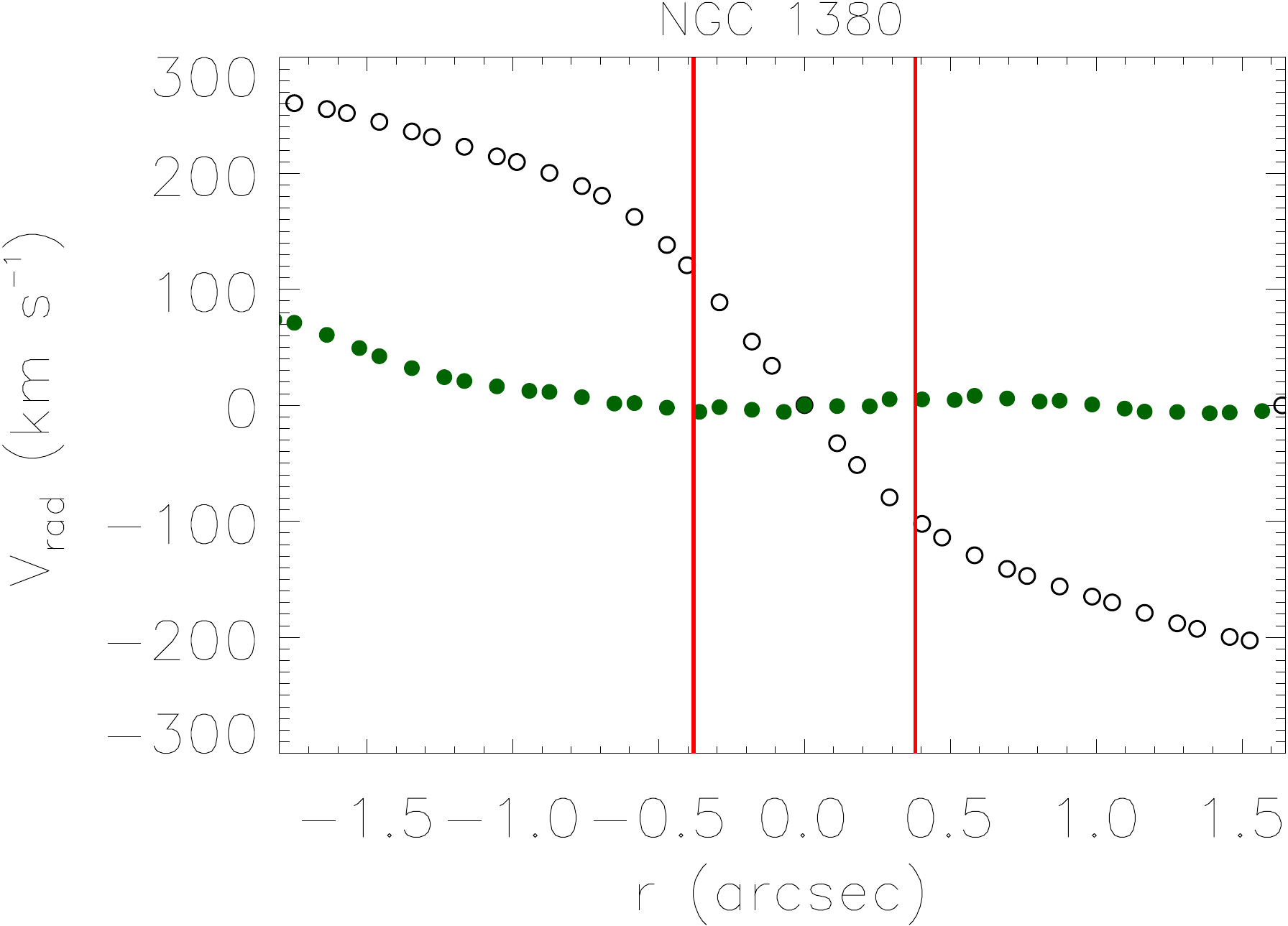}
\includegraphics[scale=0.36]{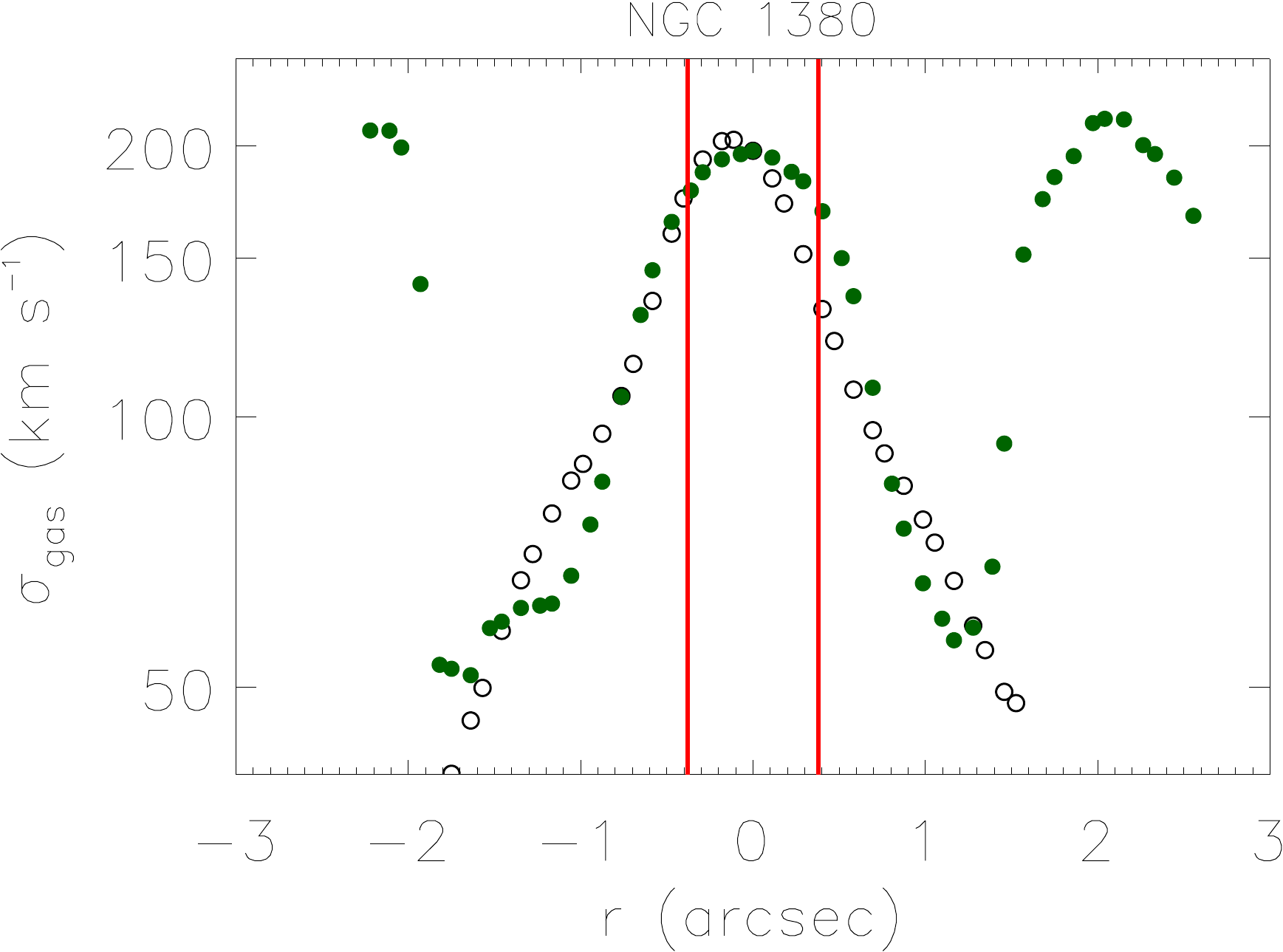}

\caption{Same as in Fig. \ref{mapa_cin_gal_1}. \label{mapa_cin_gal_2}}
\end{figure*}

\addtocounter{figure}{-1}
\addtocounter{subfigure}{1}

\begin{figure*}
\hspace{0.0cm}
\includegraphics[scale=0.32]{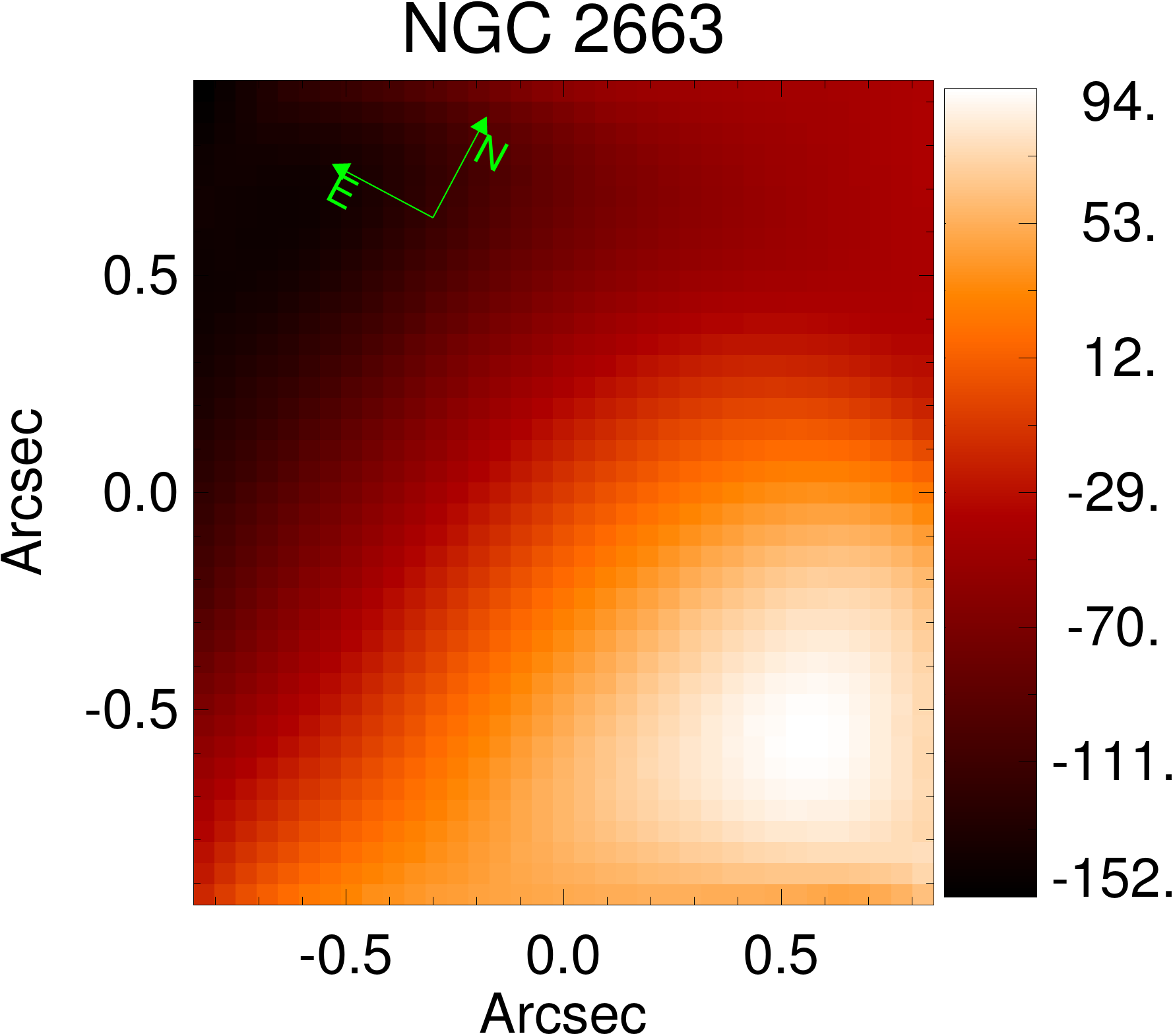}
\hspace{0.3cm}
\vspace{0.7cm}
\includegraphics[scale=0.32]{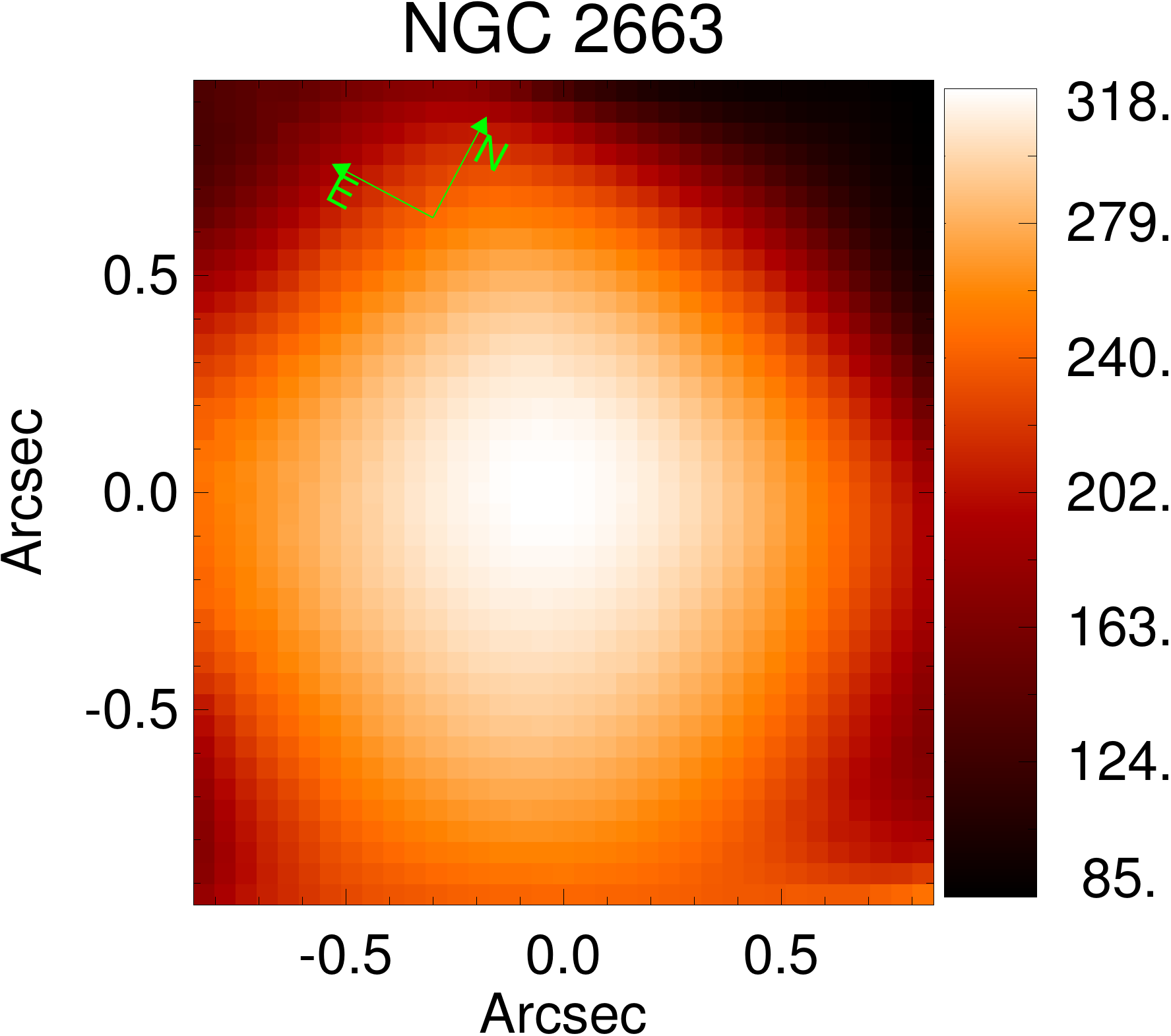}
\hspace{1.0cm}

\includegraphics[scale=0.36]{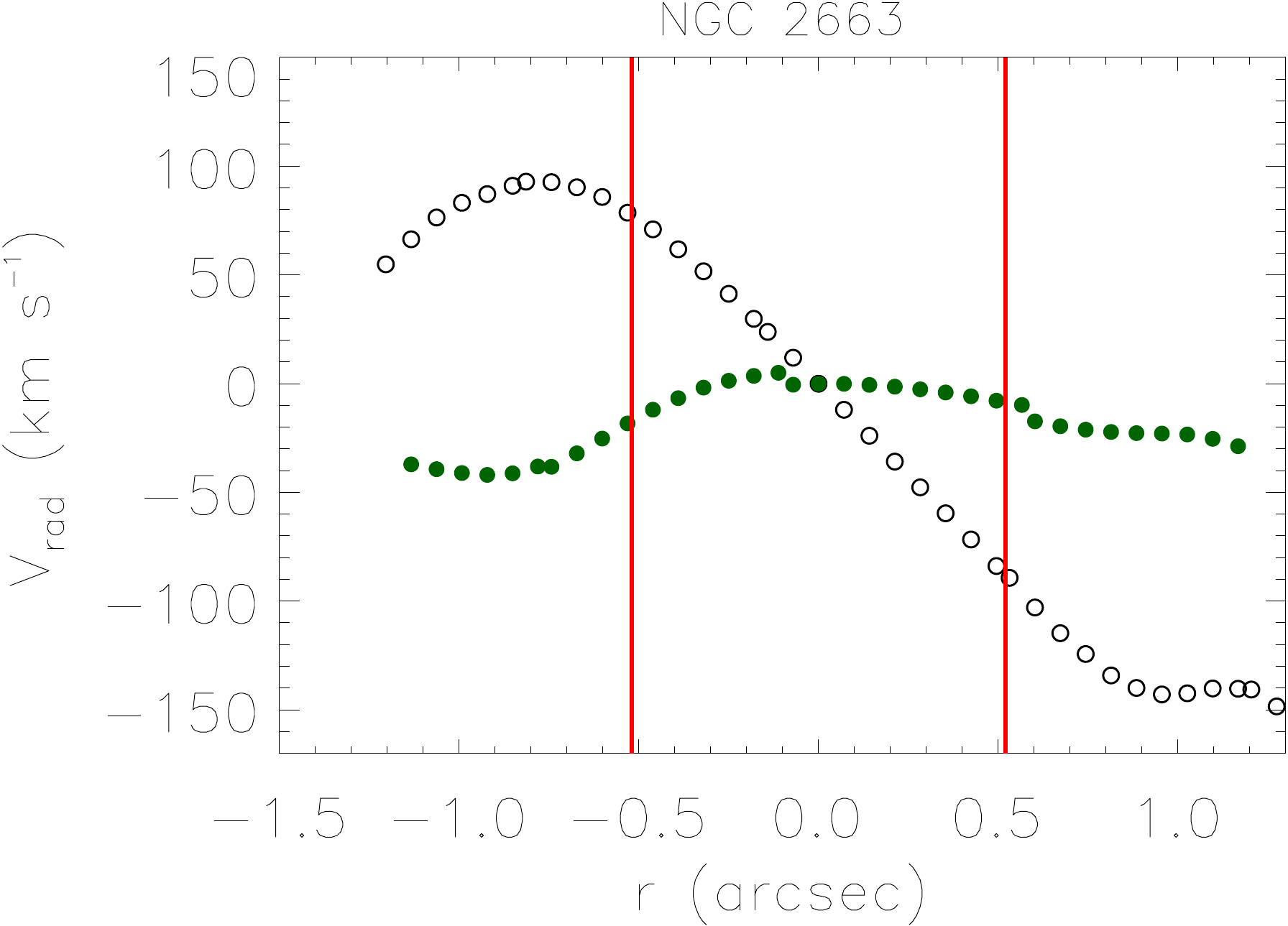}
\vspace{1.0cm}
\includegraphics[scale=0.36]{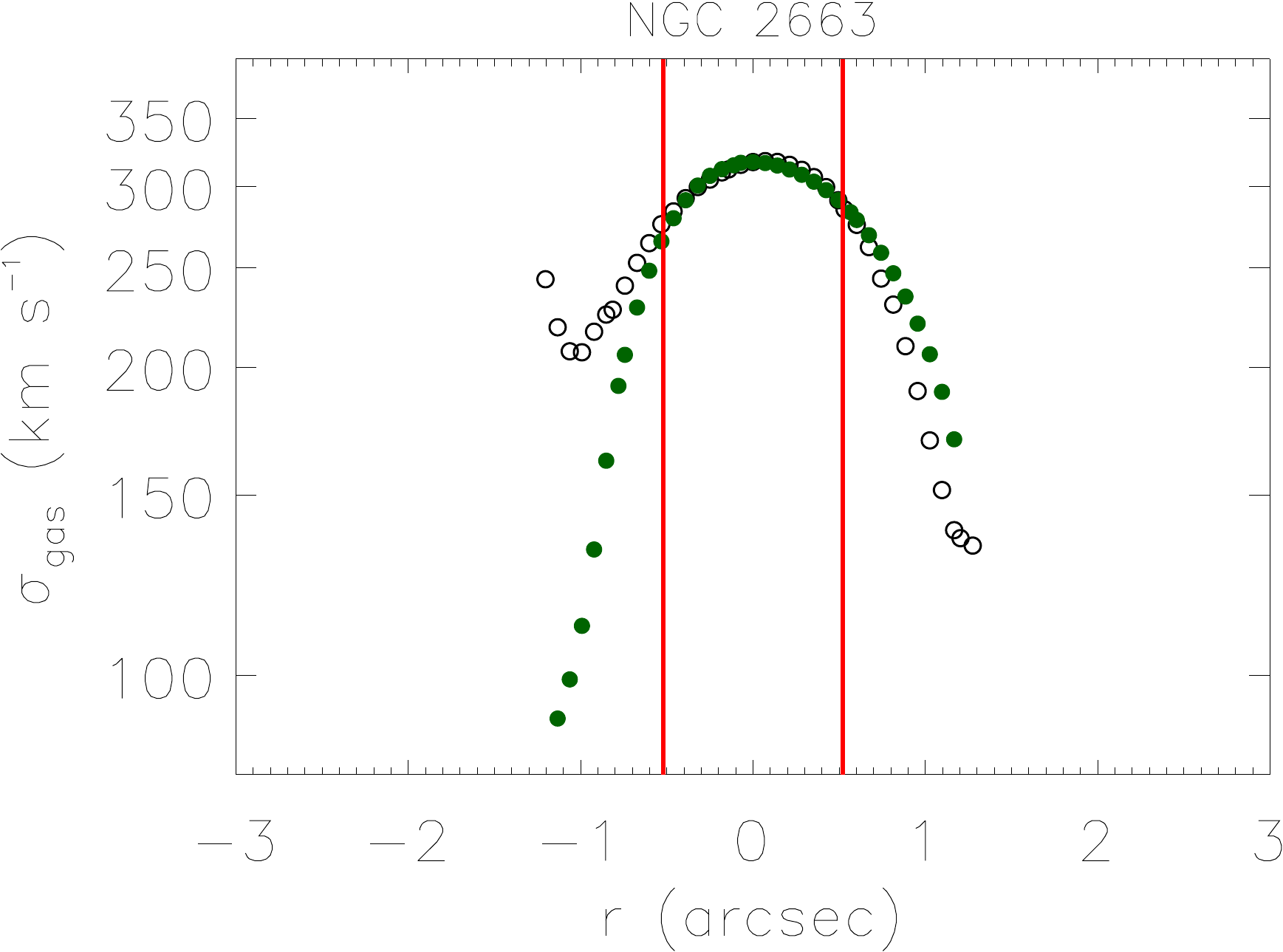}

\hspace{0.0cm}
\includegraphics[scale=0.32]{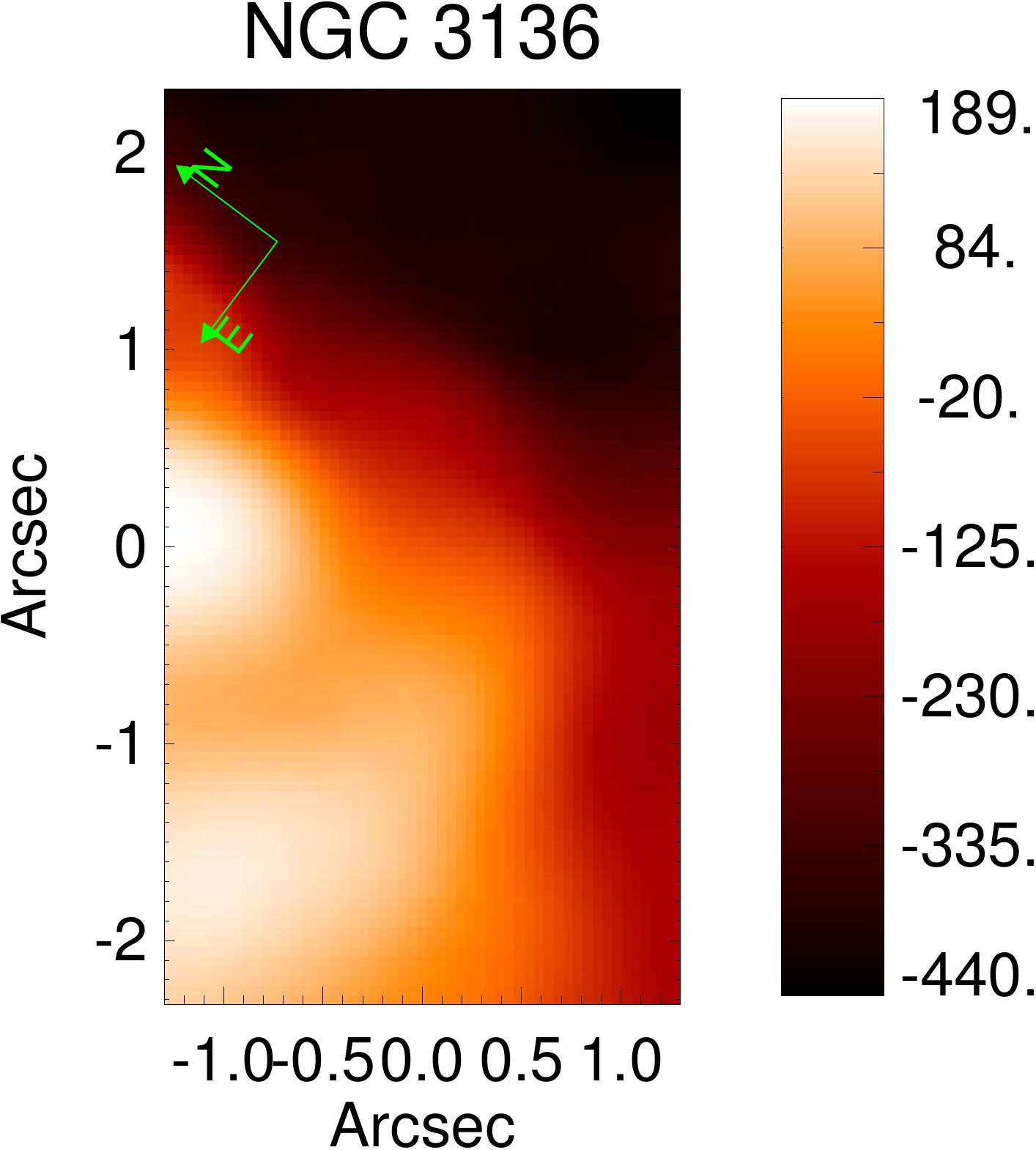}
\hspace{0.5cm}
\vspace{0.7cm}
\includegraphics[scale=0.32]{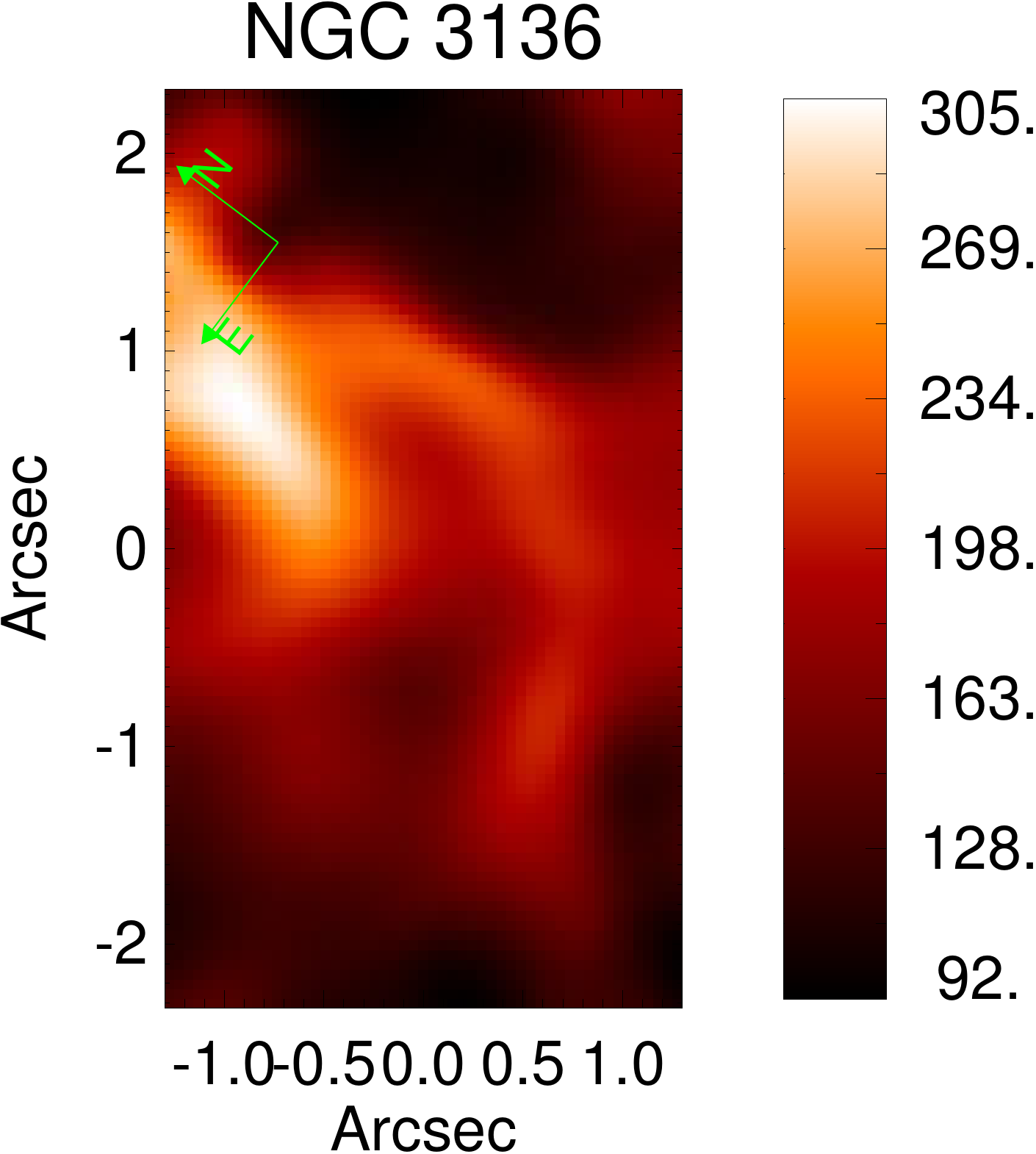}
\hspace{1.0cm}

\includegraphics[scale=0.36]{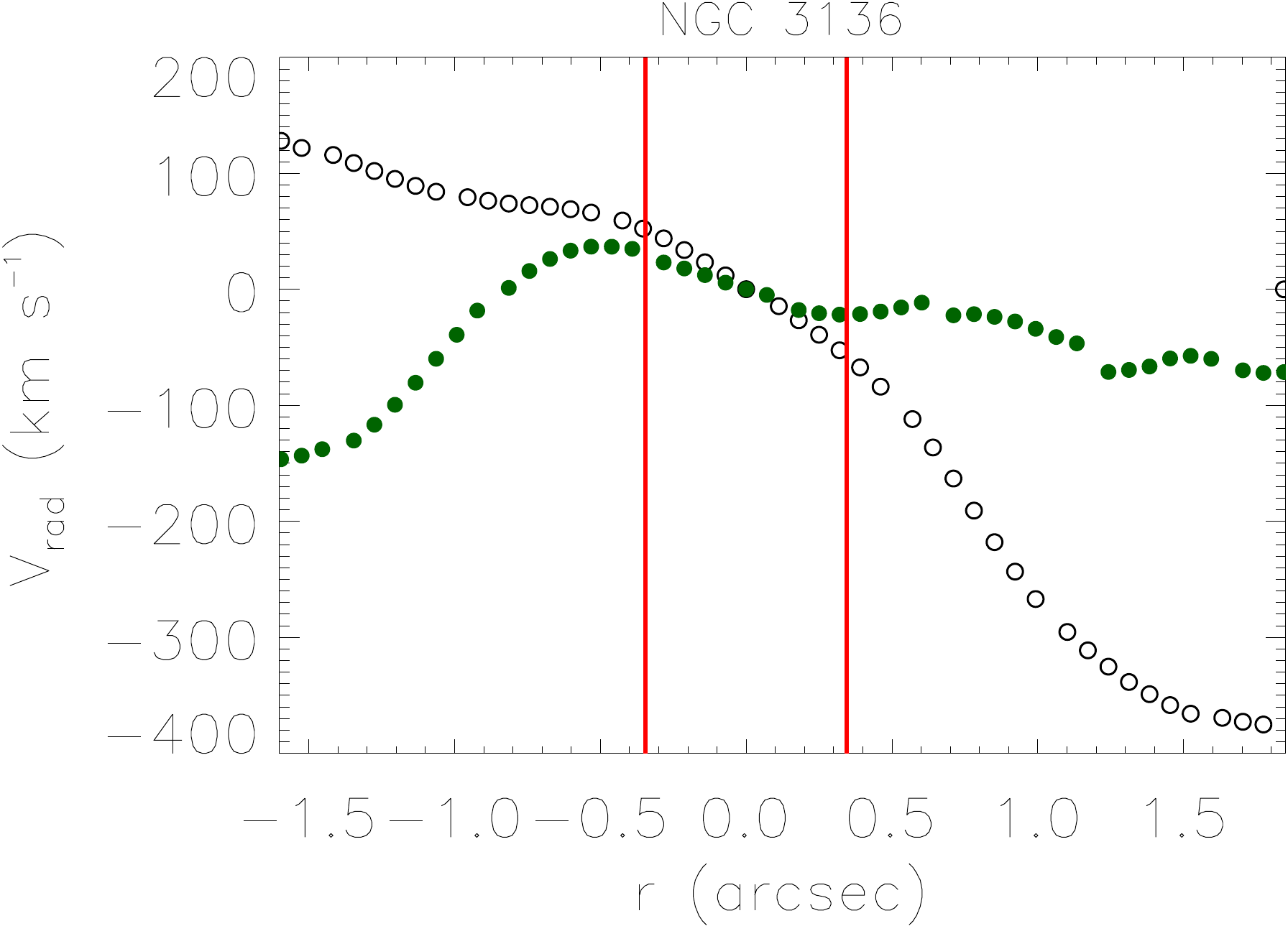}
\includegraphics[scale=0.36]{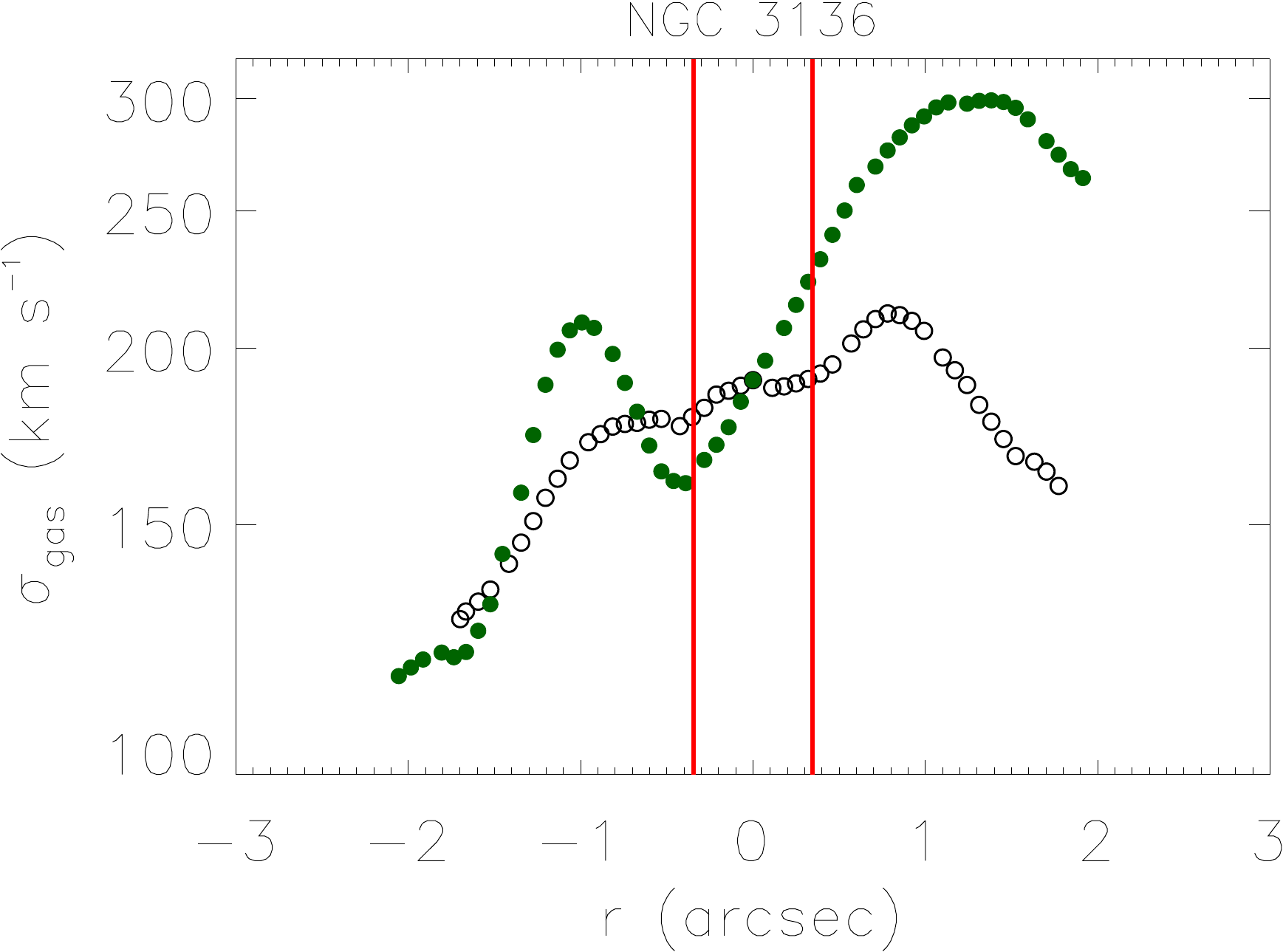}

\caption{Same as in Fig. \ref{mapa_cin_gal_1}. \label{mapa_cin_gal_3}}
\end{figure*}

\addtocounter{figure}{-1}
\addtocounter{subfigure}{1}

\begin{figure*}
\hspace{0.0cm}
\includegraphics[scale=0.32]{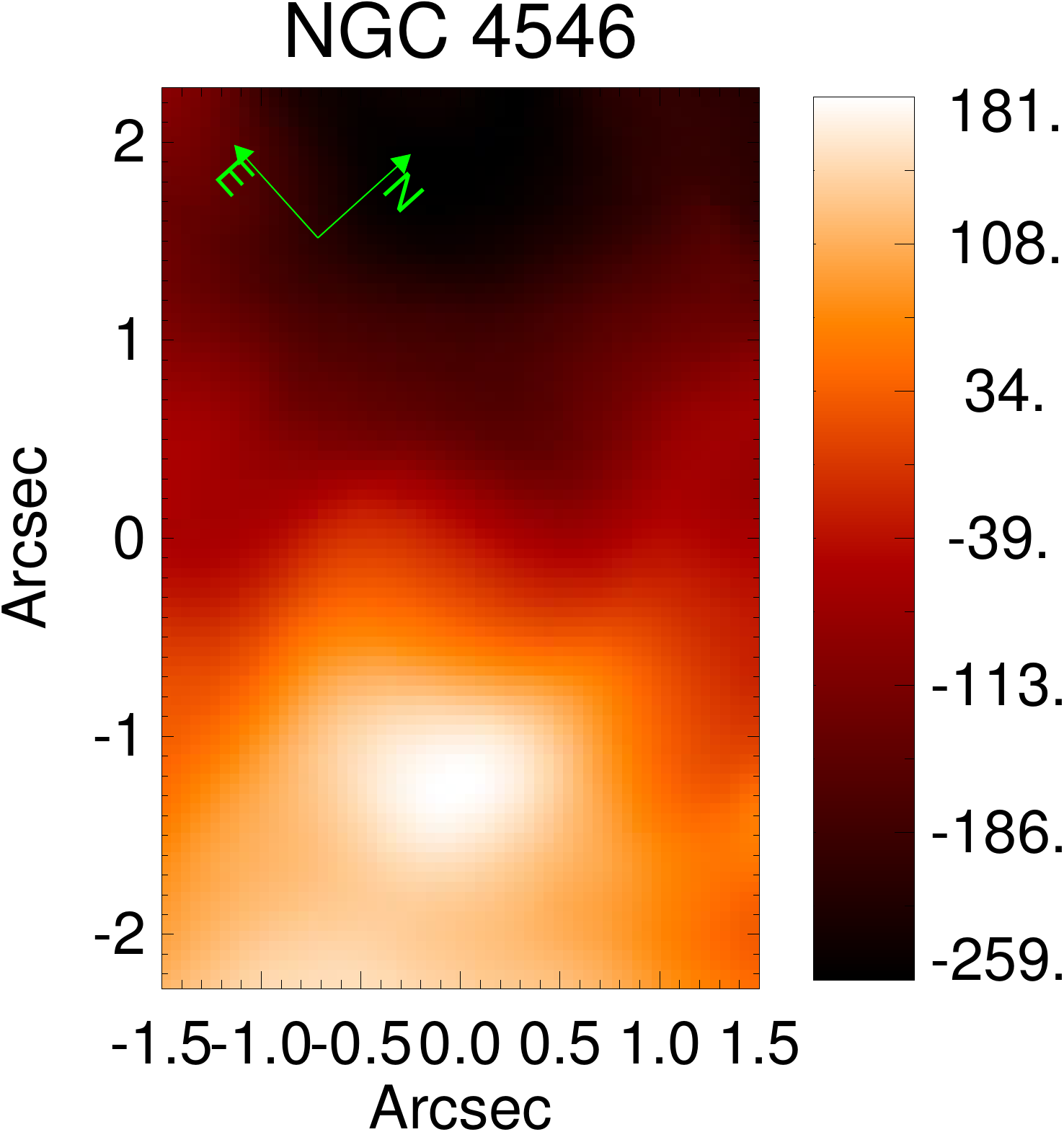}
\hspace{0.5cm}
\vspace{0.7cm}
\includegraphics[scale=0.32]{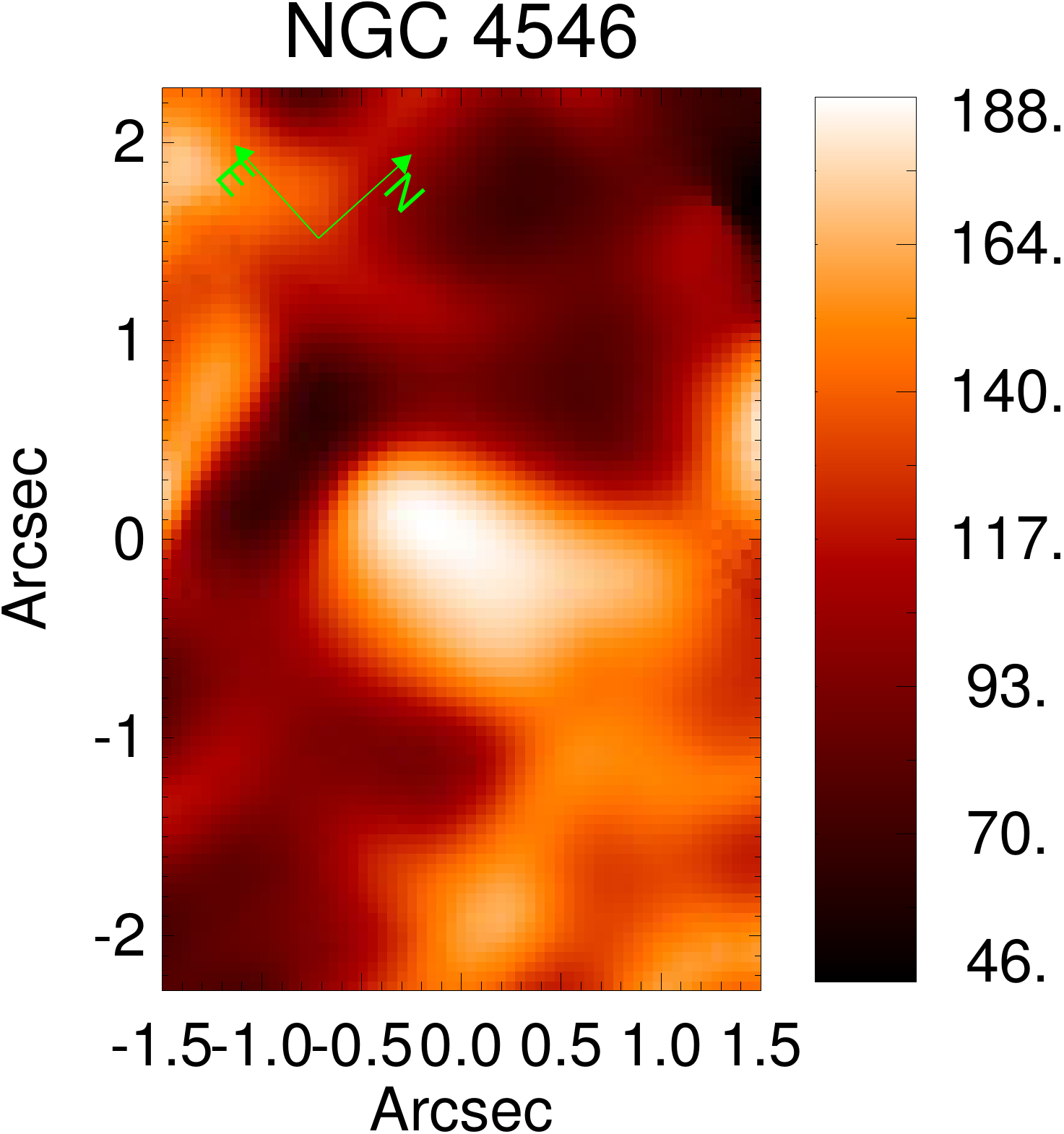}
\hspace{1.0cm}

\includegraphics[scale=0.36]{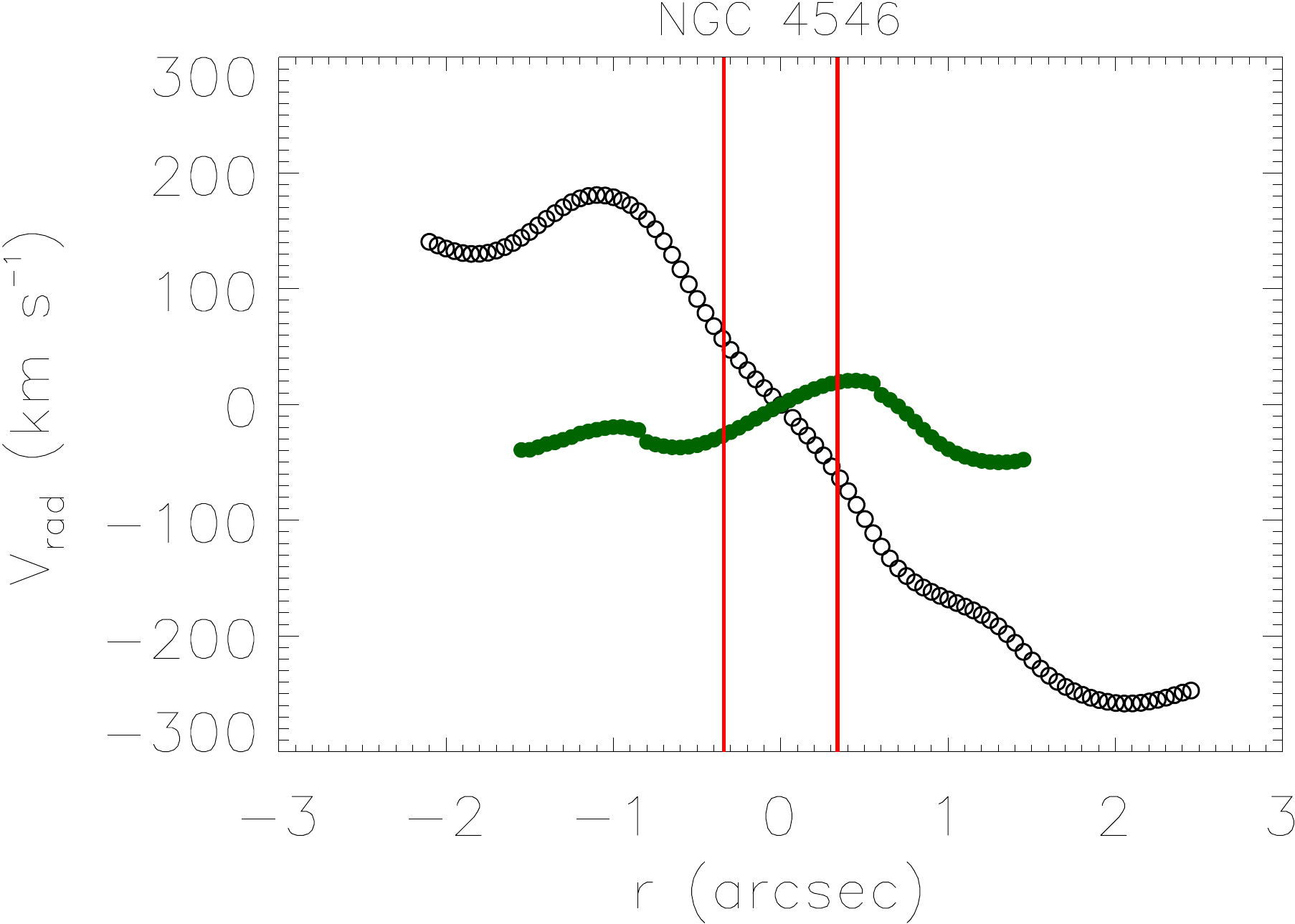}
\vspace{1.0cm}
\includegraphics[scale=0.36]{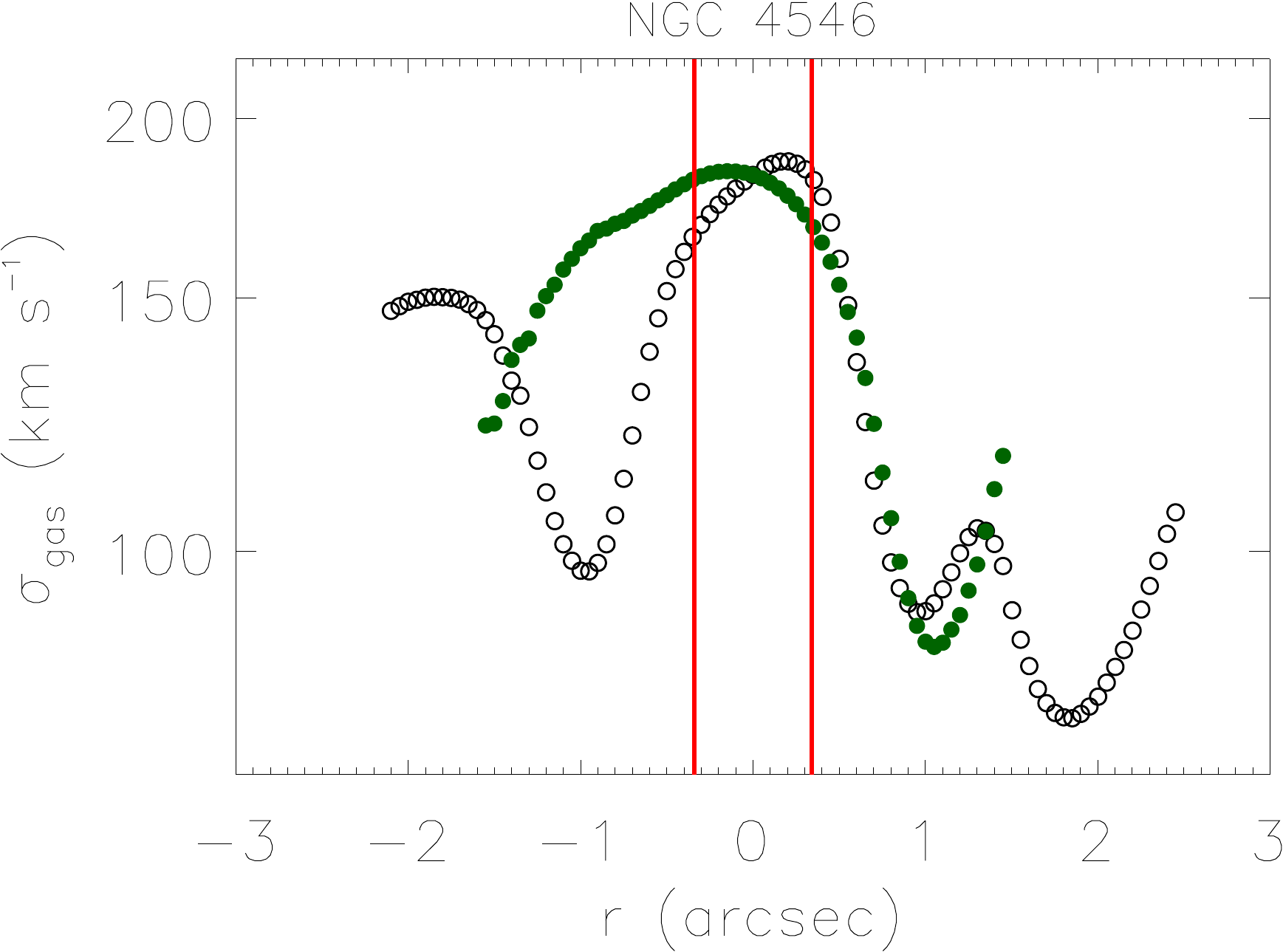}

\hspace{0.0cm}
\includegraphics[scale=0.32]{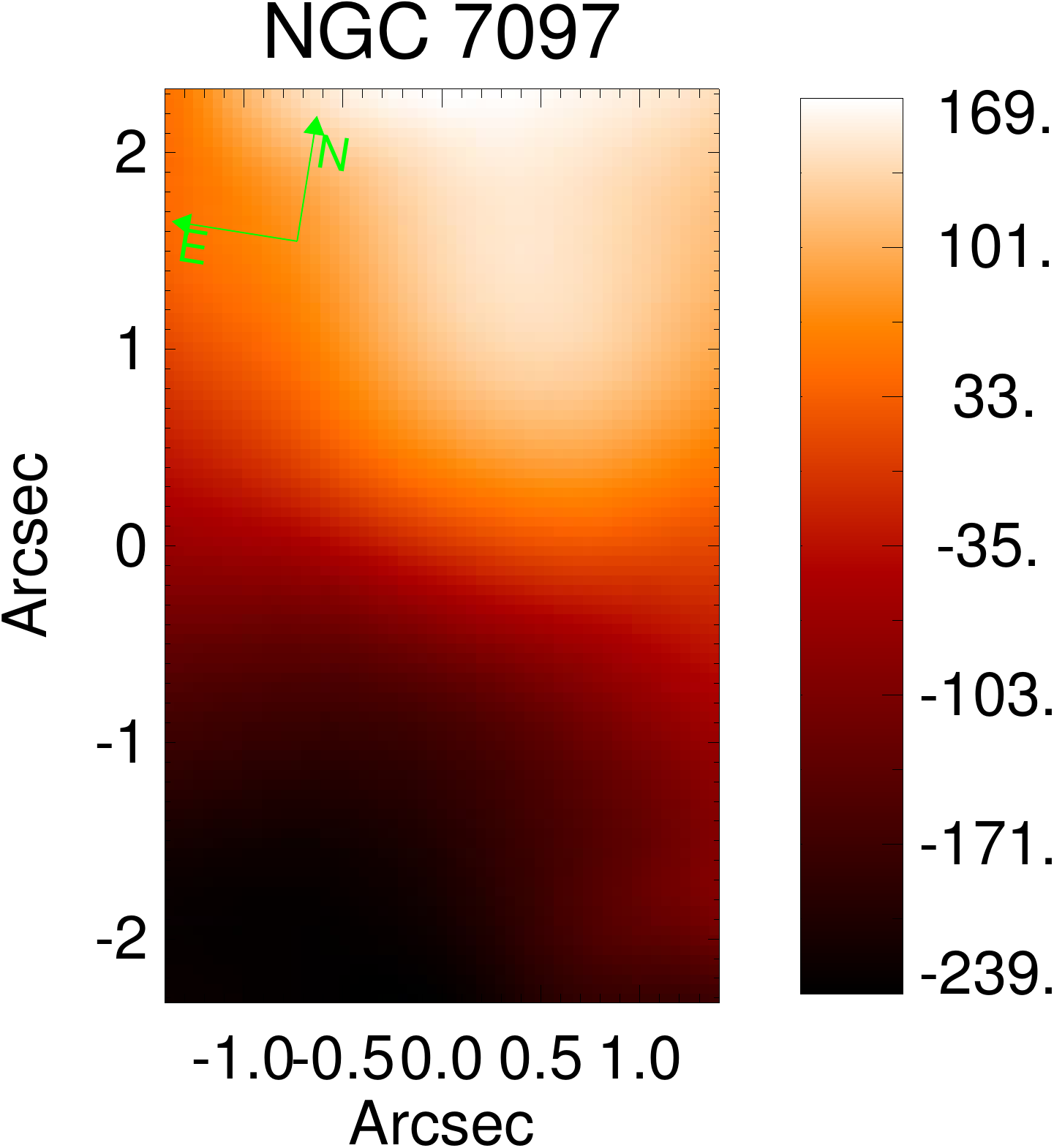}
\hspace{0.5cm}
\vspace{0.7cm}
\includegraphics[scale=0.32]{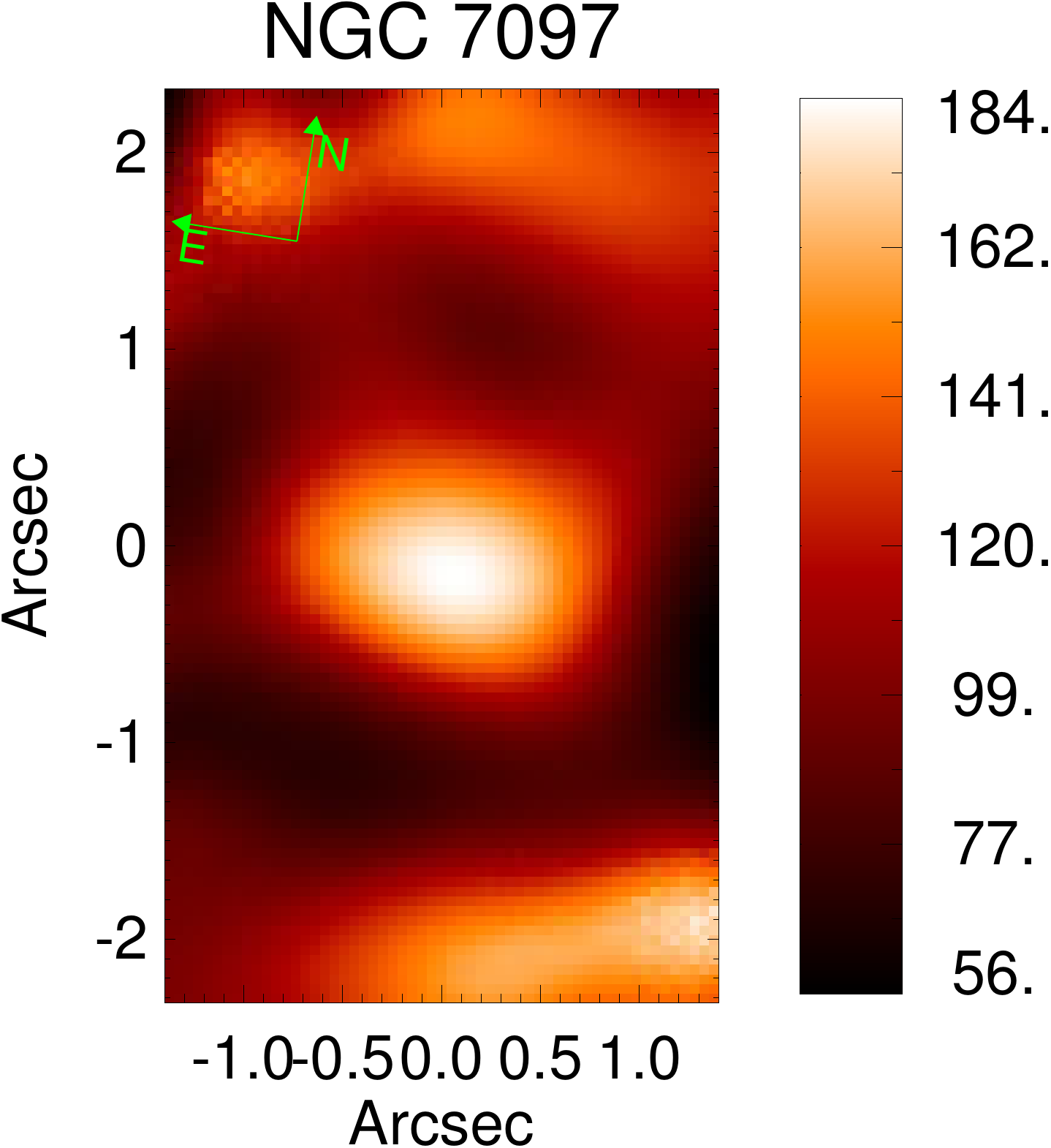}
\hspace{1.0cm}

\includegraphics[scale=0.36]{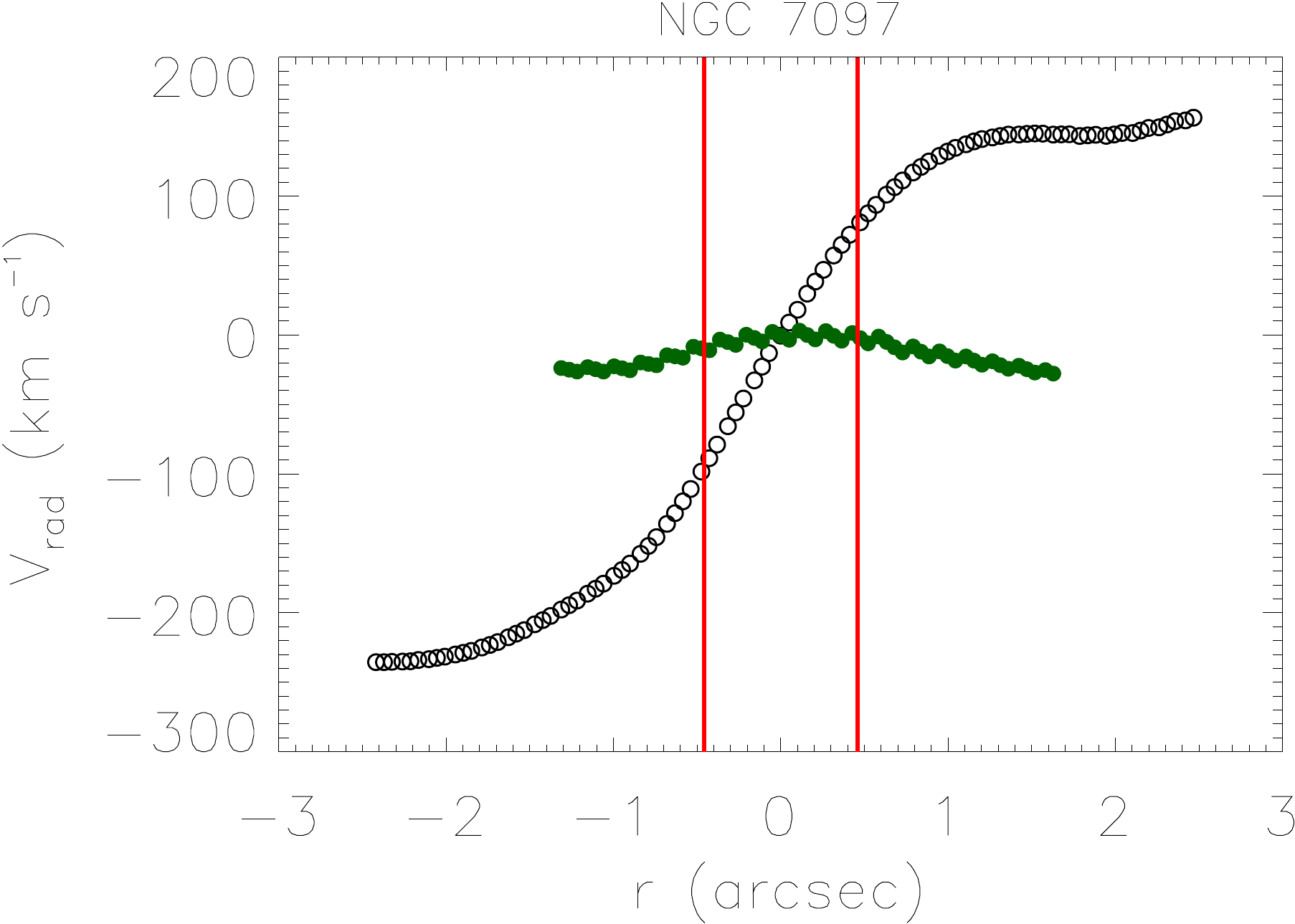}
\includegraphics[scale=0.36]{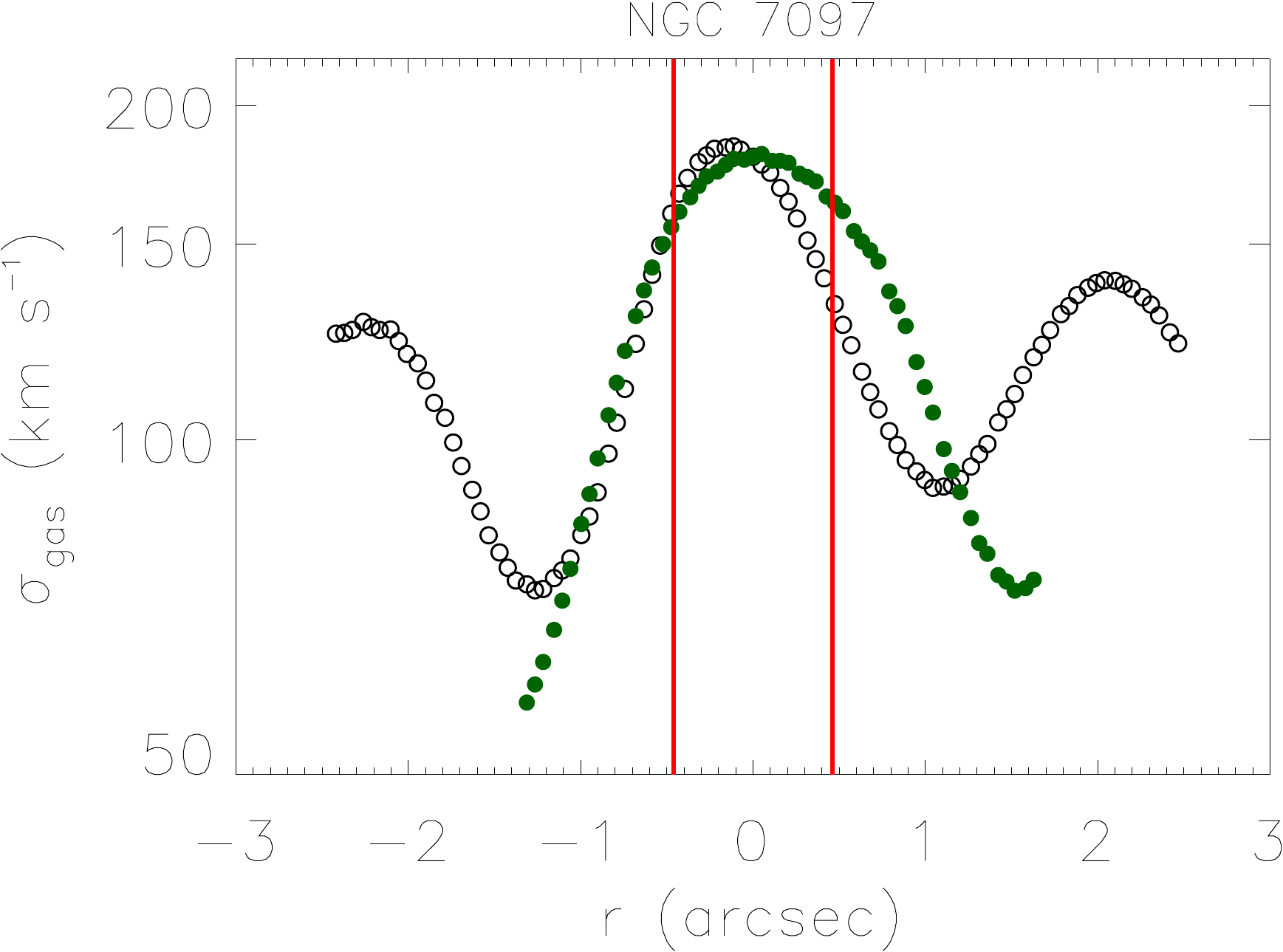}

\caption{Same as in Fig. \ref{mapa_cin_gal_1}. \label{mapa_cin_gal_4}}
\end{figure*}

\renewcommand{\thefigure}{\arabic{figure}}

In the radial velocity maps, we detected bipolar structures along the circumnuclear regions in the eight analysed galaxies. The position angles (P.A.) of these kinematic bipolar structures were measured by fitting a linear function between the points of maximum and minimum radial velocity values along this structure. The P.A. results are shown in Table \ref{tab_PA}. The errors of the P.A. were calculated by means of a Monte Carlo simulation, where we measured the P.A. a hundred times in the maps after adding Gaussian noise with $\sigma$ equal to the error of the radial velocity in each spaxel. The uncertainties associated with the radial velocity are the statistical errors taken from the fitting procedure of the emission lines. 

 \begin{table}
 \scriptsize
 \begin{center}
  \caption{Column (1): Galaxy name. Column (2): Radial velocity, in km s$^{-1}$, measured on the nucleus of the galaxies after the redshift correction applied to the galaxies using the velocities presented in paper I. It is worth mentioning that these velocities were measured for the narrow component of the emission lines, since we removed the broad component of H$\alpha$ from the gas cubes. Column (3): Position angles of the kinematic bipolar structures measured in the radial velocity maps shown in Figs. \ref{mapa_cin_gal_1}, \ref{mapa_cin_gal_2}, \ref{mapa_cin_gal_3} and \ref{mapa_cin_gal_4}. Column (4) Position angles of the kinematic bipolar structures measured in the maps of the blue and red wings of the H$\alpha$ emission line shown in Fig. \ref{RGB_Ha}.    \label{tab_PA}
}

 \begin{tabular}{@{}lccc}
  \hline
  Galaxy name & V$_{nuc}$ &P.A.$_{Vr}$ & P.A.$_{H\alpha}$ \\
  (1) & (2) & (3) & (4) \\
  \hline
  ESO 208 G-21 &13 &$-66^o \pm 2$ & $-74^o \pm 5$ \\
  NGC 1380 &-50 &$-169^o \pm 2$ & $-169^o \pm 4$\\
  IC 1459 & -65&$48^o \pm 1$ & $22^o \pm 3$\\
  NGC 7097 & -6&$-9^o \pm 3$ & $-11^o \pm 9$\\
  IC 5181 & 44&$-37^o \pm 2$ & $-24^o \pm 4$\\
  NGC 4546 & -12&$-130^o \pm 2$& $-150^o \pm 6$ \\
  NGC 2663 & 66&$-110^o \pm 3$ & $-98^o \pm 6$\\
  NGC 3136  & 69&$78^o \pm 2$ & $73^o \pm 9$\\
 
  \hline
 \end{tabular}
 \end{center}

\end{table}

In addition to the maps, we built 1D profiles along the kinematic bipolar structure and along the perpendicular direction of the same structure. In both cases, the profiles traverse the nucleus of the galaxies. The nuclear positions of the six galaxies with a BLR were estimated with the image of the red wing of the broad component of the H$\alpha$ line. In both NGC 1380 and NGC 3136, we used the [O I]$\lambda$6300 emission line to define the nuclear position because these galaxies lack a BLR. For each galaxy, the 1D profiles of the kinematic parameters are shown below their respective maps in Figs \ref{mapa_cin_gal_1}, \ref{mapa_cin_gal_2}, \ref{mapa_cin_gal_3} and \ref{mapa_cin_gal_4}. Note that the velocity values shown in Figs. \ref{mapa_cin_gal_1}, \ref{mapa_cin_gal_2}, \ref{mapa_cin_gal_3} and \ref{mapa_cin_gal_4} are presented relative to the velocity measured in the nucleus, whose values are shown in Table \ref{tab_PA}. It is worth mentioning that the velocity measured in the nucleus is relative to the redshifts of the sample galaxies shown in paper I.

In paper I, we proposed that gaseous discs were present in six galaxies of the sample. In NGC 3136, using PCA Tomography, we detected the kinematics of an extended structure, which is probably not associated with a gaseous disc. The radial velocity map of this object, shown in Fig. \ref{mapa_cin_gal_3}, reveals the overall kinematics of the structure. The P.A. of the kinematic bipolar structures detected with PCA Tomography (paper I) match the P.A. measured in the radial velocity maps of five objects. In NGC 3136, we found a difference of $\sim$ 20$^o$, but the extended structure is quite irregular. In IC 5181, the difference between both measurements is $\sim$ 16$^o$. However, a twist in the gas structure of this object was detected, which is discussed in Sections \ref{EW_extended_emission} and \ref{discs_or_cones}. In NGC 4546, the difference is $\sim$ 19$^o$. In this case, the 1D profile along the kinematic bipolar structure suggests a non-Keplerian curve, with perturbations within a projected distance of 1 arcsec from the nucleus. Although IC 1459 has a difference of $\sim$ 11$^o$ between both measurements of the P.A., the kinematic bipolar structure is affected by an outflow. This perturbation in the kinematics of IC 1459 has been previously mentioned by \citet{2002ApJ...578..787C}. 

In the gas velocity dispersion maps, we detected a peak in the nuclear region of all galaxies, except for NGC 3136, which has a velocity dispersion peak northward from the nucleus. In the circumnuclear regions of four galaxies, the variation of the velocity dispersion is the same in both parallel and perpendicular directions of the kinematic bipolar structure. In the case of ESO 208 G-21, the velocity dispersion falls more gently in the direction from the nucleus along the kinematic bipolar structure. In the case of NGC 4546 and NGC 7097, the velocity dispersion radial gradient is shallower in the direction from the nucleus along the perpendicular direction of the kinematic bipolar structure. In IC 5181, the velocity dispersion rises along the perpendicular direction of the kinematic bipolar structure. In NGC 3136, a strong positive radial gradient is present northward from the nucleus.

The gas kinematics suggest that pure gaseous discs are present only in ESO 208 G-21, NGC 1380 and NGC 7097. It is worth mentioning that in the first two objects, the gaseous disc is corotating with the stellar component (paper I). In NGC 7097, we detected a counterrotation between the gaseous disc and the stellar component (paper I). In NGC 4546 and IC 1459, gaseous discs may be present but are affected by non-Keplerian effects, possibly outflows. In NGC 3136, an ionization bicone is likely to be present. The cases of NGC 2663 and IC 5181 will be studied in Section \ref{discs_or_cones}.  

\subsection{Flux maps of the H$\alpha$ and [N II]$\lambda$6584 emission lines} \label{Halpha_NII_flux_maps}

The same Gaussian functions used to extract kinematic information from the gas structure of the sample galaxies were integrated in order to extract the flux of the H$\alpha$ and [N II]$\lambda$6583 emission lines. Figs. \ref{mapa_fluxo_gal_1} and \ref{mapa_fluxo_gal_2} show the results. The same figures present the 1D profiles of the H$\alpha$ flux maps along the parallel and the perpendicular directions of the kinematic bipolar structures. 

\renewcommand{\thefigure}{\arabic{figure}\alph{subfigure}}
\setcounter{subfigure}{1}

\begin{figure*}
\hspace{-2.5cm}
\includegraphics[scale=0.28]{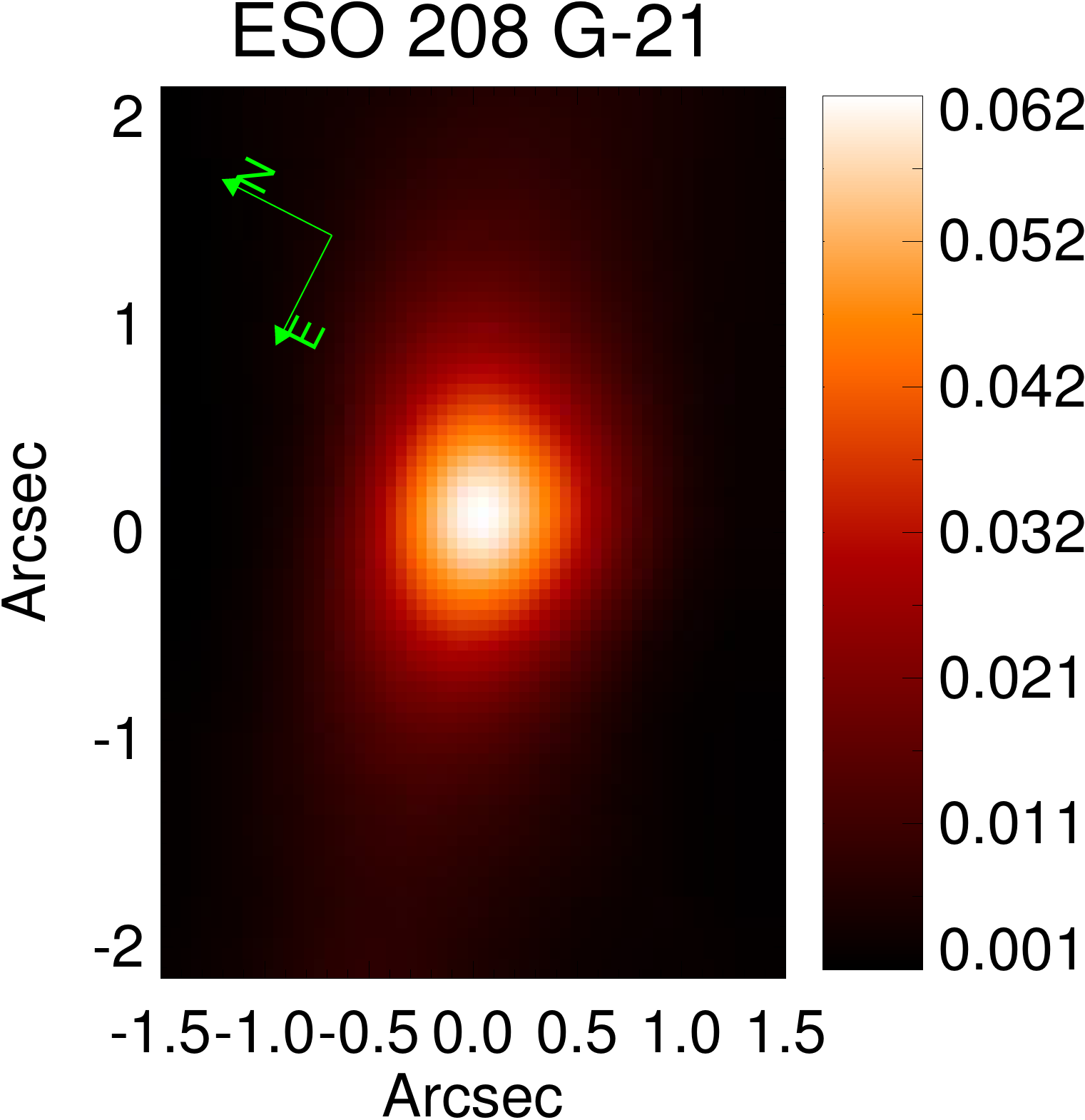}
\hspace{0.5cm}
\vspace{0.8cm}
\includegraphics[scale=0.28]{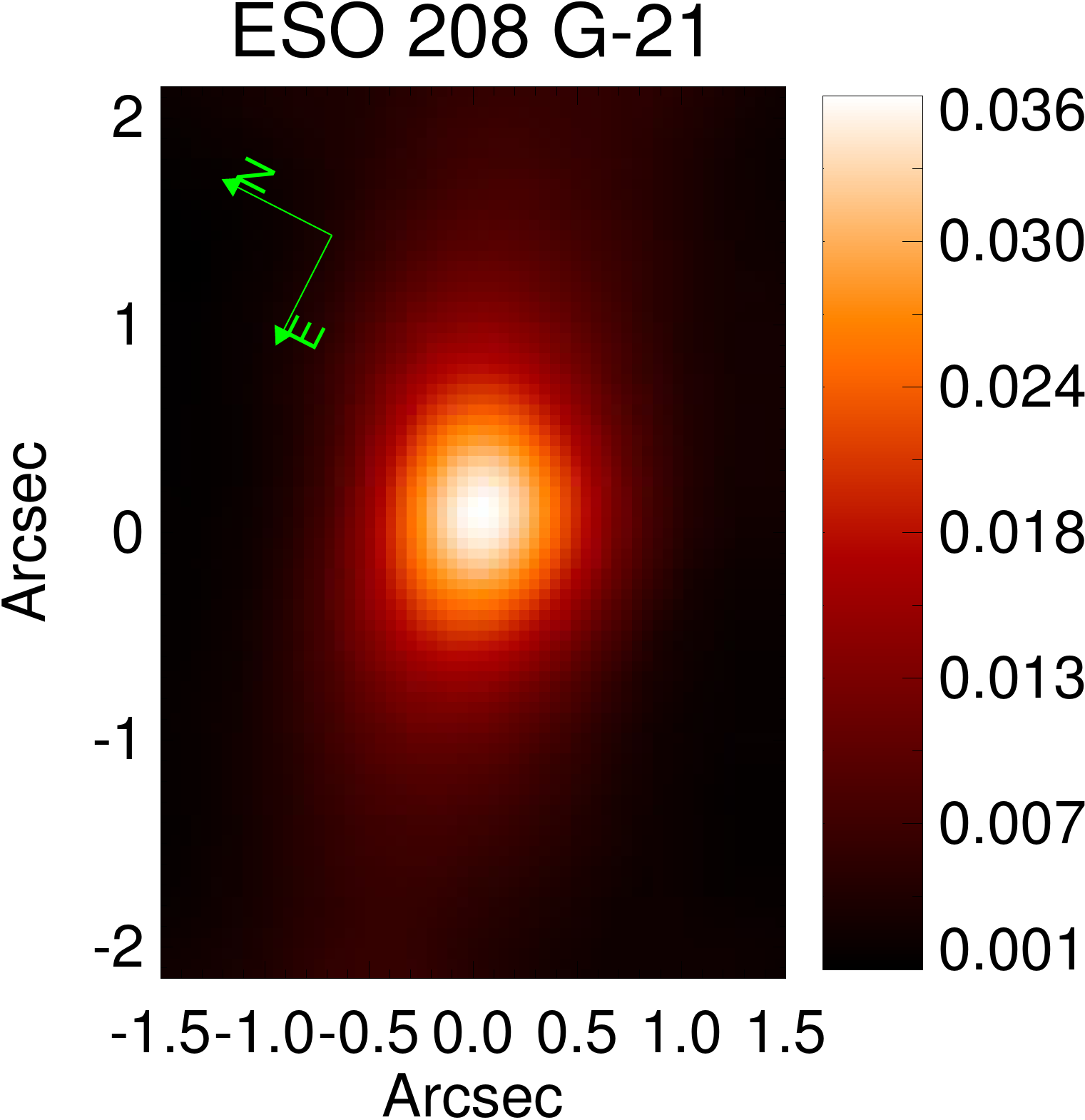}
\hspace{0.5cm}
\includegraphics[scale=0.34]{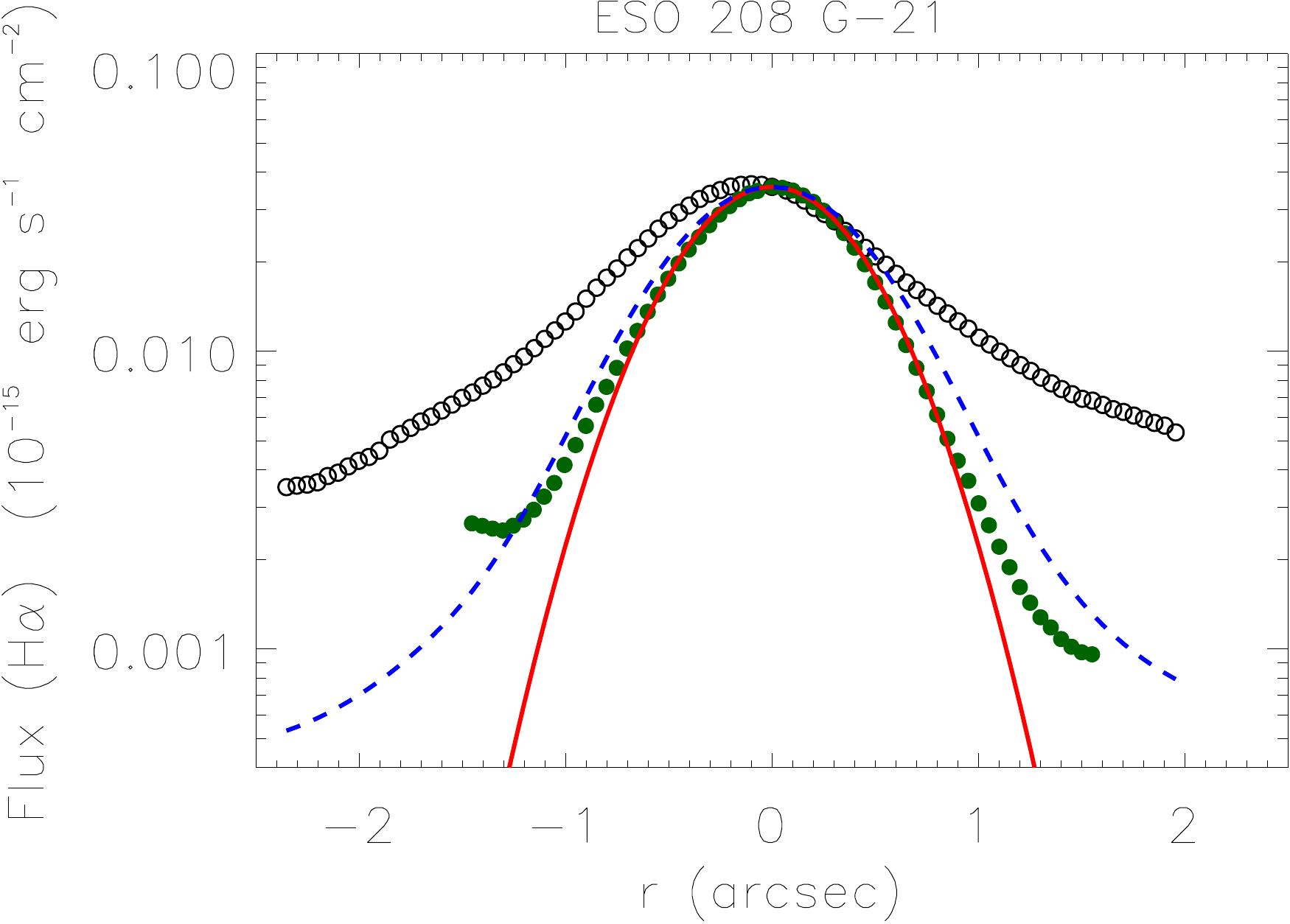}
\hspace{-0.8cm}

\hspace{-2.5cm}
\includegraphics[scale=0.28]{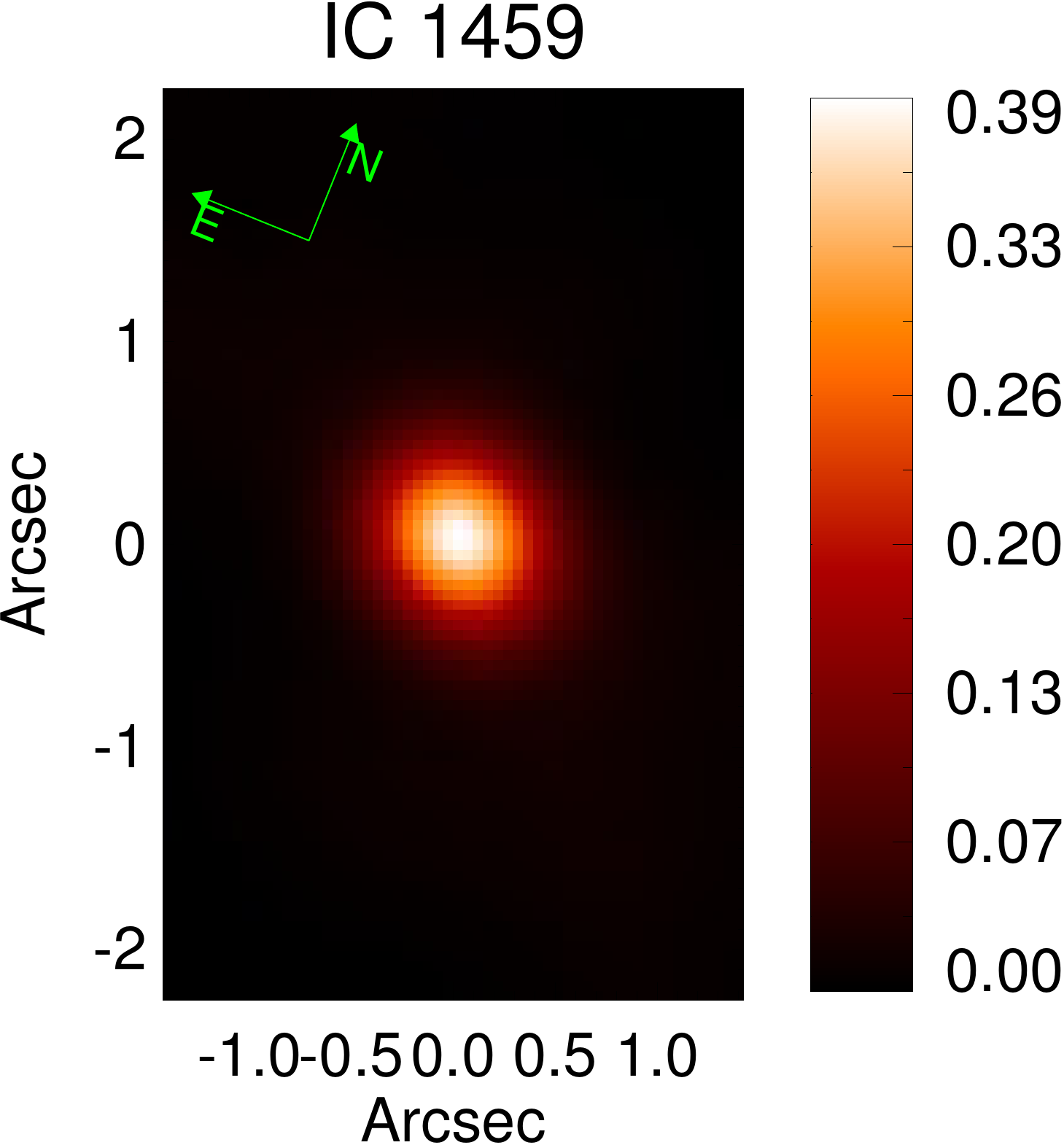}
\hspace{0.5cm}
\vspace{0.8cm}
\includegraphics[scale=0.28]{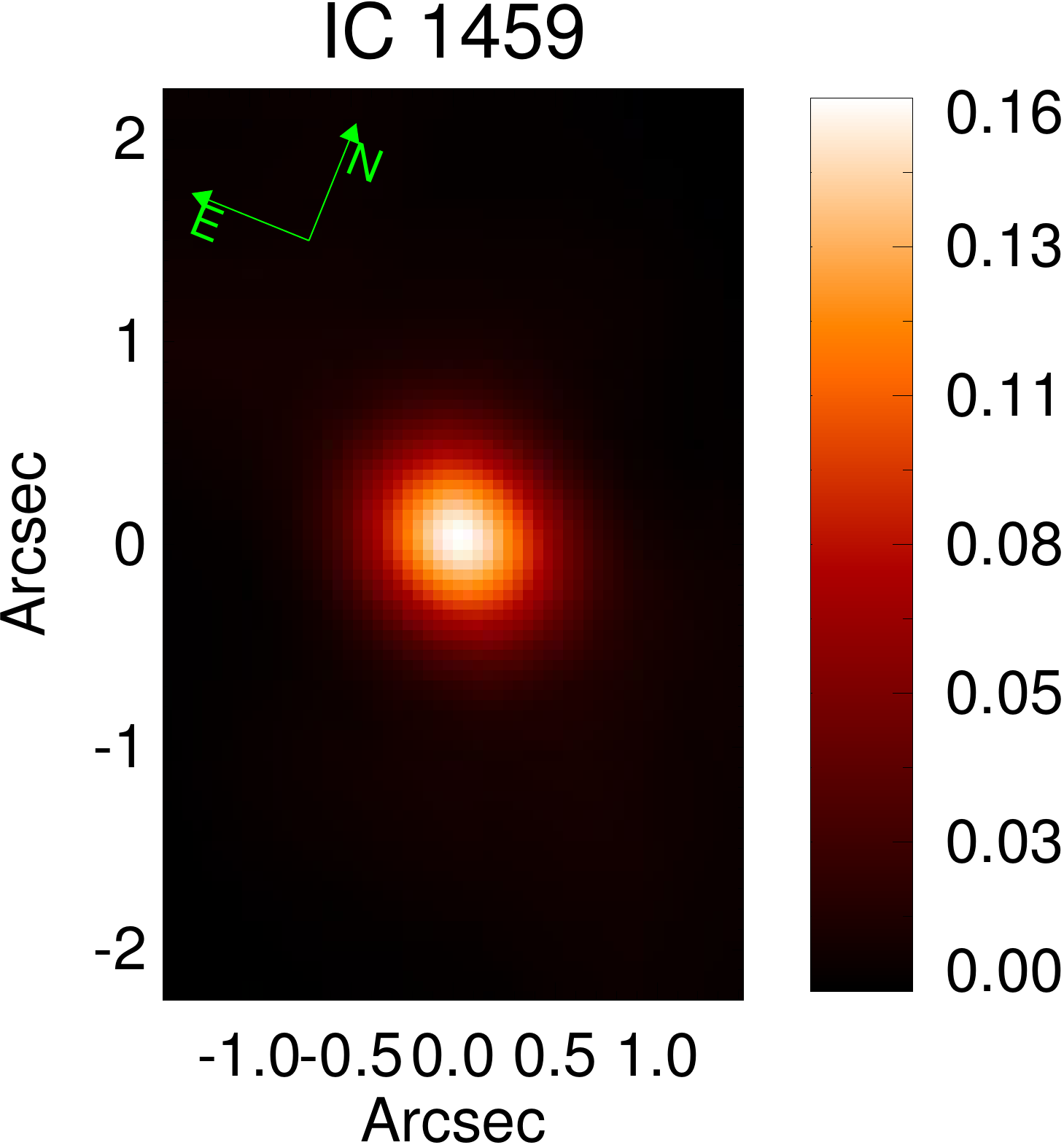}
\hspace{0.5cm}
\includegraphics[scale=0.34]{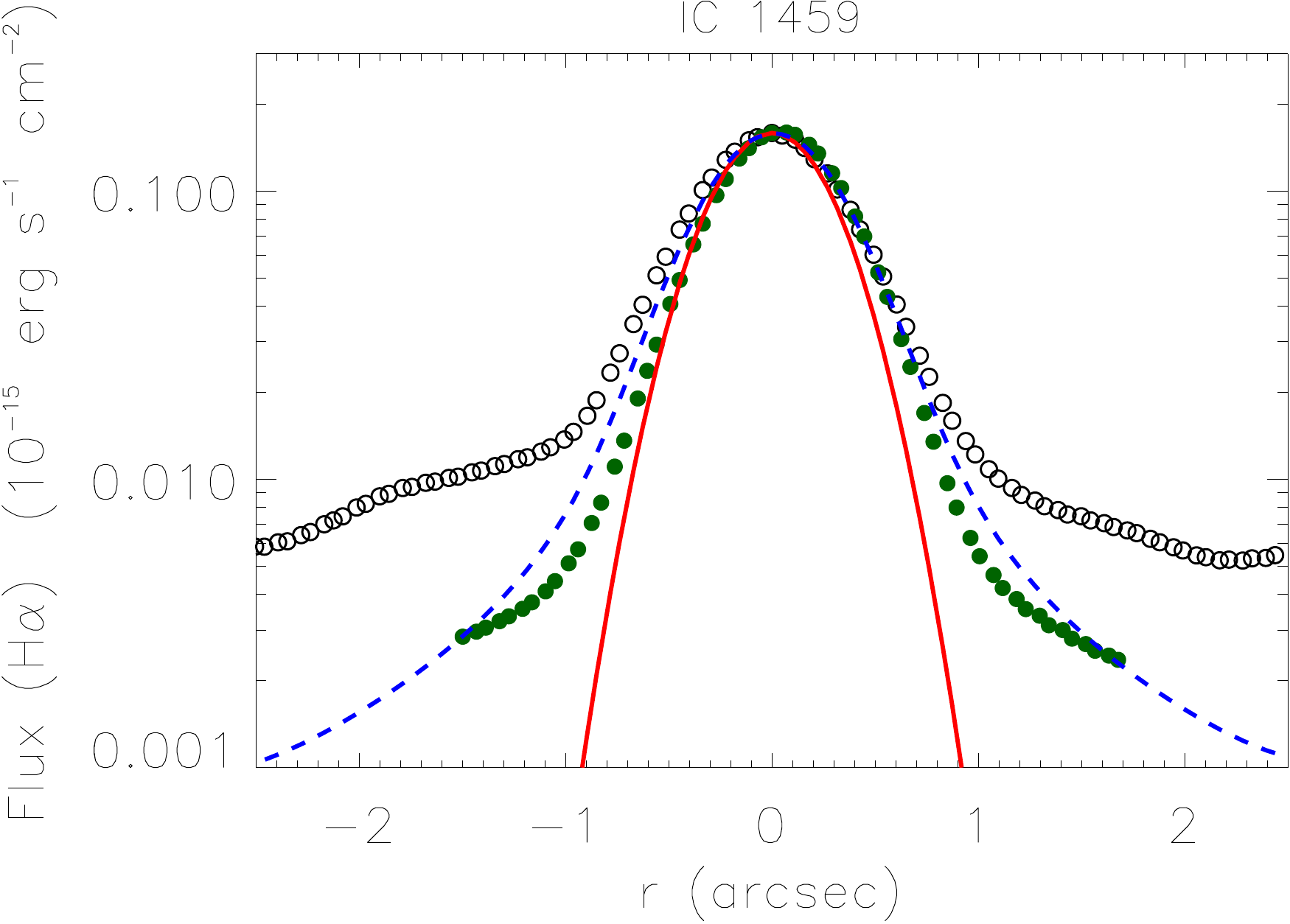}
\hspace{-0.8cm}

\hspace{-2.5cm}
\includegraphics[scale=0.28]{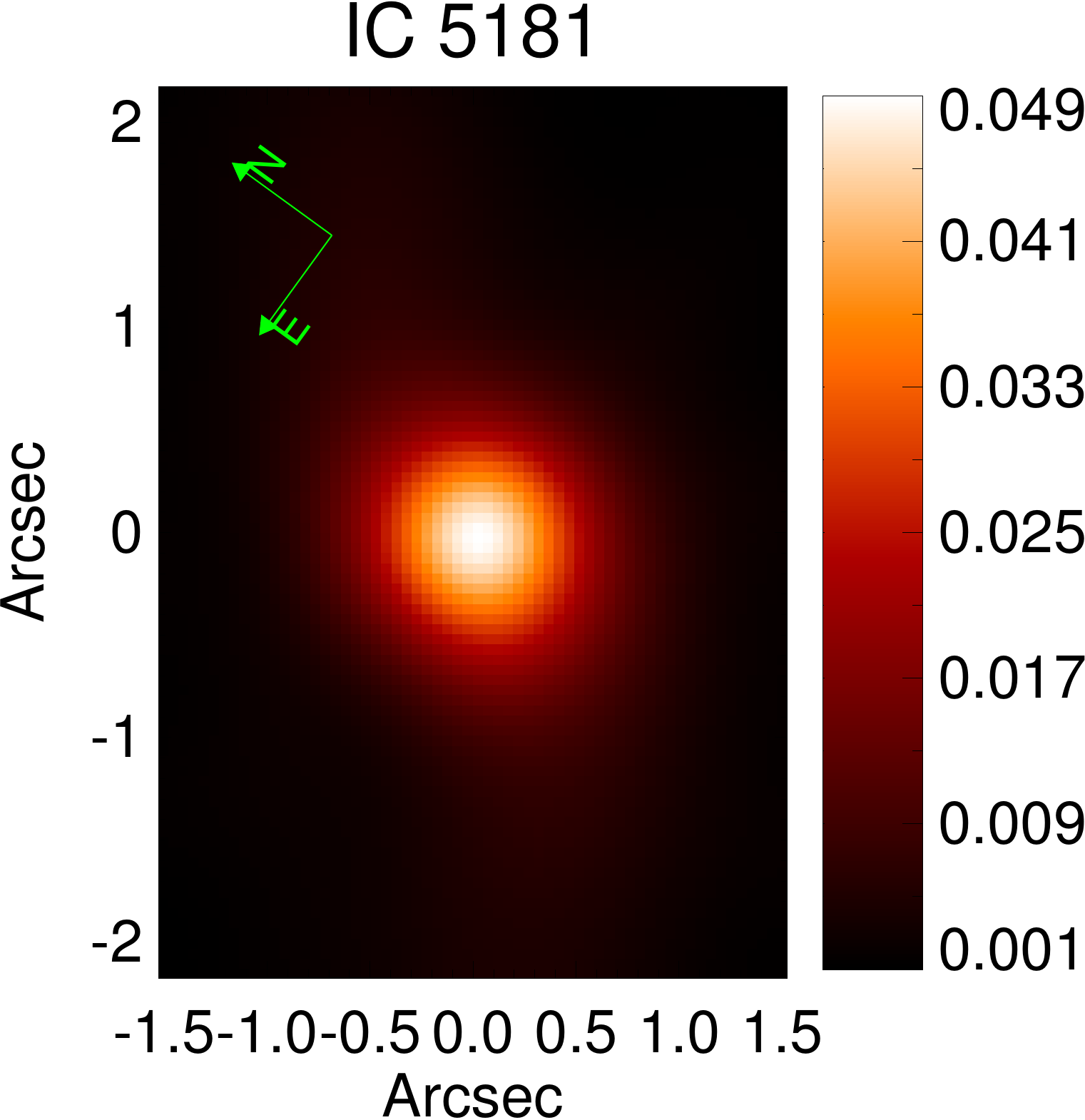}
\hspace{0.5cm}
\vspace{0.8cm}
\includegraphics[scale=0.28]{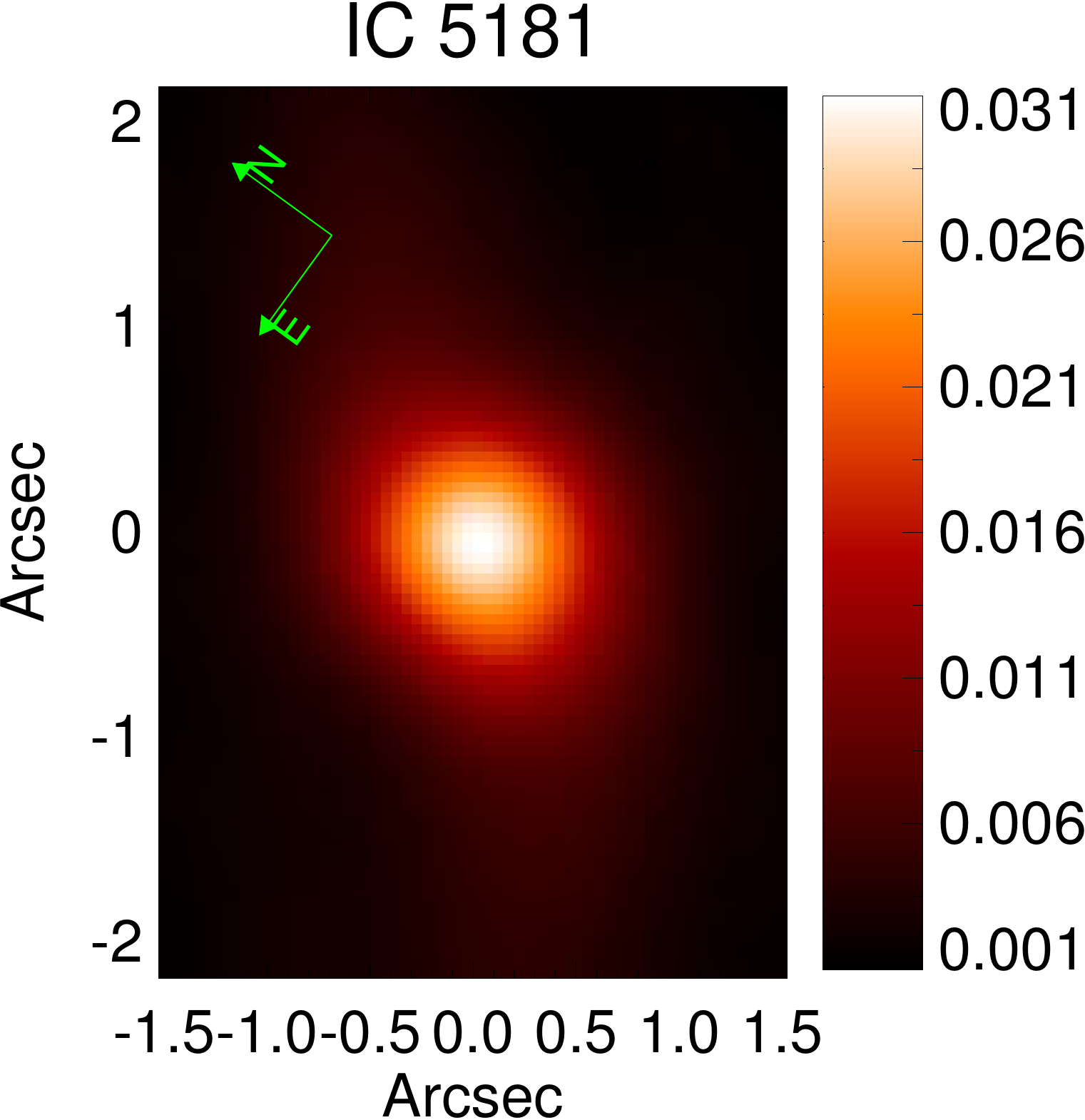}
\hspace{0.5cm}
\includegraphics[scale=0.34]{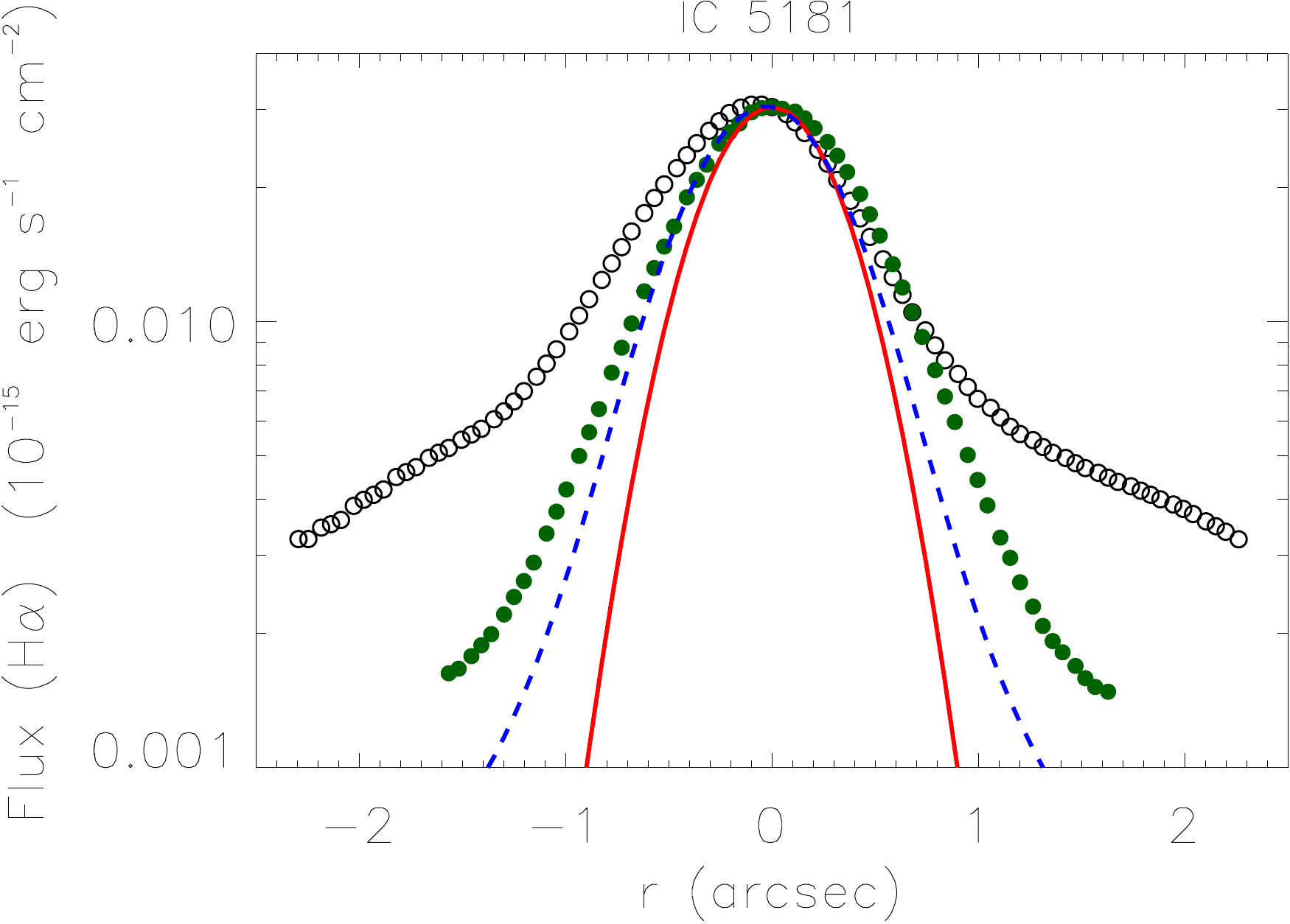}
\hspace{-0.8cm}

\hspace{-2.5cm}
\includegraphics[scale=0.28]{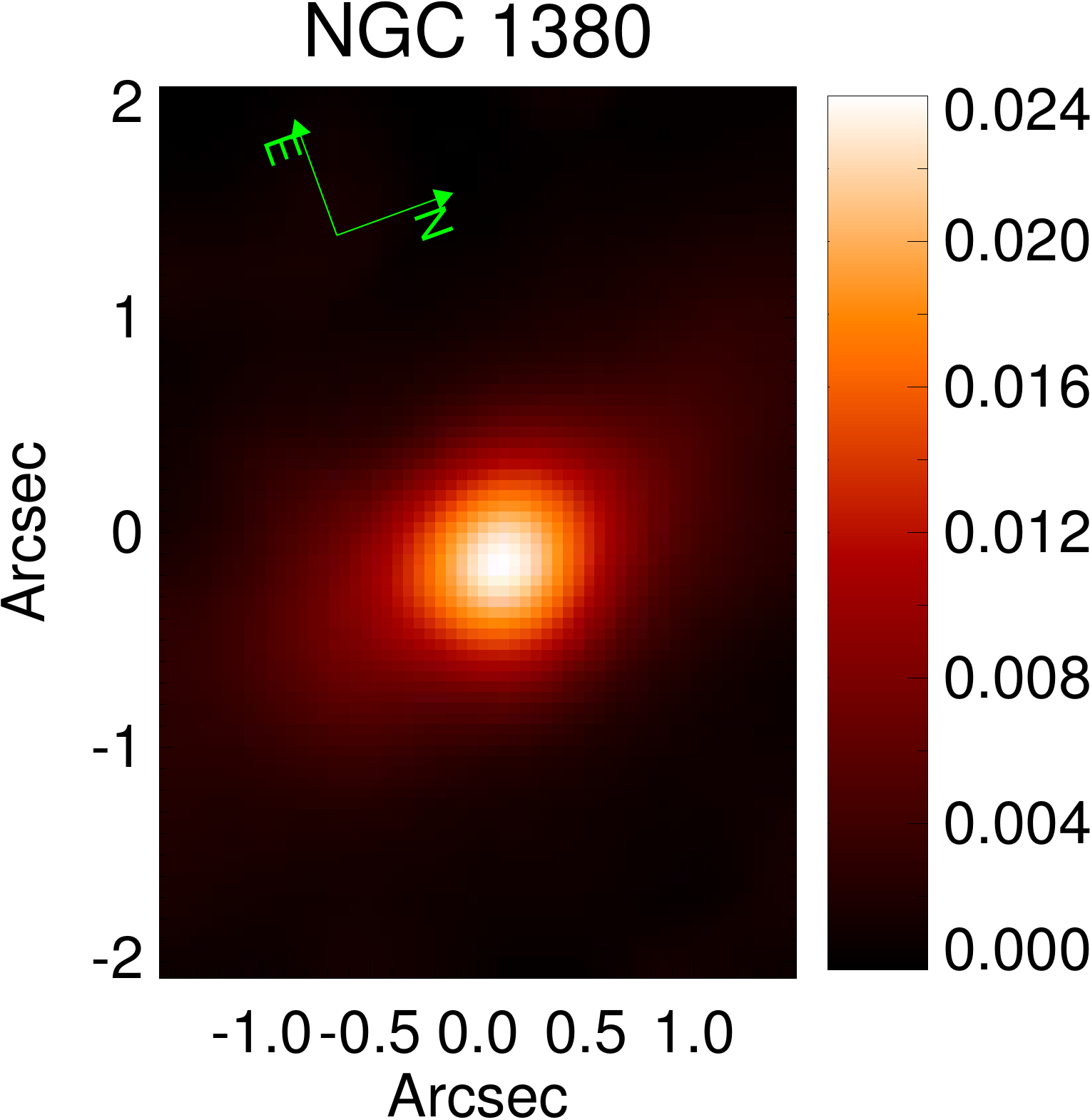}
\hspace{0.5cm}
\vspace{0.6cm}
\includegraphics[scale=0.28]{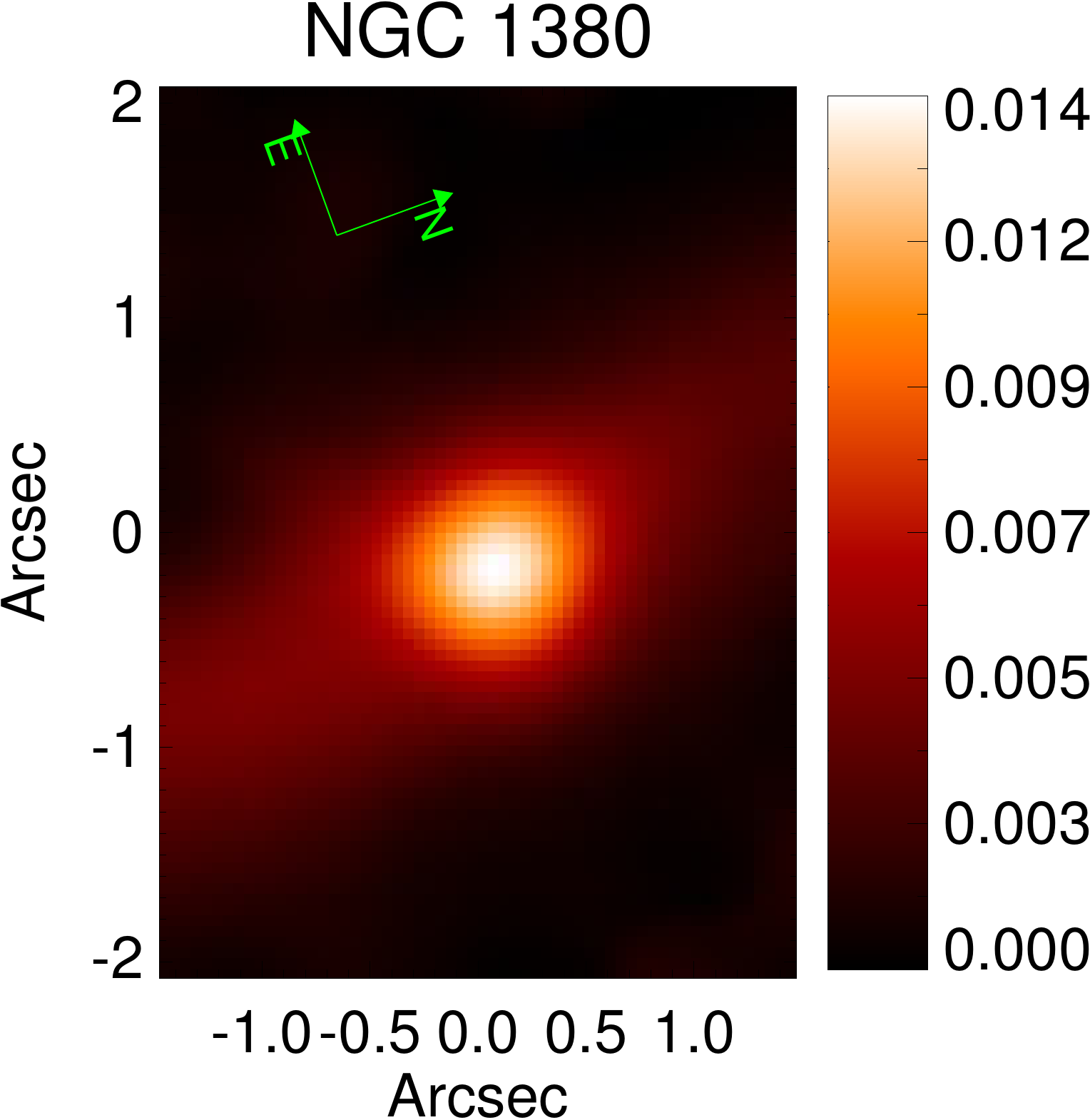}
\hspace{0.5cm}
\includegraphics[scale=0.34]{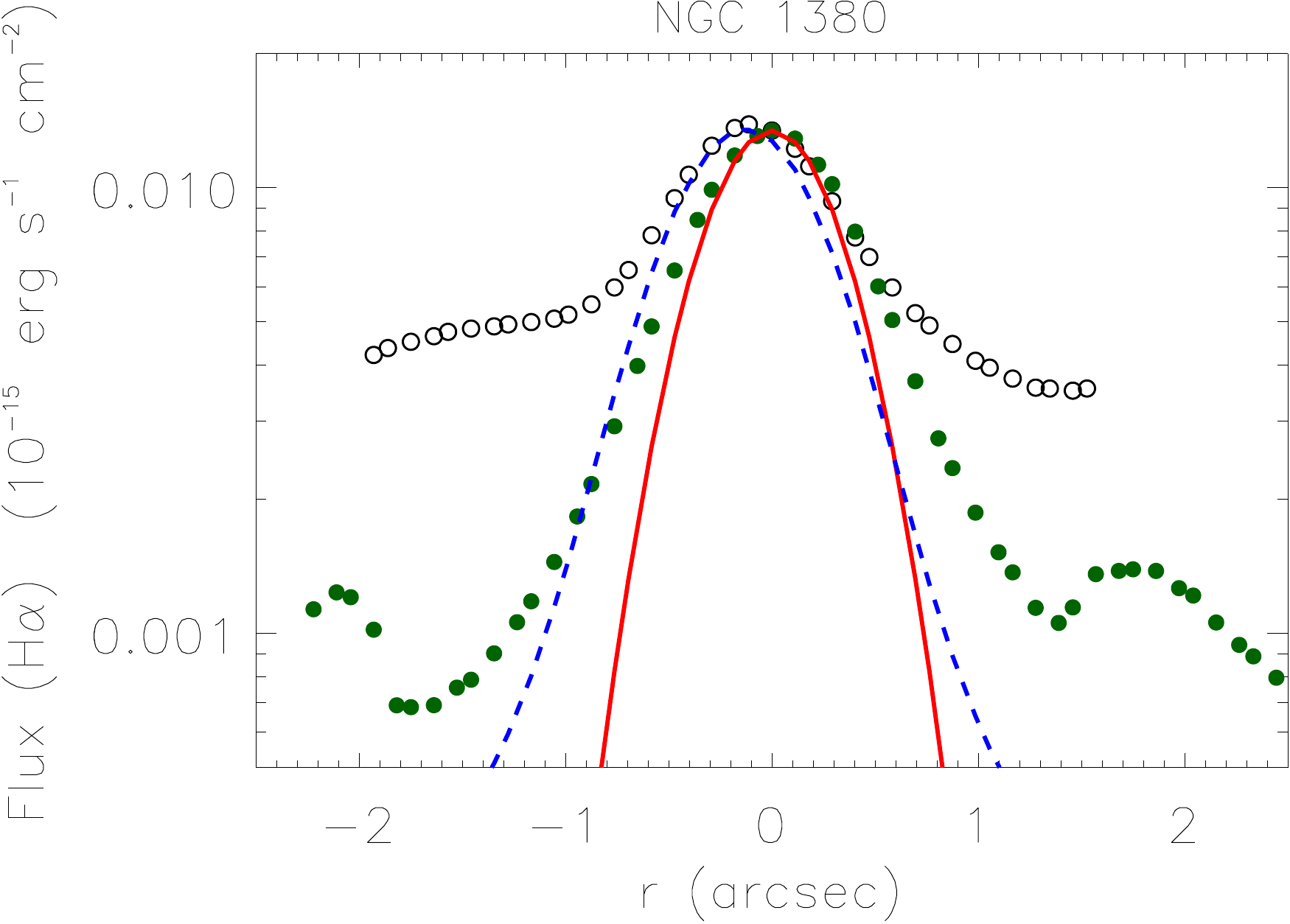}
\hspace{-0.8cm}

\caption{Left: [N II]$\lambda$6583 flux map, in erg s$^{-1}$ cm$^{-2}$. Centre:  H$\alpha$ flux map, in erg s$^{-1}$ cm$^{-2}$. Right: 1D profile of the H$\alpha$ flux maps. It is worth mentioning that the results shown above correspond to the NLR only, since we removed the broad components of H$\alpha$ before the construction of the maps. The hollow black circles were extracted along the kinematic bipolar structures, while the filled green circles were extracted from the perpendicular direction. The red lines correspond to the PSF of the gas cubes. The blue dashed lines are the simple model of a spatially infinite, optically and geometrically thin gas structure with constant density and filling factor, which is photoionized by a central AGN. This model falls with R$^{-2}$. In each galaxy, the model was convolved with the PSF of the gas cubes.  \label{mapa_fluxo_gal_1} }
\end{figure*}

\addtocounter{figure}{-1}
\addtocounter{subfigure}{1}

\begin{figure*}

\hspace{-0.7cm}
\includegraphics[scale=0.25]{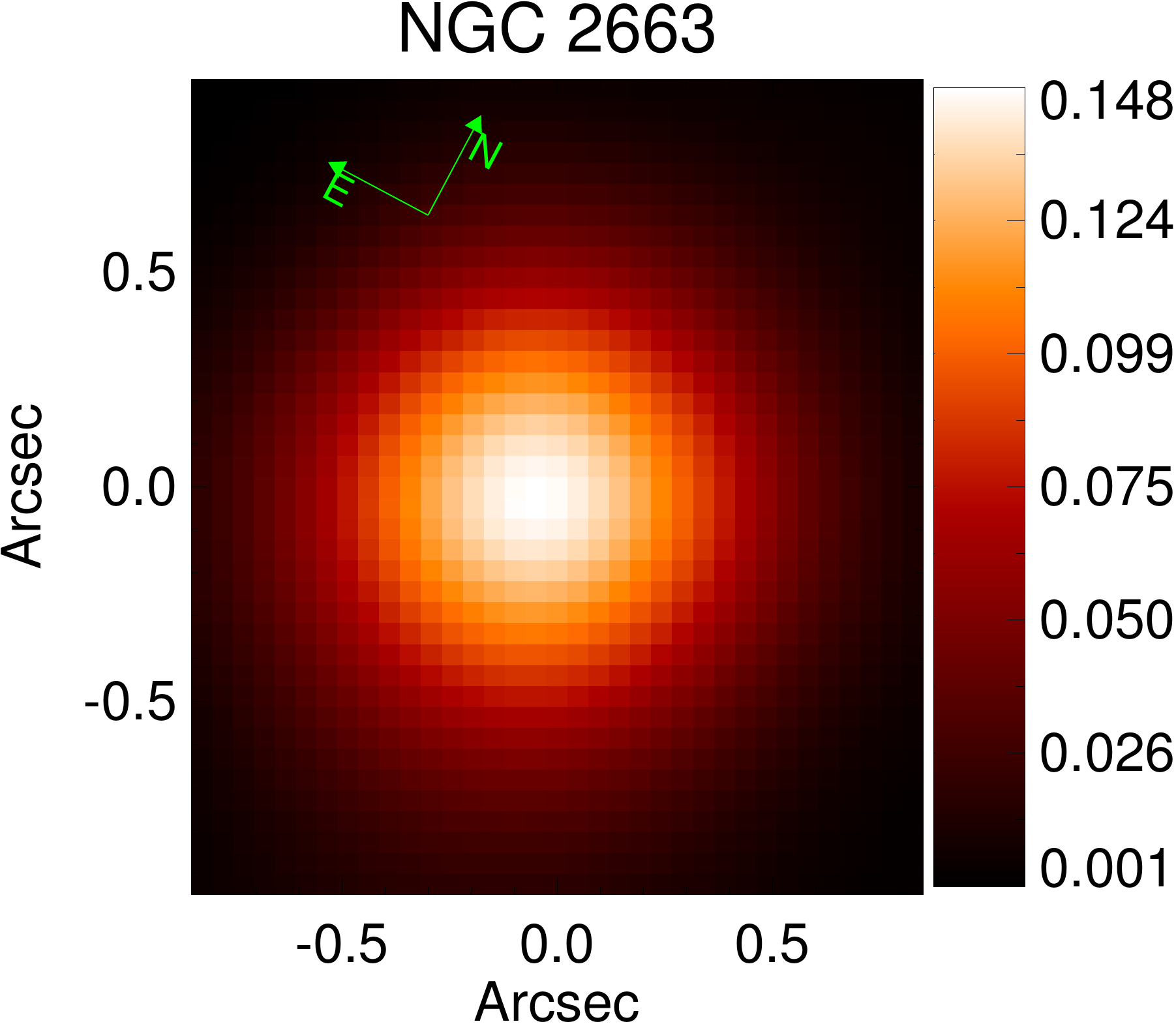}
\hspace{0.15cm}
\vspace{1.0cm}
\includegraphics[scale=0.25]{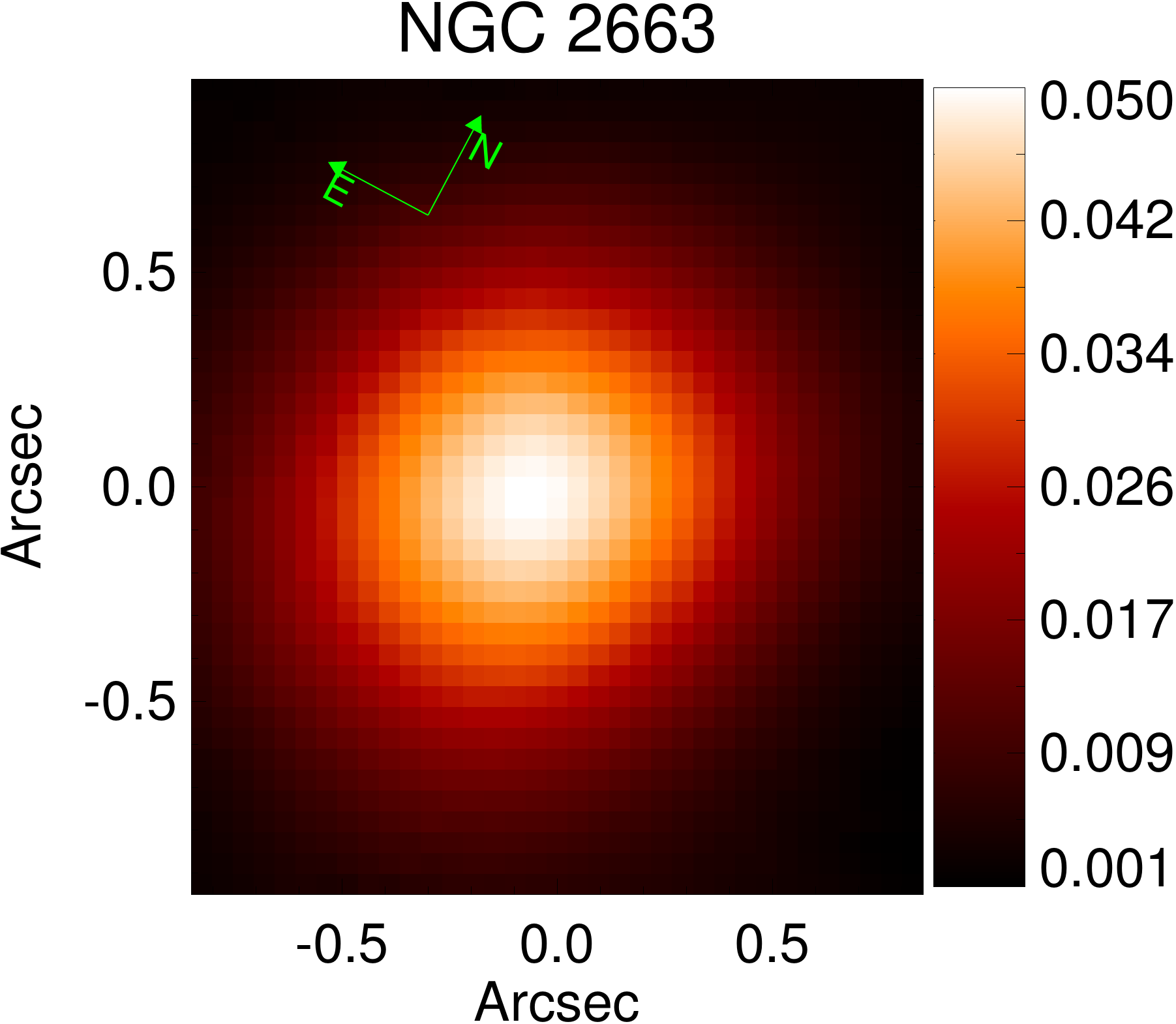}
\hspace{0.55cm}
\includegraphics[scale=0.29]{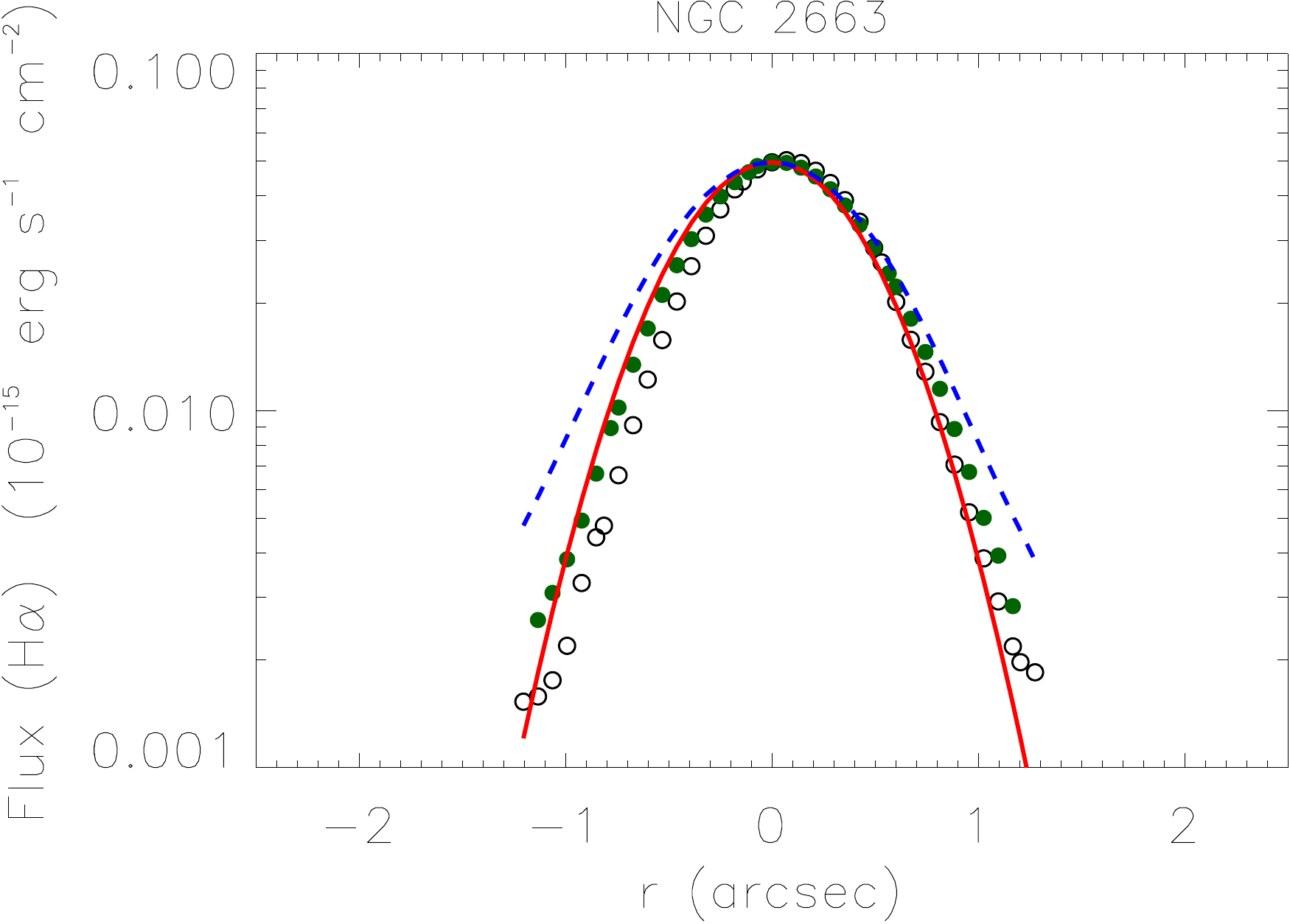}
\hspace{0.45cm}

\hspace{-2.5cm}
\includegraphics[scale=0.28]{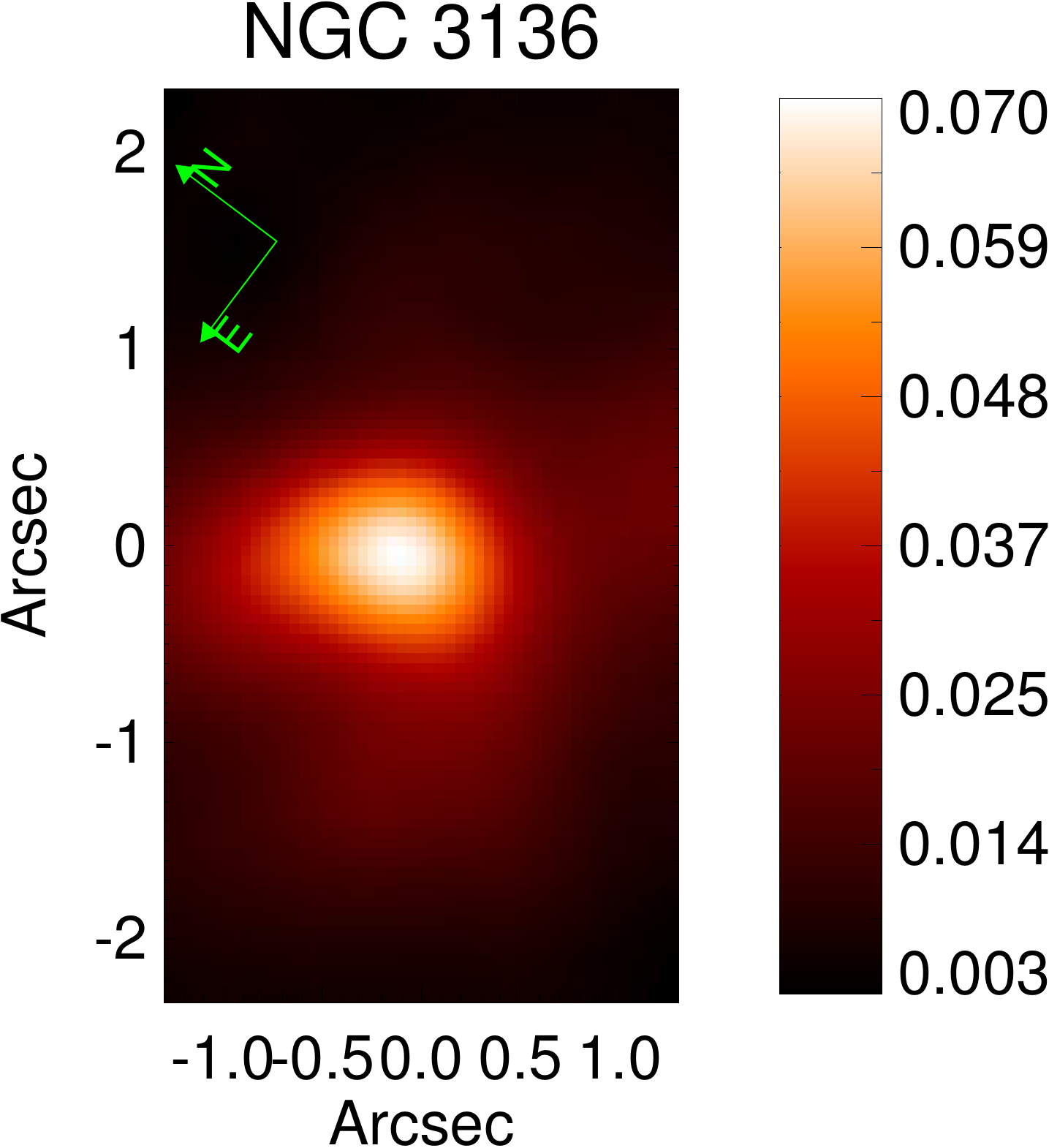}
\hspace{0.5cm}
\vspace{0.9cm}
\includegraphics[scale=0.28]{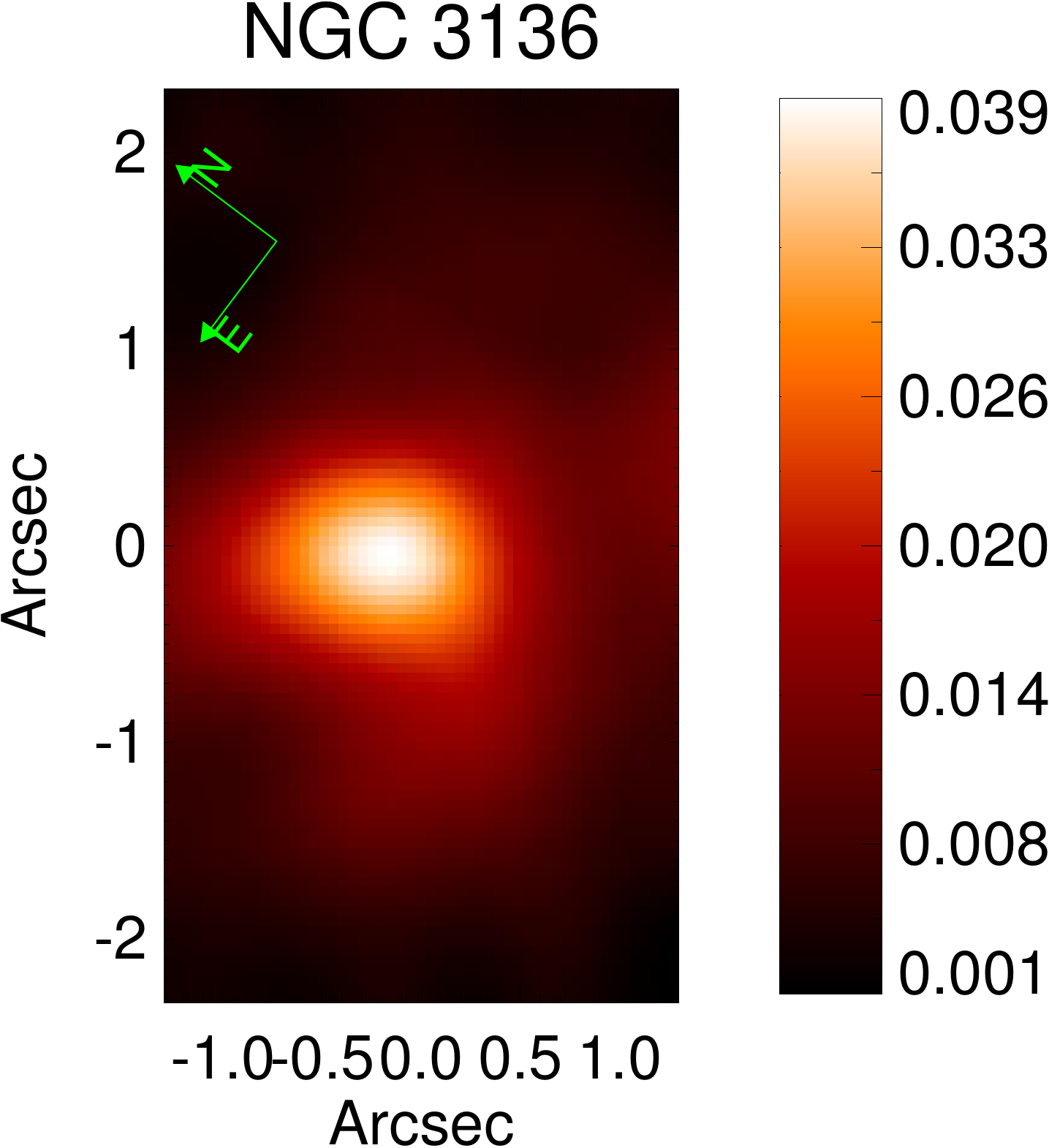}
\hspace{0.5cm}
\includegraphics[scale=0.34]{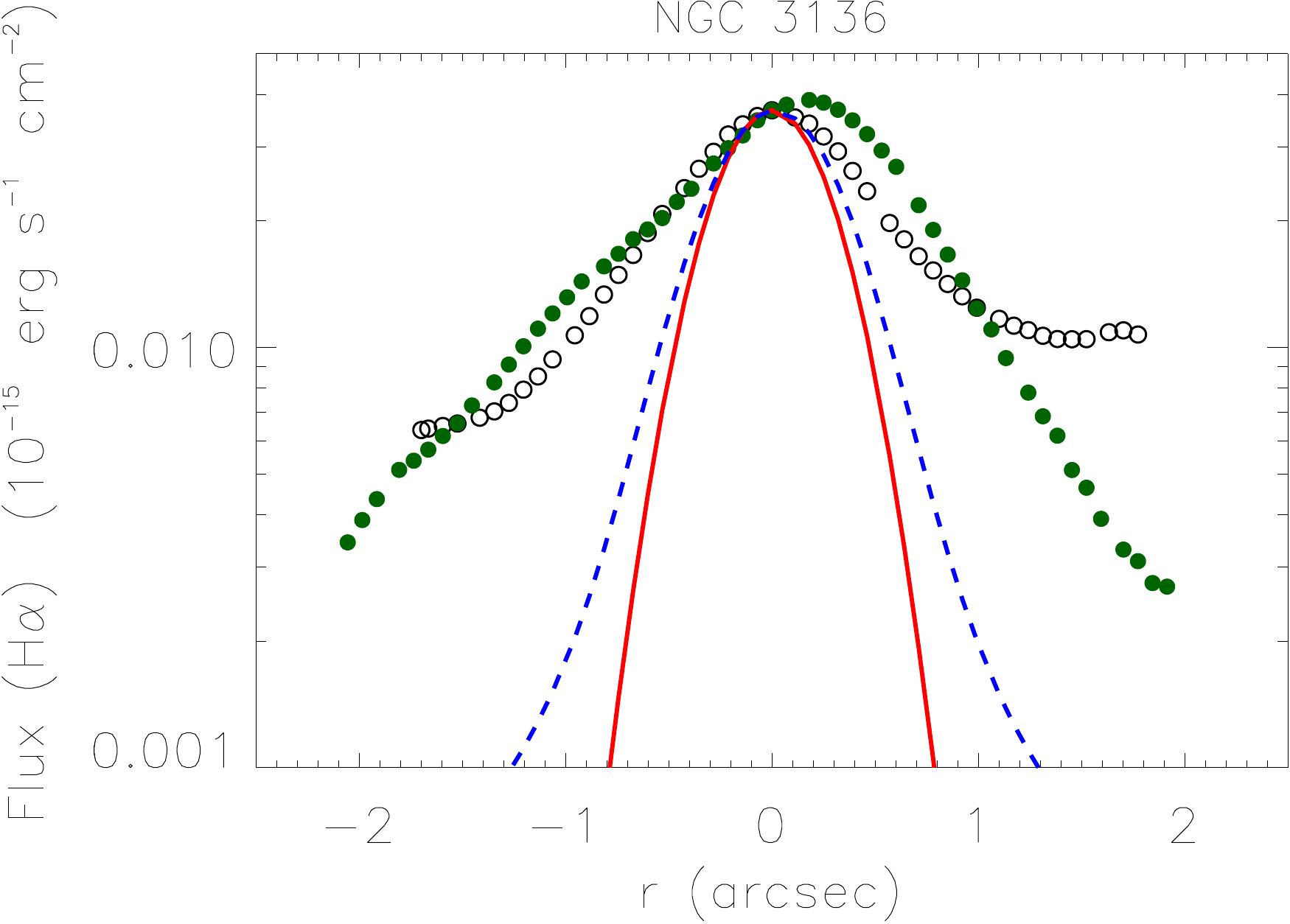}
\hspace{-0.8cm}

\hspace{-2.5cm}
\includegraphics[scale=0.28]{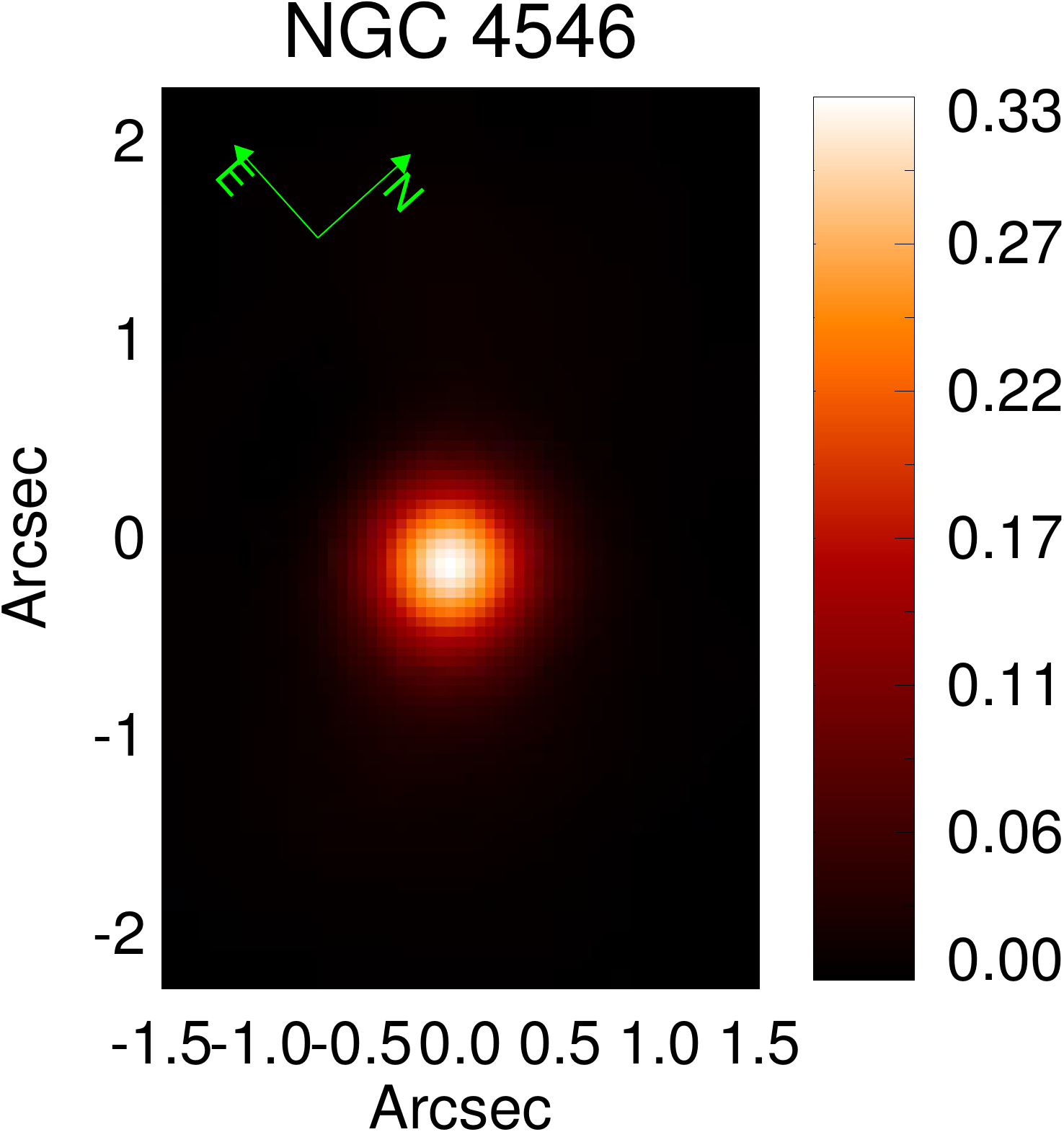}
\hspace{0.5cm}
\vspace{0.9cm}
\includegraphics[scale=0.28]{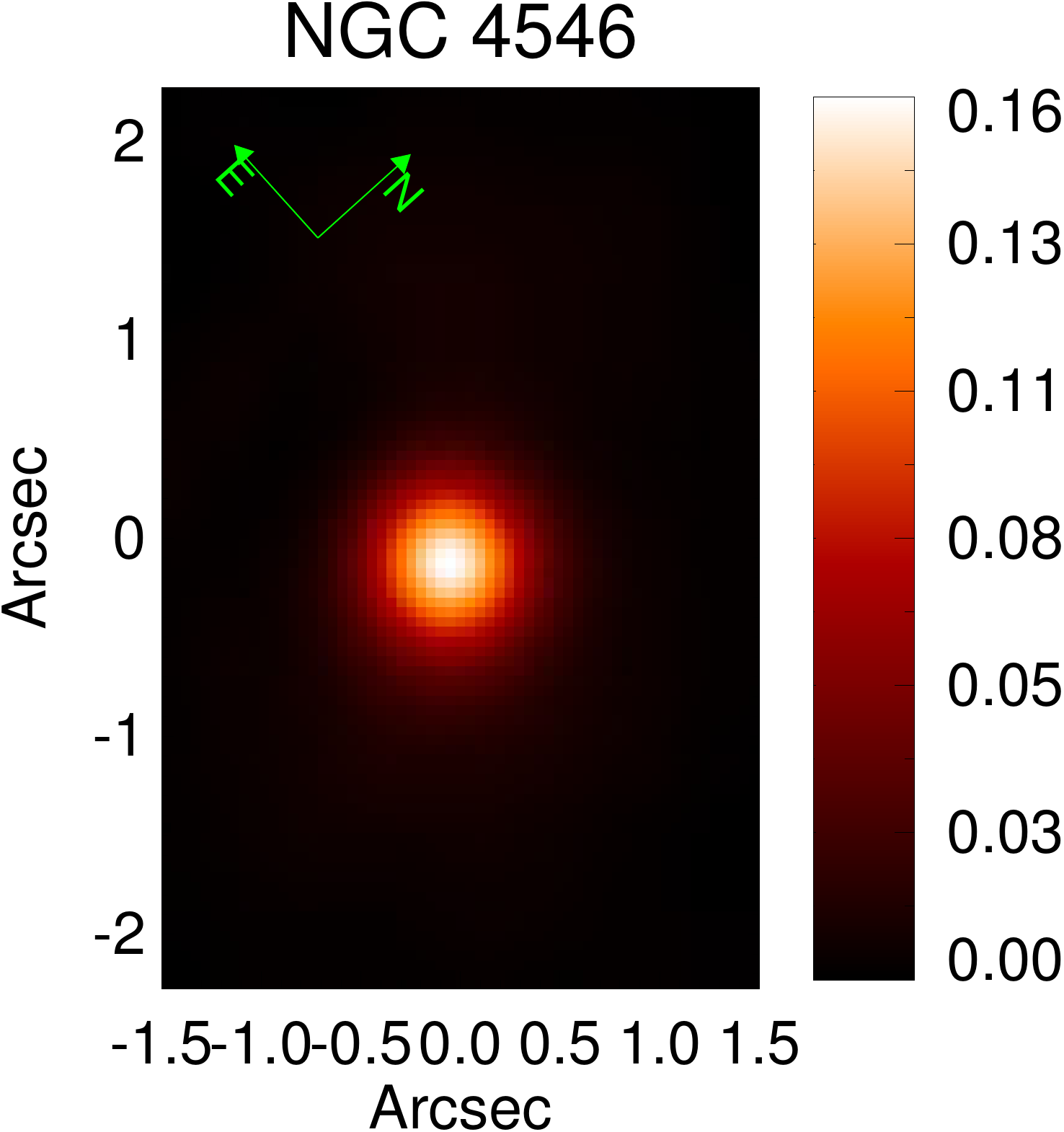}
\hspace{0.5cm}
\includegraphics[scale=0.34]{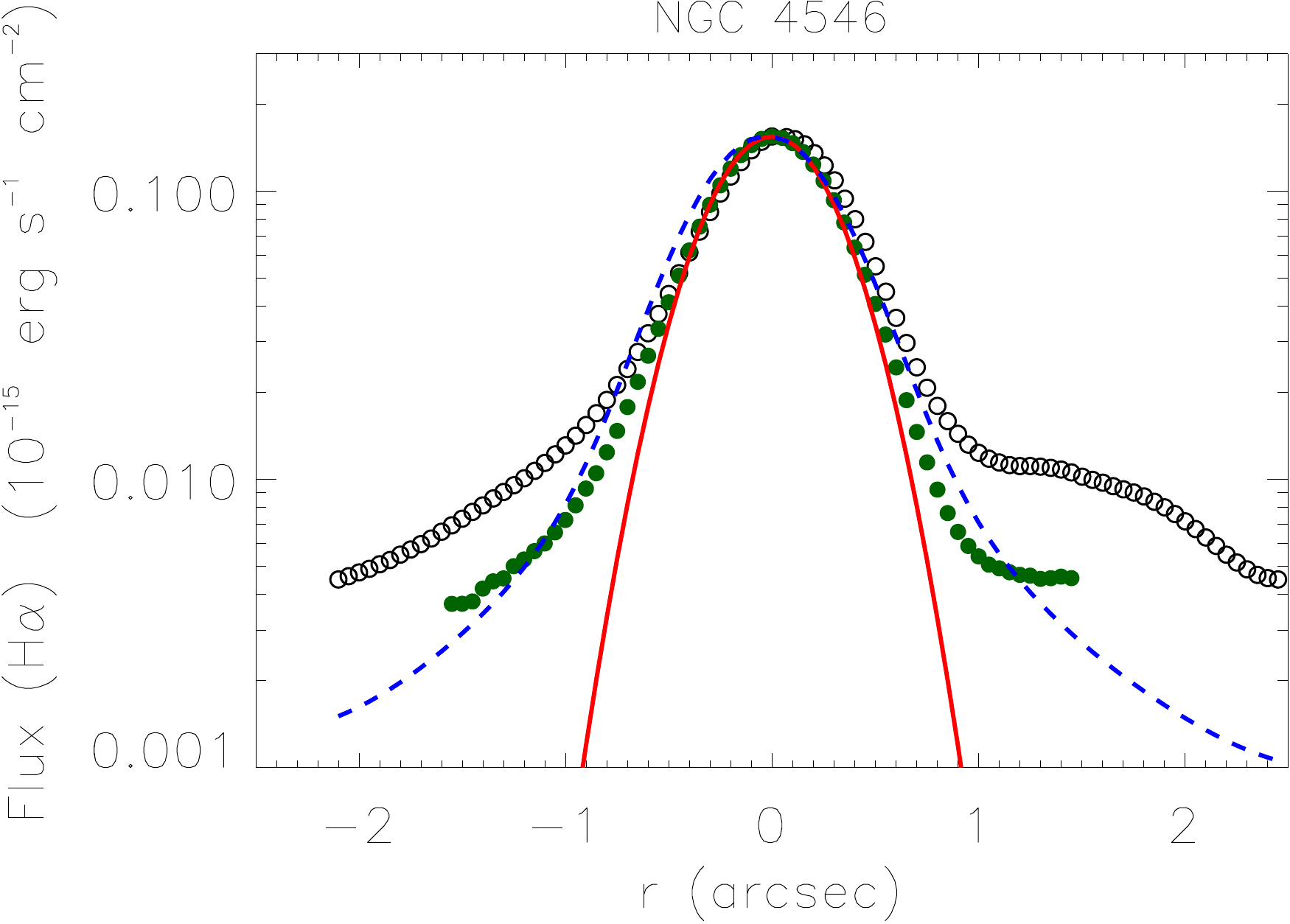}
\hspace{-0.8cm}

\hspace{-2.5cm}
\includegraphics[scale=0.28]{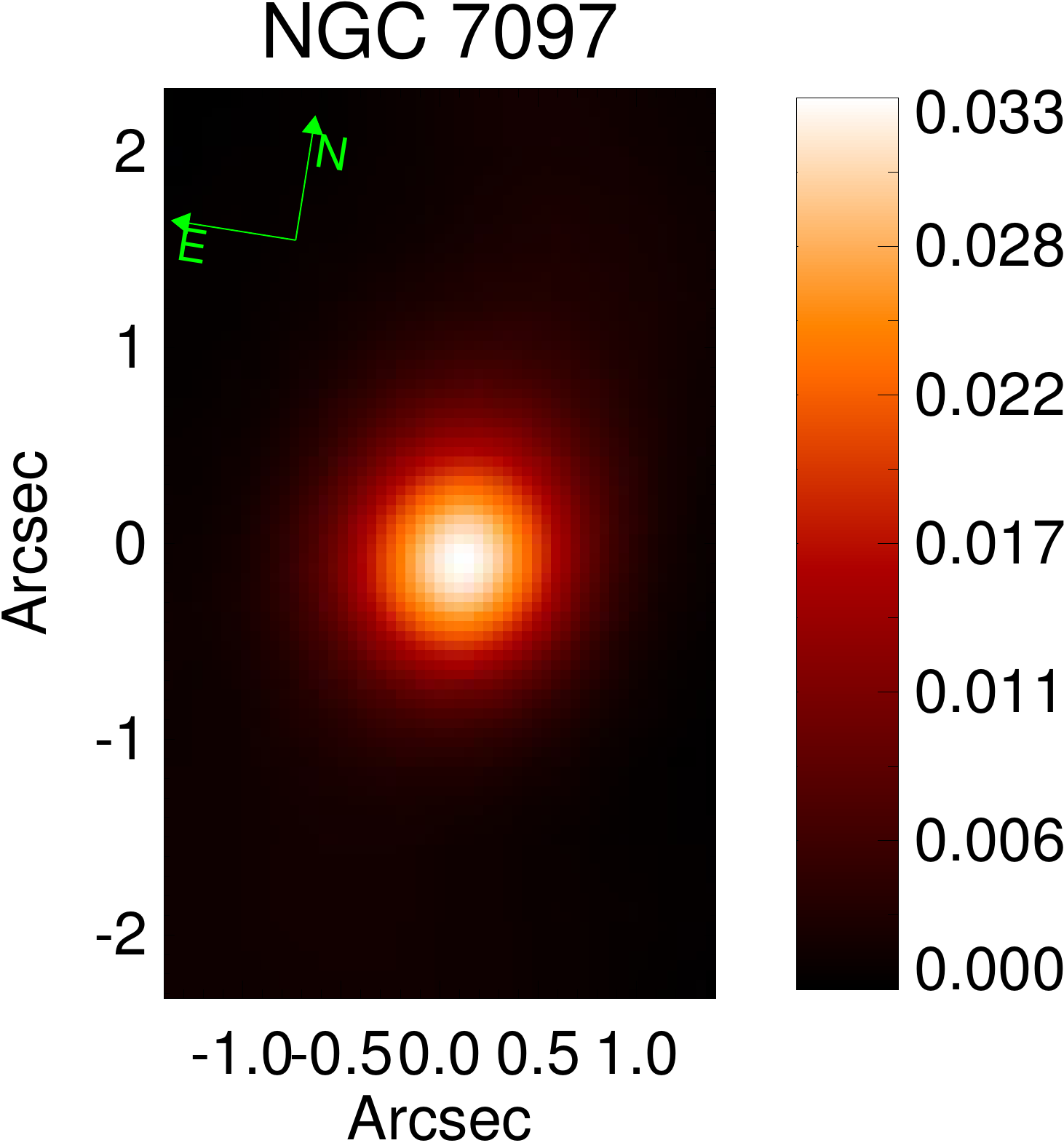}
\hspace{0.5cm}
\vspace{0.9cm}
\includegraphics[scale=0.28]{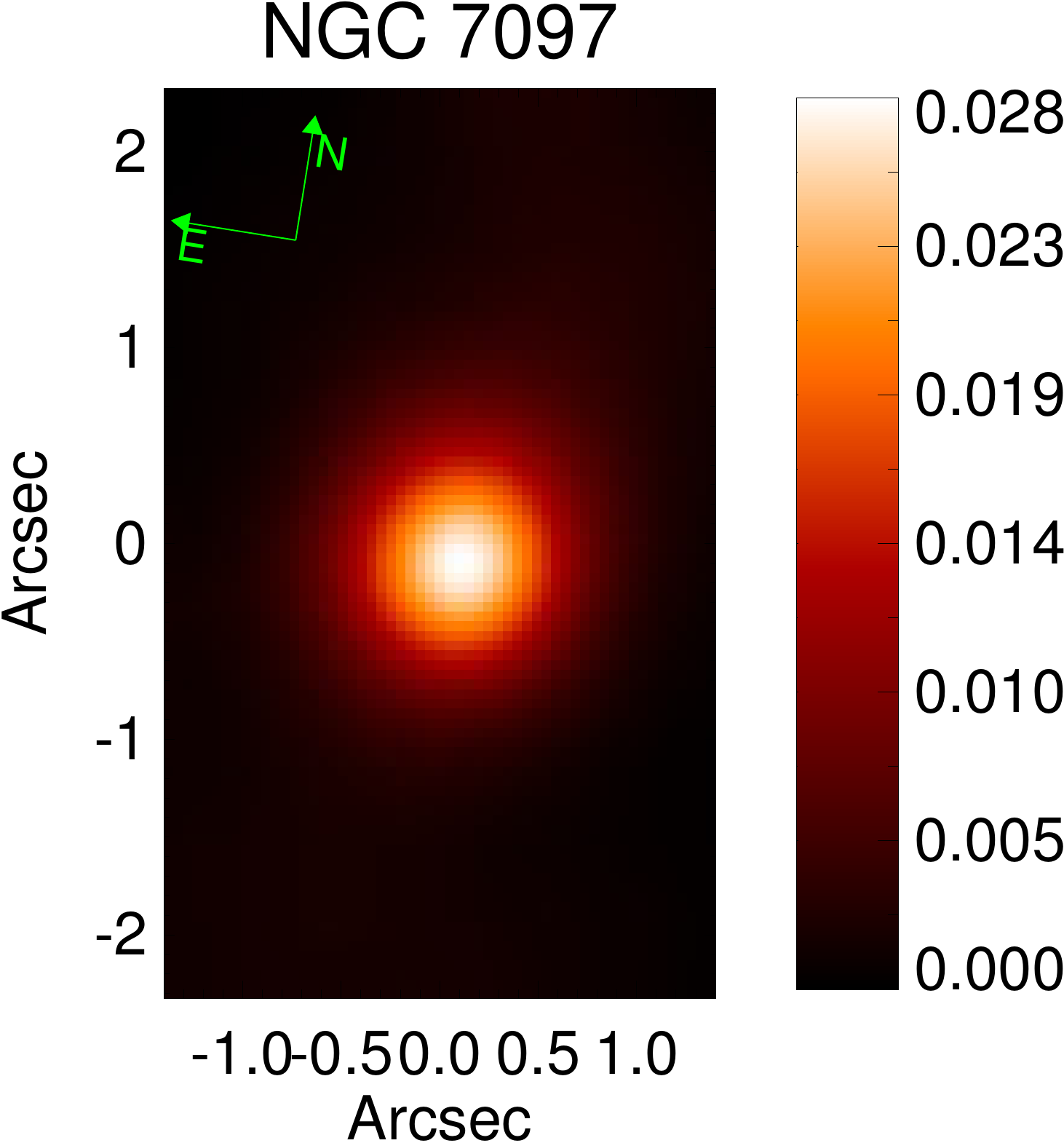}
\hspace{0.5cm}
\includegraphics[scale=0.34]{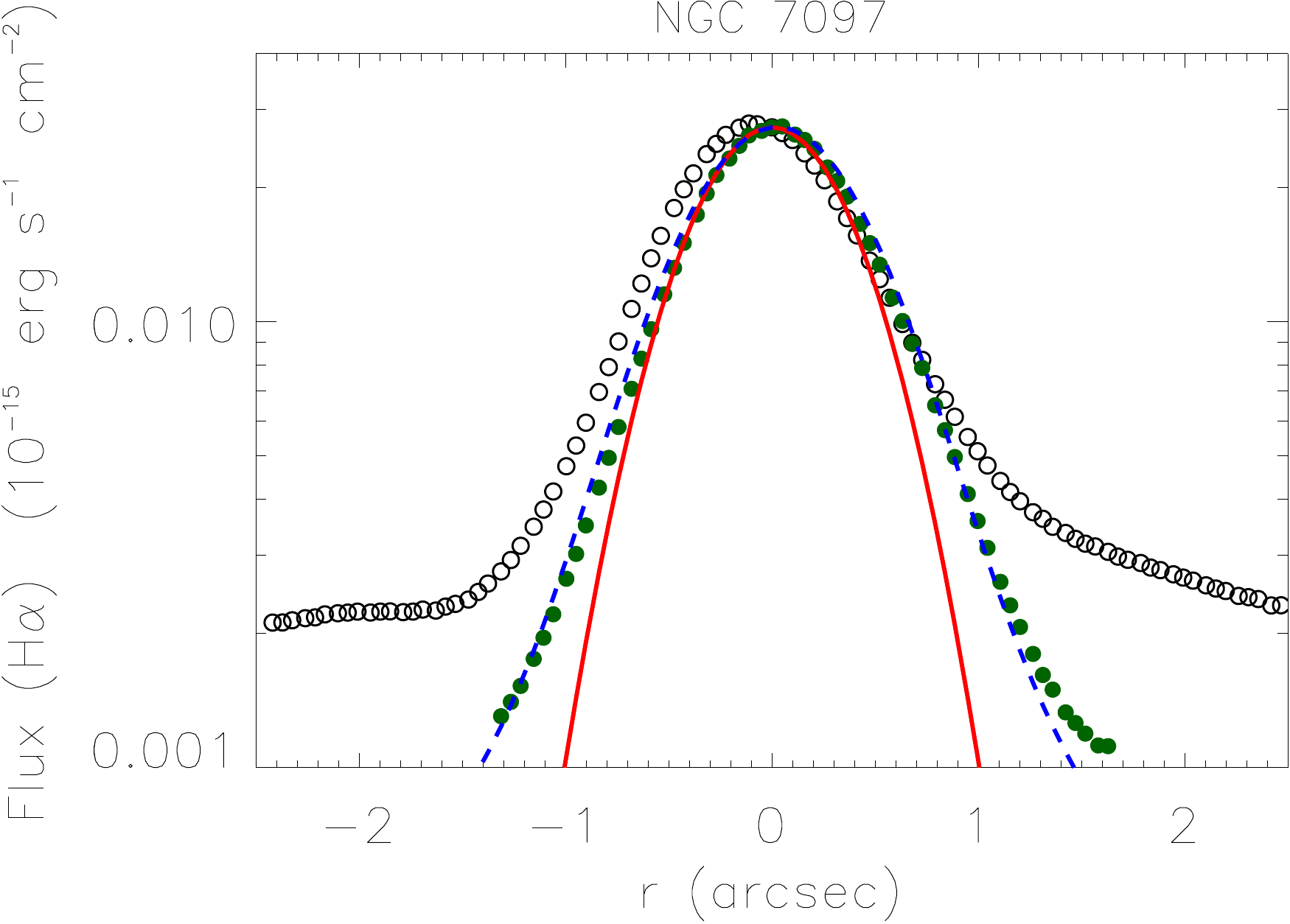}
\hspace{-0.8cm}

\caption{Same as in Fig. \ref{mapa_fluxo_gal_1} \label{mapa_fluxo_gal_2}}
\end{figure*}

\renewcommand{\thefigure}{\arabic{figure}}

In the flux maps, we see a concentrated emission in the nuclear region of the galaxies. Indeed, \citet{2007ApJ...654..125S} showed that galaxies containing AGNs have a large concentration of the H$\alpha$ flux in their nucleus. This is observed in the six galaxies containing a BLR and also in NGC 1380 and NGC 3136. This latter galaxy also has a slightly extended emission in the NE-SW direction. This feature was also reported by \citet{1994A&AS..105..341G}. The extended emission is in the same position as the two compact objects detected with PCA Tomography applied to the spectral range 6250-6800 \AA\ (paper I). 

The 1D profiles of the H$\alpha$ flux reveal that, in six galaxies of the sample, the circumnuclear regions are more intense along the kinematic bipolar structures. Moreover, in the direction perpendicular to the kinematic bipolar structures, we always detect emission and, from now on, we will call this ``the low-velocity emission'', as can be clearly seen in Figs. \ref{mapa_cin_gal_1}, \ref{mapa_cin_gal_2}, \ref{mapa_cin_gal_3} and \ref{mapa_cin_gal_4}. In NGC 3136, the 1D profiles show that both directions have, approximately, the same behaviour. This is in agreement with the hypothesis that the geometry of the circumnuclear region of this galaxy is not disc-like. In NGC 2663, the 1D profiles in both directions are consistent with a point-like source. This is expected, since we did not detect any emission lines in the spectrum of the circumnuclear region of this galaxy.

\begin{figure*}
\begin{center}
\includegraphics[scale=0.34]{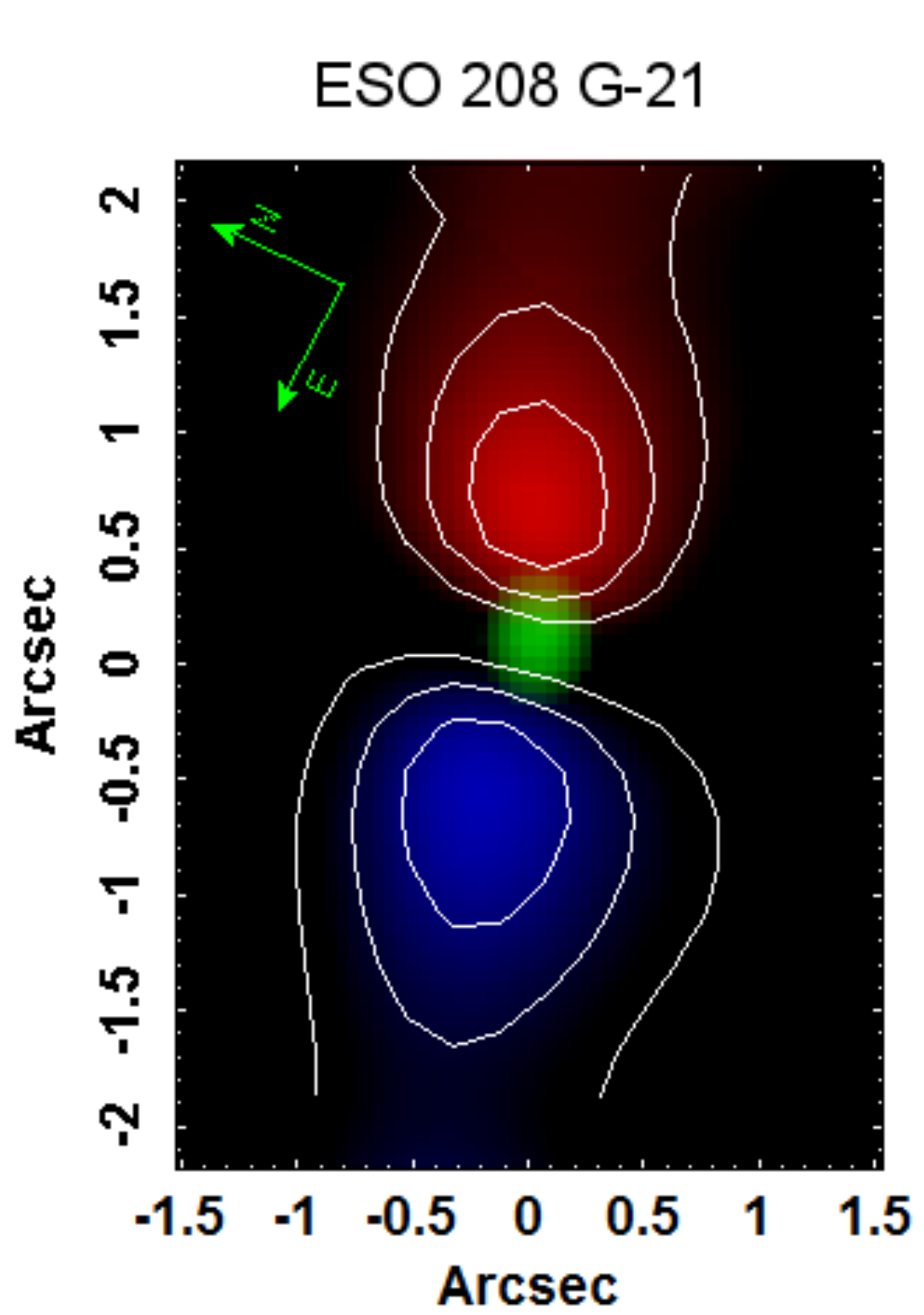}
\vspace{0cm}
\includegraphics[scale=0.34]{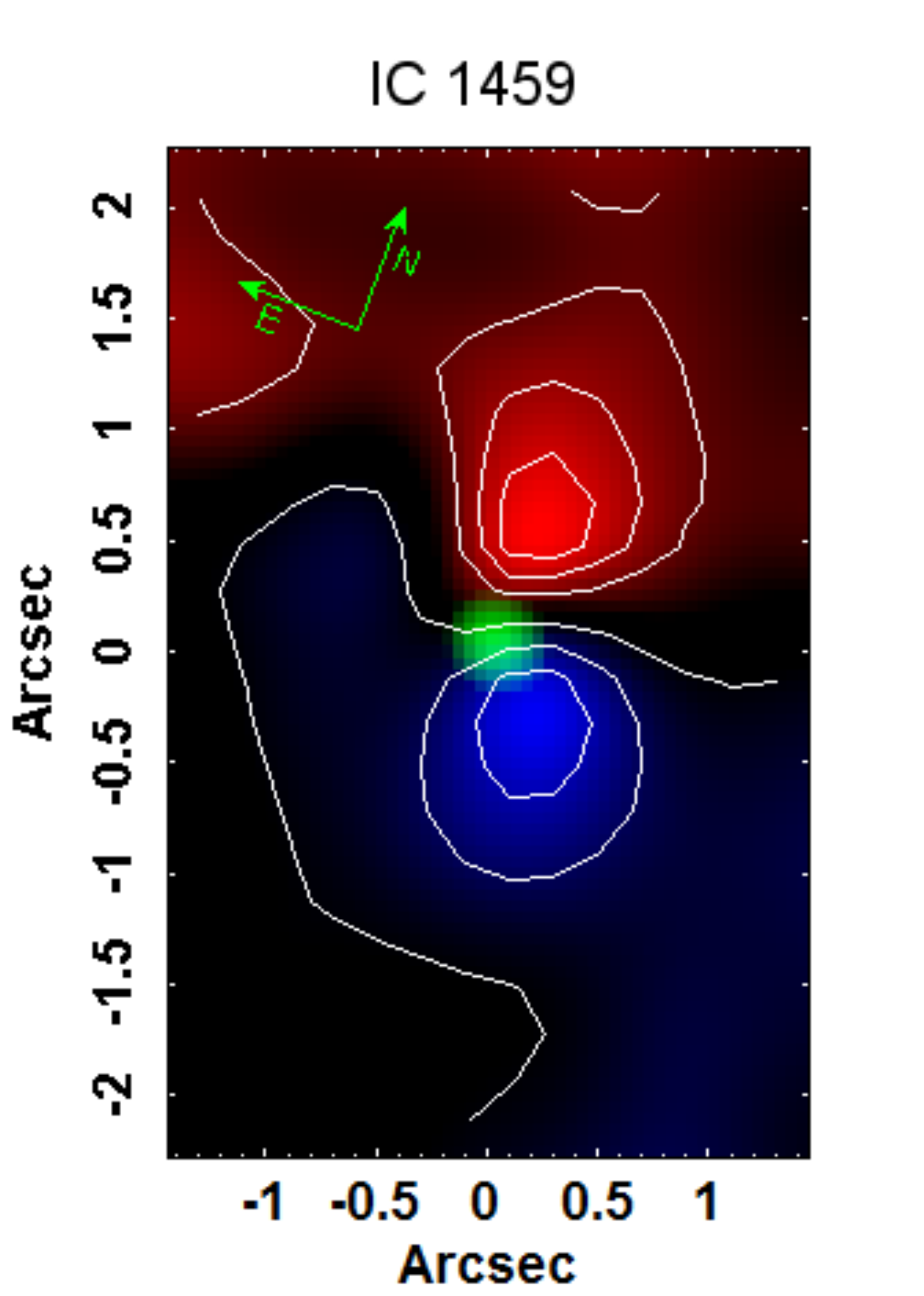}
\includegraphics[scale=0.34]{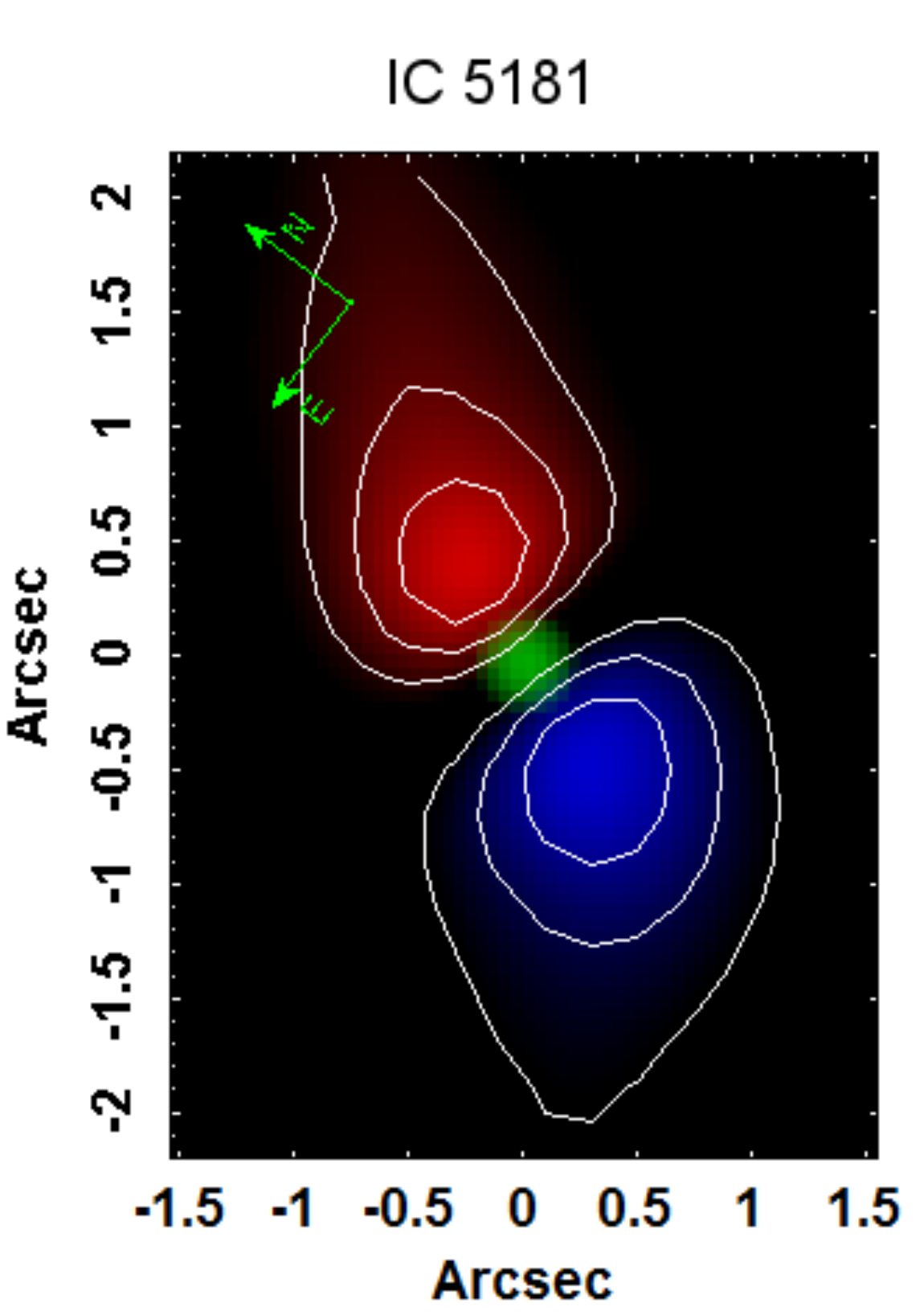}
\vspace{0cm}
\includegraphics[scale=0.34]{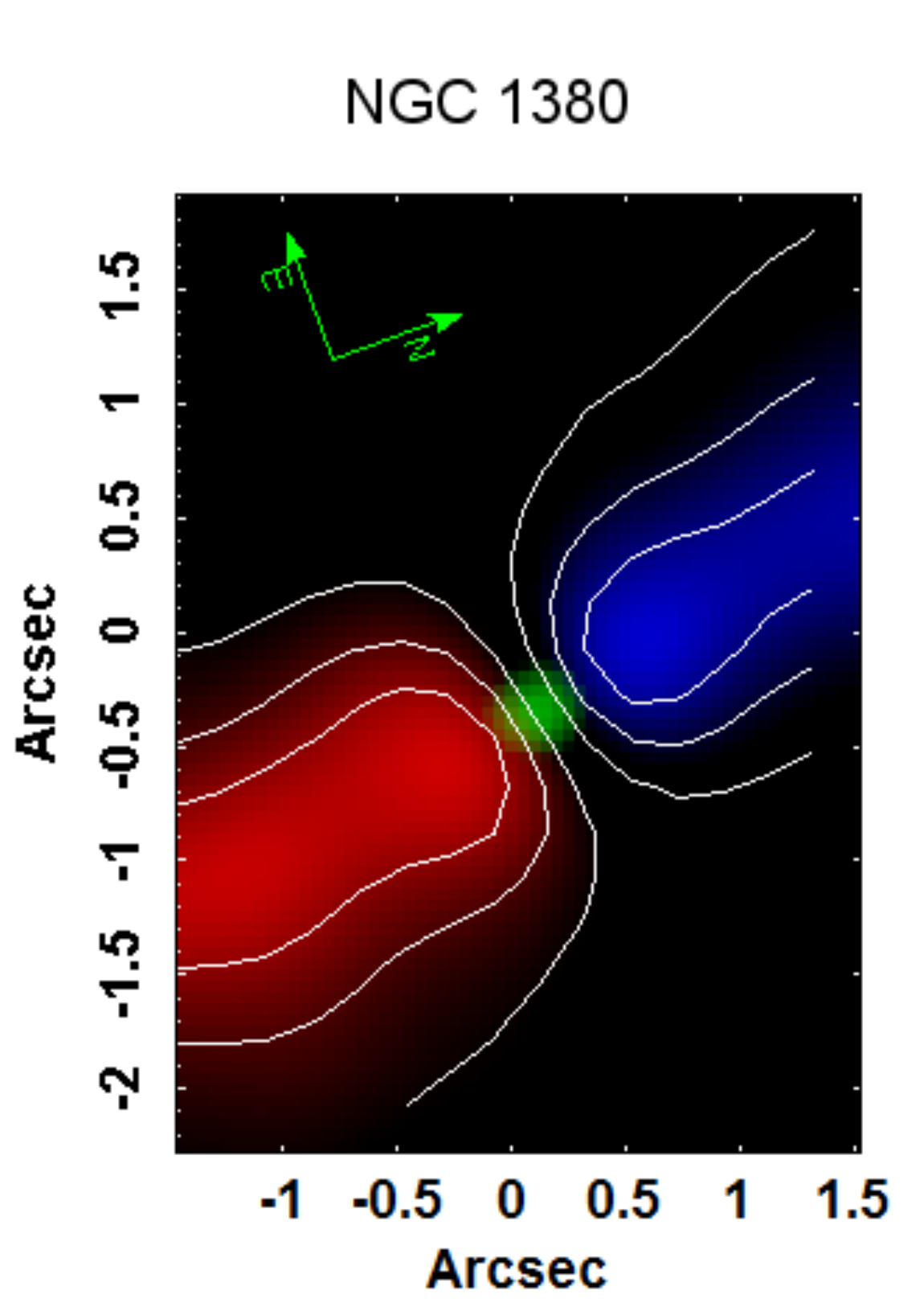}
\includegraphics[scale=0.34]{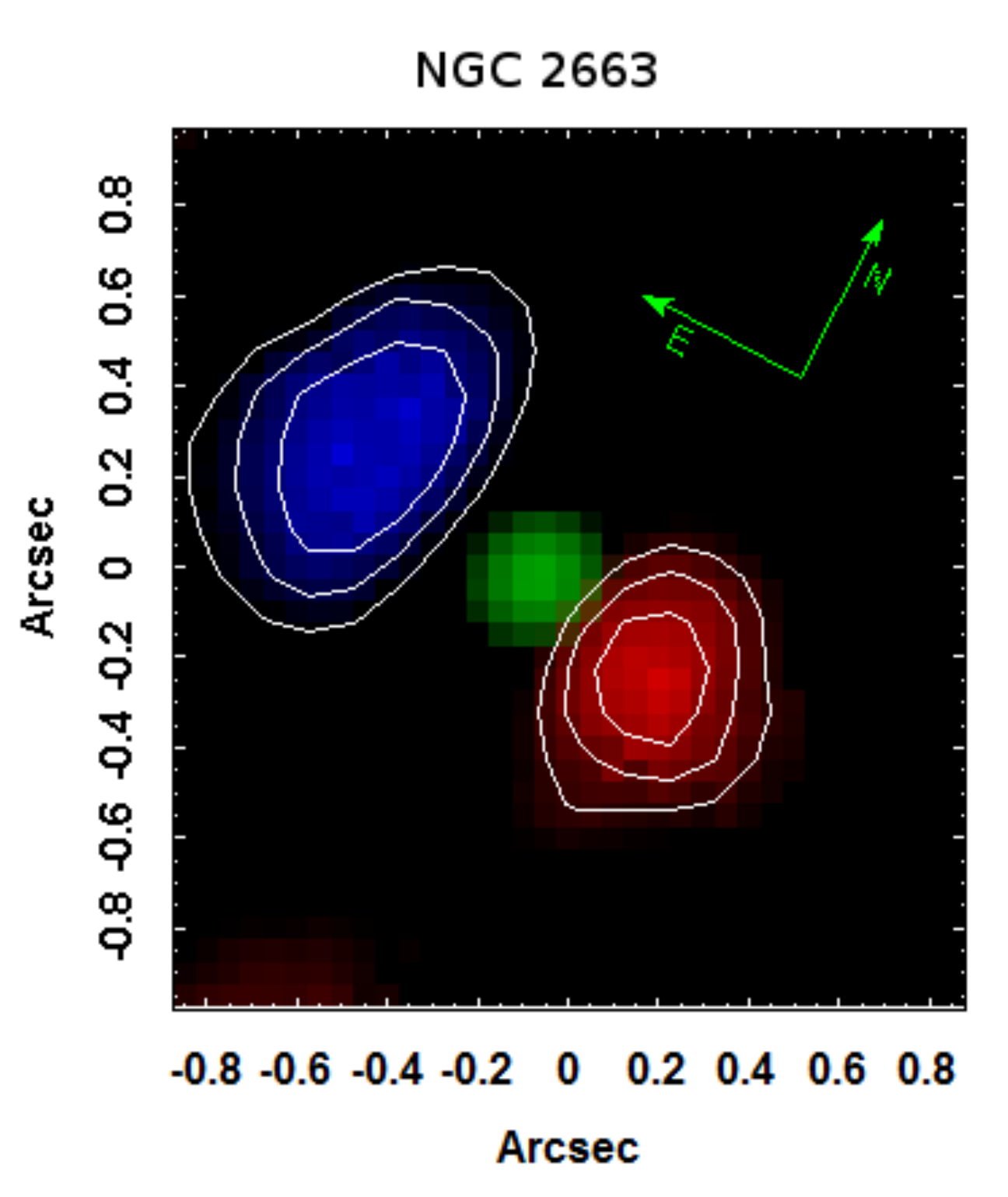}
\vspace{0cm}
\includegraphics[scale=0.34]{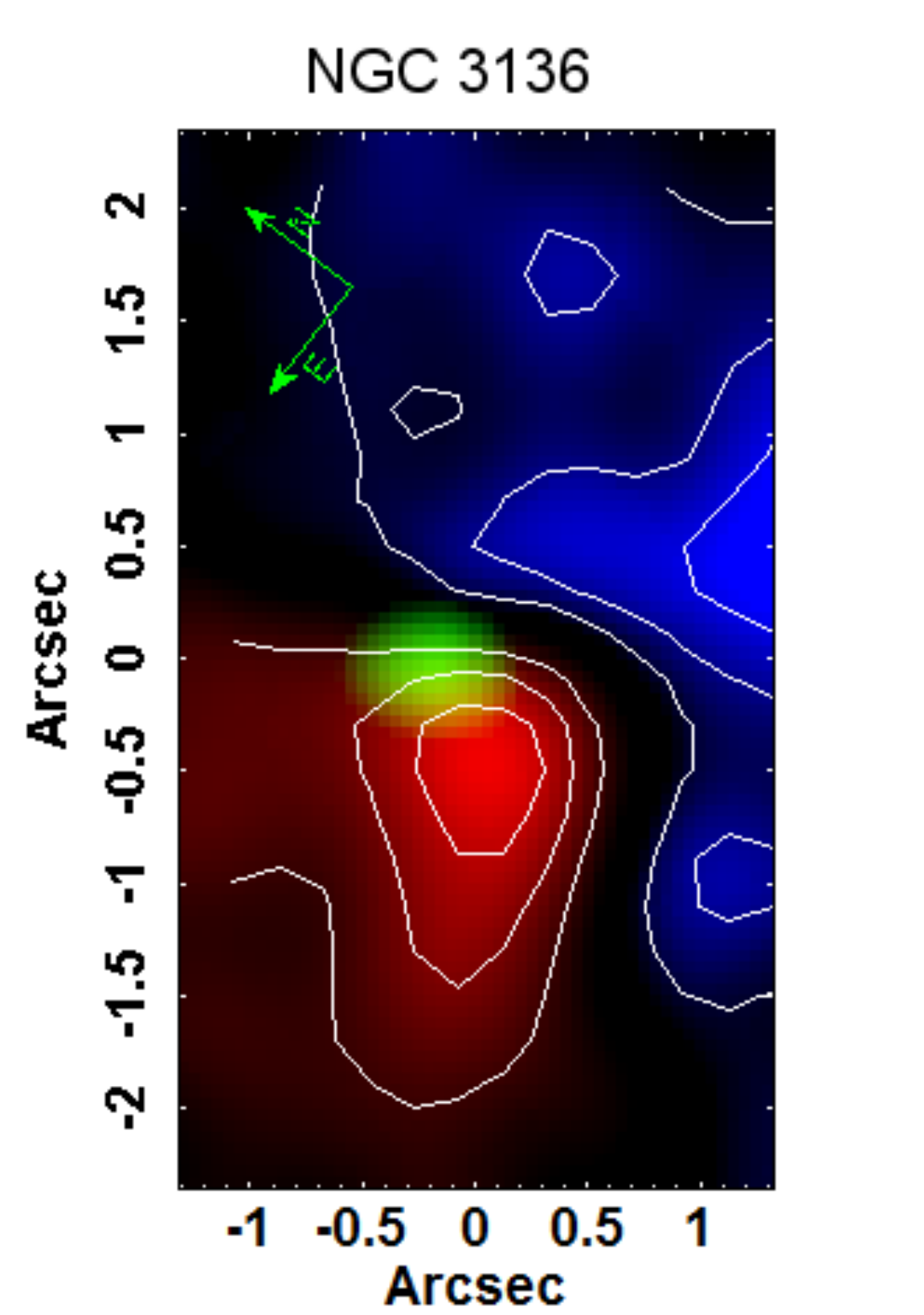}
\includegraphics[scale=0.34]{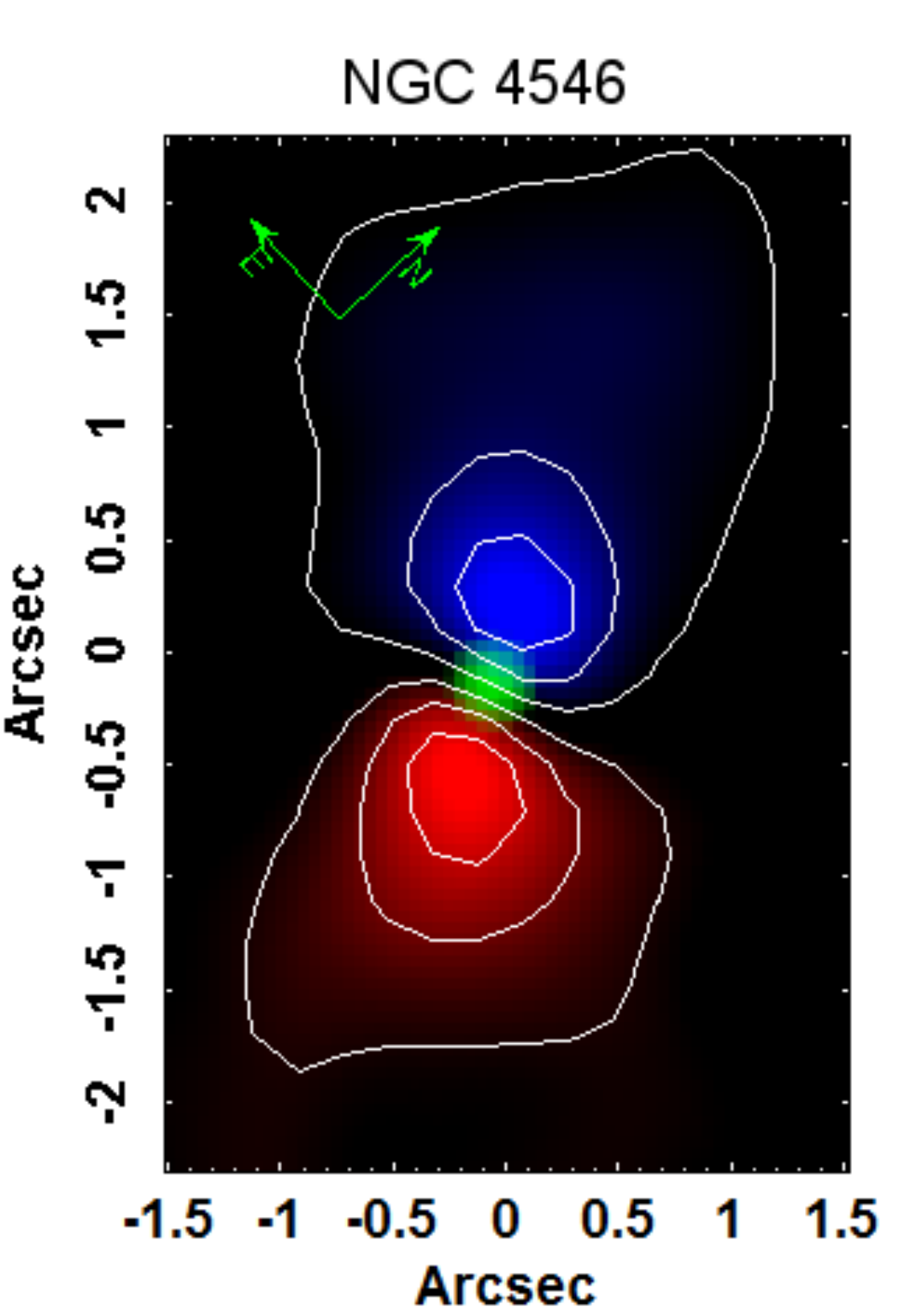}
\vspace{0cm}
\includegraphics[scale=0.34]{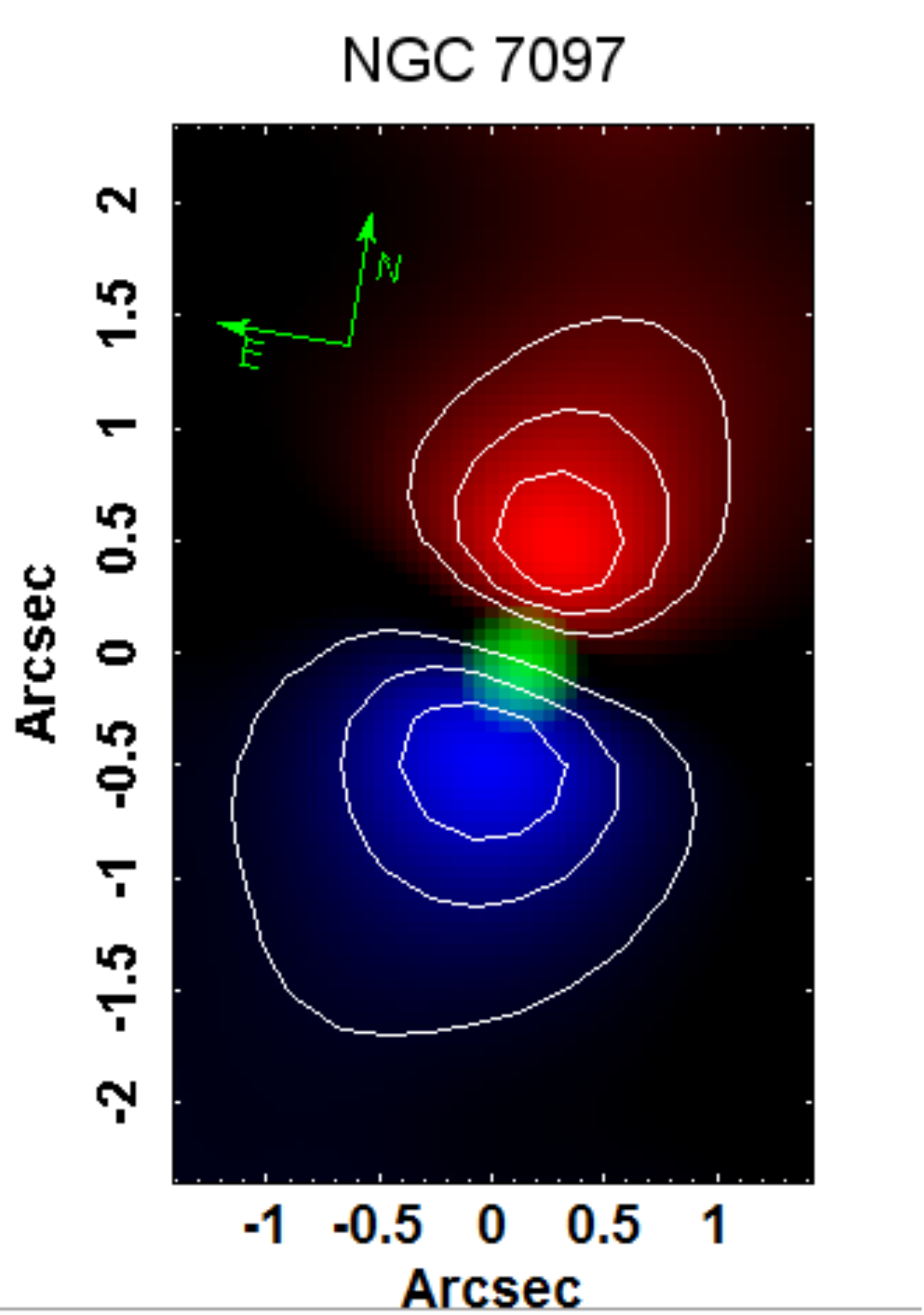}

\caption{RGB images corresponding to the blue and red wings of the H$\alpha$ emission lines. It is worth mentioning that the results shown above correspond to the NLR only, since we removed the broad components of H$\alpha$ before the construction of the maps. The image of the blue (red) wing minus the image of the red (blue) wing is shown in blue (red). The tip of the PSF corresponding to the flux maps in Figs. \ref{mapa_fluxo_gal_1} and \ref{mapa_fluxo_gal_2} is shown in green. We also added isophotes for each structure, which are shown as white contours. \label{RGB_Ha}
}
\end{center}
\end{figure*}

In addition to the flux maps, we show in Fig. \ref{RGB_Ha} images corresponding to the red and the blue wings of the narrow component of the H$\alpha$ emission line (remembering that the H$\alpha$ broad component was already subtracted). Again, we see bipolar structures in the circumnuclear region of the galaxies. Moreover, the images shown in Fig. \ref{RGB_Ha} are very similar to the tomograms related to the gas kinematics, which were obtained with PCA Tomography (paper I). We measured the P.A. of these structures and the results are shown in Table \ref{tab_PA}. The P.A. measured in these images are in agreement with the P.A. measured in their corresponding tomograms shown in paper I within 3$\sigma$, except for IC 1459, whose difference between both measurements is $\sim$ 4$\sigma$. The isophotes presented in Fig. \ref{RGB_Ha} of ESO 208 G-21, NGC 1380 and NGC 7097 show narrow structures, which is compatible with the hypothesis of gaseous discs. In IC 1459, the outer isophotes are irregular, probably affected by the outflows. This is also seen, at a lower intensity, in NGC 4546. Although the isophotes of IC 5181 and NGC 2663 are disc-like, these cases are discussed in Section \ref{discs_or_cones}. For NGC 3136, the structure seen in Fig. \ref{RGB_Ha} also resembles an ionization bicone.

\subsection{[N II]/H$\alpha$ ratio maps} \label{nii_halpha_ratio_section}

The maps of the [N II]$\lambda$6583/H$\alpha$ ratio, shown in Figs. \ref{perfil_razao_1} and \ref{perfil_razao_2}, were drawn by dividing the flux maps of both emission lines. The 1D profiles along the kinematic bipolar structures and along the low-velocity emission are also shown in Figs. \ref{perfil_razao_1} and \ref{perfil_razao_2}.

The [N II]/H$\alpha$ ratio has higher values in the nucleus of seven galaxies in the sample. In the circumnuclear region of galaxies ESO 208 G-21, IC 5181, NGC 4546 and NGC 7097, [N II]/H$\alpha$ $>$ 1.0 along the kinematic bipolar structure. This is typical of LINER emission. In the low-velocity emission of these galaxies, [N II]/H$\alpha$ $<$ 1.0. In IC 1459, the 1D profiles along both directions have similar behaviours, also seen in NGC 2663.

In NGC 1380, the 1D profile along the kinematic bipolar structure reveals that [N II]/H$\alpha$ $\sim$ 1.5 in a region located 1 arcsec south from the nucleus. In the same direction, at 1.8 arcsec from the nucleus, [N II]/H$\alpha$ $\sim$ 0.6, which is typical of an H II region. Northward, also within a projected distance of 1.8 arcsec, [N II]/H$\alpha$ $\sim$ 0.8. Probably, a second H II region is to be found here. In NGC 3136, a few compact structures are seen in the map with [N II]/H$\alpha$ $>$ 1.8. One of the objects is likely to be the central AGN studied in papers I and II. In this galaxy, in the position of the second compact structure detected with PCA Tomography (paper I), we have [N II]/H$\alpha$ $\sim$ 1.6.

\renewcommand{\thefigure}{\arabic{figure}\alph{subfigure}}
\setcounter{subfigure}{1}

\begin{figure*}
\includegraphics[scale=0.32]{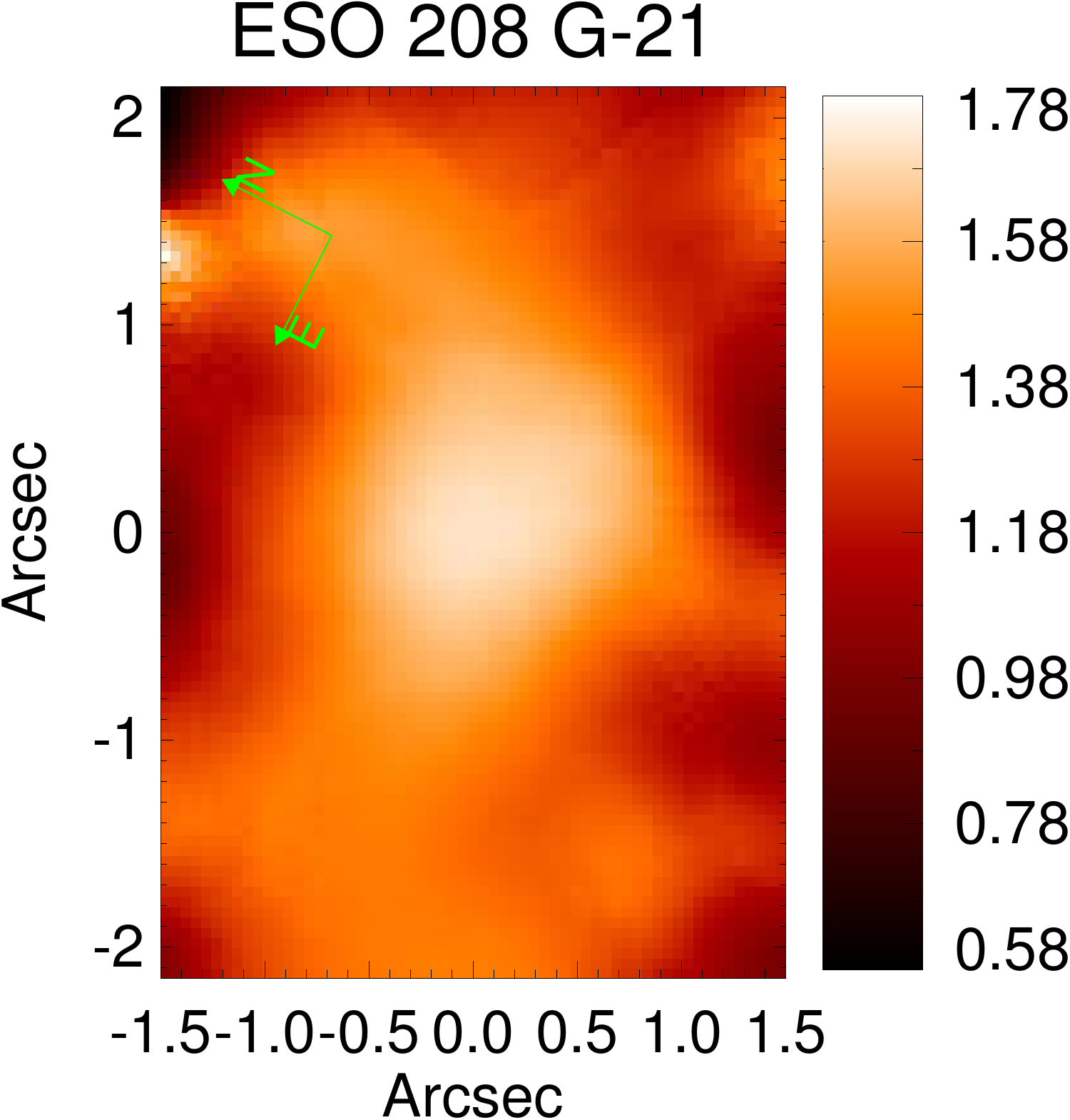}
\vspace{0.2cm}
\hspace{0.5cm}
\includegraphics[scale=0.34]{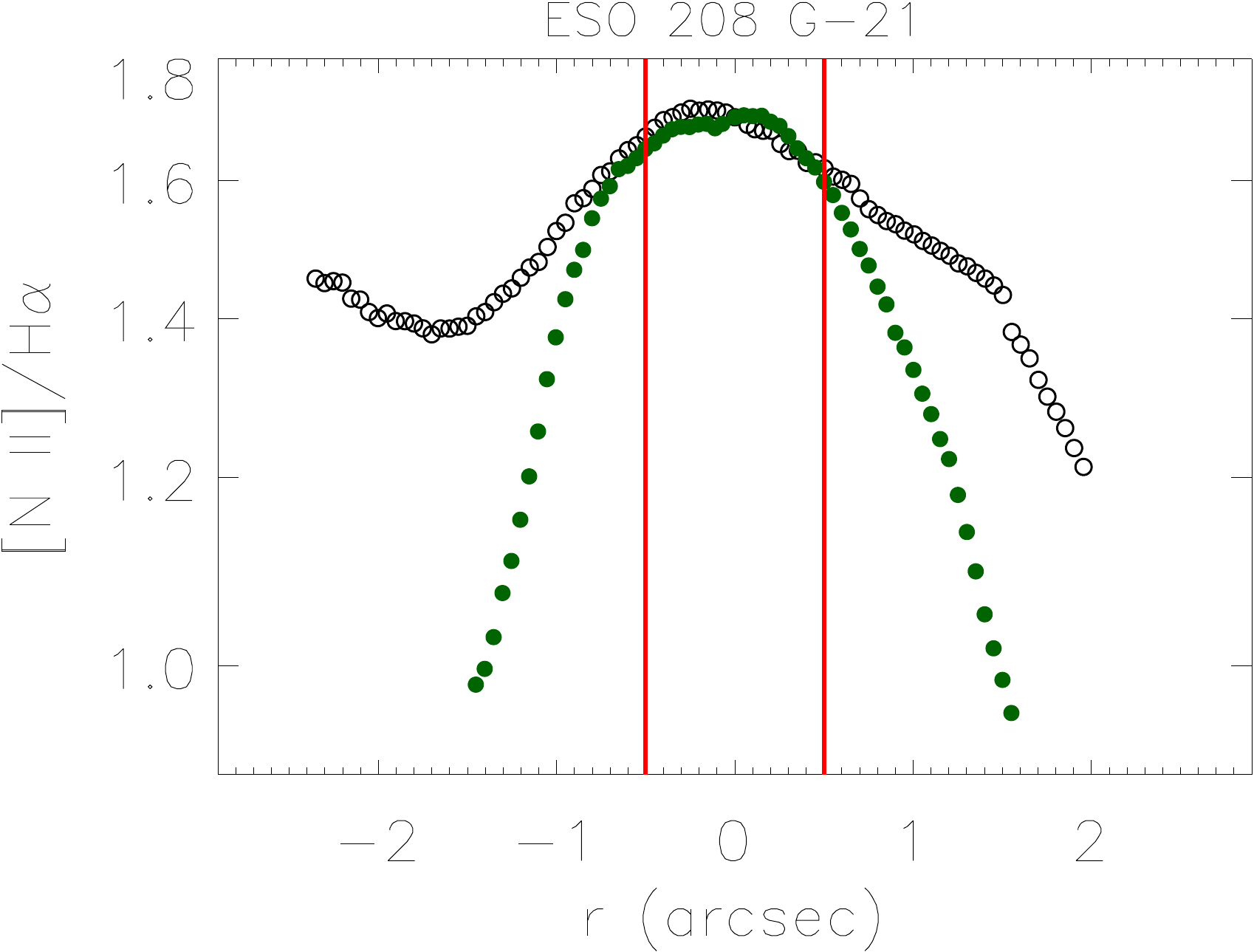}
\hspace{0.5cm}

\includegraphics[scale=0.32]{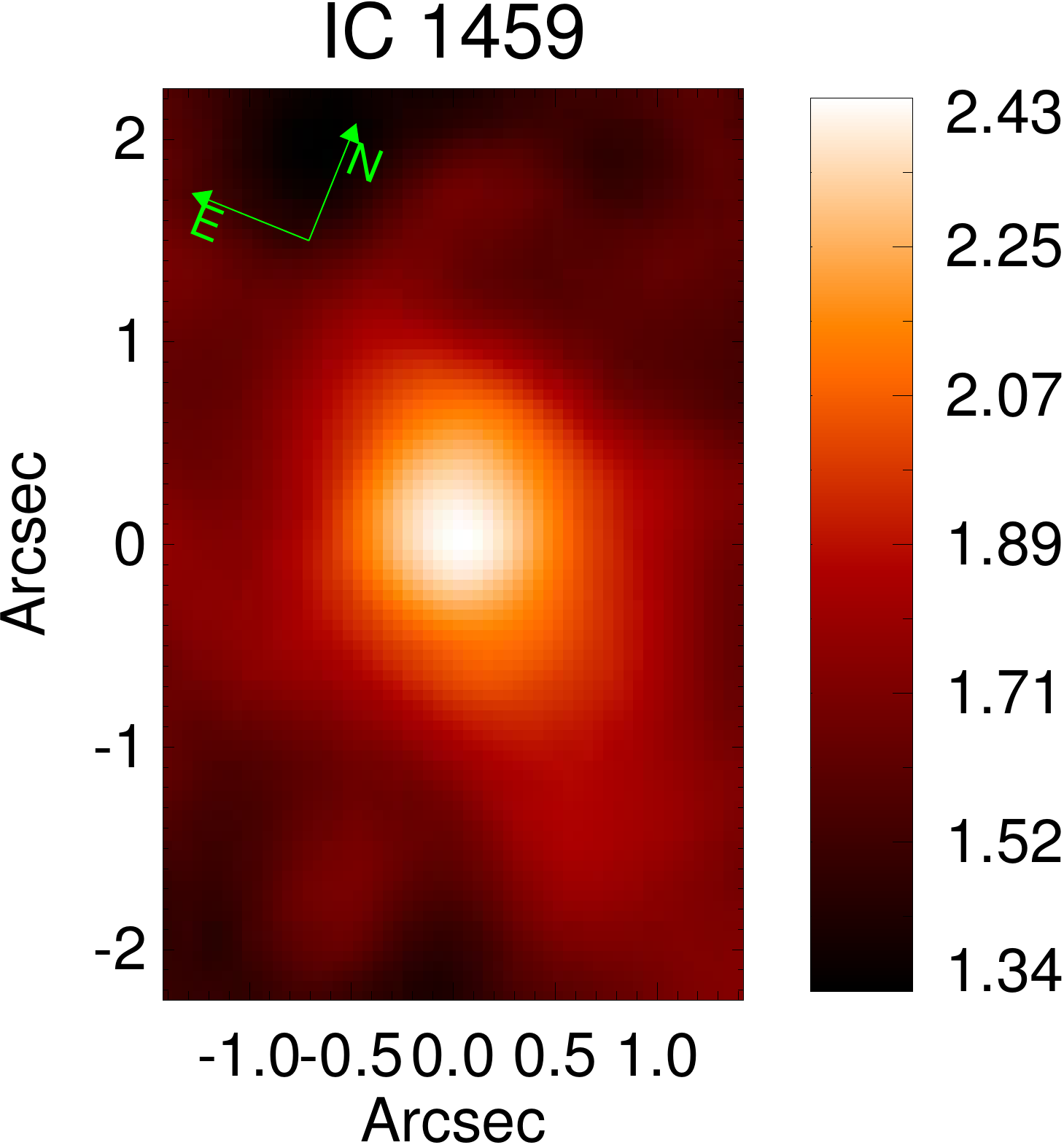}
\vspace{0.2cm}
\hspace{0.5cm}
\includegraphics[scale=0.34]{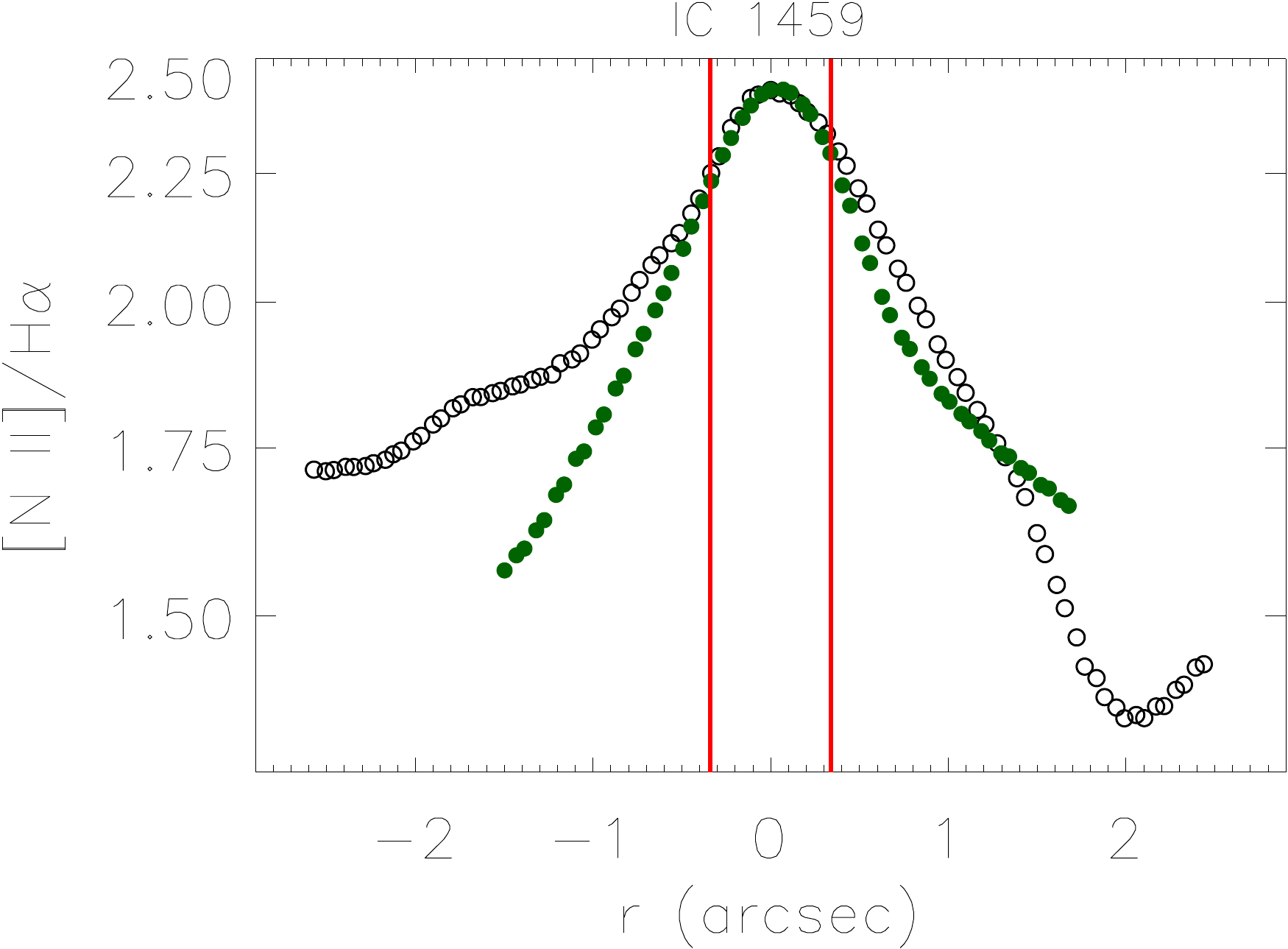}
\hspace{0.5cm}

\includegraphics[scale=0.32]{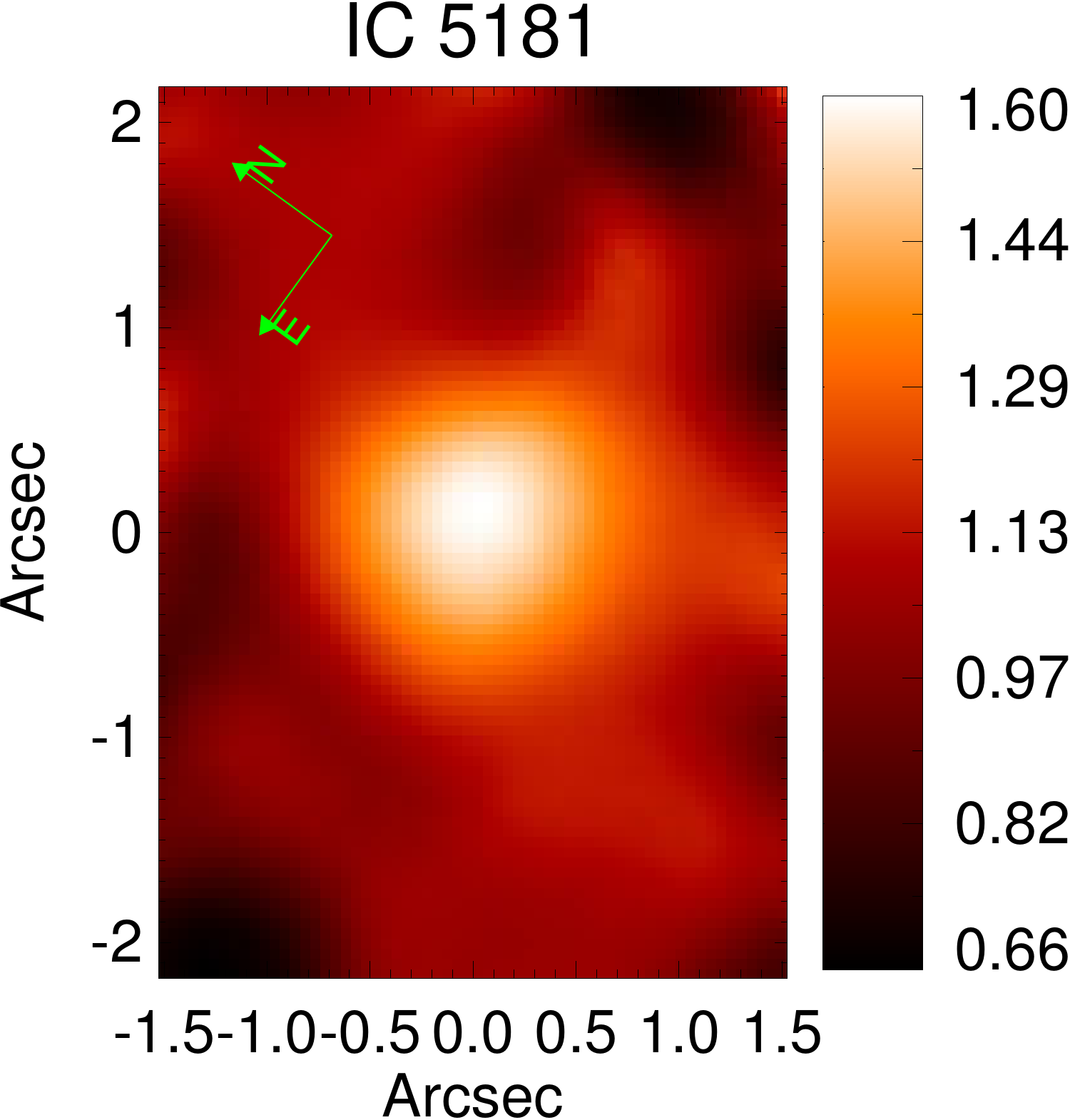}
\vspace{0.2cm}
\hspace{0.5cm}
\includegraphics[scale=0.34]{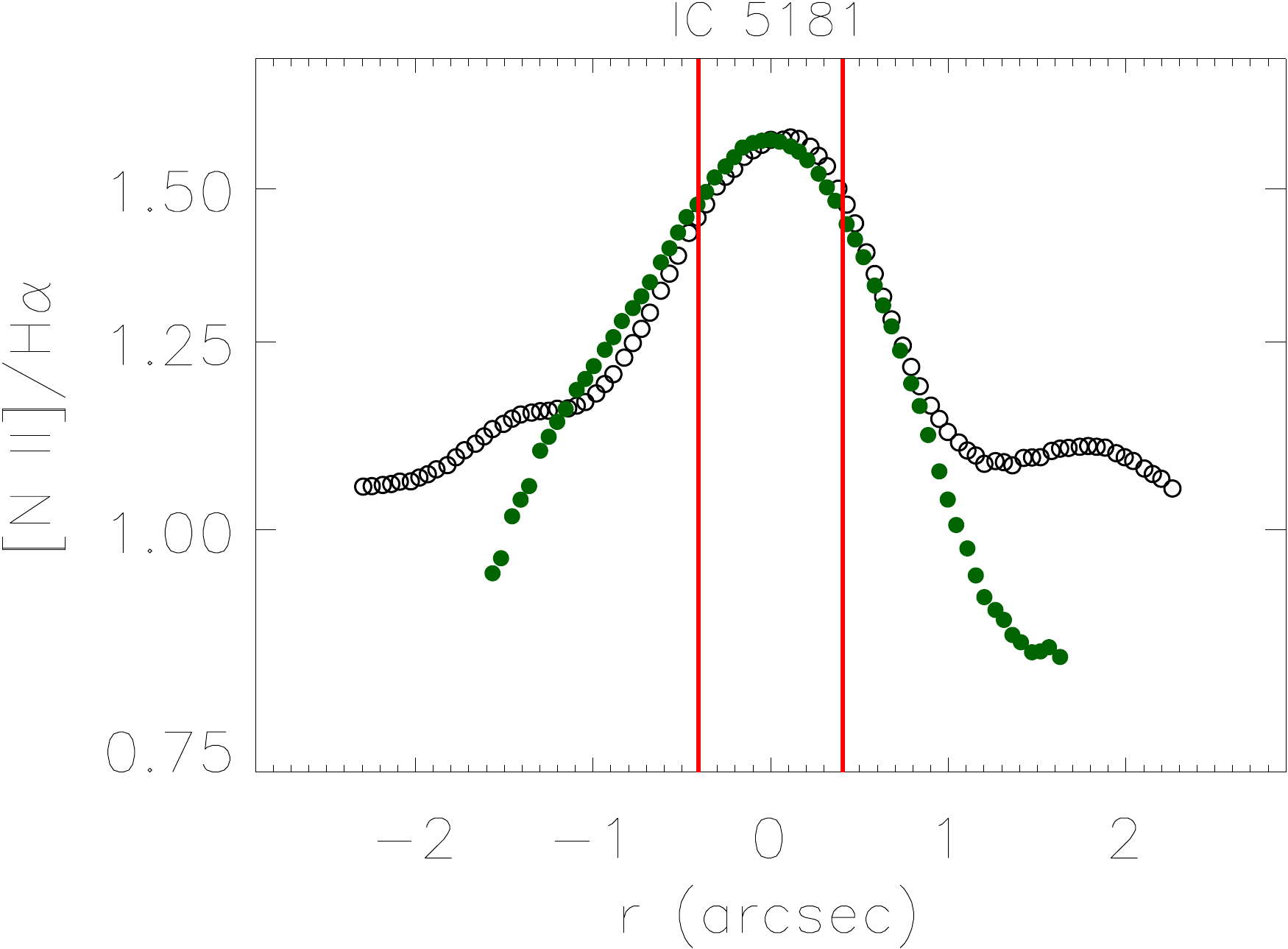}
\hspace{0.5cm}

\includegraphics[scale=0.32]{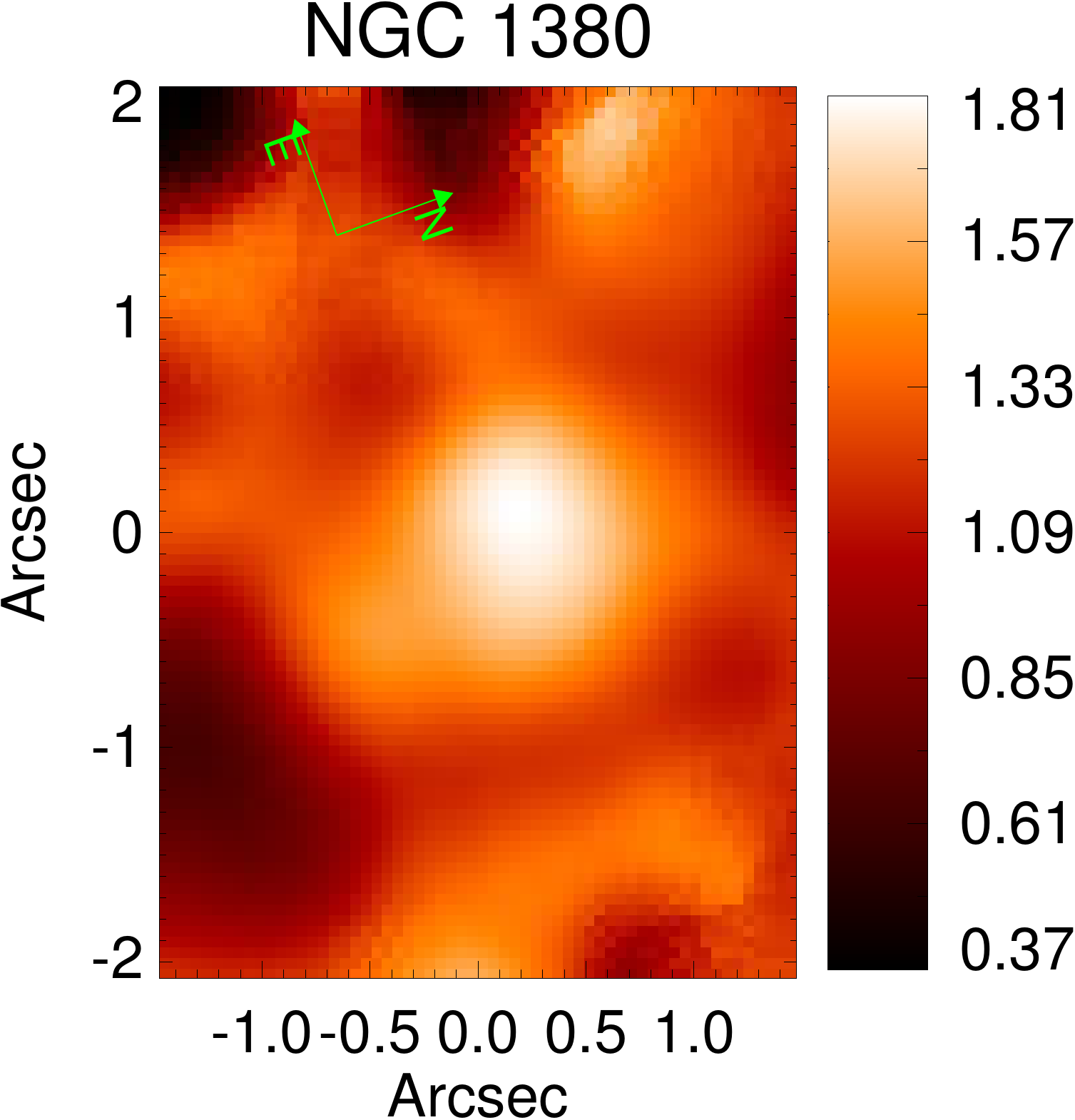}
\vspace{0.2cm}
\hspace{0.5cm}
\includegraphics[scale=0.34]{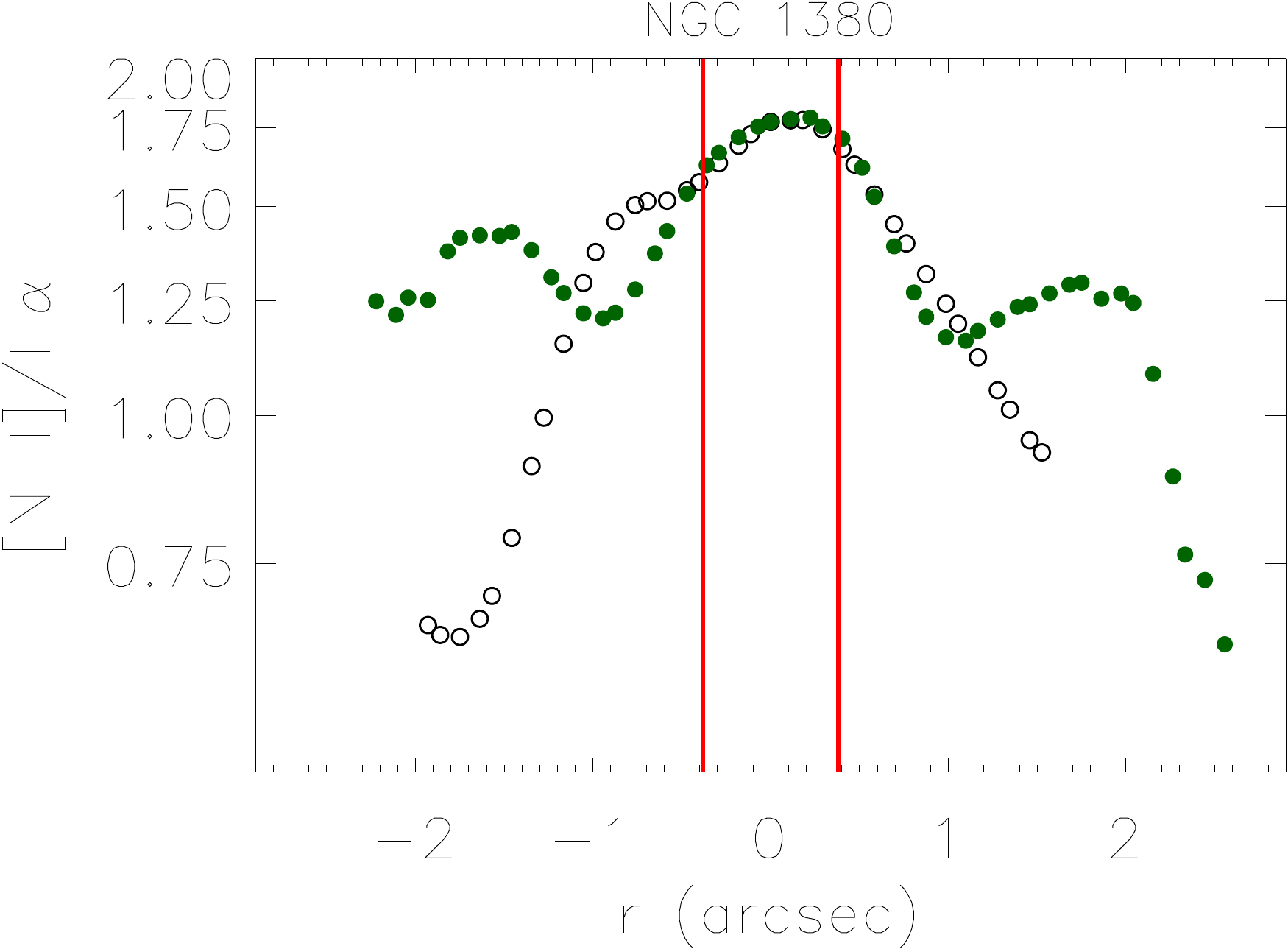}
\hspace{0.5cm}

\caption{Left: [N II]/H$\alpha$ ratio maps. Right: 1D profiles of the [N II]/H$\alpha$ ratio. It is worth mentioning that the results shown above correspond to the NLR only, since we removed the broad components of H$\alpha$ before the construction of the maps. The hollow black circles correspond to the observed ratio along the kinematic bipolar structure, while the filled green circles are related to the observed ratio along the low-velocity emission. The vertical red lines delimit the FWHM of the PSFs of the data cubes. \label{perfil_razao_1}} 
\end{figure*}

\addtocounter{figure}{-1}
\addtocounter{subfigure}{1}

\begin{figure*}
\includegraphics[scale=0.32]{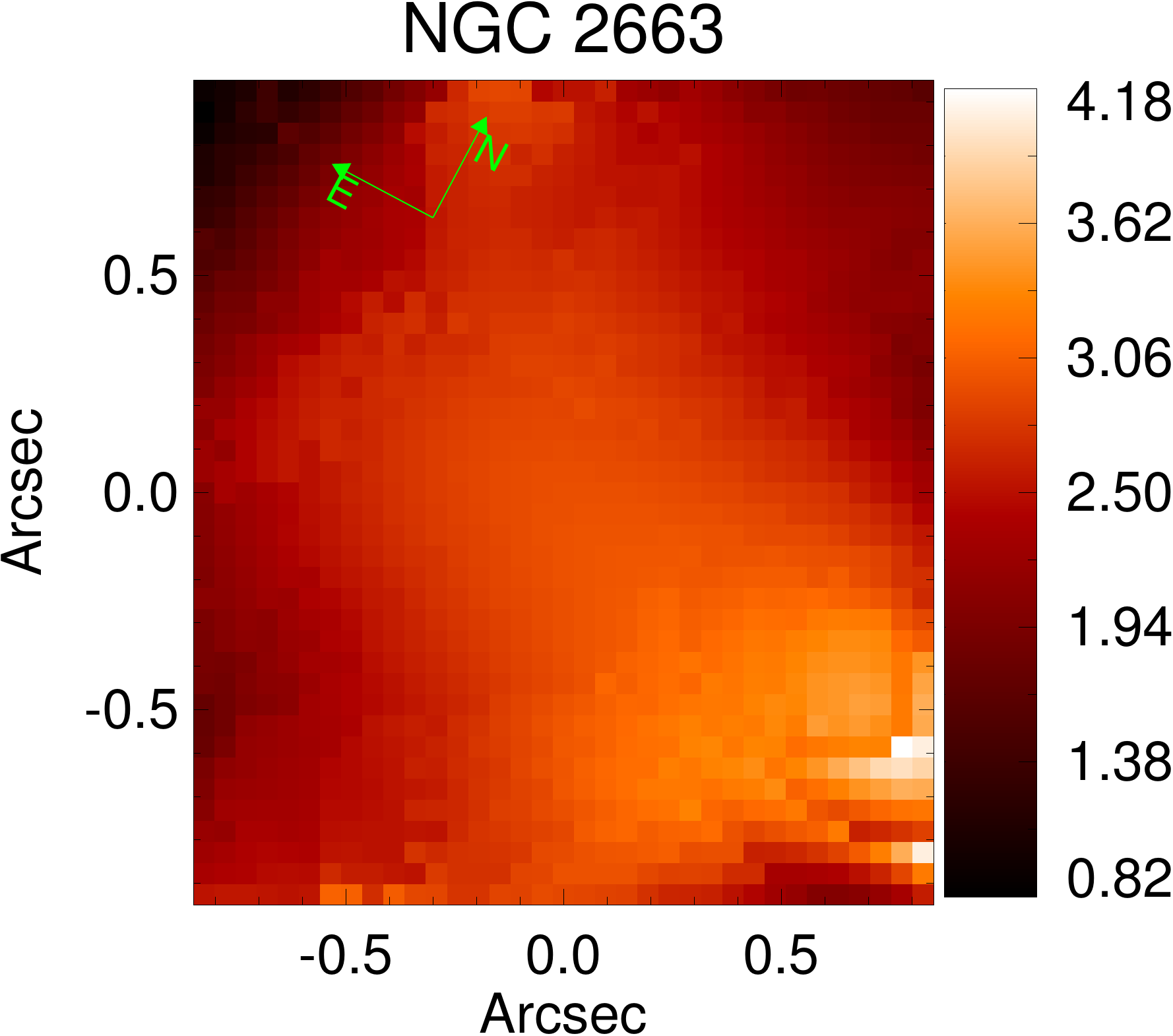}
\vspace{0.2cm}
\hspace{1cm}
\includegraphics[scale=0.34]{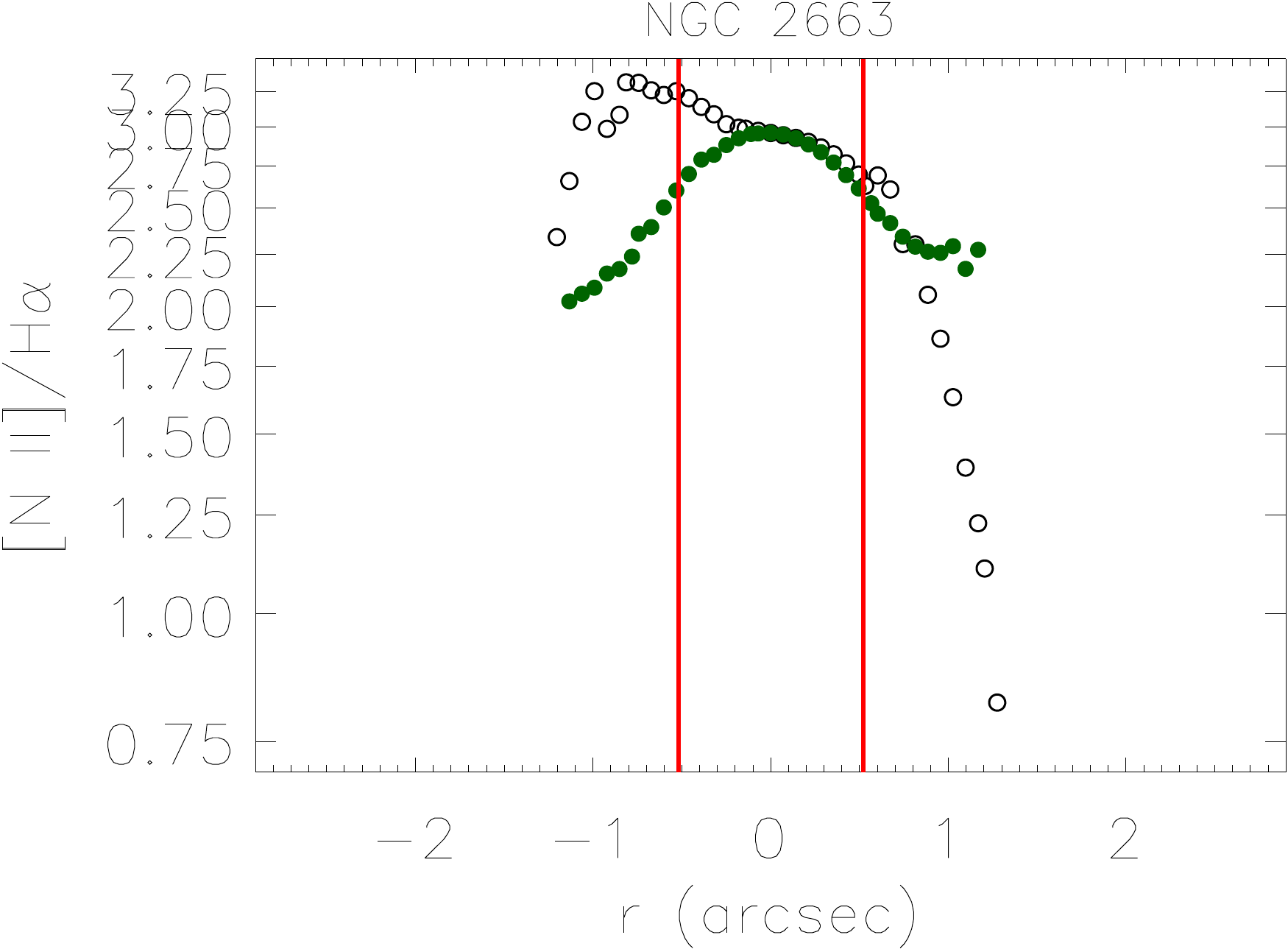}
\hspace{0.5cm}

\includegraphics[scale=0.32]{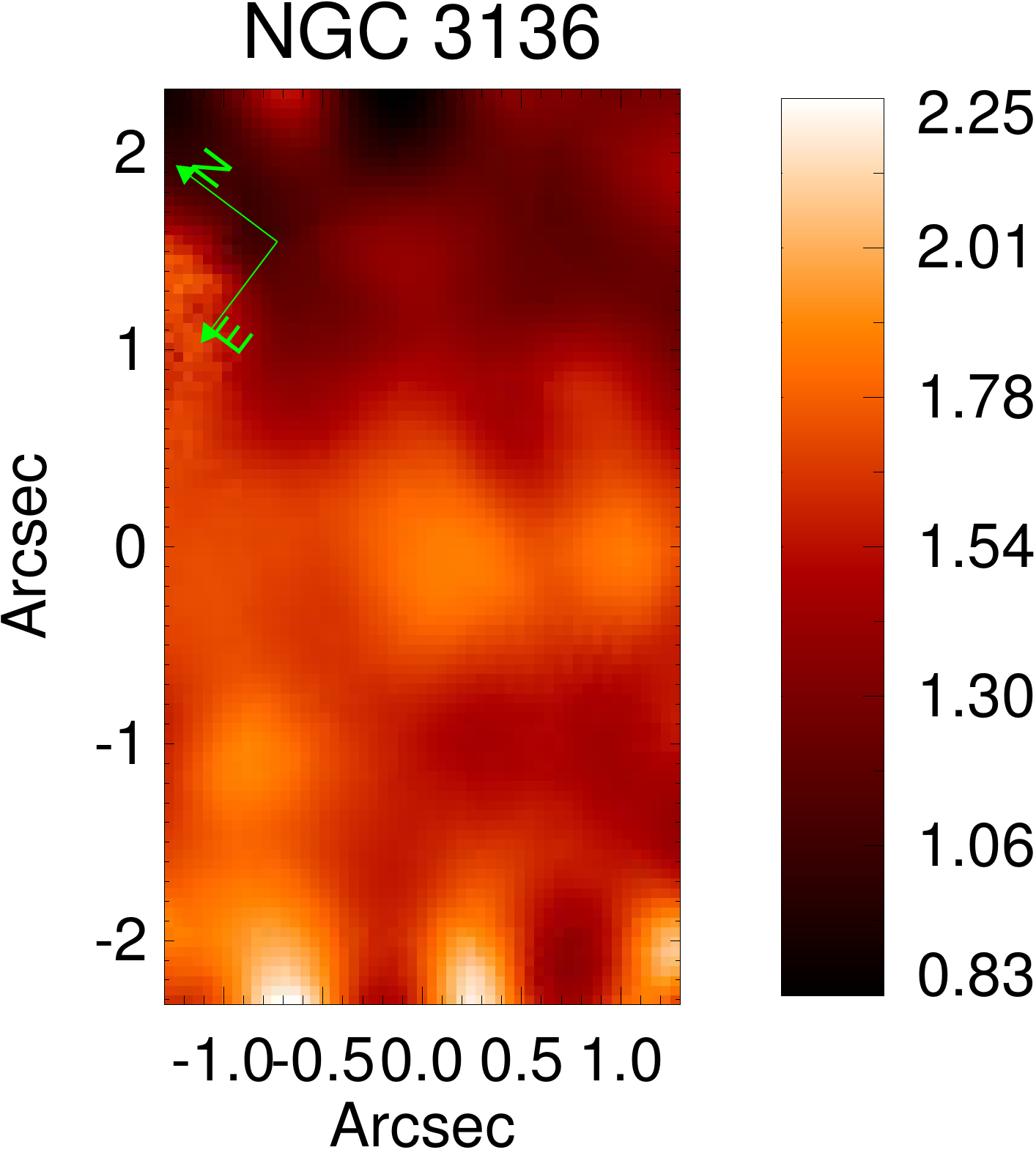}
\vspace{0.2cm}
\hspace{0.5cm}
\includegraphics[scale=0.34]{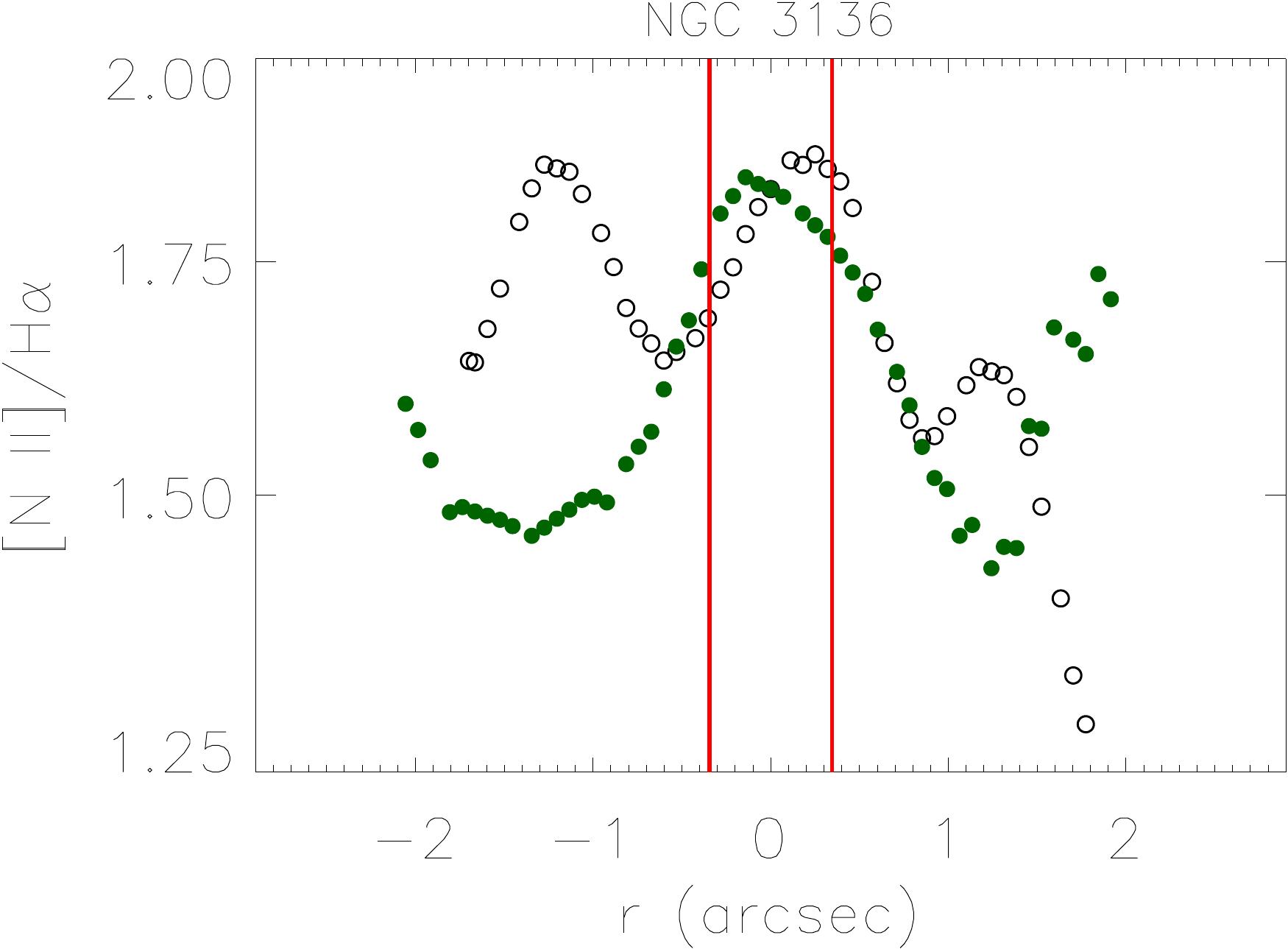}
\hspace{0.5cm}

\includegraphics[scale=0.32]{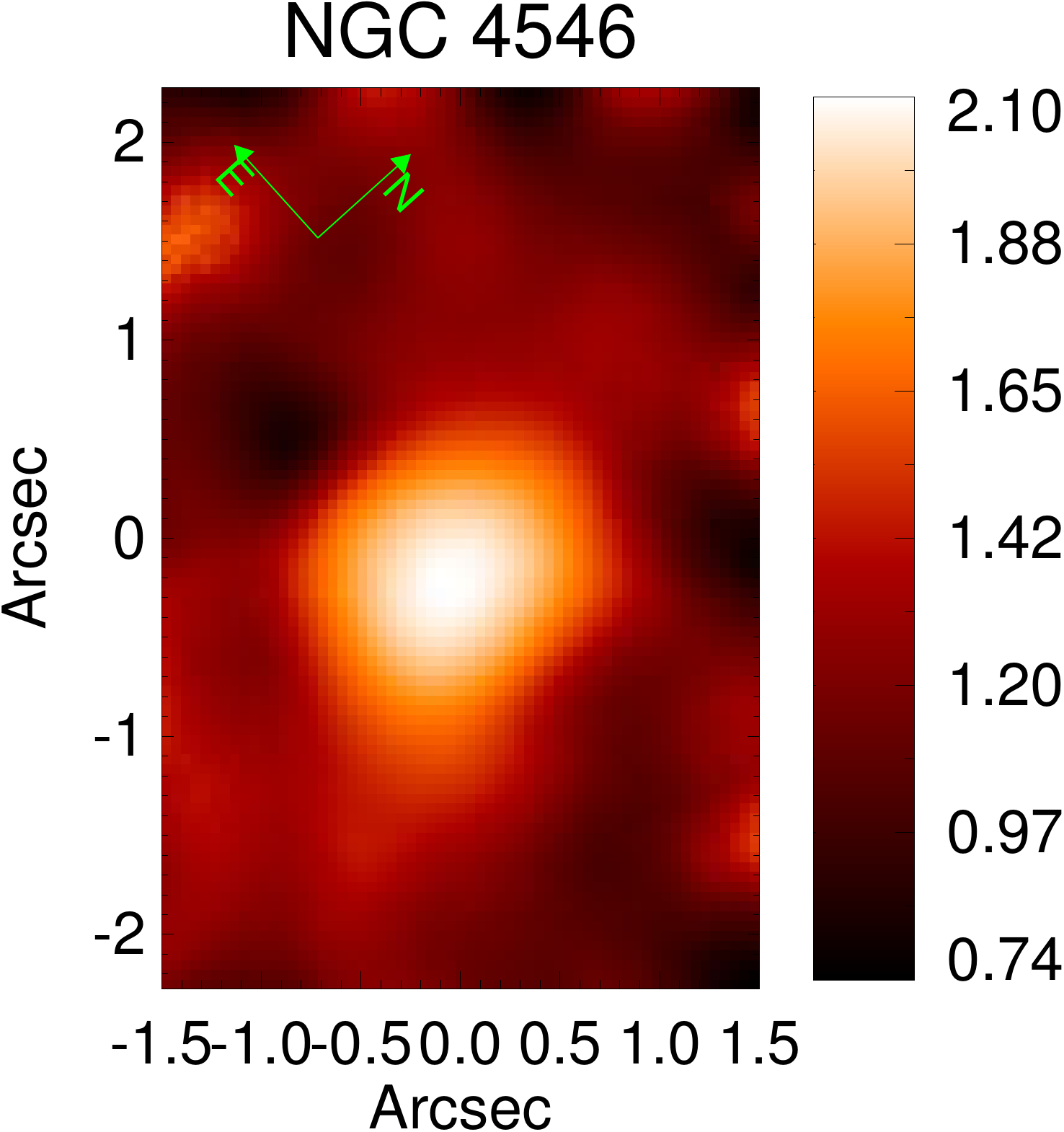}
\vspace{0.2cm}
\hspace{0.5cm}
\includegraphics[scale=0.34]{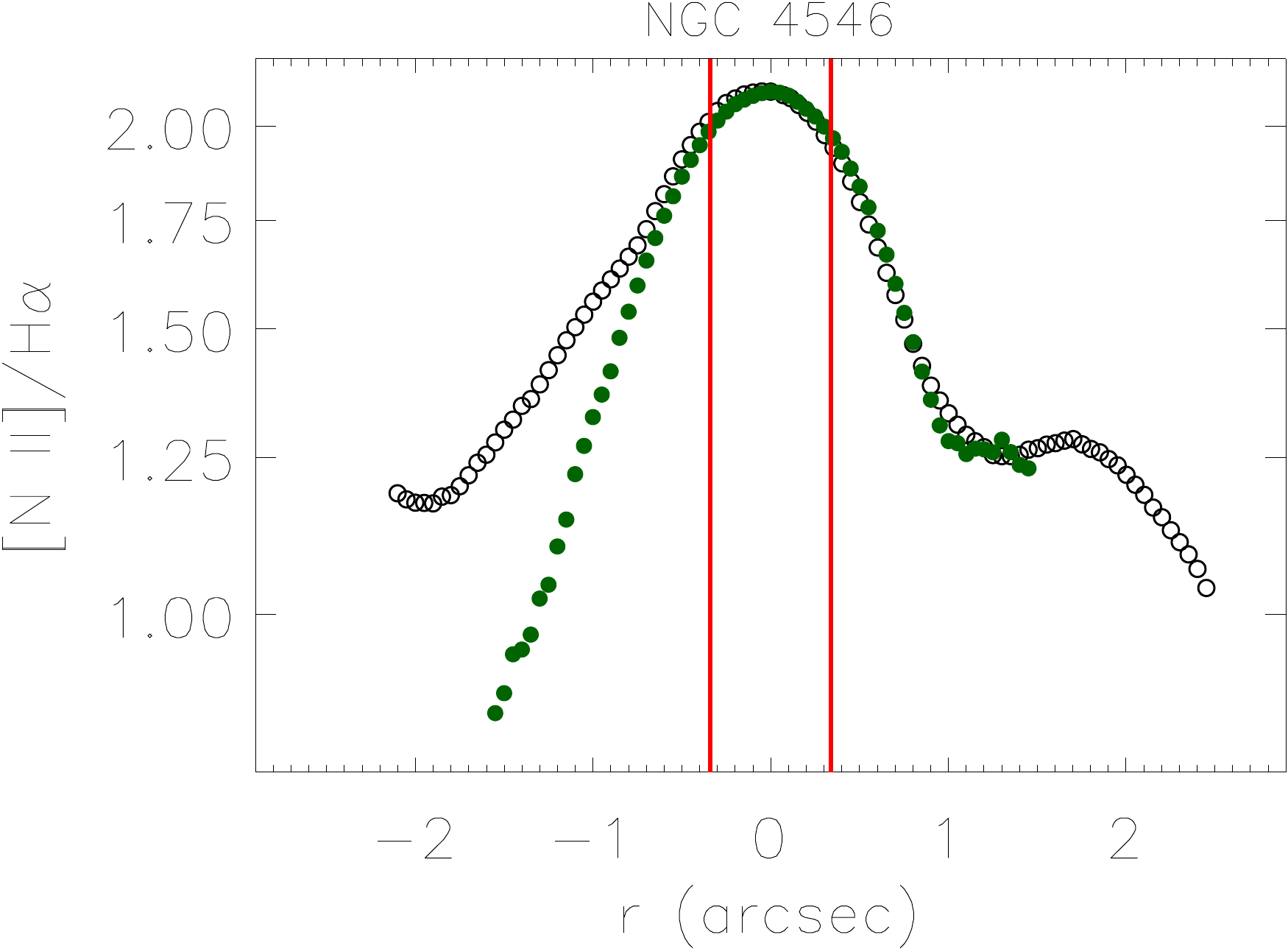}
\hspace{0.5cm}

\includegraphics[scale=0.32]{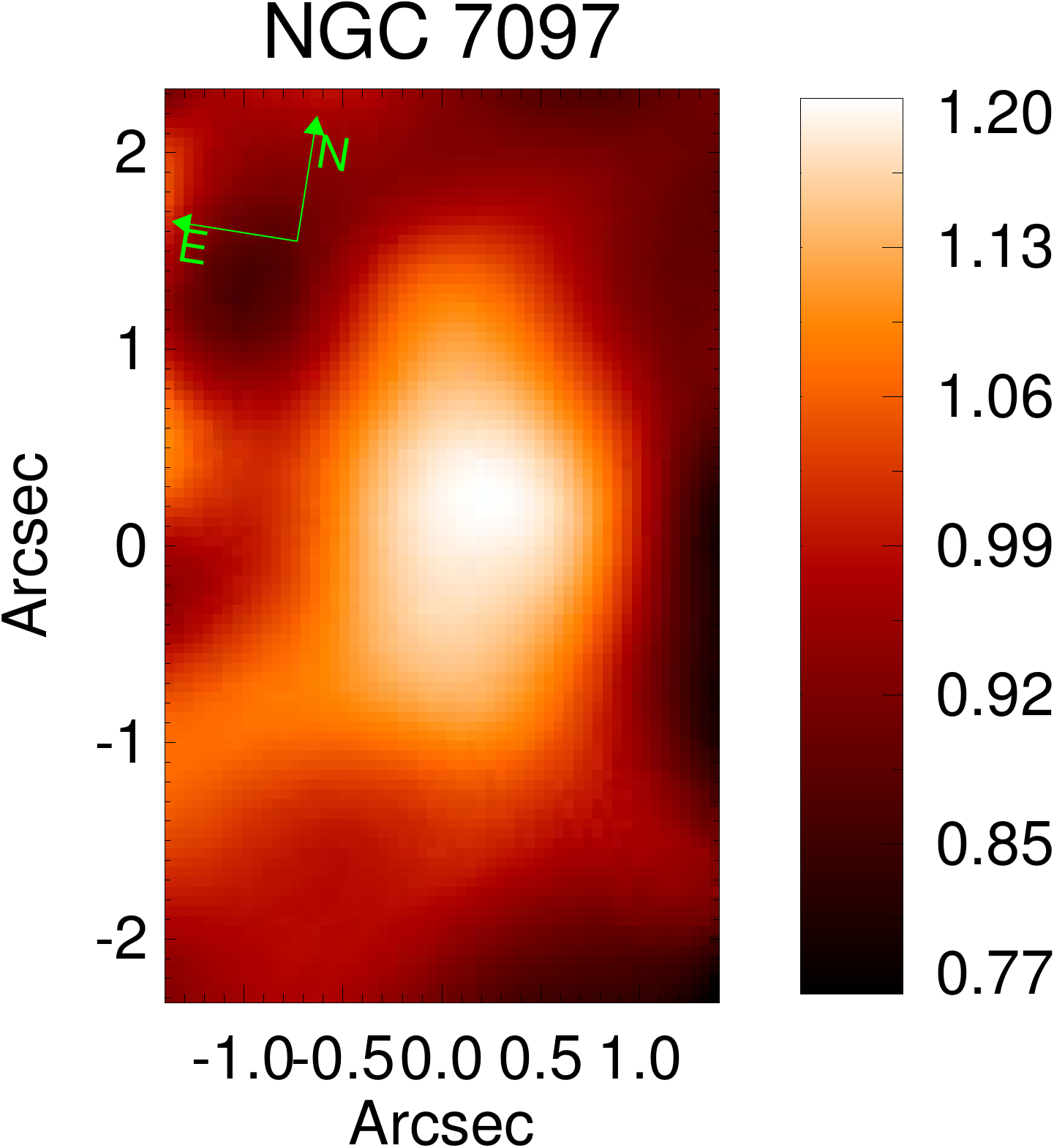}
\vspace{0.2cm}
\hspace{0.5cm}
\includegraphics[scale=0.34]{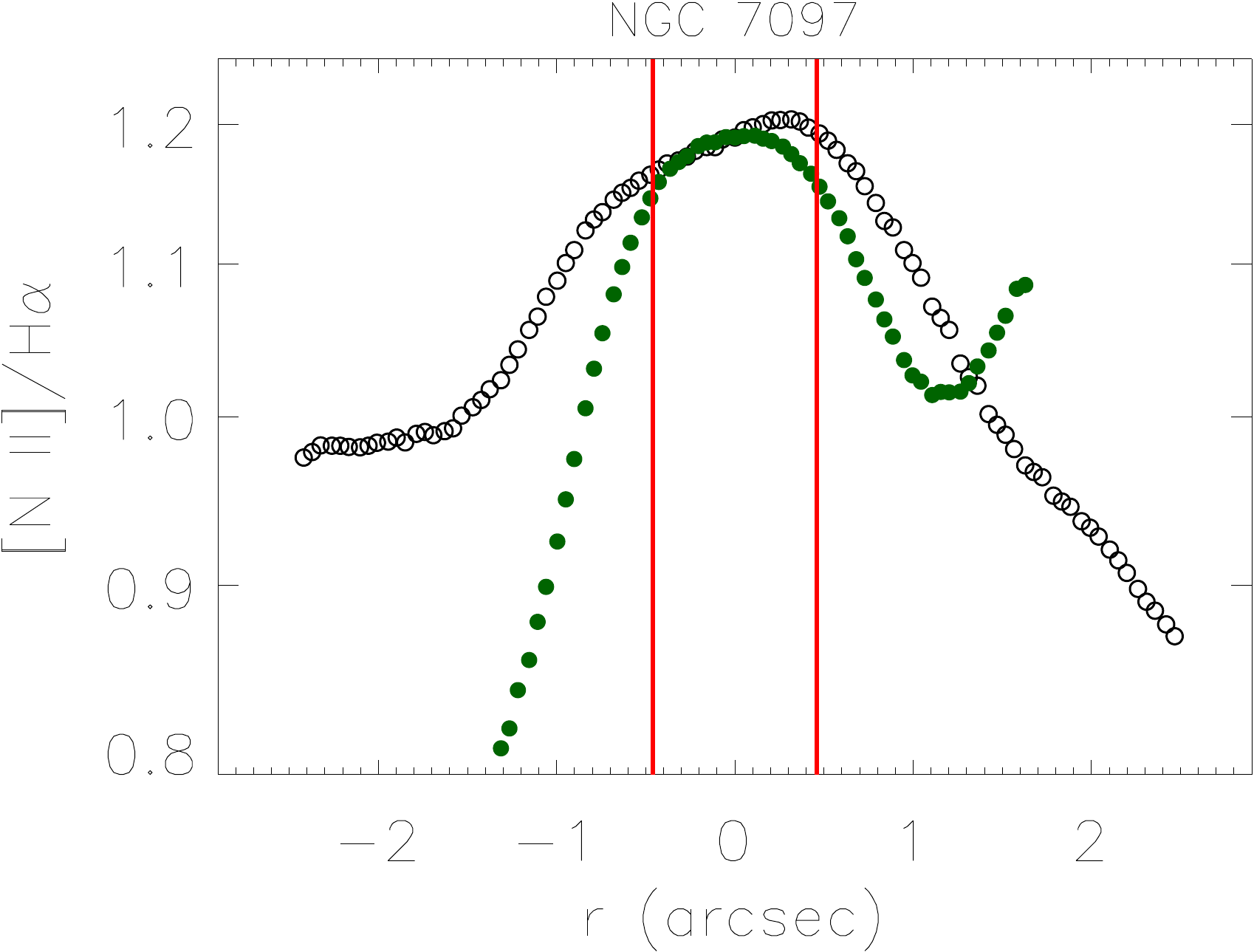}
\hspace{0.5cm}

\caption{Same as in Fig. \ref{perfil_razao_1}. \label{perfil_razao_2}} 
\end{figure*}

\renewcommand{\thefigure}{\arabic{figure}}



\subsection{Maps of the equivalent widths of the [N II] and H$\alpha$ emission lines} \label{EW_extended_emission}

In order to construct the EW maps, each spectrum of the gas cubes was divided by the average of the continuum regions adjacent to the H$\alpha$ absorption line in the spectrum taken from the stellar component. Subsequently, we integrated the [N II] and H$\alpha$ emission lines in the same way described in Section \ref{Halpha_NII_flux_maps}. The EW(H$\alpha$) and EW([N II]) maps are presented in Figs. \ref{mapa_ew_gal_1} and \ref{mapa_ew_gal_2}. 

\renewcommand{\thefigure}{\arabic{figure}\alph{subfigure}}
\setcounter{subfigure}{1}

\begin{figure*}
\hspace{-1.3cm}
\includegraphics[scale=0.32]{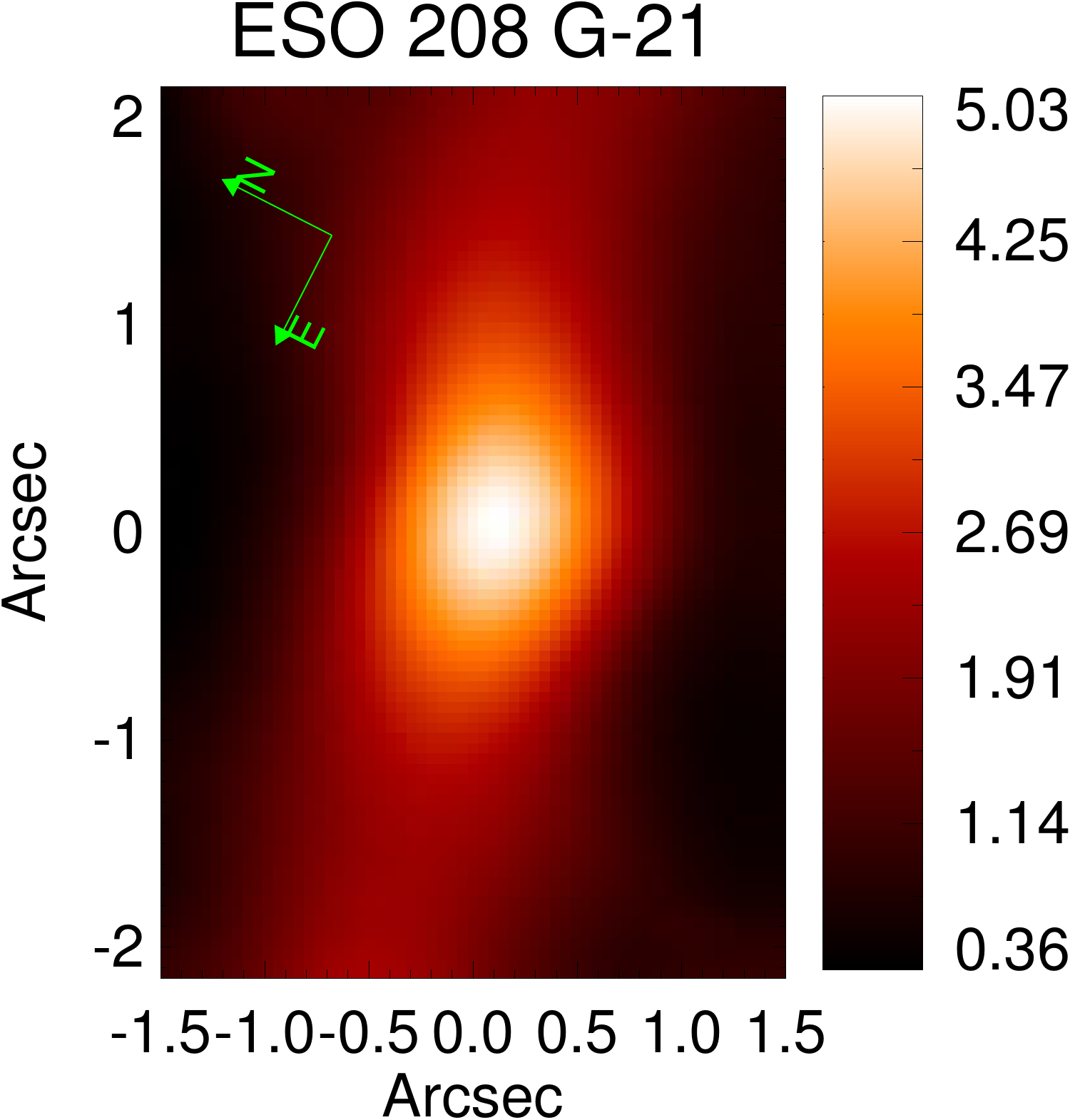}
\hspace{0.5cm}
\vspace{0.2cm}
\includegraphics[scale=0.32]{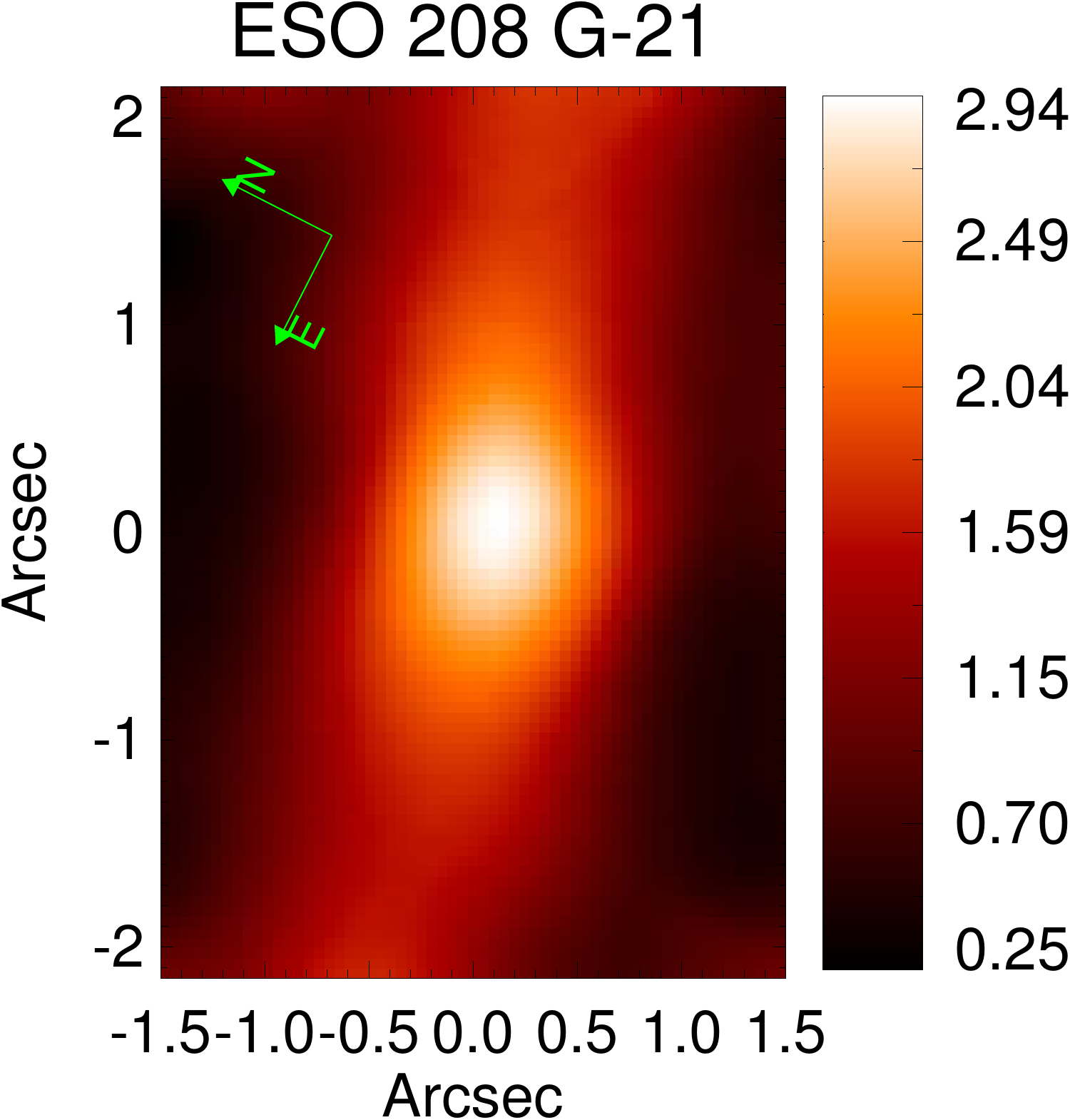}
\hspace{0.2cm}
\includegraphics[scale=0.32]{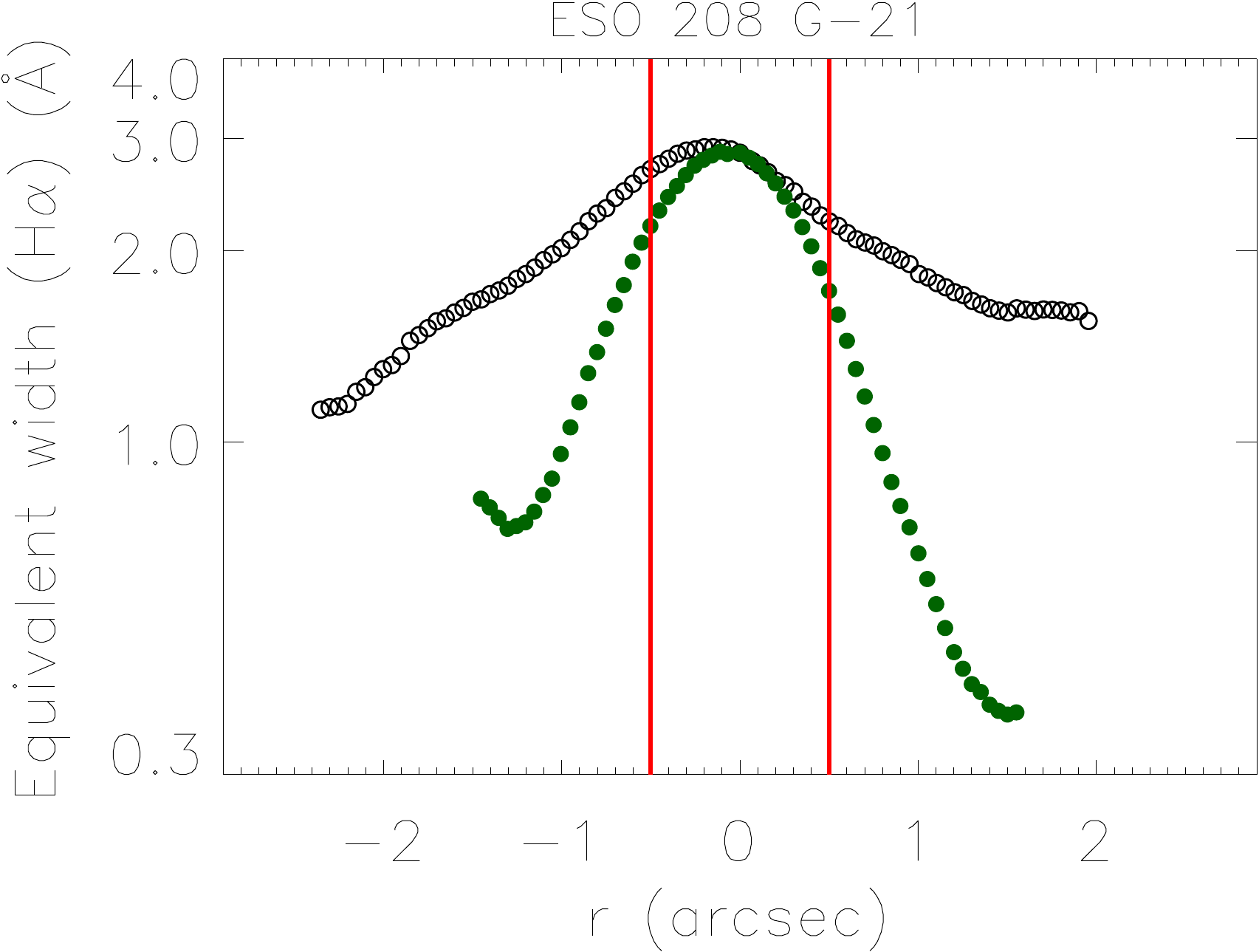}
\hspace{-0.8cm}

\hspace{-1.3cm}
\includegraphics[scale=0.32]{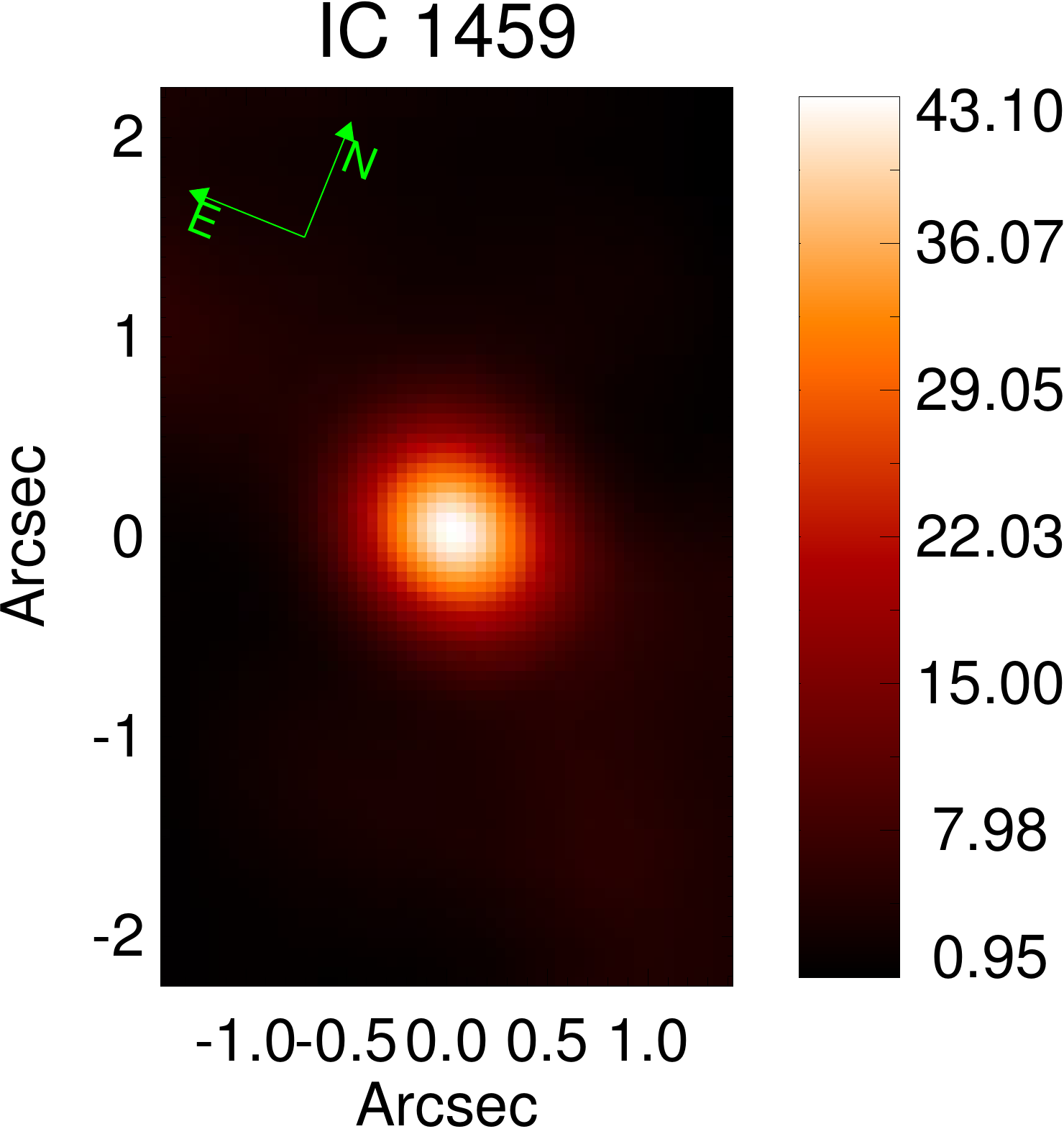}
\hspace{0.5cm}
\vspace{0.2cm}
\includegraphics[scale=0.32]{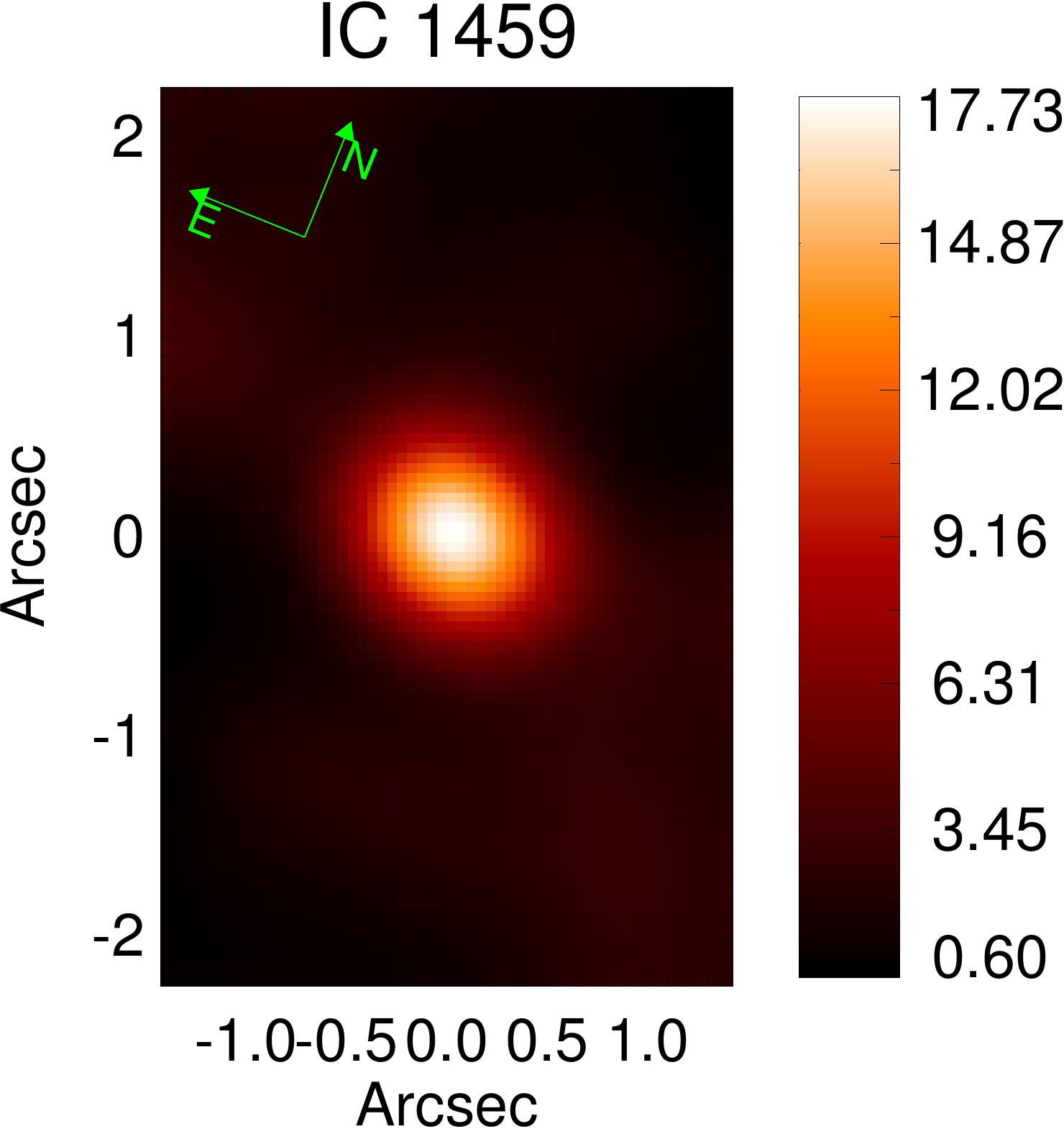}
\hspace{0.2cm}
\includegraphics[scale=0.32]{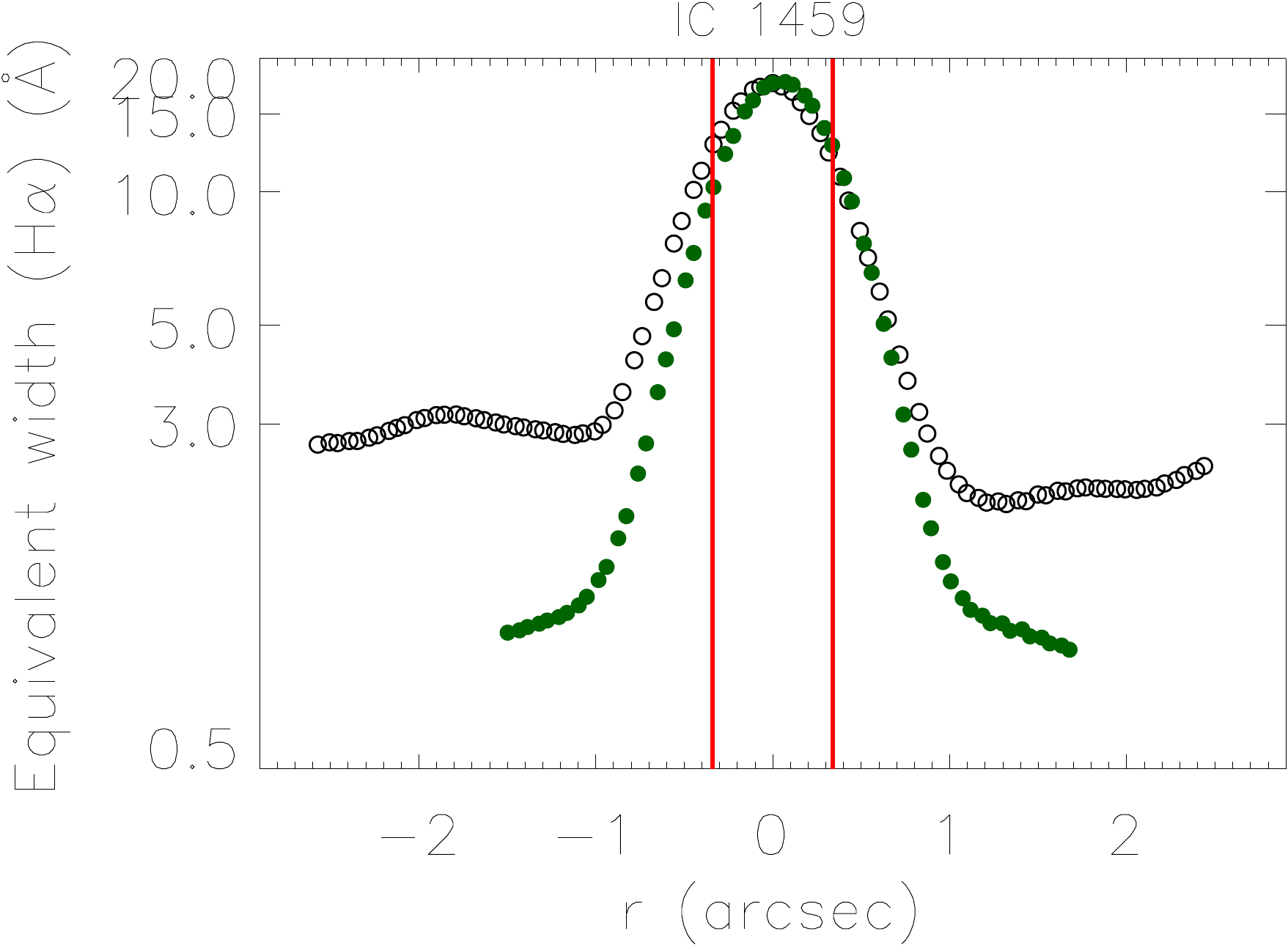}
\hspace{-0.8cm}

\hspace{-1.3cm}
\includegraphics[scale=0.32]{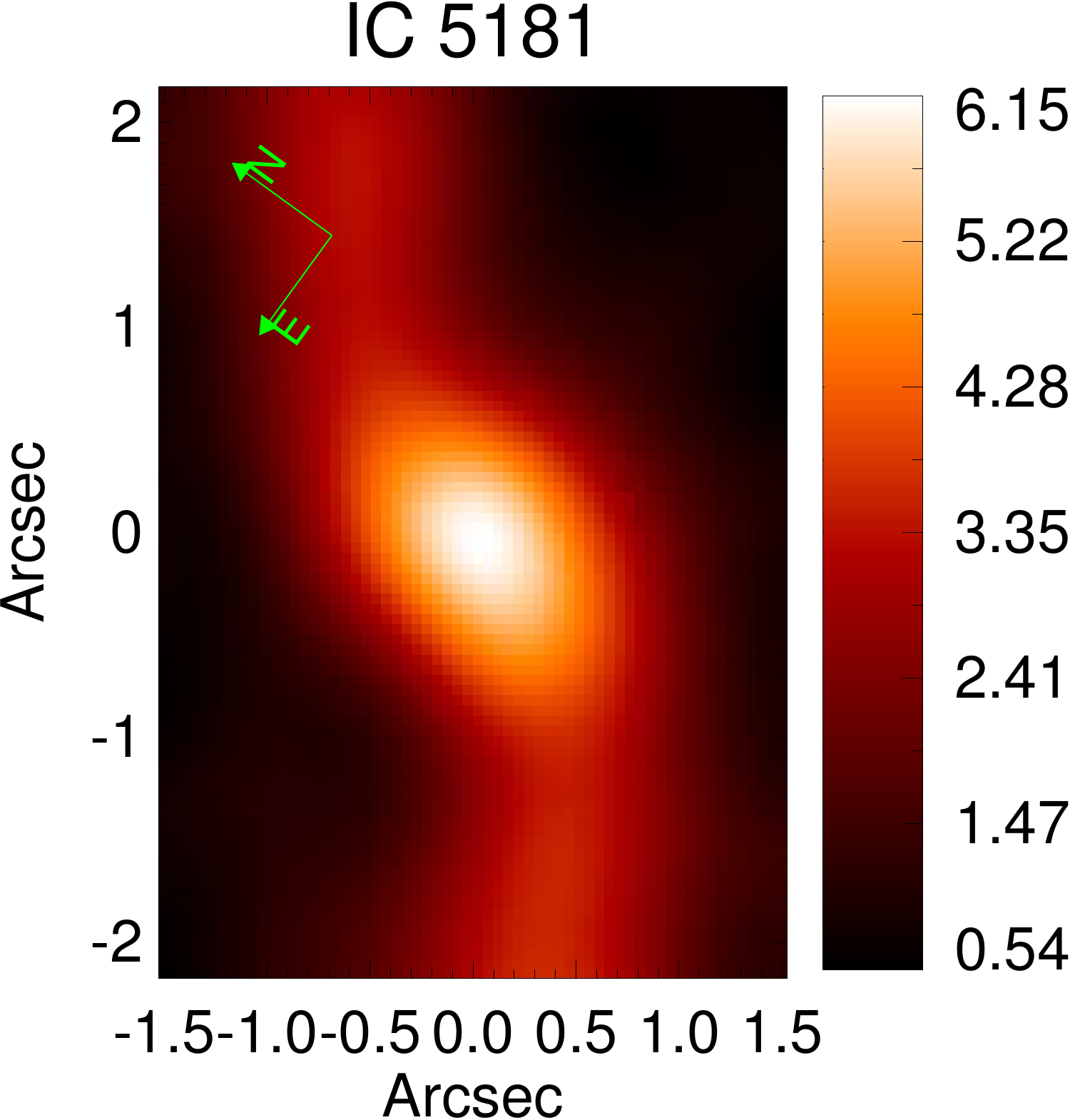}
\hspace{0.5cm}
\vspace{0.2cm}
\includegraphics[scale=0.32]{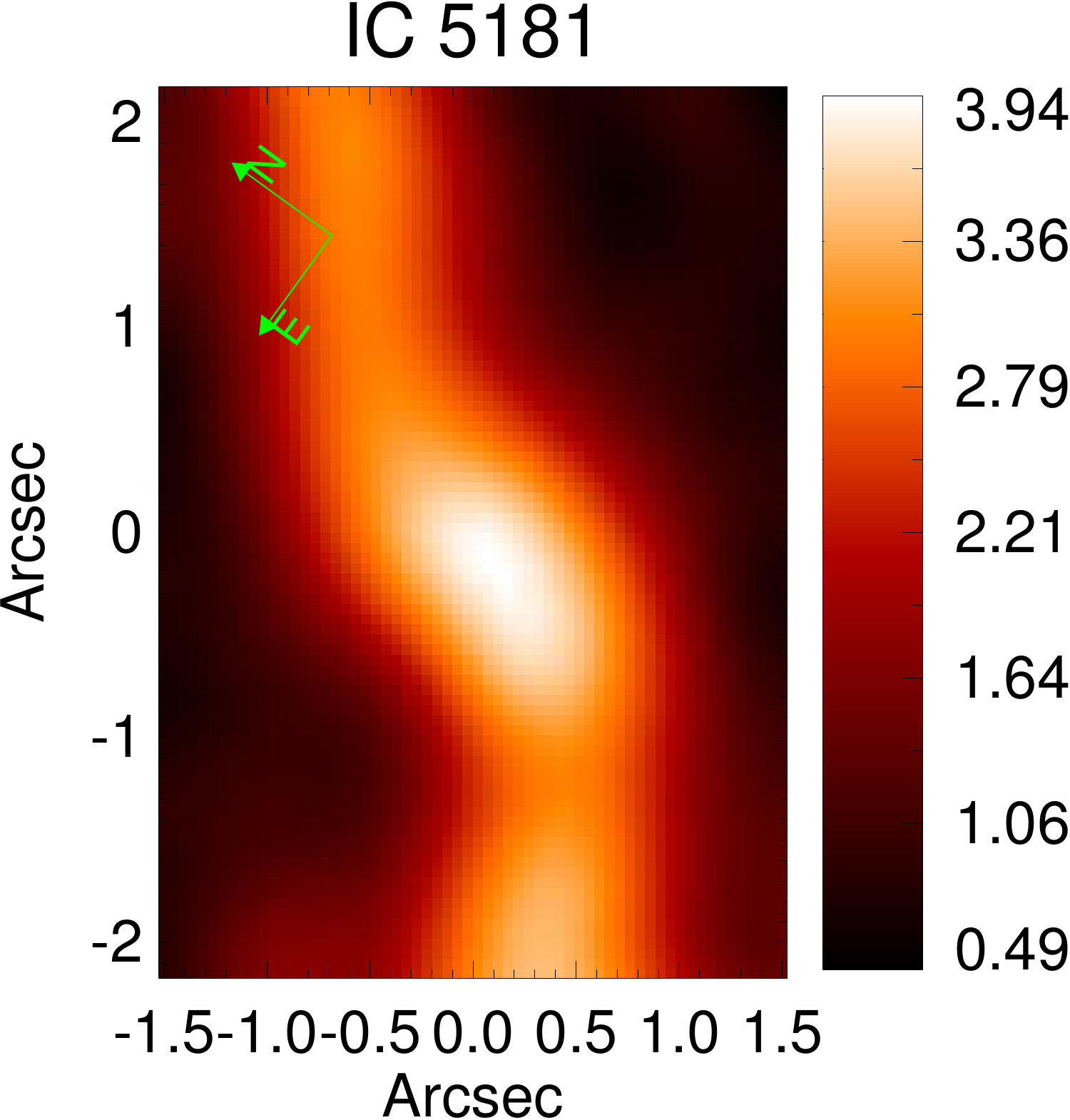}
\hspace{0.2cm}
\includegraphics[scale=0.32]{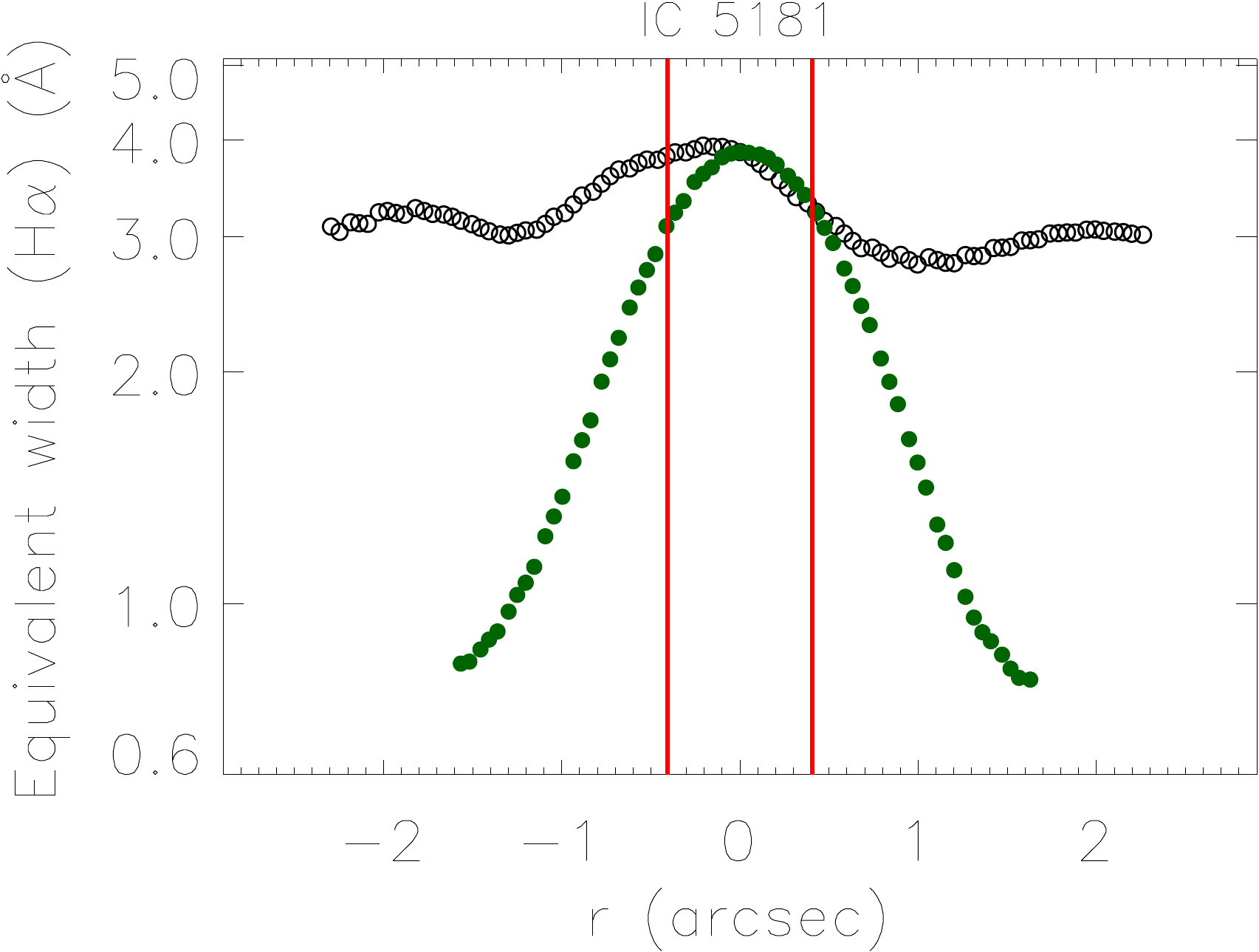}
\hspace{-0.8cm}

\hspace{-1.3cm}
\includegraphics[scale=0.32]{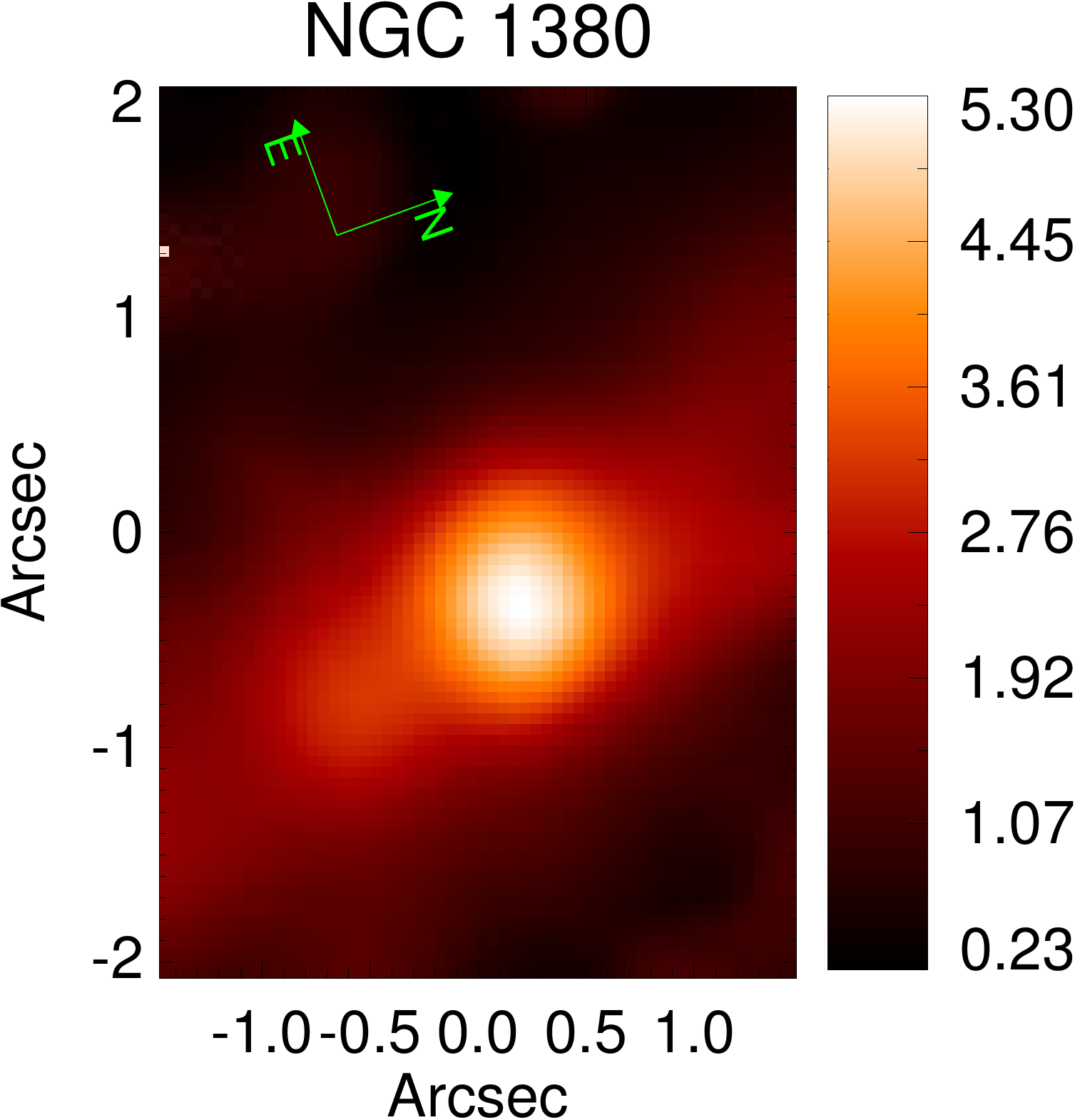}
\hspace{0.5cm}
\vspace{0.2cm}
\includegraphics[scale=0.32]{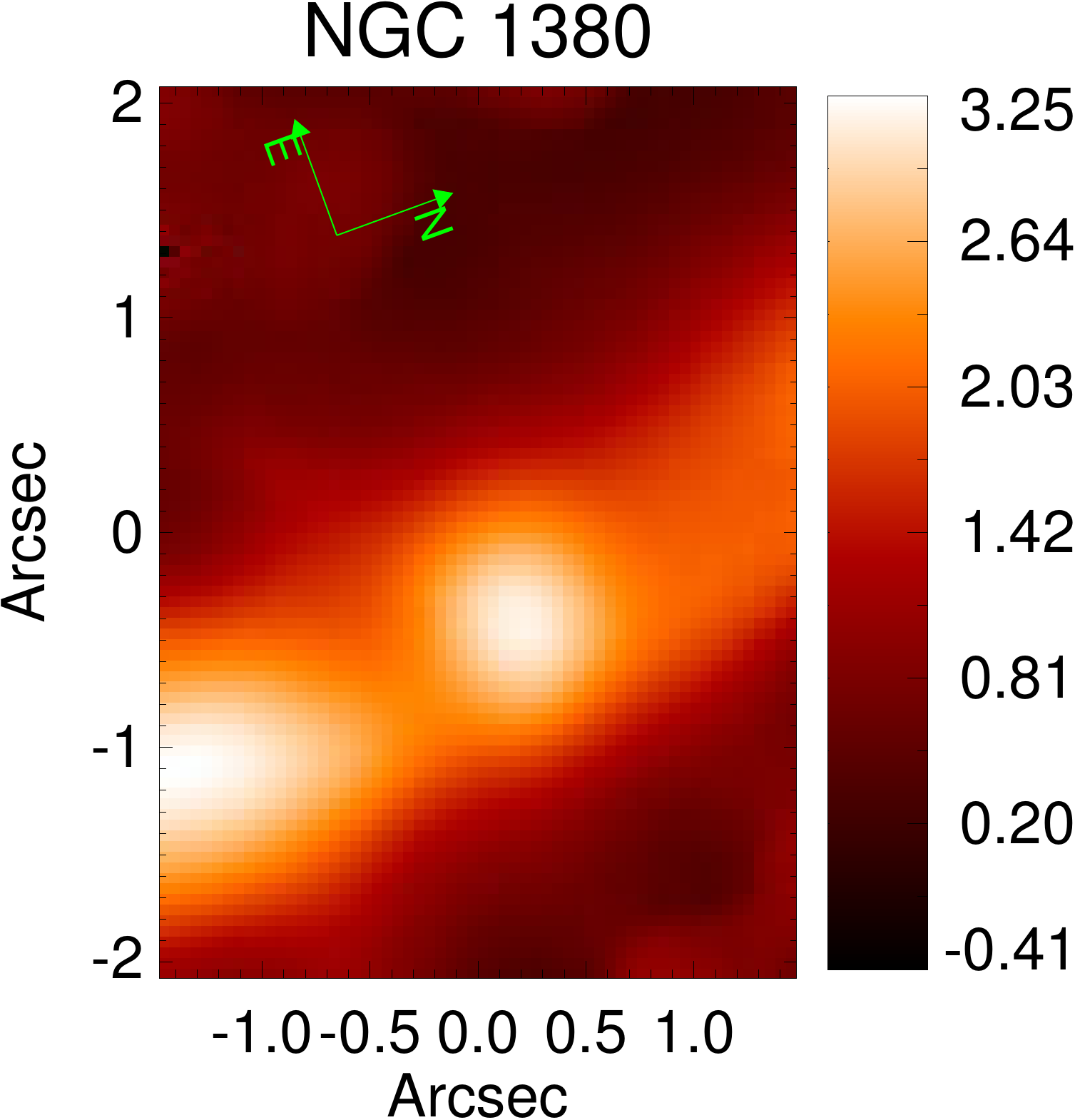}
\hspace{0.2cm}
\includegraphics[scale=0.32]{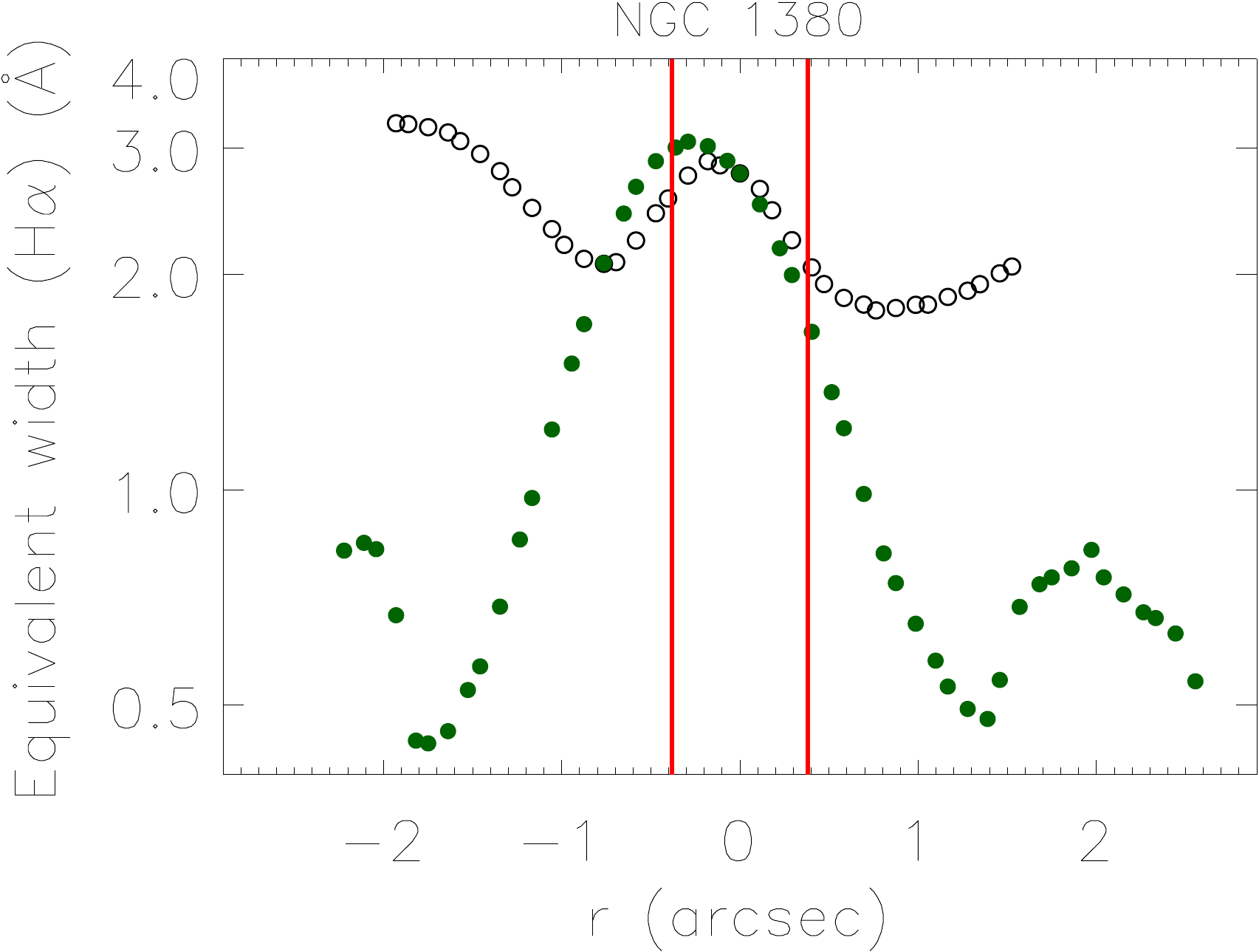}
\hspace{-0.8cm}

\caption{Left: [N II] EW maps, in \AA. Centre: H$\alpha$ EW, in \AA. Right: 1D profiles of the H$\alpha$ EW. The hollow black circles were extracted along the kinematic bipolar structures, while the filled green circles were extracted from the low-velocity emission. The vertical red lines delimit the FWHM of the PSFs of the data cubes. \label{mapa_ew_gal_1}}
\end{figure*}

\addtocounter{figure}{-1}
\addtocounter{subfigure}{1}

\begin{figure*}

\hspace{-1.3cm}
\includegraphics[scale=0.3]{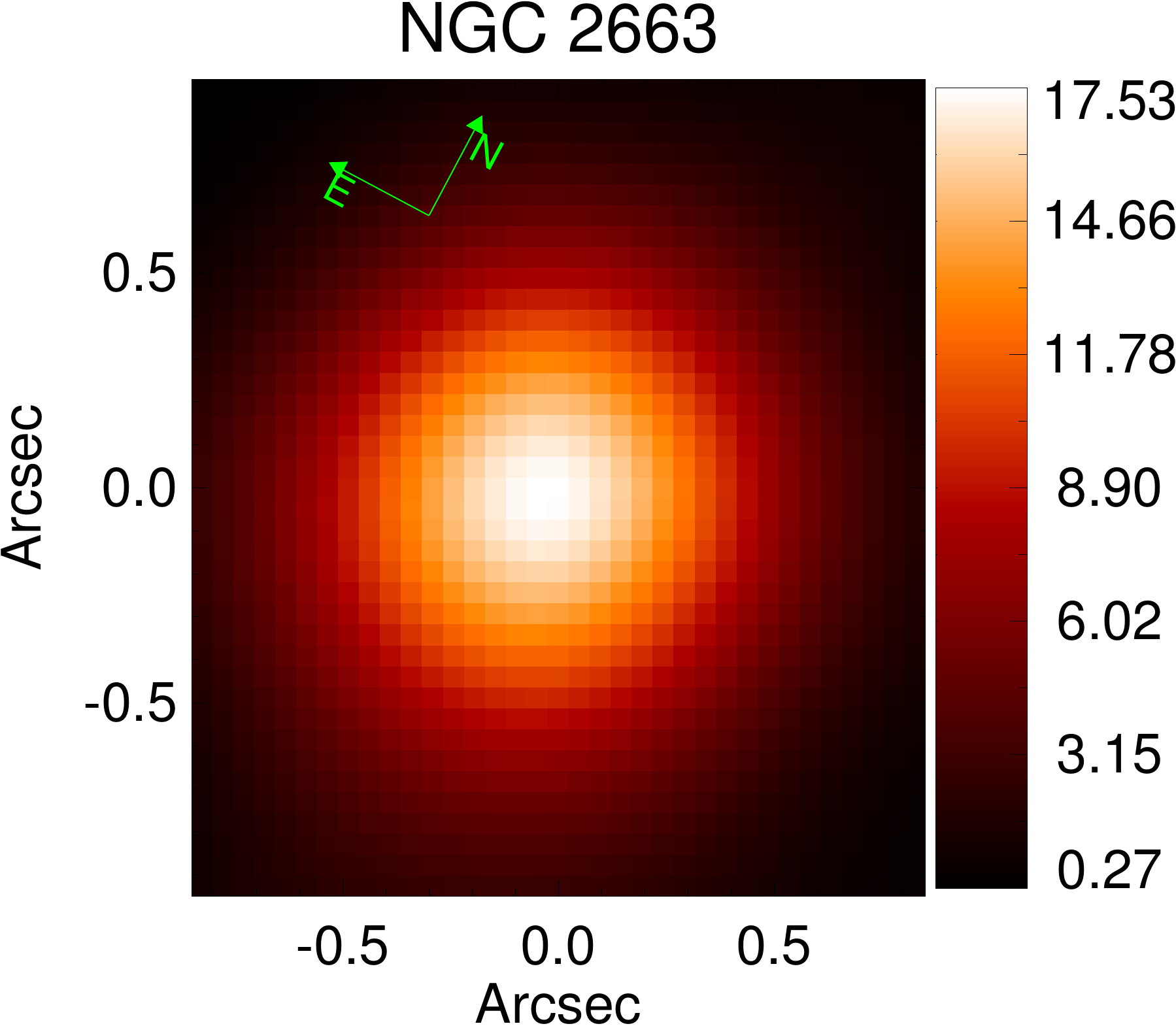}
\hspace{0.45cm}
\vspace{0.2cm}
\includegraphics[scale=0.3]{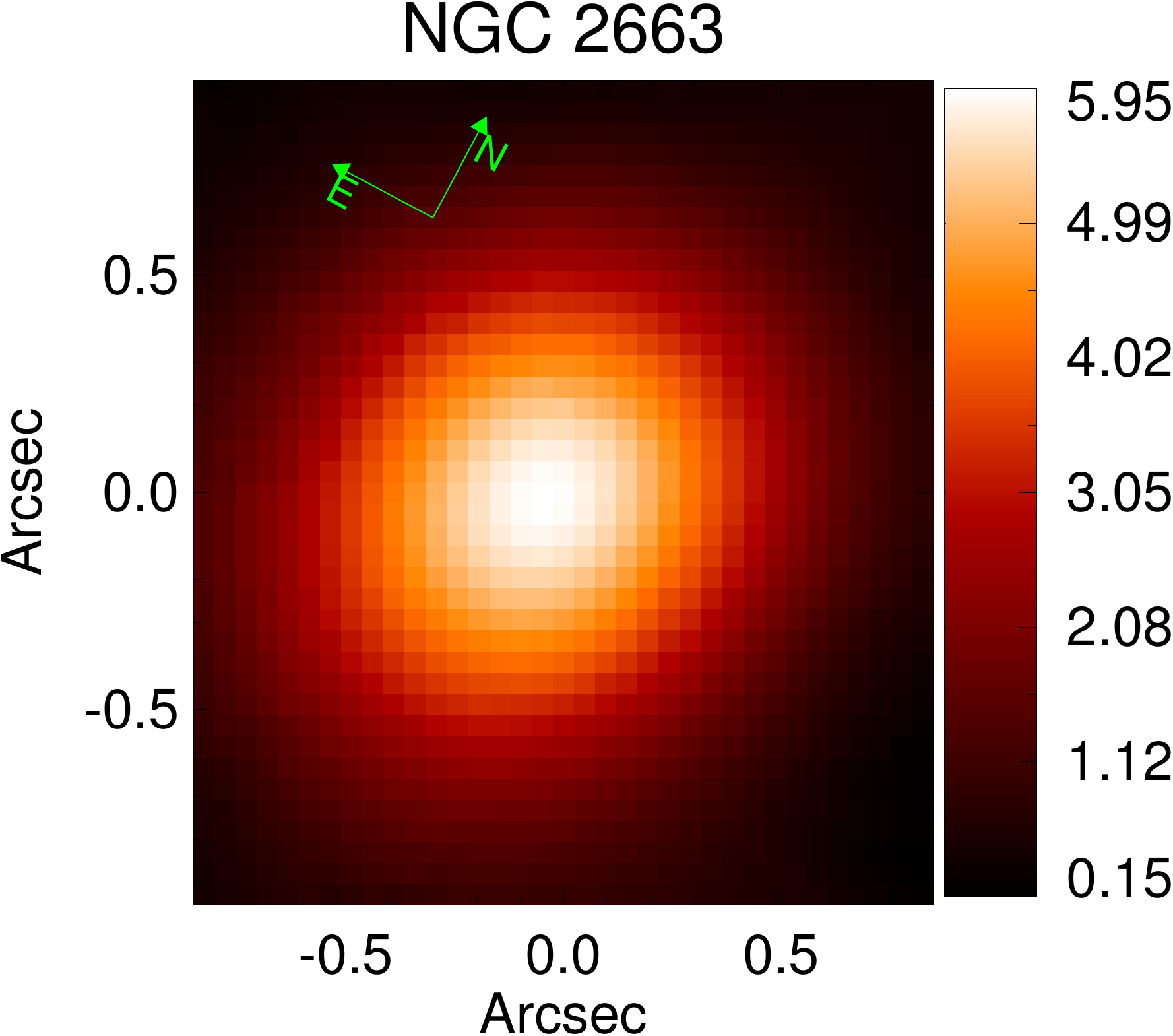}
\hspace{0.45cm}
\includegraphics[scale=0.3]{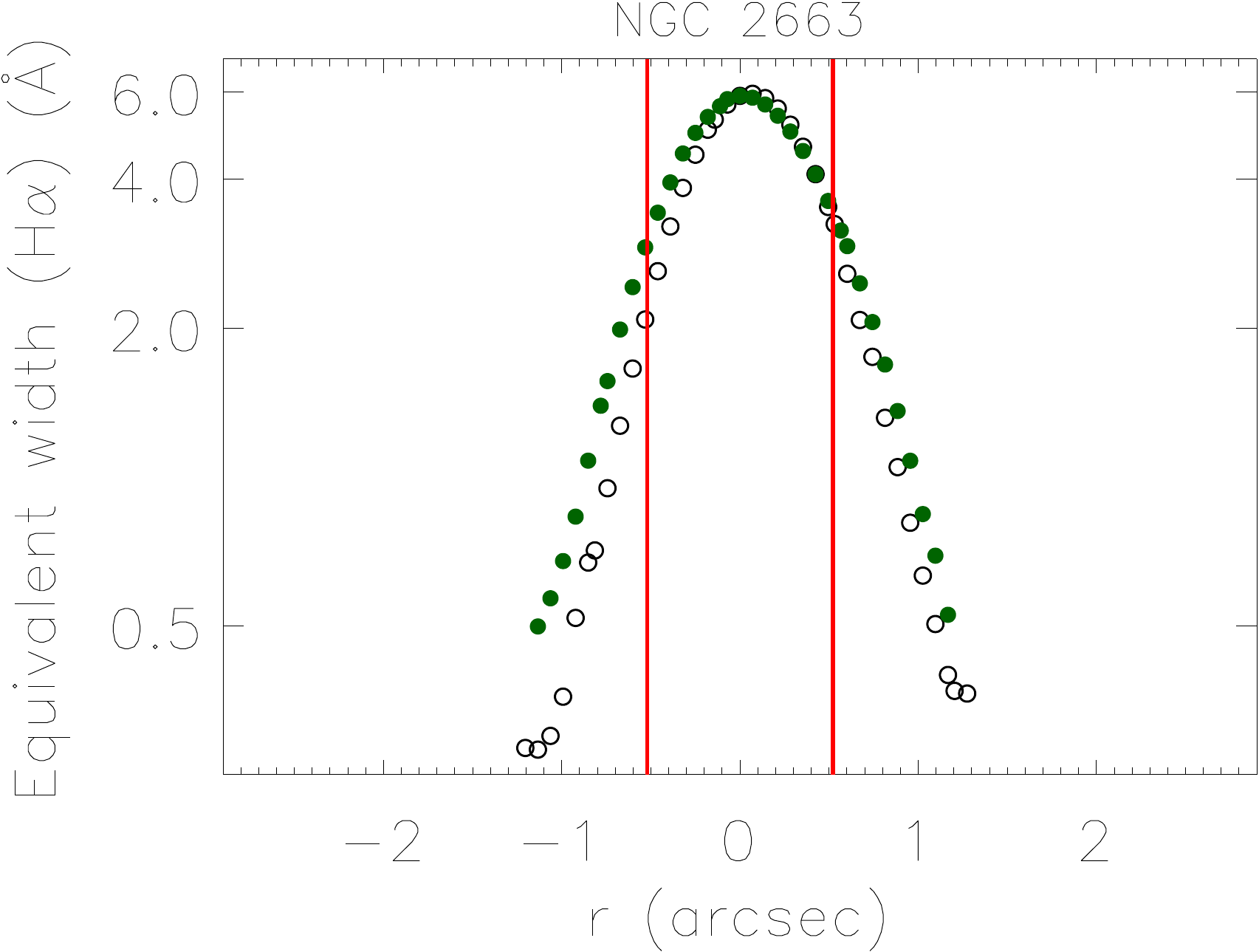}
\hspace{0.45cm}

\hspace{-1.3cm}
\includegraphics[scale=0.32]{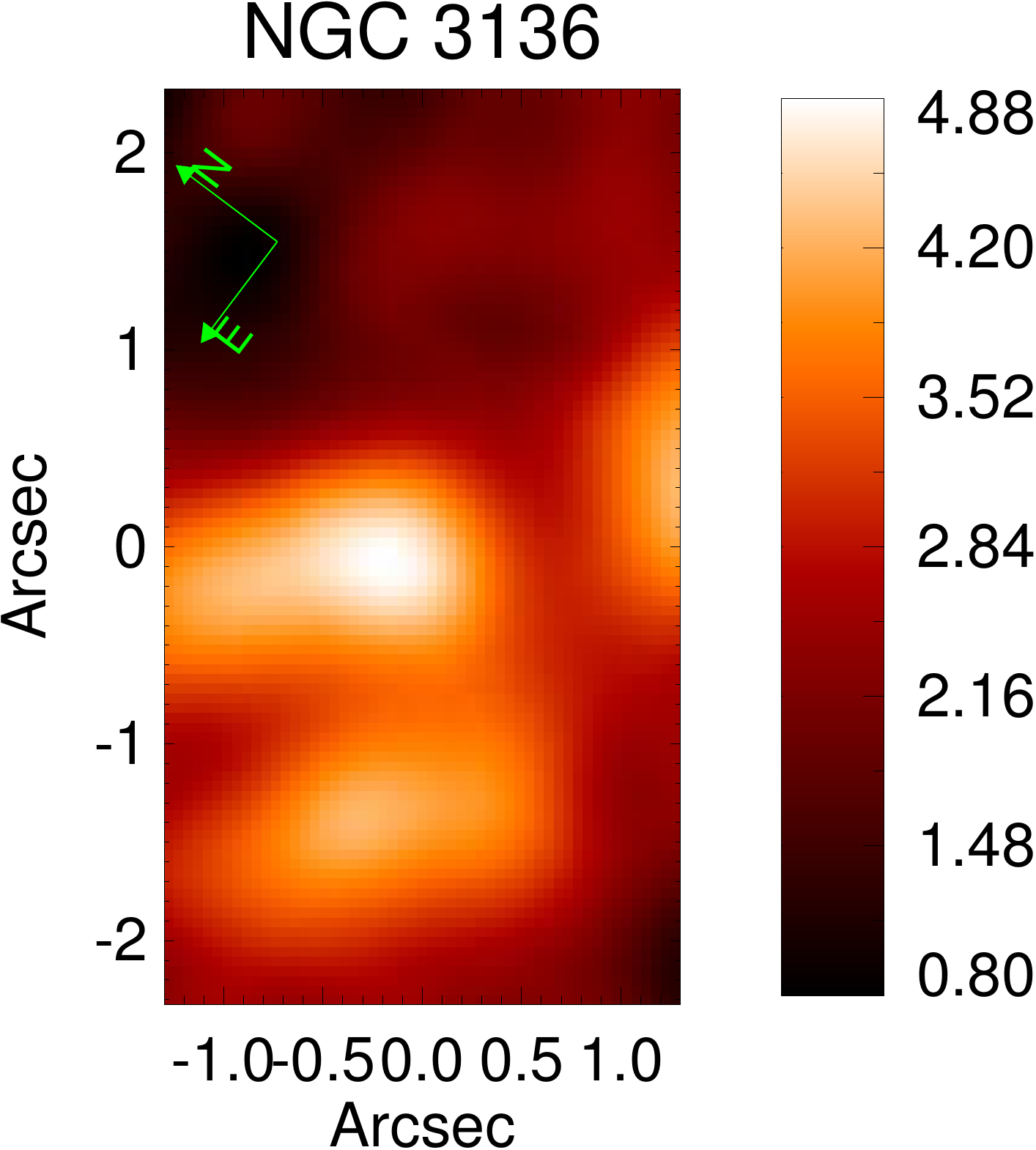}
\hspace{0.5cm}
\vspace{0.2cm}
\includegraphics[scale=0.32]{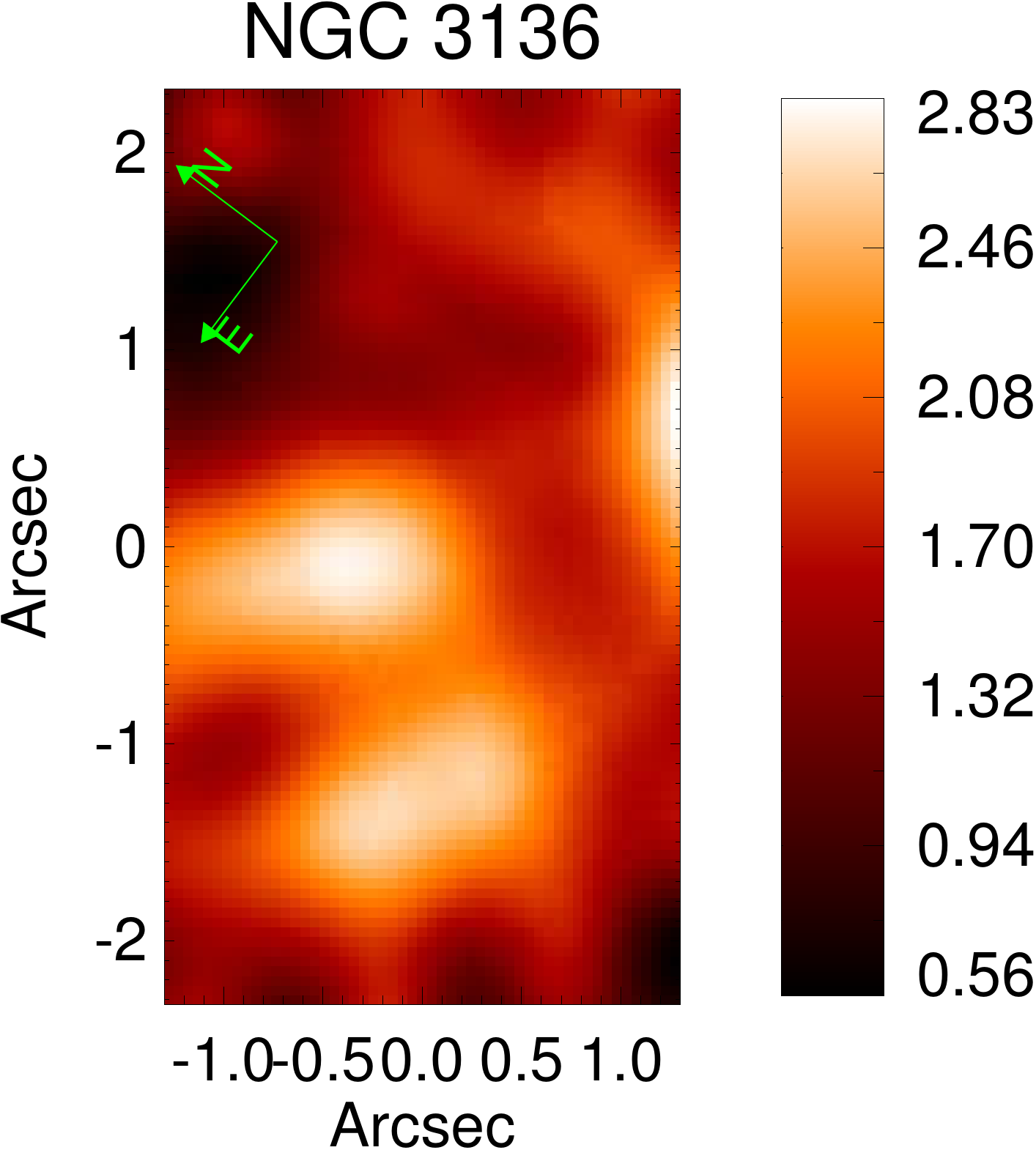}
\hspace{0.2cm}
\includegraphics[scale=0.32]{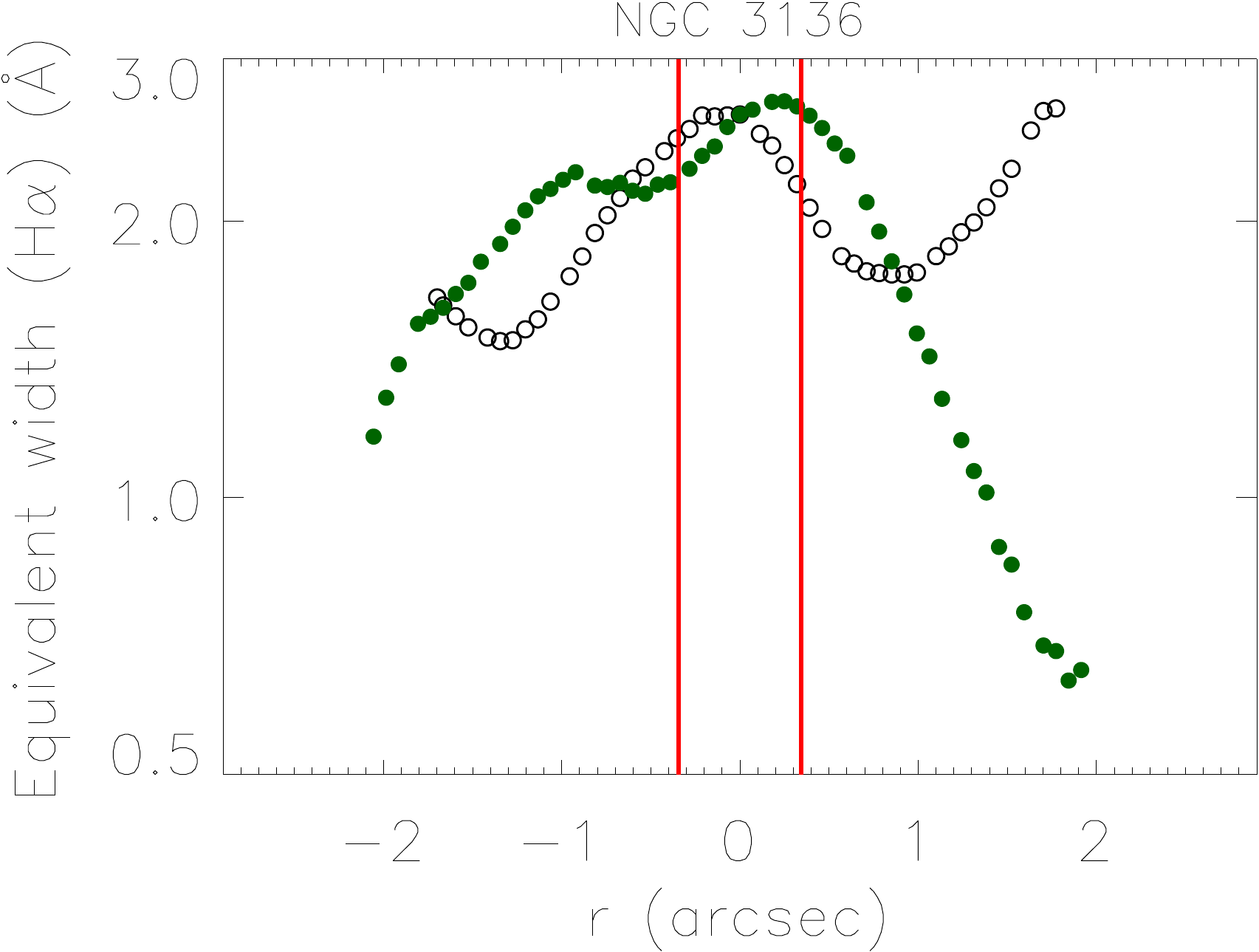}
\hspace{-0.8cm}

\hspace{-1.3cm}
\includegraphics[scale=0.32]{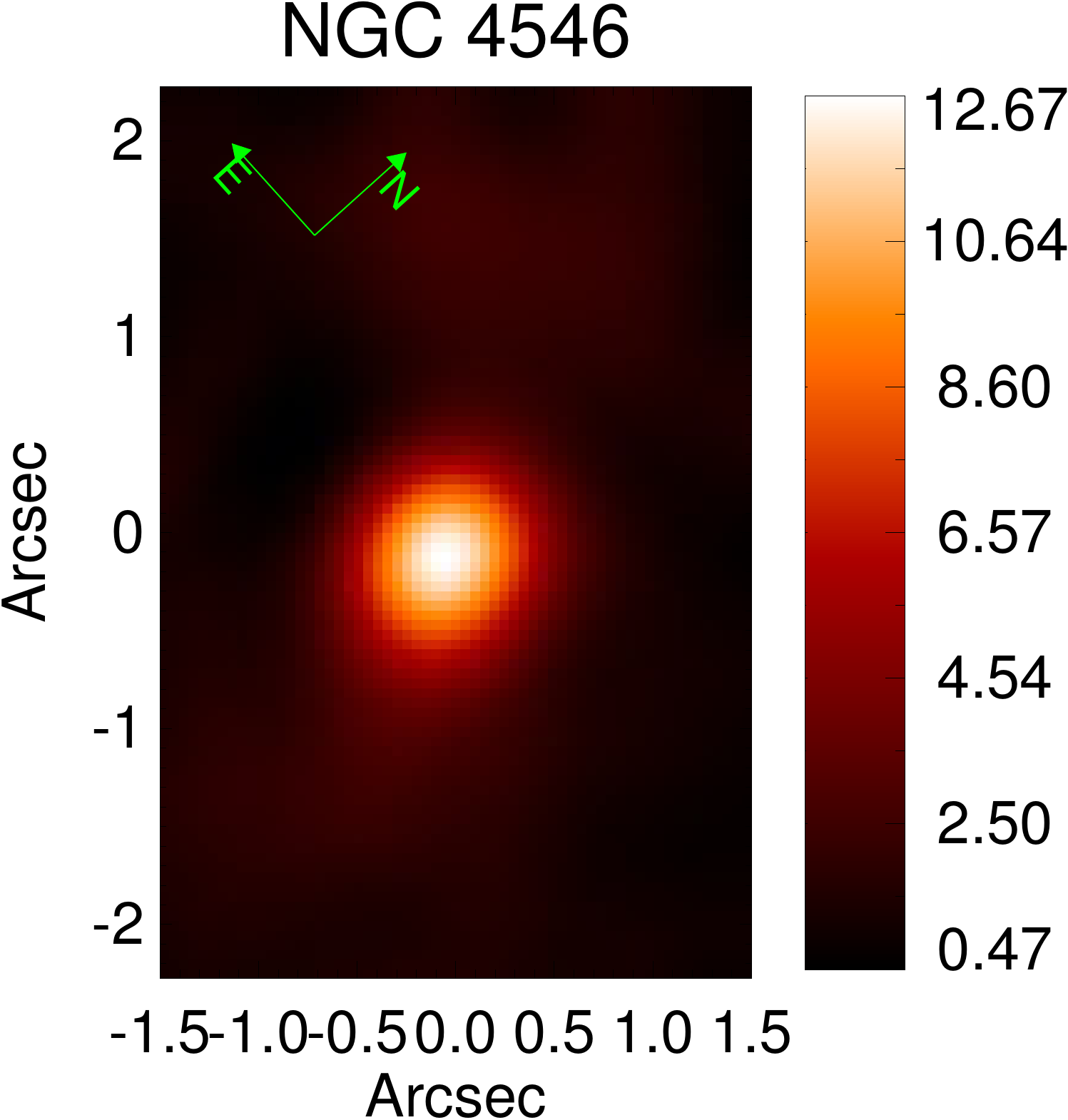}
\hspace{0.5cm}
\vspace{0.2cm}
\includegraphics[scale=0.32]{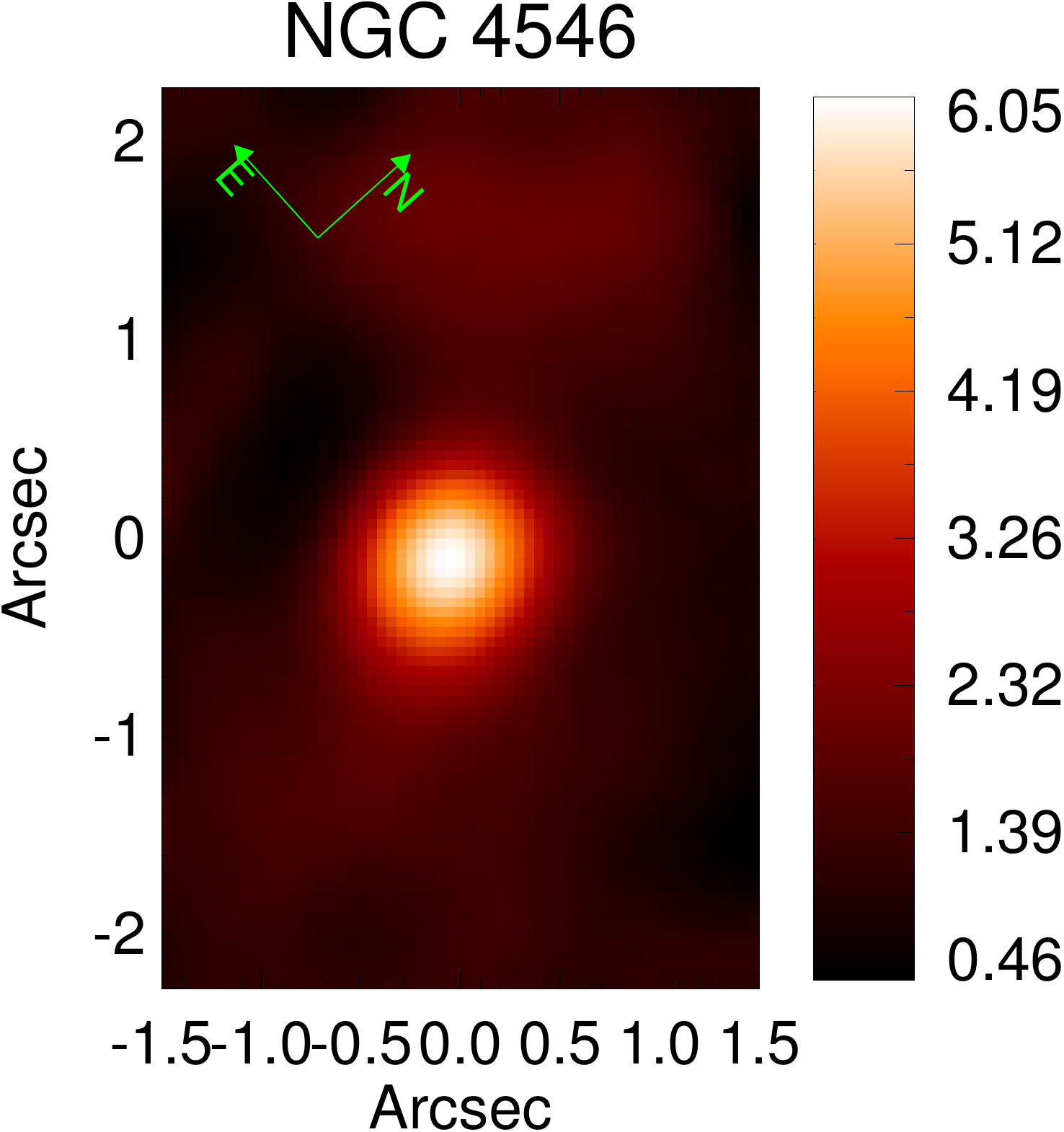}
\hspace{0.2cm}
\includegraphics[scale=0.32]{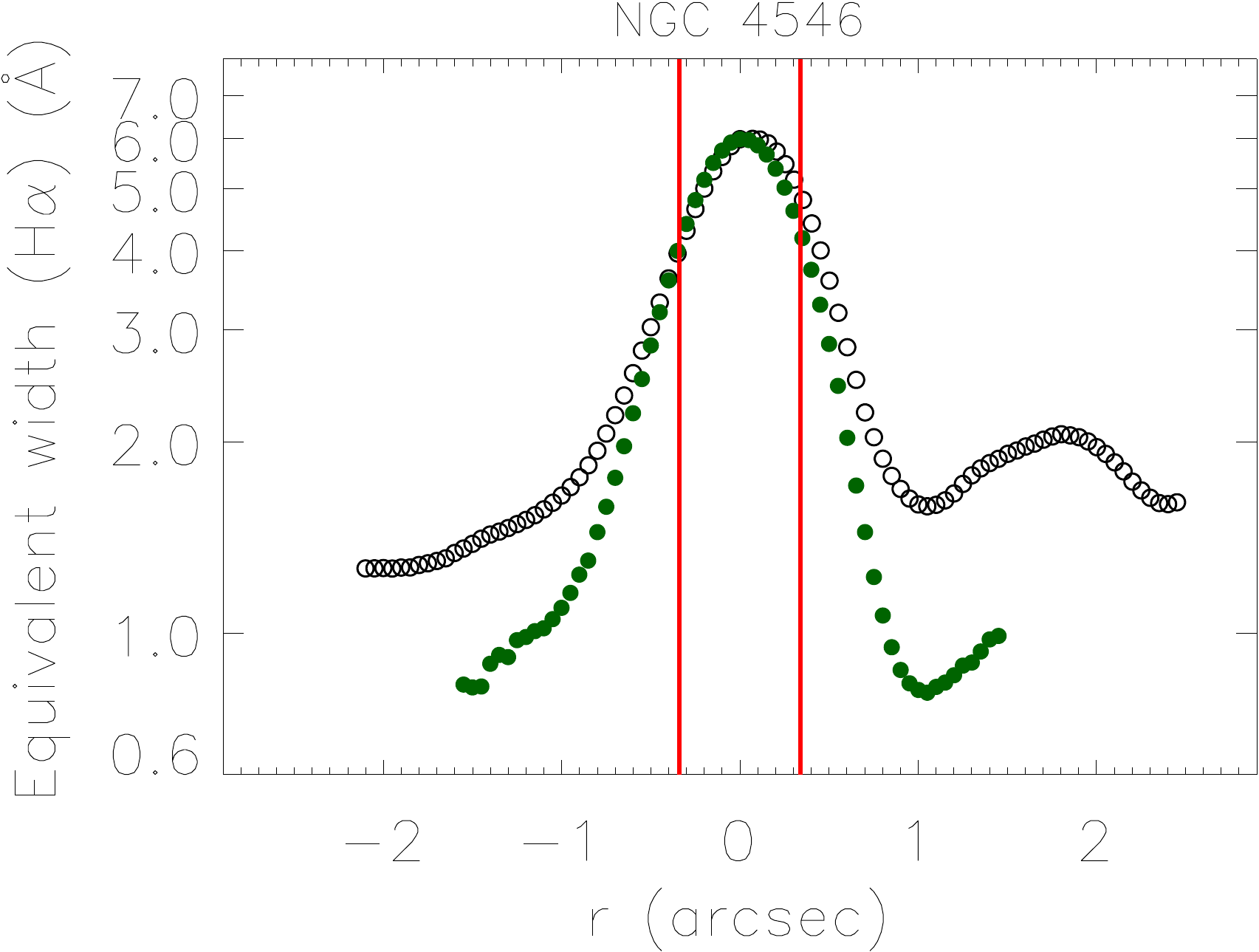}
\hspace{-0.8cm}

\hspace{-1.3cm}
\includegraphics[scale=0.32]{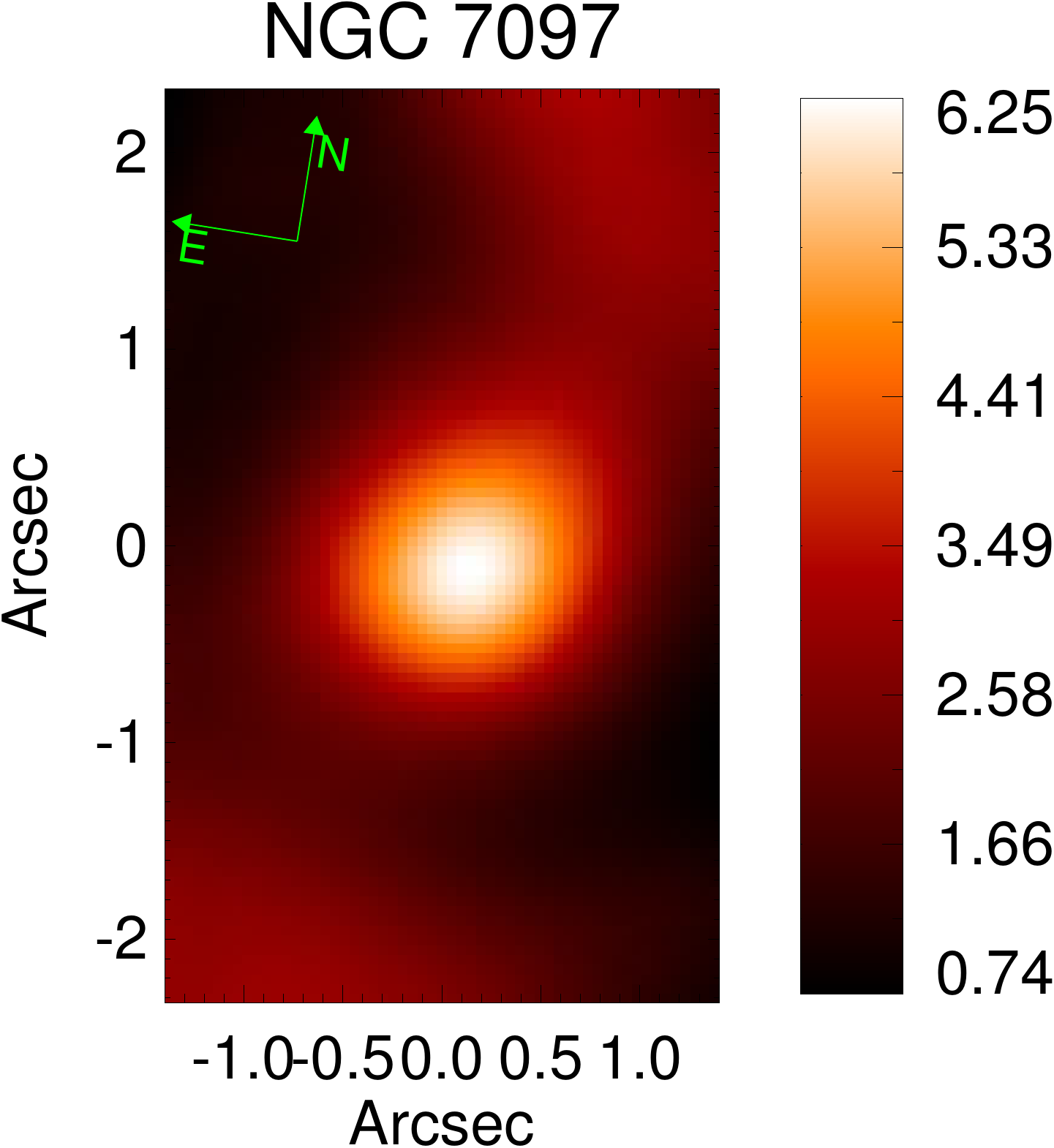}
\hspace{0.5cm}
\vspace{0.2cm}
\includegraphics[scale=0.32]{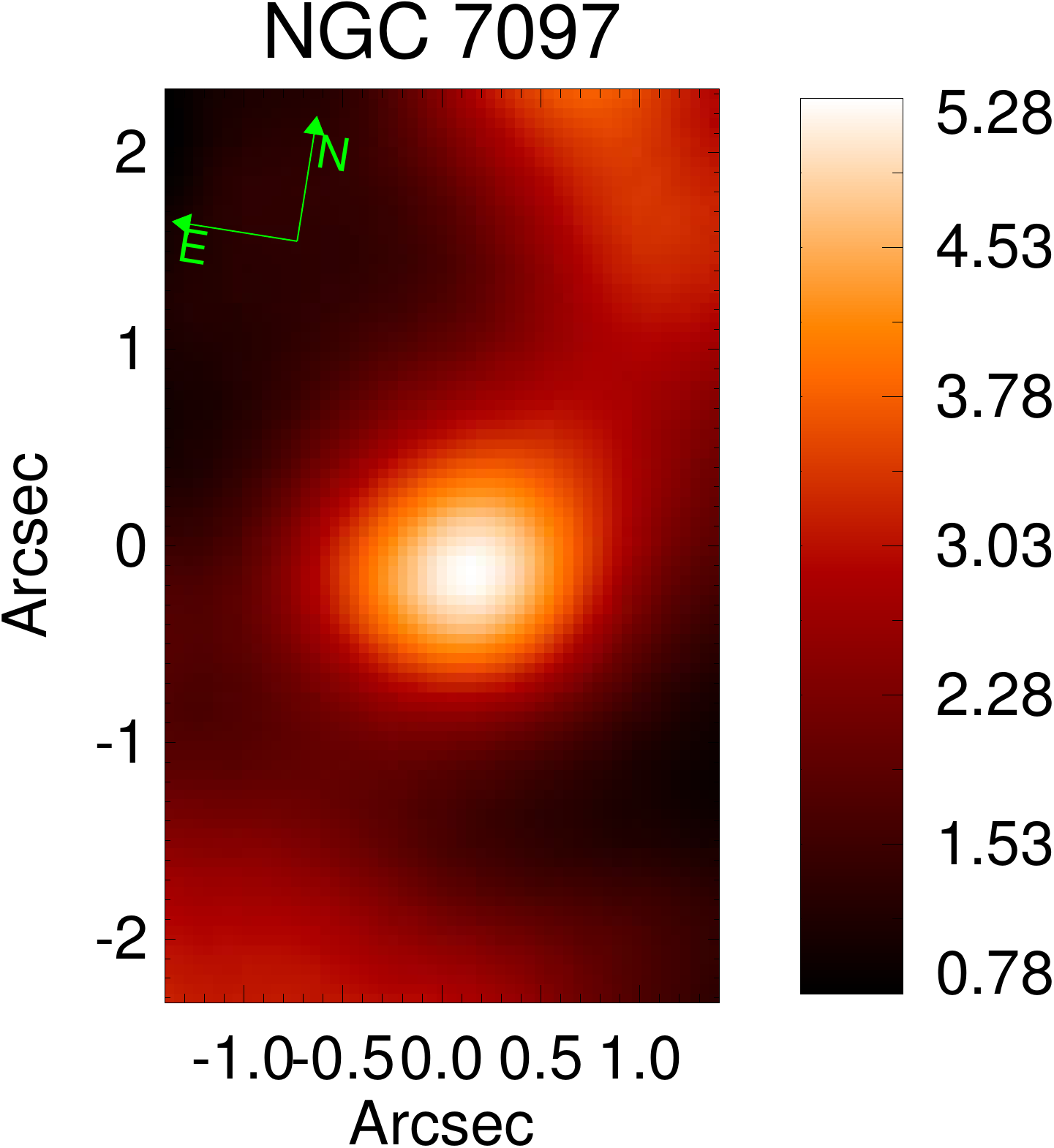}
\hspace{0.2cm}
\includegraphics[scale=0.32]{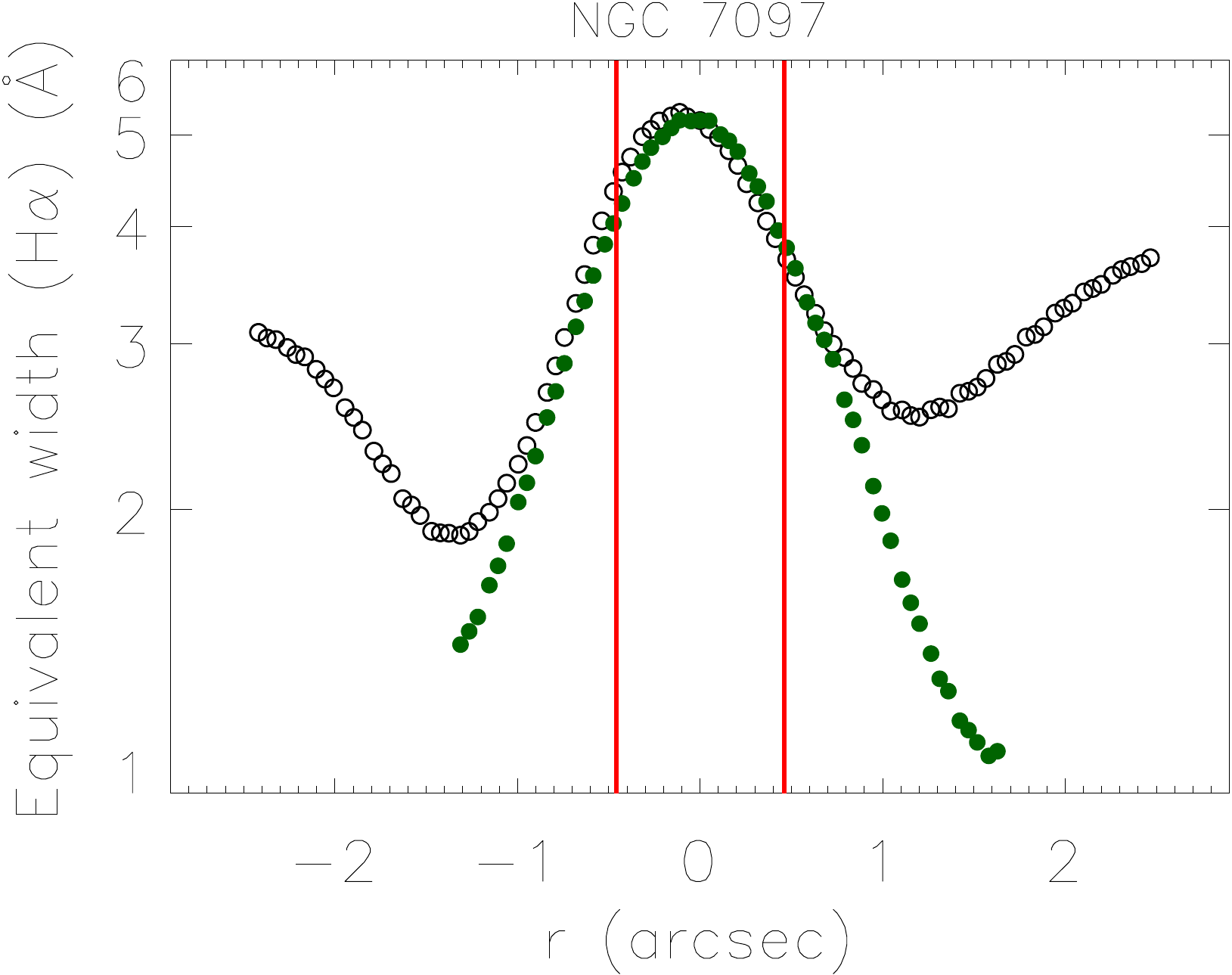}
\hspace{-0.8cm}

\caption{Same as in Fig. \ref{mapa_ew_gal_1}. \label{mapa_ew_gal_2}}
\end{figure*}

\renewcommand{\thefigure}{\arabic{figure}}

In five of the eight galaxies shown in Figs. \ref{mapa_ew_gal_1} and \ref{mapa_ew_gal_2}, the EW(H$\alpha$) is higher along the kinematic bipolar structure than in the low-velocity emission. In NGC 2663, the 1D profiles along both directions have the same behaviour. In all galaxies, the nucleus has the highest values of the EWs of both emission lines, except for NGC 1380, whose case is discussed below. In IC 5181, both EW maps reveal a spiral structure, similar to an integral sign. This structure is revisited in Section \ref{discs_or_cones}.

In NGC 1380, the EW(H$\alpha$) map revealed an object located 1.8 arcsec south from the nucleus. In the same position, we have [N II]/H$\alpha$ $\sim$ 0.6. The EW([N II]) map does not show any structure in this region. These results reinforce the hypothesis that an H II region is present at this position. At 1.8 arcsec northward from the nucleus, we detected an extended object in the EW(H$\alpha$) map. The [N II]/H$\alpha$ ratio values in this position also suggest an H II region. In addition to these star-forming regions, the EW([N II]) map shows a compact object located at 0.7 arcsec south from the nucleus, in the same position where [N II]/H$\alpha$ $\sim$ 1.5. In the EW(H$\alpha$) map, this object is not seen. This may be the consequence of a second LINER located in the central region of NGC 1380. This discussion is revisited in Section \ref{hst_nuclear} and also in a future paper.

Maps of the EW of both lines of NGC 3136 show at least three extended structures. One corresponds to the structure detected in the flux maps of both lines, which may be related to two compact objects detected with PCA Tomography (paper I). The other two structures, one located 1.4 arcsec southeastward from the nucleus and the other one located 1.5 arcsec southwestward from the nucleus have [N II]/H$\alpha$ $\sim$ 1.5 and $\sim$ 1.3, respectively. Both ratios are typical of LINERs.

\subsection{Density maps} \label{densitymapsection}

Following procedures in Section \ref{FOV_properties}, we used the [S II]$\lambda$6716 / [S II]$\lambda$6731 ratio to calculate the electron density in each spaxel of the gas cube. The [S II] fluxes were extracted using the same procedure described in Section \ref{Halpha_NII_flux_maps}. However, only the amplitudes of the Gaussian functions were assumed as free parameters. Radial velocity and velocity dispersion were constrained to the values found for the [N II] and H$\alpha$ emission lines. The electron density was calculated by means of the relation proposed by \citet{2014A&A...561A..10P}, assuming $T_e$ = 10000K. Note that in Section \ref{FOV_properties} we used the {\sc NEBULAR} package under the {\sc IRAF} environment to calculate $n_e$. As we used this package to estimate the nuclear densities in paper II, we were able to compare the densities of both the nuclear and circumnuclear regions. On the other hand, \citet{2014A&A...561A..10P} proposed an empirical numerical function to calculate $n_e$, which was easier to implement in the code that we used to extract the flux from the emission lines. Within the errors, both calculation methods provide the same results. The density maps are shown in Figs. \ref{perfil_dens_1} and \ref{perfil_dens_2}.

\renewcommand{\thefigure}{\arabic{figure}\alph{subfigure}}
\setcounter{subfigure}{1}

\begin{figure*}
\begin{center}
\includegraphics[scale=0.35]{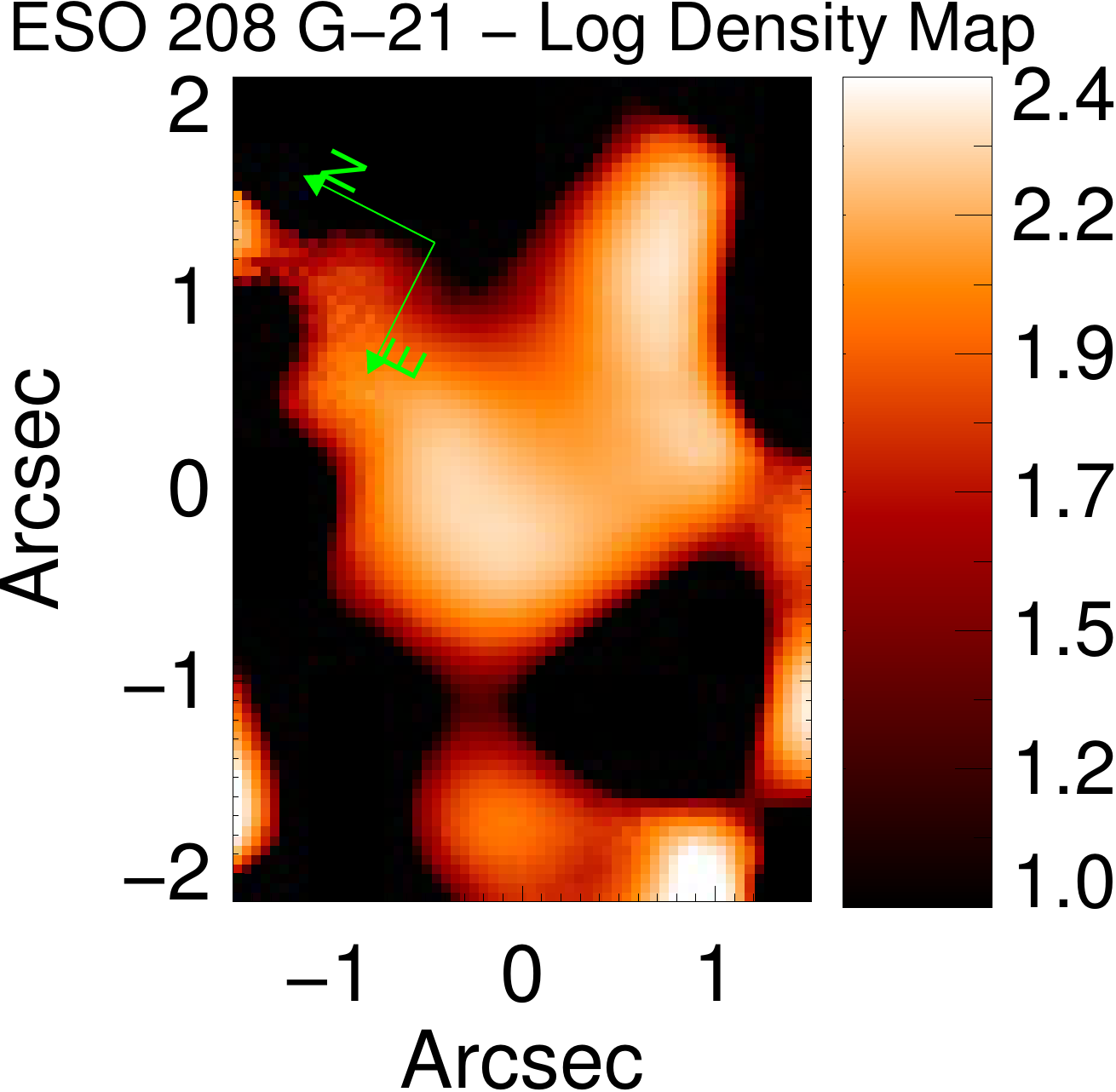}
\vspace{1cm}
\hspace{0.5cm}
\includegraphics[scale=0.35]{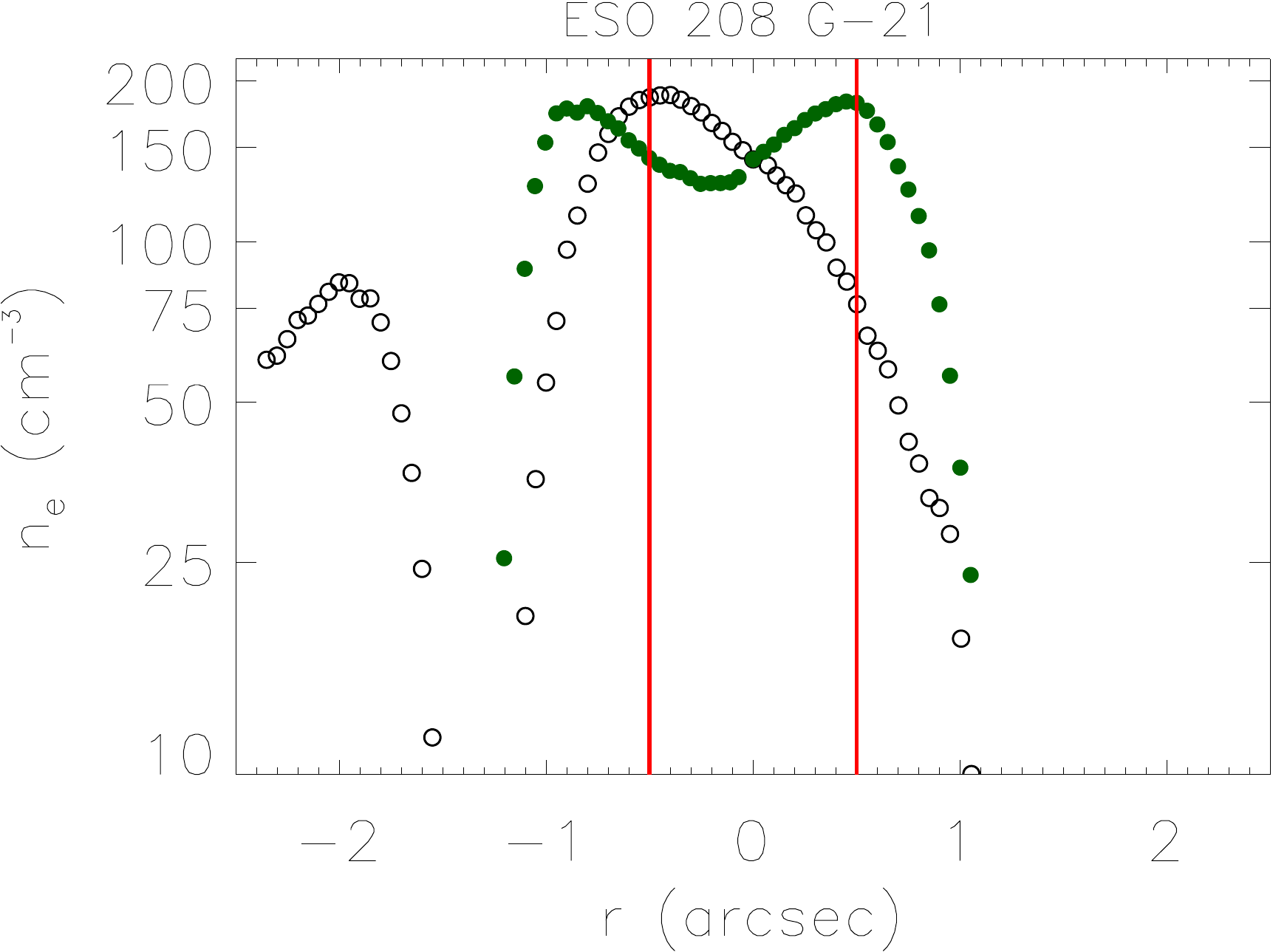}
\hspace{0.5cm}

\includegraphics[scale=0.35]{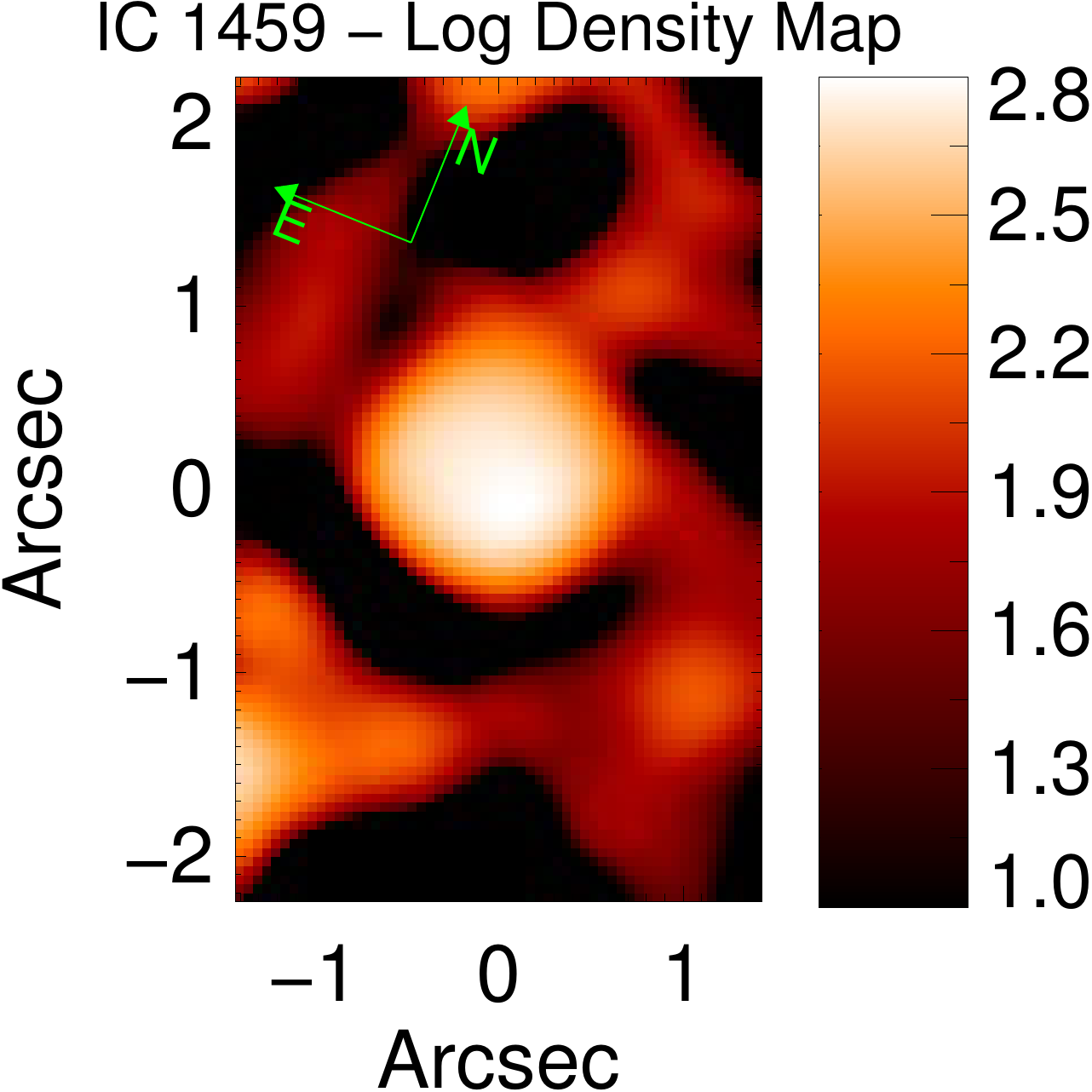}
\vspace{1cm}
\hspace{0.5cm}
\includegraphics[scale=0.35]{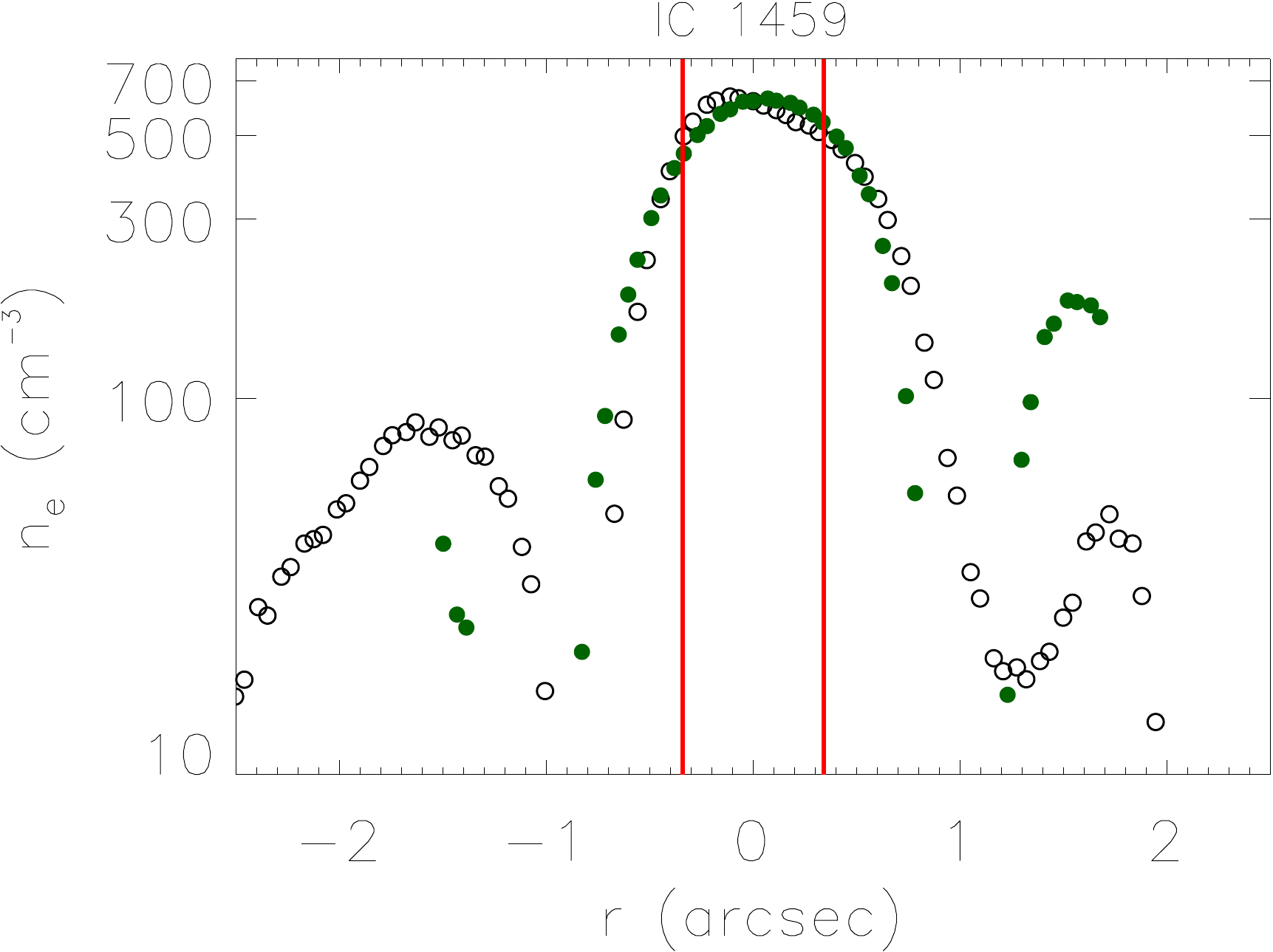}
\hspace{0.5cm}

\includegraphics[scale=0.35]{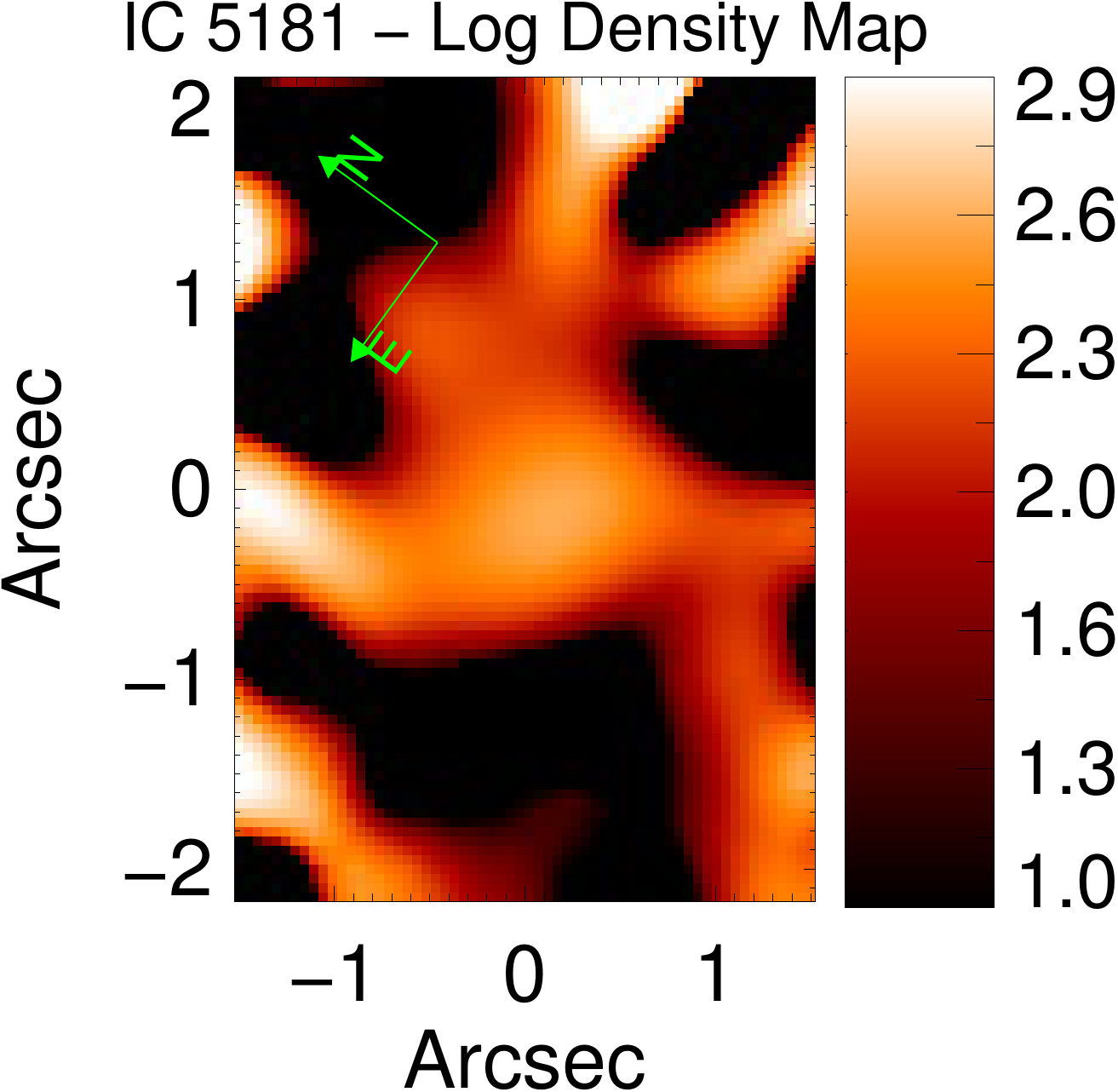}
\vspace{1cm}
\hspace{0.5cm}
\includegraphics[scale=0.35]{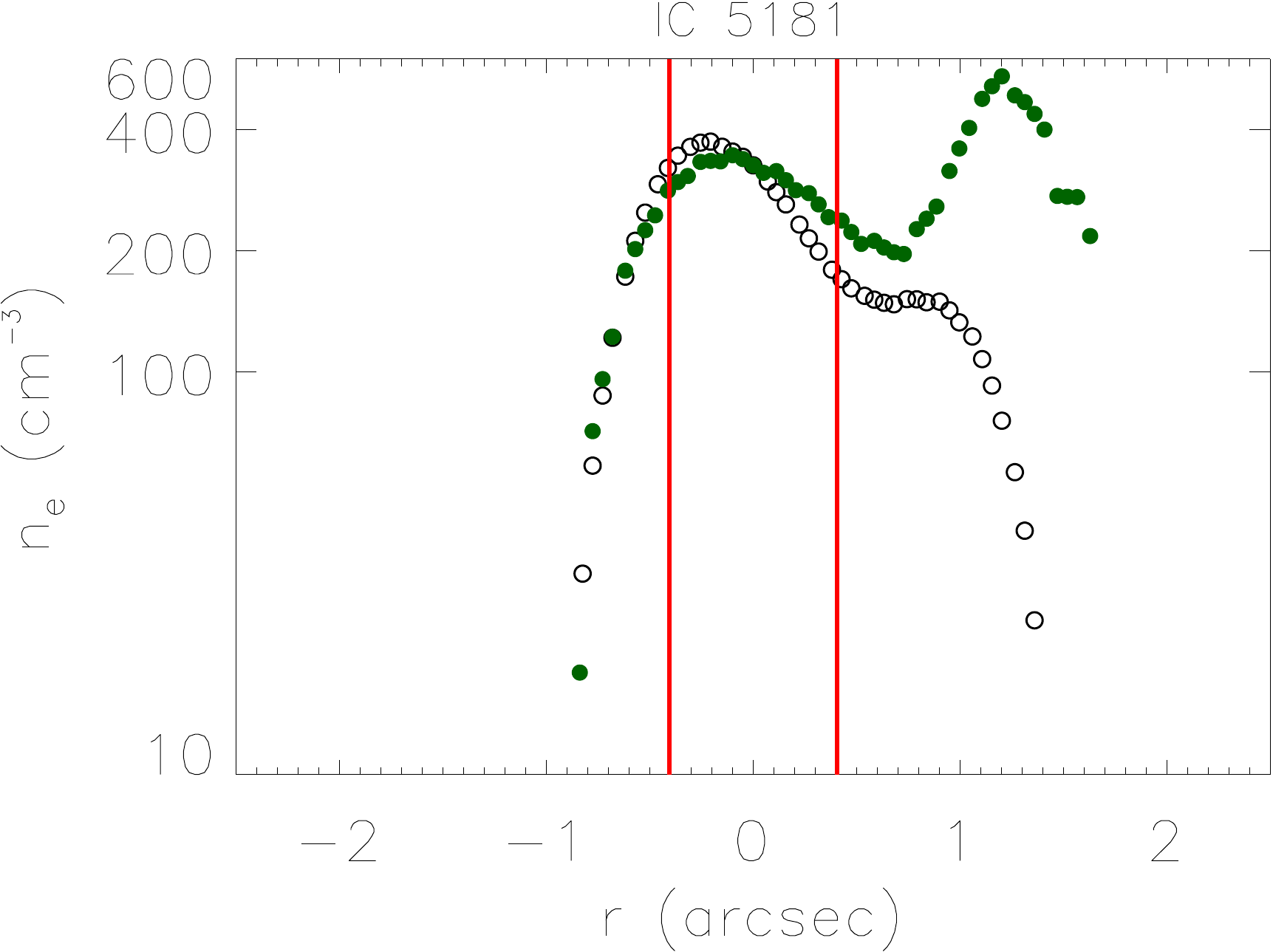}
\hspace{0.5cm}

\includegraphics[scale=0.35]{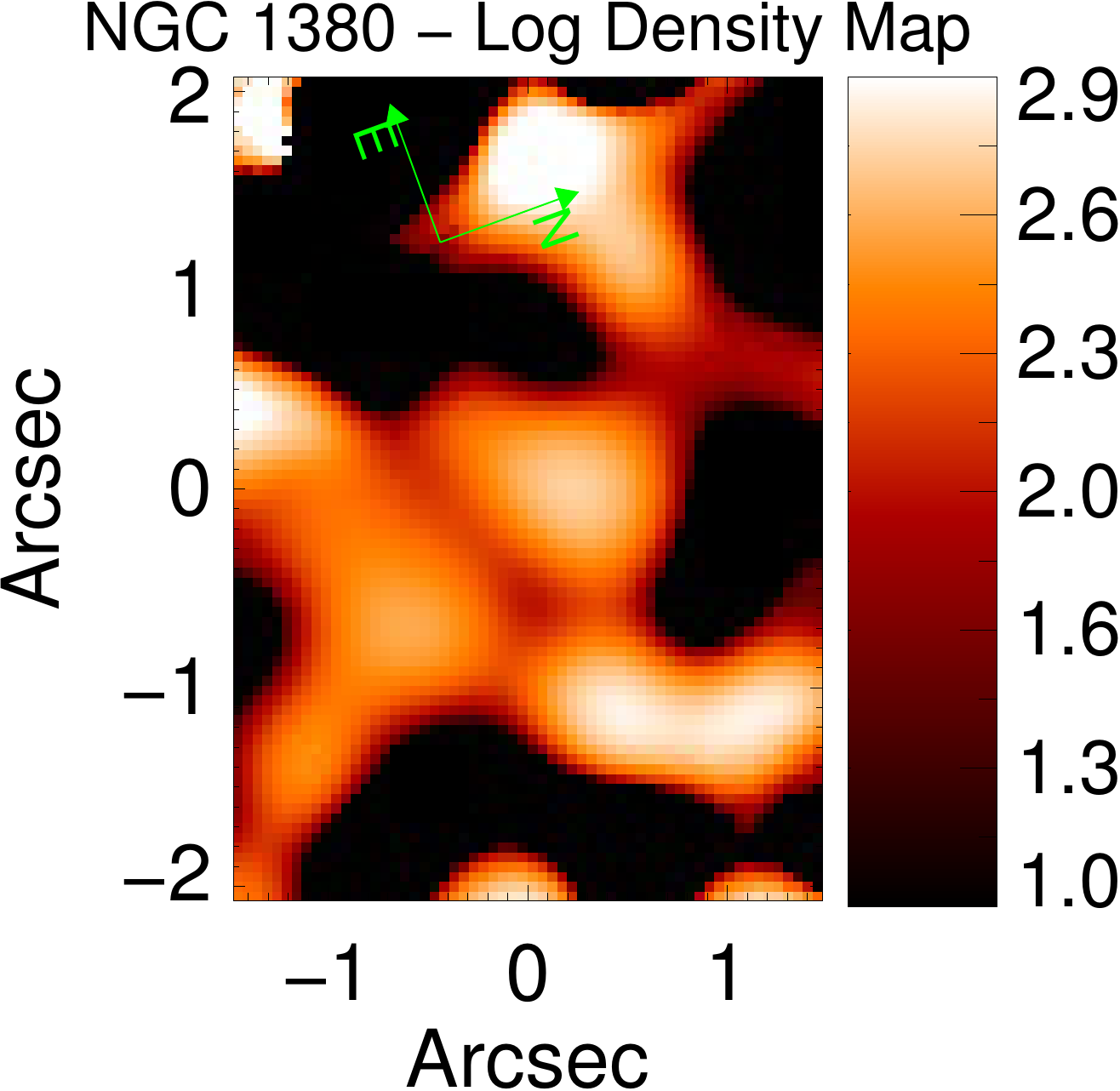}
\vspace{1cm}
\hspace{0.5cm}
\includegraphics[scale=0.35]{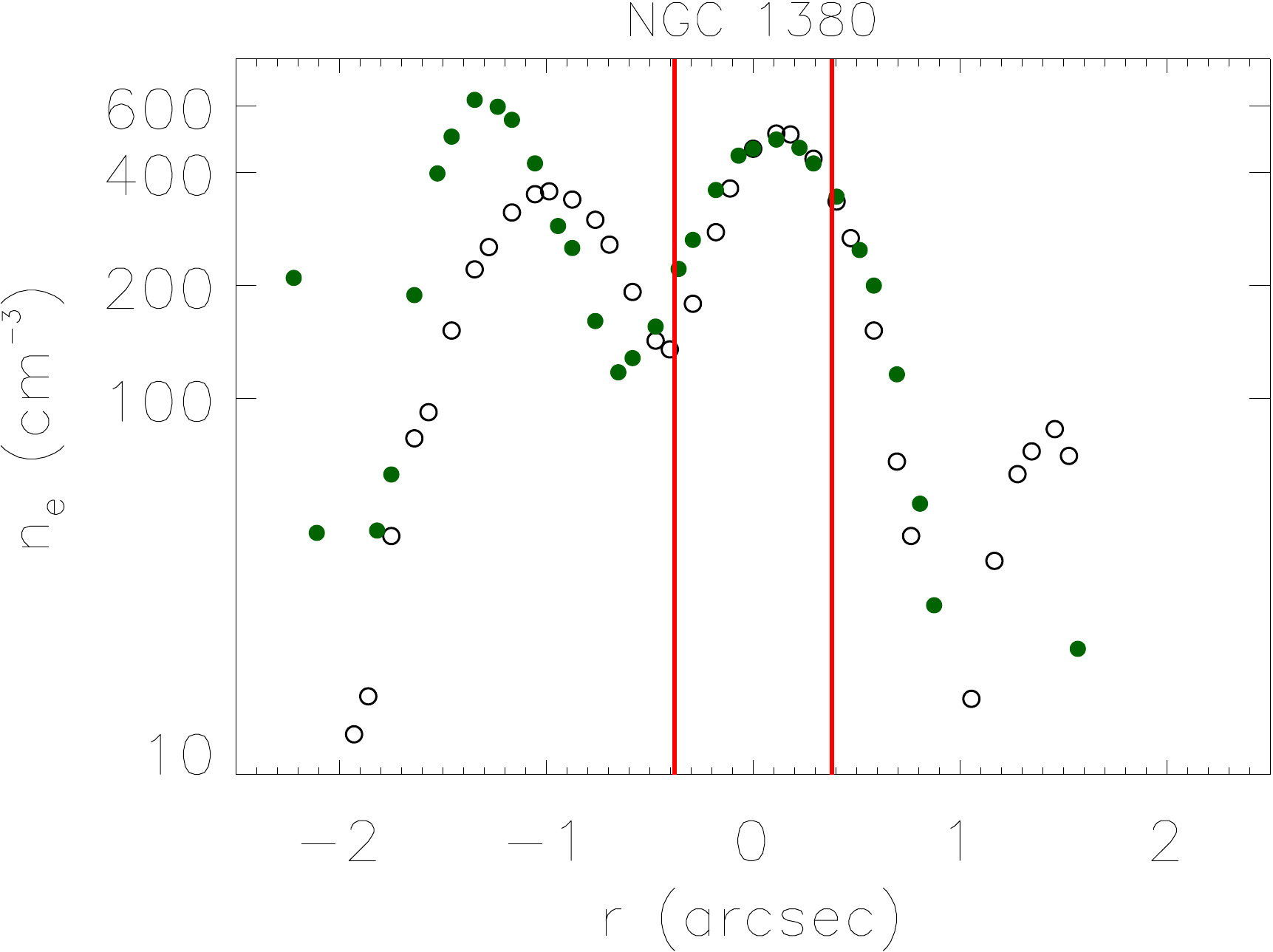}
\hspace{0.5cm}

\caption{Left: Density maps in cm$^{-3}$. Right: 1D profiles of the density along the parallel and perpendicular directions of bipolar gas structure. The hollow black circles were extracted along the kinematic bipolar structures, while the filled green circles were extracted from the low-velocity emission. The vertical red lines delimit the FWHM of the PSFs of the data cubes. \label{perfil_dens_1}} 
\end{center}
\end{figure*}

\addtocounter{figure}{-1}
\addtocounter{subfigure}{1}

\begin{figure*}
\begin{center}
\includegraphics[scale=0.35]{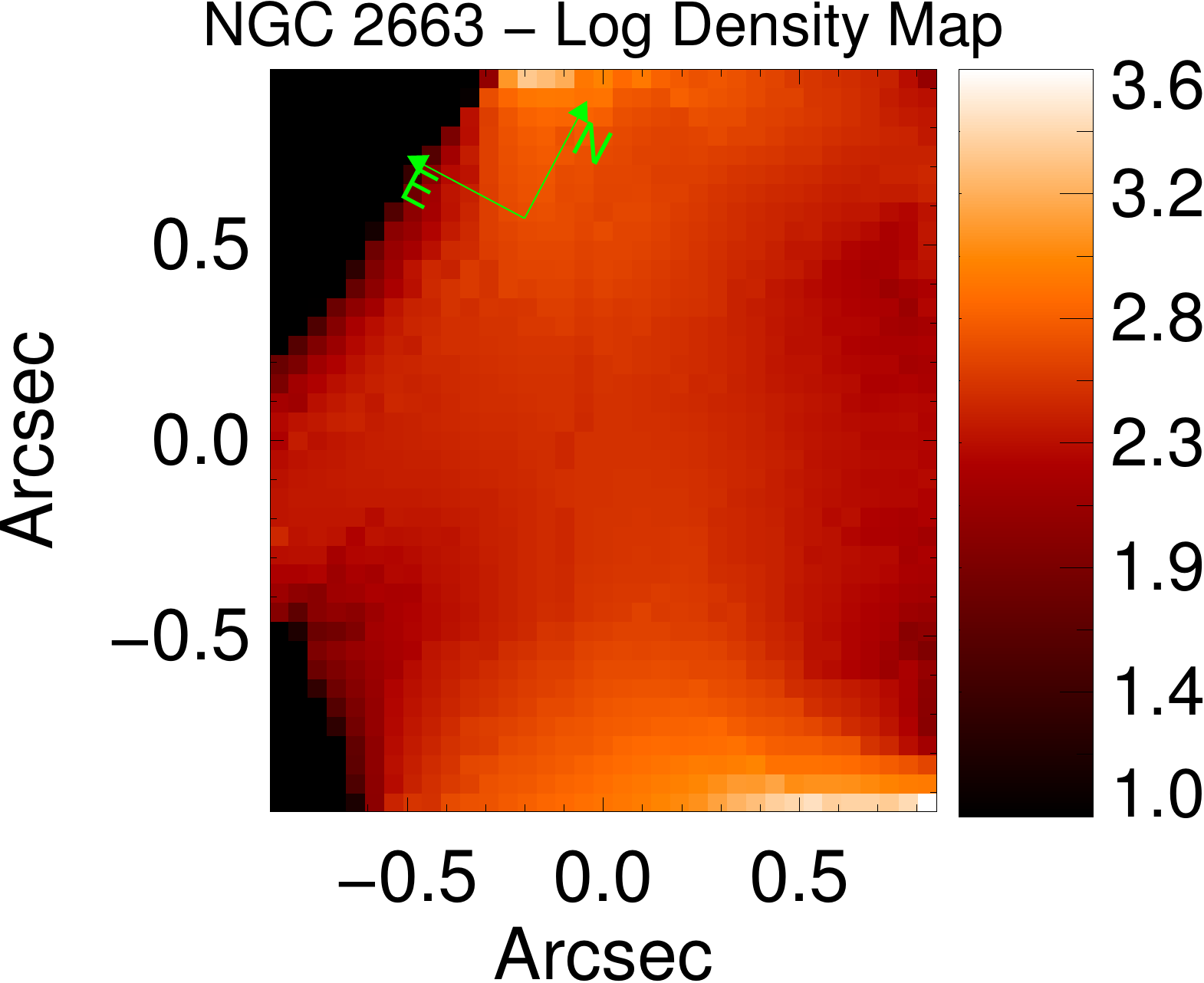}
\vspace{1cm}
\hspace{1cm}
\includegraphics[scale=0.35]{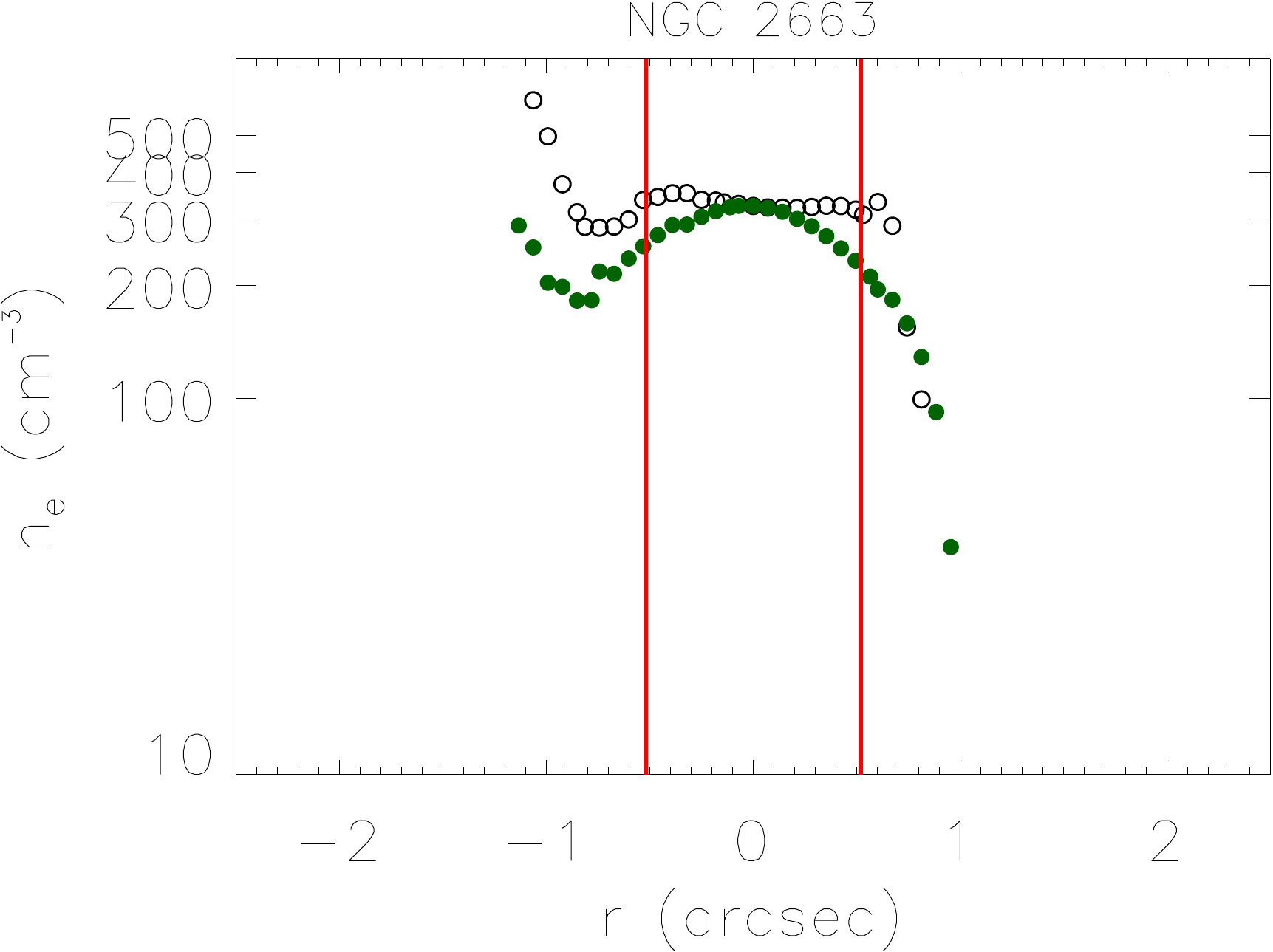}
\hspace{1.5cm}

\includegraphics[scale=0.35]{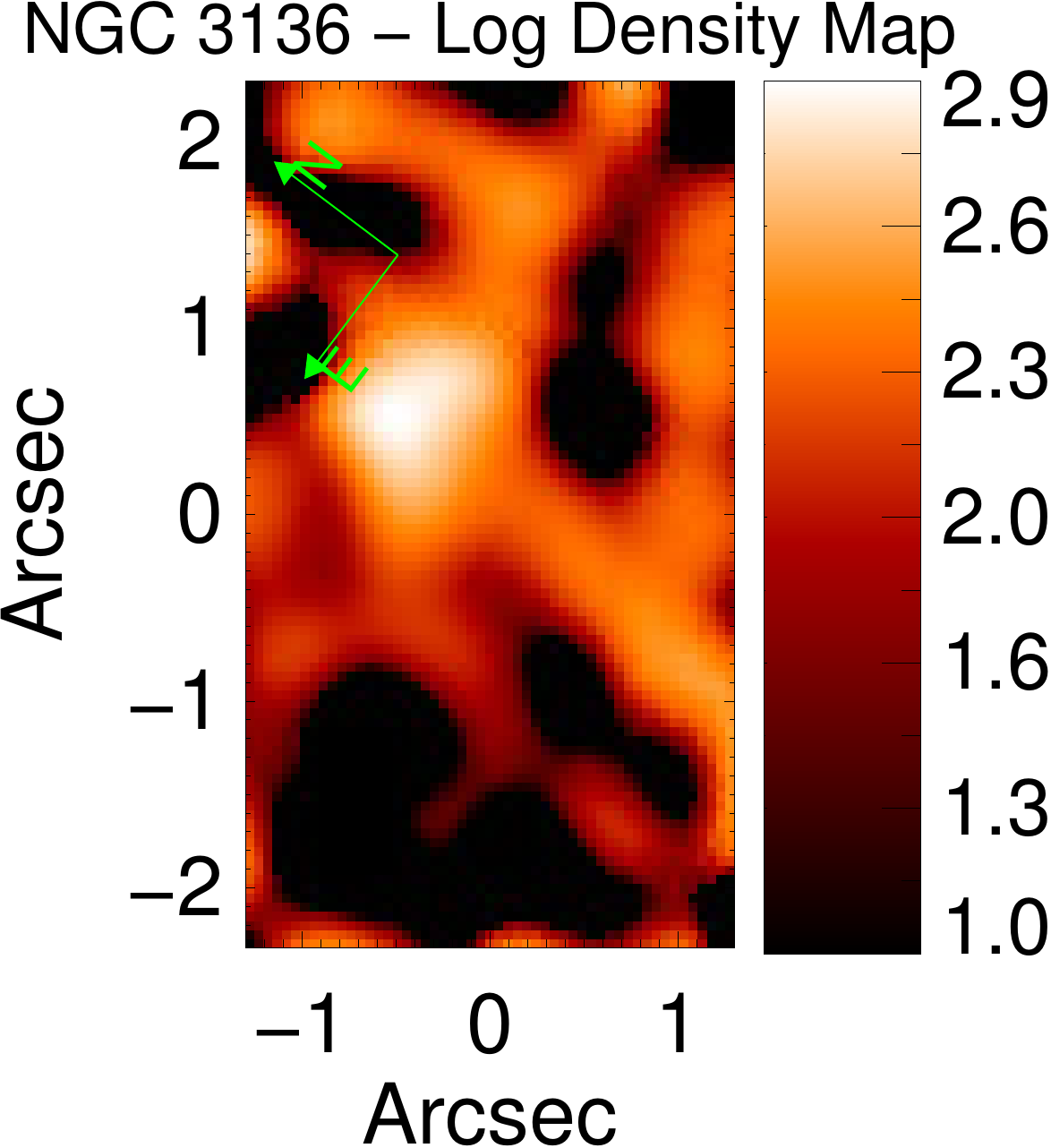}
\vspace{1cm}
\hspace{0.5cm}
\includegraphics[scale=0.35]{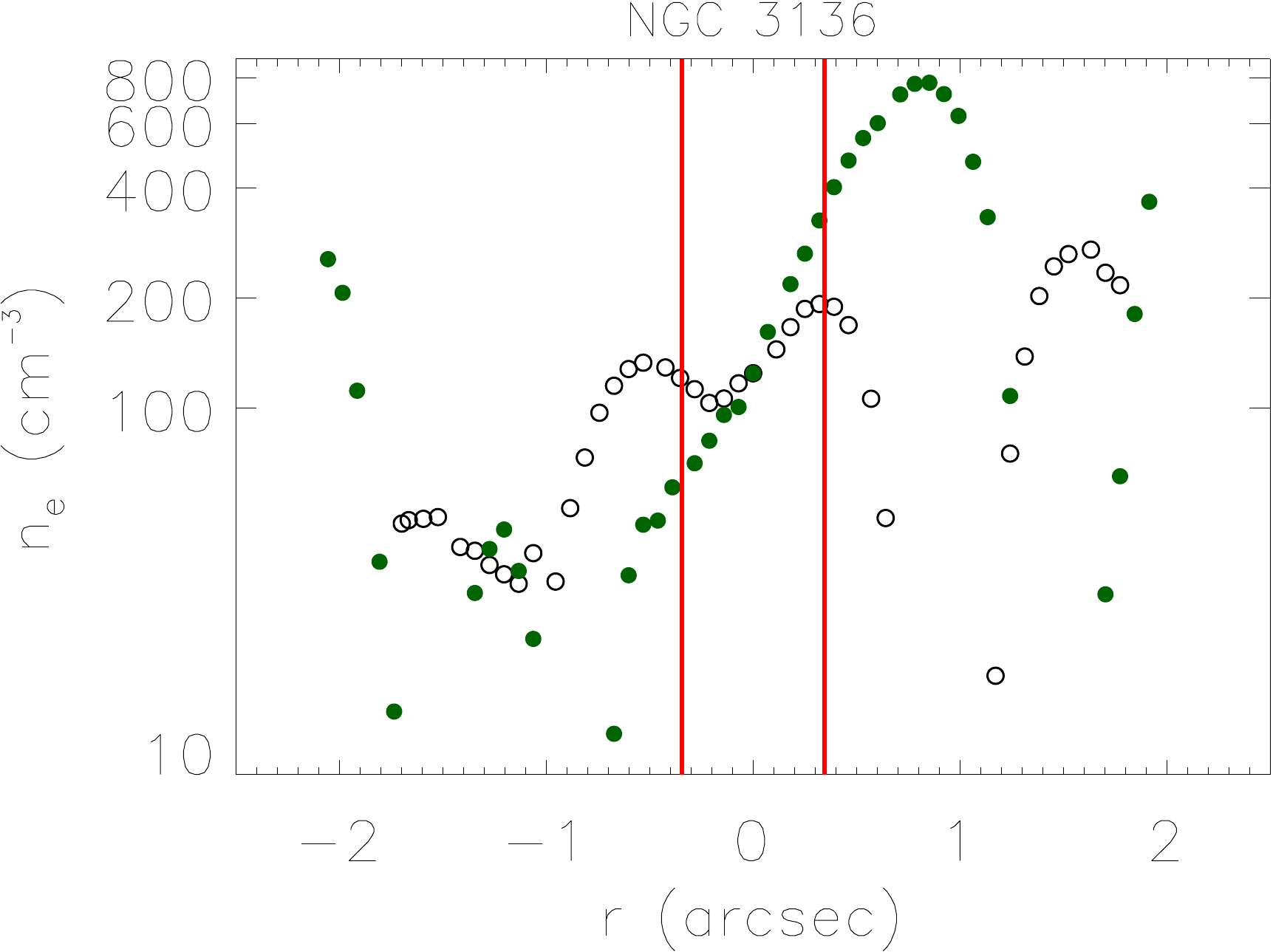}
\hspace{1.5cm}

\includegraphics[scale=0.35]{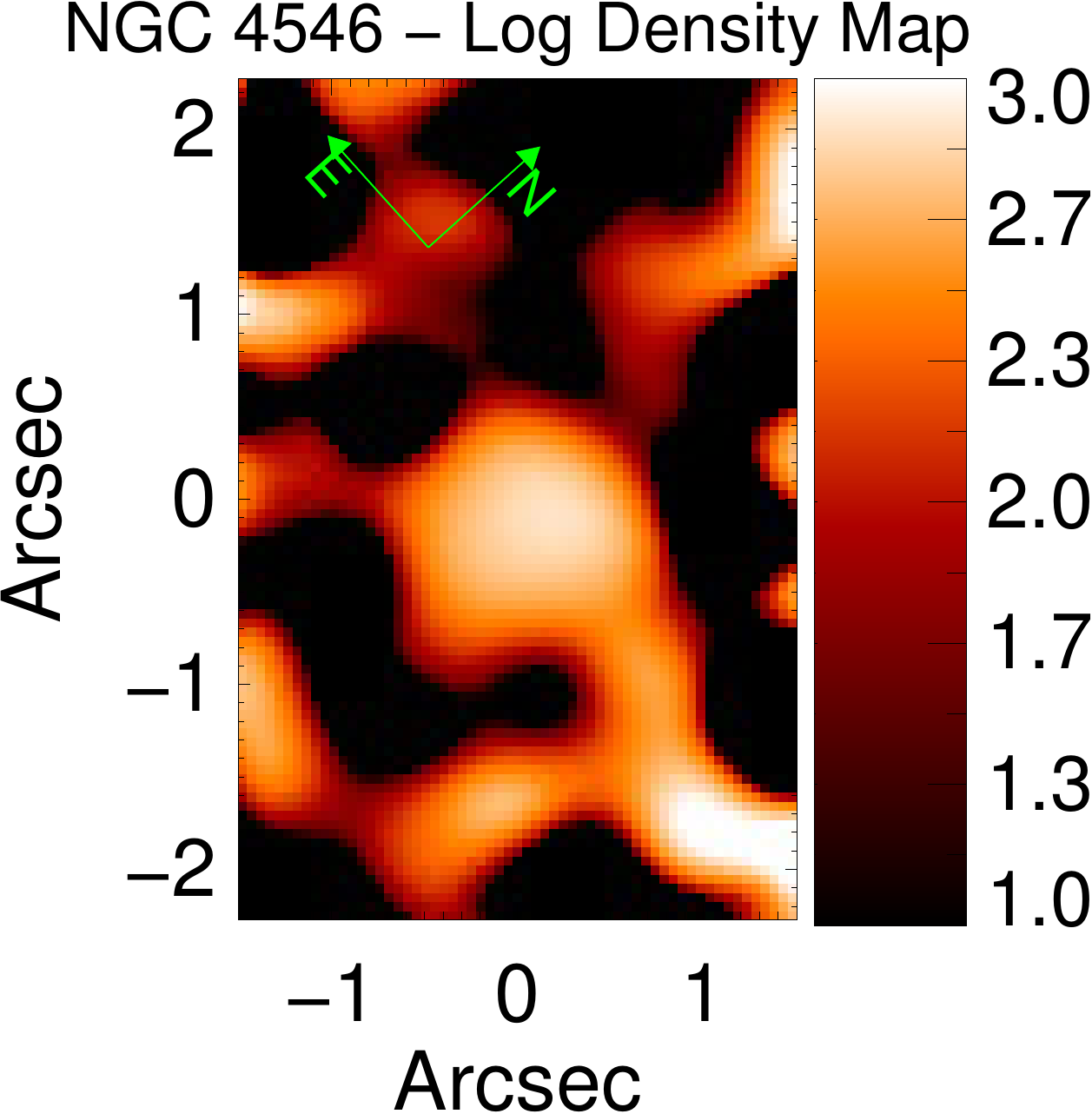}
\vspace{1cm}
\hspace{0.5cm}
\includegraphics[scale=0.35]{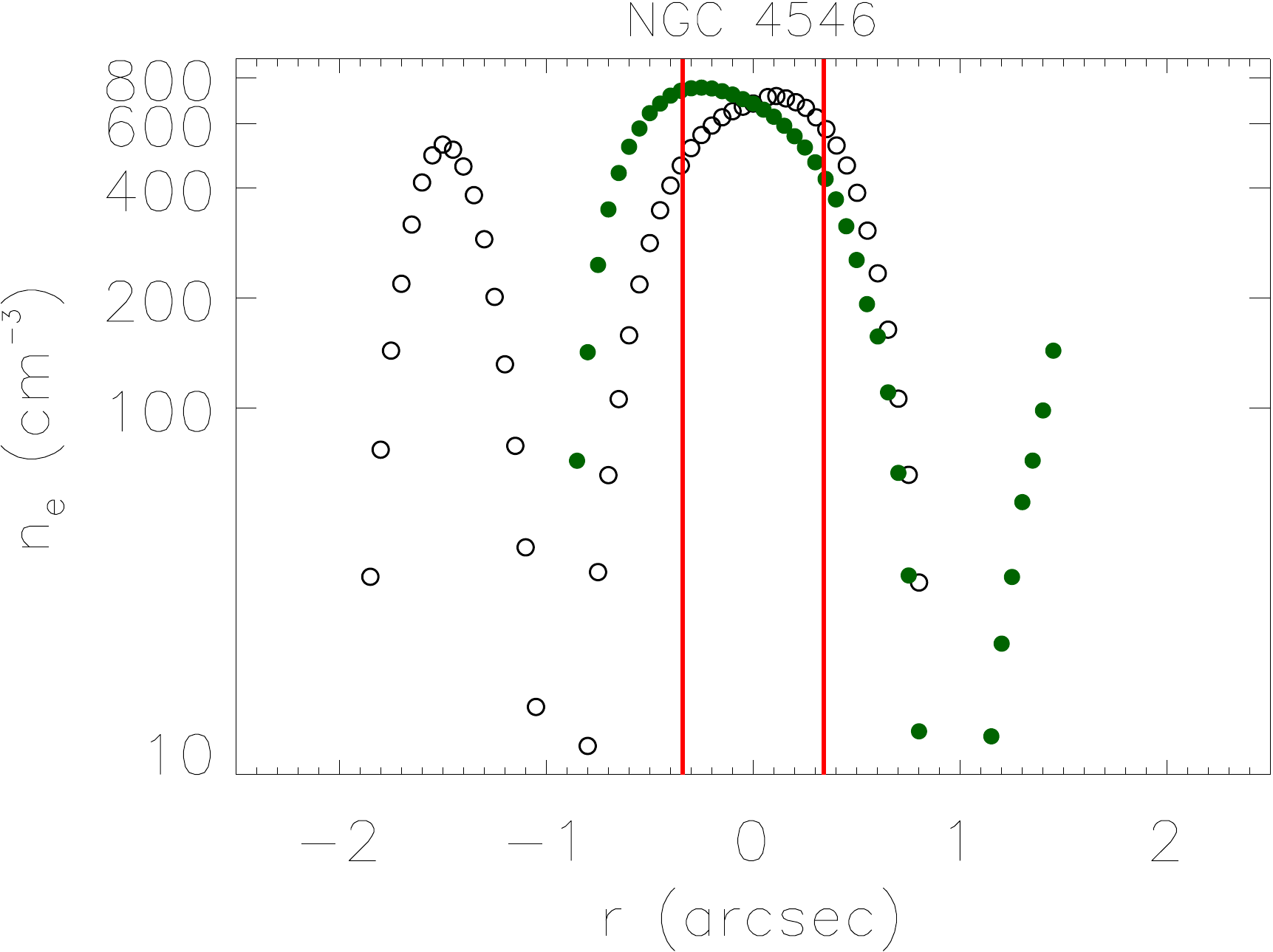}
\hspace{0.5cm}

\includegraphics[scale=0.35]{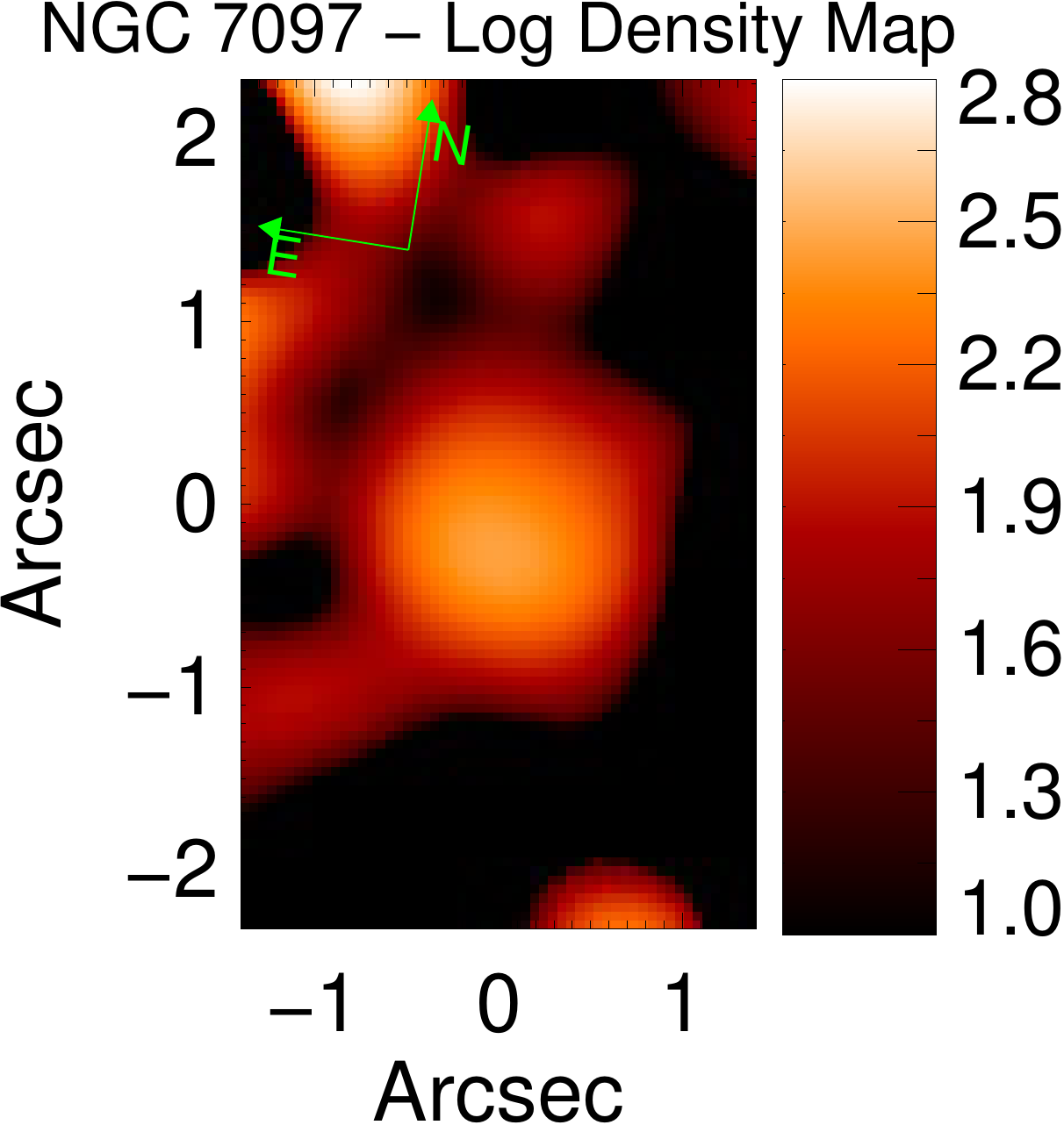}
\vspace{1cm}
\hspace{0.5cm}
\includegraphics[scale=0.35]{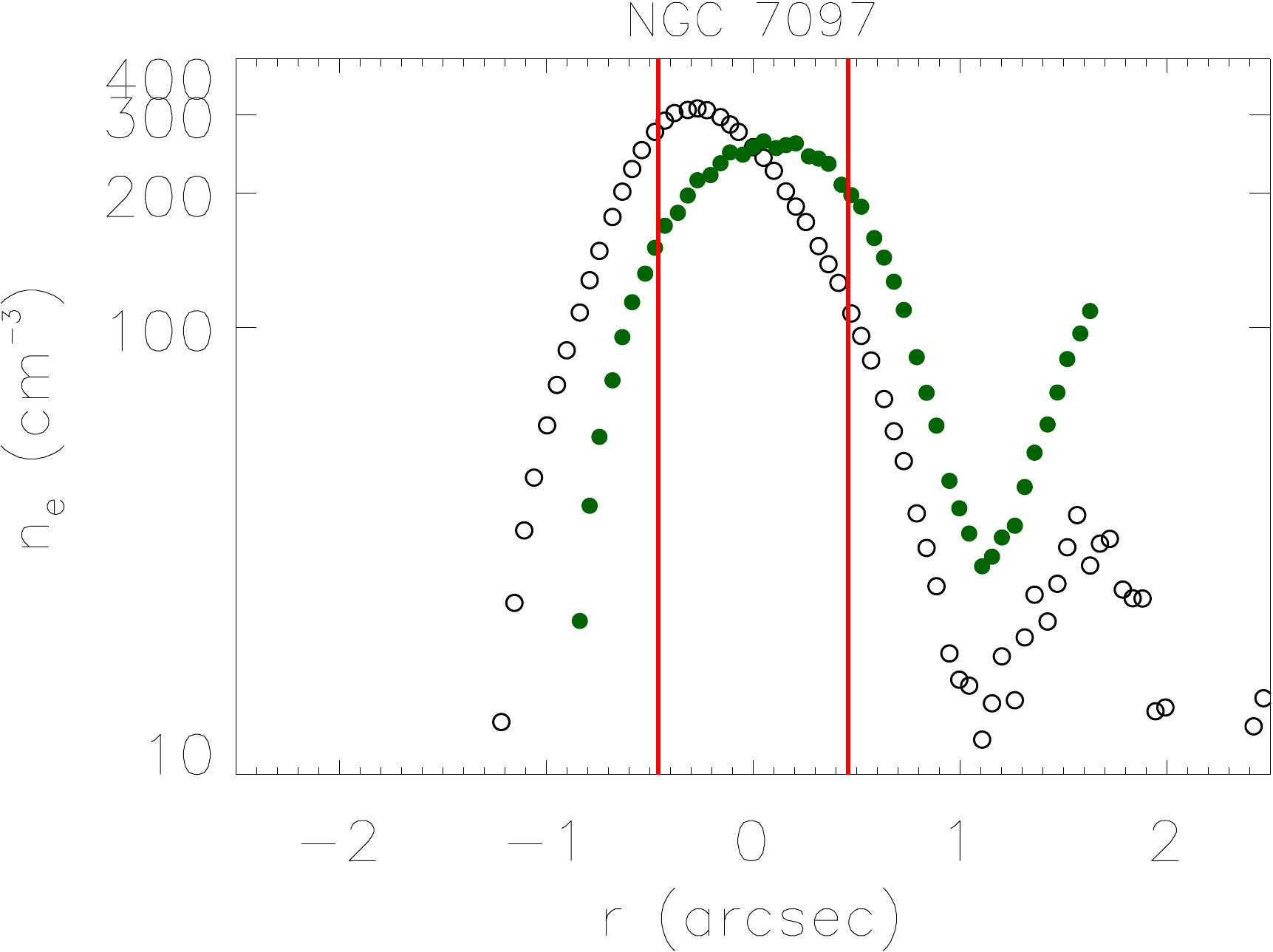}
\hspace{0.5cm}

\caption{Same as in Fig. \ref{perfil_dens_1}. \label{perfil_dens_2}} 
\end{center}
\end{figure*}

\renewcommand{\thefigure}{\arabic{figure}}

In the outer regions of the FOV, the signal-to-noise ratio of the [S II] lines is quite low. This may be the cause of several artefacts seen in Figs. \ref{perfil_dens_1} and \ref{perfil_dens_2}, since the measurement of the density is very uncertain. Thus, it is quite hard to interpret these maps. Still, some things may be worth mentioning. For instance, in NGC 2663, the density seems to be constant along the FOV analysed in this paper. In NGC 3136, the densest regions do not coincide with any object mentioned before, not even the nucleus. In ESO 208 G-21, two structures are seen in the respective density map, wherein none of them are in the same position as the AGN.

\section{Is AGN photoionization enough to explain the circumnuclear emission of the sample galaxies?} \label{agn_flux_distribution}

We want to check if the circumnuclear emission of the sample galaxies can be explained by photoionization from their AGN. As a first approach, we will adopt the same model as \citet{2010MNRAS.402.2187S} and \citet{2013A&A...558A..43S}, that is the case where the flux distribution of the H$\alpha$ emission line is caused by a point-like ionizing source embedded in an infinitesimally thin gas disc of constant density and filling factor being optically thin to Lyman continuum radiation. In this simplified model, the H$\alpha$ emission line flux is expected to fall as R$^{-2}$, where R = 0 sets the position of the AGN. In fact, the flux distribution of H$\alpha$ follows the strength of the radiation field emitted by an AGN. It is a simplistic model, since an infinitesimally geometrically thin and optically thin gas distribution may not apply for real ETGs. Moreover, the assumptions of a constant density and filling factor do not hold for real objects. For instance, a negative radial density distribution was observed for some galaxies \citep{1988ApJ...324..134F,2008AJ....136.1677W}. Given the limitations of the model, it is quite helpful to verify if the photoionization caused by an AGN is enough to explain the circumnuclear emission. For each galaxy, this model was convolved by the PSF of the respective gas cube. The results are shown together with the 1D profiles of the observed H$\alpha$ emission line flux in Figs. \ref{mapa_fluxo_gal_1} and \ref{mapa_fluxo_gal_2}.  

Along the low-velocity emission, the observed profiles are somehow consistent with the model results. In some cases, as ESO 208 G-21, IC 1459 and NGC 4546, the observed 1D profile along this direction is steeper than the model predictions. One possible reason is that these objects have a negative radial gradient of densities. Another hypothesis is that the nebula is not optically thin around the regions closest to the AGN. In NGC 1380 and NGC 3136, the observed 1D profile along the low-velocity emission is shallower than the model predictions. Probably, the presence of other compact ionization sources apart from the central AGN of both NGC 1380 and NGC 3136 may be affecting the observed 1D profiles in both directions (see Section \ref{hst_nuclear}). It is worth mentioning that in NGC 3136, the kinematic bipolar structure is probably related to an ionization bicone while for the other objects we found a disc-like geometry. In NGC 7097, the results from the model match the observed 1D profile along the low-velocity emission. However, it does not mean that this region of NGC 7097 follows the very simplistic assumptions of the model. For this galaxy, it is possible that the gas distribution along the low-velocity emission is not geometrically thin, which would result in a shallower profile caused by projection effects. Indeed, this geometry may be true for the other galaxies of the sample, except for NGC 3136 whose geometry is probably associated with an ionization bicone. On the other hand, the 1D profiles of the flux distribution of H$\alpha$ along the kinematic bipolar structures are much shallower that the model predictions. Although projection effects may also be happening here, this is probably not enough to account for the missing photons emerging from the gas structure (see e.g., the results of \citealt{2013A&A...558A..43S}). Therefore, other photoionization sources may be important to explain part of the circumnuclear emission of these galaxies. 

Another way to test the importance of AGNs in photoionizing the circumnuclear regions is by means of photoionization models. To do so, we used the nebular spectral-simulation program {\sc cloudy}\footnote{Calculations were performed with version 07.02.00 of {\sc cloudy}, last described by \citet{1998PASP..110..761F}.}. We assumed a plane-parallel geometry, a power-law continuum $f_\nu$ $\propto$ $\nu^\alpha$, with $\alpha$ $\sim$ -1.5, which is typical for LINER-like AGNs \citep{2008ARA&A..46..475H}, a lower cut in the energy of the ionizing photons of 27 eV (photons with less energy than 27 eV have already been absorbed by the most internal regions of the nucleus), a filling factor of 10$^{-3}$ and a gas density of 200 cm$^{-3}$. We also used the ionization parameter $U$ for the calculations, which is the ratio between the ionizing photon density and the gas density. Since an AGN is a compact source, we may calculate $U$ as

\begin{equation}
	U = \frac{n_\gamma}{n_e} = \frac{Q(H)}{4\pi r^2n_ec},
	\label{ionization_parameter_eq}
\end{equation}
where $n_\gamma$ is the ionizing photon density, $c$ is the speed of light and $Q(H)$ is the number of ionizing photons (i.e. with energy $>$ 13.6 eV) emitted by the AGN. It is previously known that LINER-like AGNs have log $U$ $\sim$ -3.5 \citep{1983ApJ...264..105F,1983ApJ...269L..37H,2008ARA&A..46..475H}. For our calculations, we found that -3.5 $<$ log $U$ $<$ -3.4 produces 1.81 $<$ [O III]/H$\beta$ $<$ 2.98, [N II]/H$\alpha$ $\sim$ 1.10, [S II]/H$\alpha$ $\sim$ 0.77, and [O I]/H$\alpha$ $\sim$ 0.11. These results match the observed line ratios shown in Table \ref{tab_f_NLR_ext}. For log $U$ = -3.4 and $n_e$ = 200 cm$^{-3}$, an AGN must have n$_\gamma$ = 0.079 cm$^{-3}$ in order to produce a LINER-like spectrum.

In order to verify how far an AGN may ionize the circumnuclear region, we have to estimate $Q(H)$ for the sample galaxies. With the nuclear H$\alpha$ luminosity (paper II and erratum) and assuming that the nebulae are radiation bounded, it is possible to estimate the number of ionizing photons emitted by the AGNs. This is calculated as
\begin{equation}
	Q(H) = \frac{L_{H\alpha}}{h\nu_{H\alpha}} \frac{\alpha_B(H^0,T)}{\alpha_{H\alpha}(H^0,T)},
	\label{number_ionizing_photons}
\end{equation}
where $\alpha_B$ is the recombination coefficient for all levels of the H atom, $\alpha_{H\alpha}$ is the recombination coefficient for the H$\alpha$ line, $h$ is Planck's constant and $\nu_{H\alpha}$ is the wave frequency of the photon related to the transition of the H$\alpha$ line. We used  $\alpha_B$ = 2.59$\times$10$^{-13}$ cm$^3$ s$^{-1}$ and $\alpha_{H\alpha}$ = 9.39$\times$10$^{-14}$ cm$^3$ s$^{-1}$ \citep{2006agna.book.....O}. Table \ref{tab:ionizing_photons} shows the results of the observed $Q(H)$ of the sample galaxies. Assuming log $U$ = -3.4 and $n_e$ = 200 cm$^{-3}$, we estimated the projected distance where the observed $Q(H)$ for the sample galaxies produce the value for $n_\gamma$ predicted by the photoionization model. The results are shown in Table \ref{tab:ionized_radius}. Note that all projected distances are smaller than 1.25 arcsec. Figs. \ref{mapa_fluxo_gal_1} and \ref{mapa_fluxo_gal_2} suggest that all sample galaxies have ionized gas beyond 1.25 arcsec along the gaseous discs. Also, if we assume that an AGN is photoionizing the region with a projected distance of 2 arcsec, then to maintain n$_\gamma$ = 0.079 cm$^{-3}$ the number of ionizing photons should be 10 times higher than the observed values for the sample galaxies. These results reinforce the need for alternative photoionization sources. 

\begin{table}
 \scriptsize
 \begin{center}
  \caption{Number of ionizing photons emitted by the AGN of the sample galaxies. \label{tab:ionizing_photons}}
 \begin{tabular}{@{}lc}
  \hline
  Galaxy name & log($Q(H)$)  \\
  \hline
  ESO 208 G-21 & 51.15 $\pm$ 0.22\\
  
  IC 1459 & 51.74 $\pm$ 0.13\\
  
  IC 5181 & 51.06 $\pm$ 0.39\\
  
  NGC 1380 & 50.61 $\pm$ 0.24\\
    
  NGC 2663 & 51.88 $\pm$ 0.44\\
  
  NGC 3136 & 50.66 $\pm$ 0.19\\
  
  NGC 4546 & 51.14 $\pm$ 0.40\\
  
  NGC 7097 &  51.38 $\pm$ 0.13\\
  \hline
 \end{tabular}
 \end{center}

\end{table}

\begin{table}
 \scriptsize
 \begin{center}
  \caption{Column (1): galaxy name. Column (2): Projected distances at which the ionizing photons emitted by the nucleus of the sample galaxies have a density of 0.079 cm$^{-3}$. This value for the ionizing photon density is produced by an AGN with log $U$ $\sim$ -3.4, $n_e$ $\sim$ 200 cm$^{-3}$ that, according to photoionization models, is able to explain the line ratios shown in Table \ref{tab:ionized_radius}. Column (3): same projected distances presented in column (2), but in pc. \label{tab:ionized_radius}}

 \begin{tabular}{@{}lcc}
  \hline
  Name &  $r$ & $r$  \\
  &  (arcsec) & (pc)\\
  (1) & (2) & (3)   \\
  \hline
  ESO 208 G-21 &  0.85 & 70\\
  
  IC 1459 &  1.06 & 139 \\
  
  IC 5181 & 0.53 & 64 \\
  
  NGC 1380 &  0.43 & 38 \\
    
  NGC 2663 &  1.25 & 163 \\
  
  NGC 3136 &  0.34 & 40 \\
  
  NGC 4546 &  0.84 & 69 \\
  
  NGC 7097 &  0.61 & 92\\
  \hline
 \end{tabular}
 \end{center} 

\end{table}

\section{Discs or outflows? The cases of IC 5181 and NGC 2663.} \label{discs_or_cones}

In three galaxies of the sample, we propose that the kinematic bipolar structures are pure gaseous discs. In NGC 1380 and ESO 208 G-21, the P.A. of the gaseous disc and the P.A. of the stellar disc are the same (paper I) and, probably, the gaseous disc had an internal origin. Although the gas and the stellar component are counterrotating in NGC 7097 (paper I, \citealt{1986ApJ...305..136C}), the radial velocity map and 1D profiles (Fig. \ref{mapa_cin_gal_4}), in addition to the image of the red and blue wings of the H$\alpha$ emission line, suggest that the gas component has a disc-like geometry. In IC 1459 and NGC 4546, a gaseous disc may exist, but they are affected by non-Keplerian motions, probably outflows (see Section \ref{kinematics_extended_emission}). In NGC 3136, an ionization bicone seems very likely. 

An interesting case is IC 5181. In this galaxy, the planes defined by the kinematics of the stellar and gas components are perpendicular (paper I, \citealt{2013A&A...560A..14P}). Although an external gas accretion is able to produce a gaseous disc perpendicular to a stellar disc \citep{2006MNRAS.366.1151S}, which is the scenario proposed by \citet{2013A&A...560A..14P}, an outflow is likely to be present in this object. The velocity dispersion map shows that lower values are seen along the kinematic bipolar structure, which is compatible with a geometric thin disc. On the other hand, the maps of EW(H$\alpha$) and EW([N II]) show a spiral structure, similar to an integral sign, in the inner regions ($r$ $<$ 0.7 arcsec). Two possible scenarios are suggested. One is that outflows are happening in regions very close to the AGN. The other possibility is that the gas is moving through a non-axisymmetric potential. This would create a concentration of gas in certain regions that would be outlined by the EW images of the emission lines \citep{2006MNRAS.366.1151S}. 
 
Another curious case is NGC 2663. The radial velocity map reveals a bipolar structure (Fig. \ref{mapa_cin_gal_3}). The isophotes shown in the map of the red and blue wings of the H$\alpha$ emission line (Fig. \ref{RGB_Ha}) are regular. Although both properties are typical of a gaseous disc, these kinematics are seen within the unresolved region of the nucleus. This may be the reason why this galaxy has high values for the FWHM of the Gaussian profiles representing the emission lines of the spectrum extracted from the nucleus (paper II). Although we may not discount the possibility that NGC 2663 has a gaseous disc, the hypothesis of an ionization bicone is also plausible here. This is the only case for which the AGN has enough photons to photoionize the region along the bipolar structure (see Fig. \ref{mapa_fluxo_gal_2}). The spatial extent is smaller when compared to the other galaxies (the galaxies IC 1459, IC 5181 and NGC 7097 have, approximately, the same spatial scale as NGC 2663). Although these two characteristics could be applied to a gaseous disc, an ionization bicone seems very probable in this galaxy.

\section{HST observations of the sample galaxies} \label{hst_nuclear}

\begin{figure*}
\begin{center}
\includegraphics[scale=0.26]{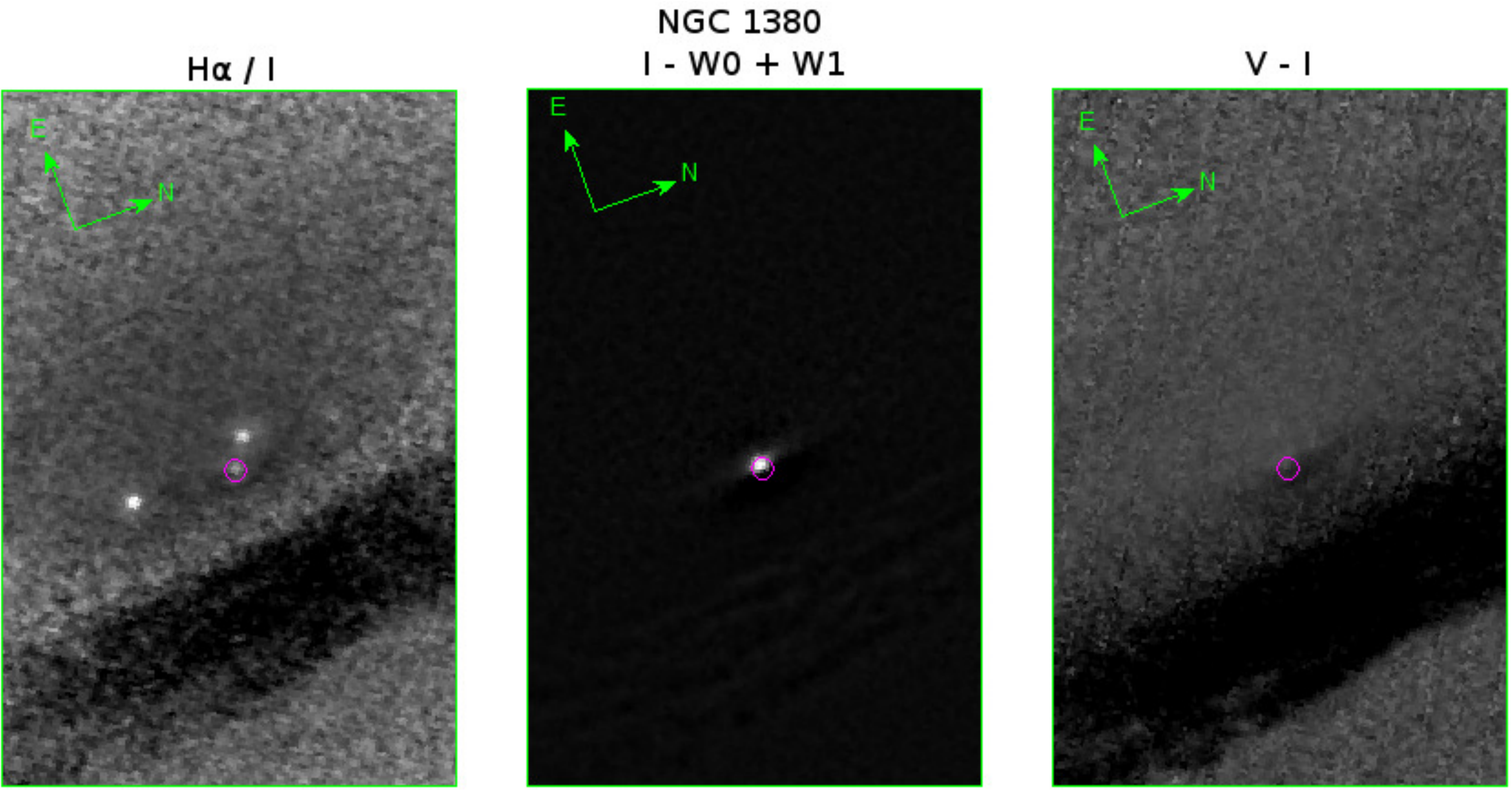}
\hspace{1.0cm}
\vspace{1.0cm}
\includegraphics[scale=0.26]{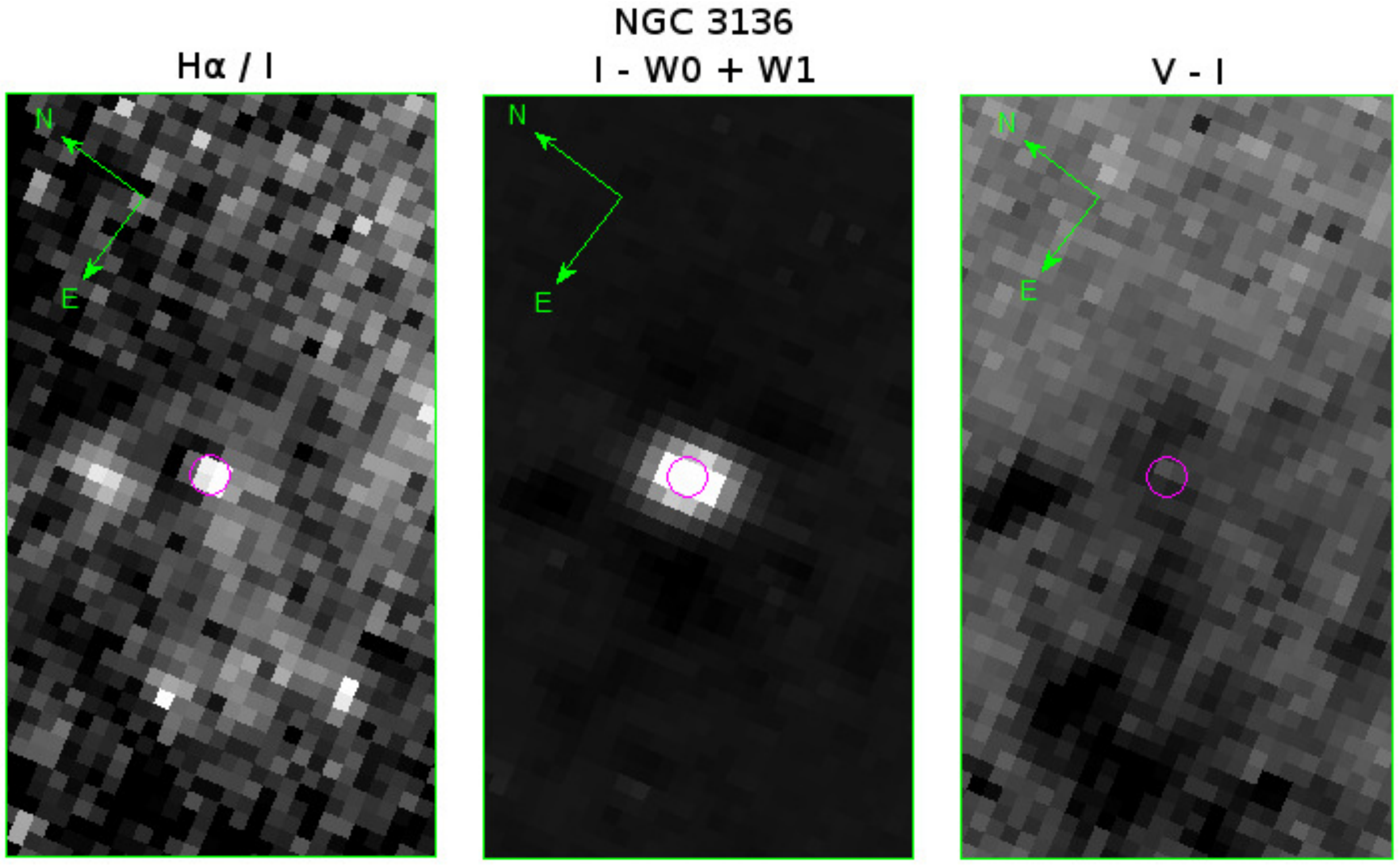}
\includegraphics[scale=0.26]{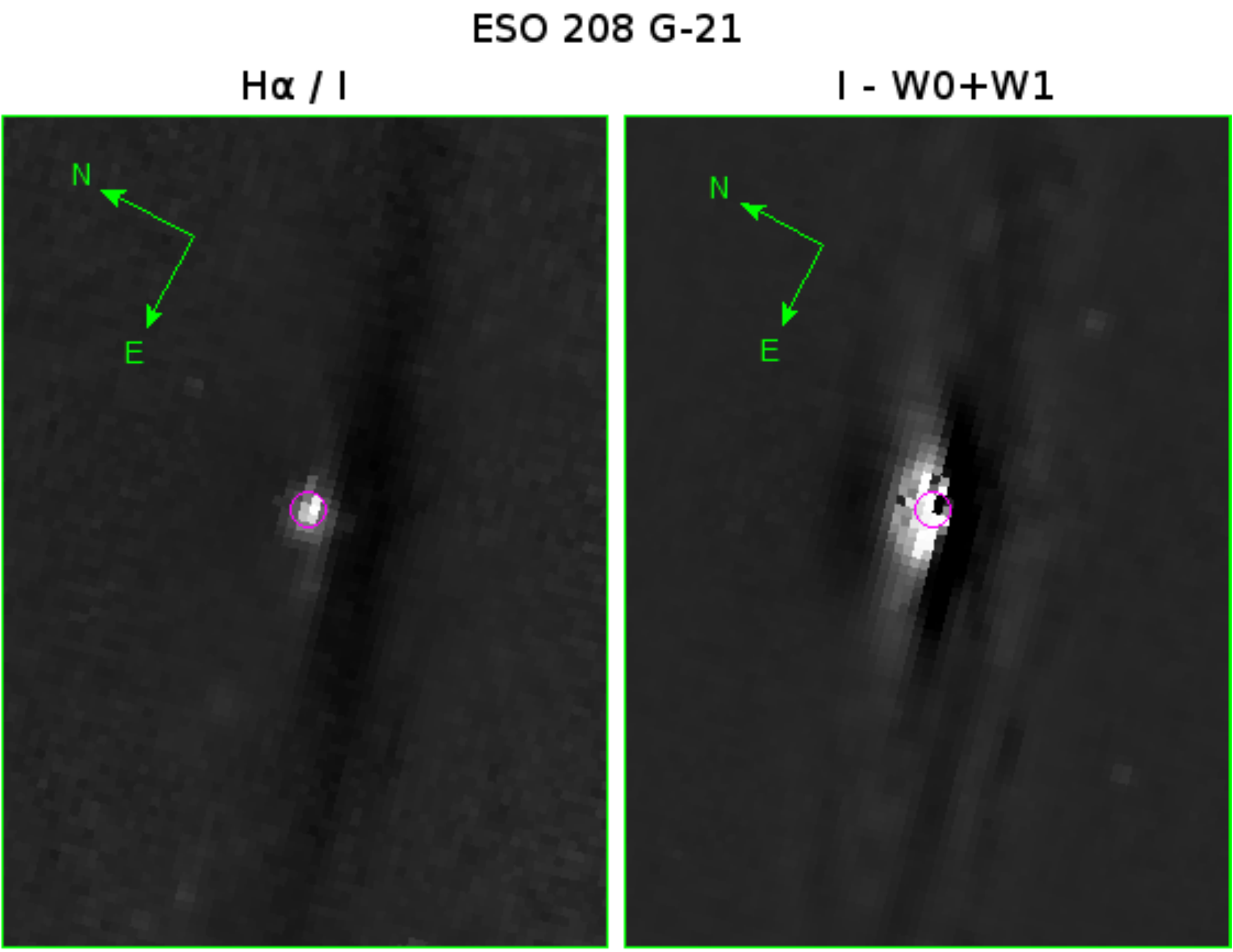}
\hspace{0.5cm}
\includegraphics[scale=0.26]{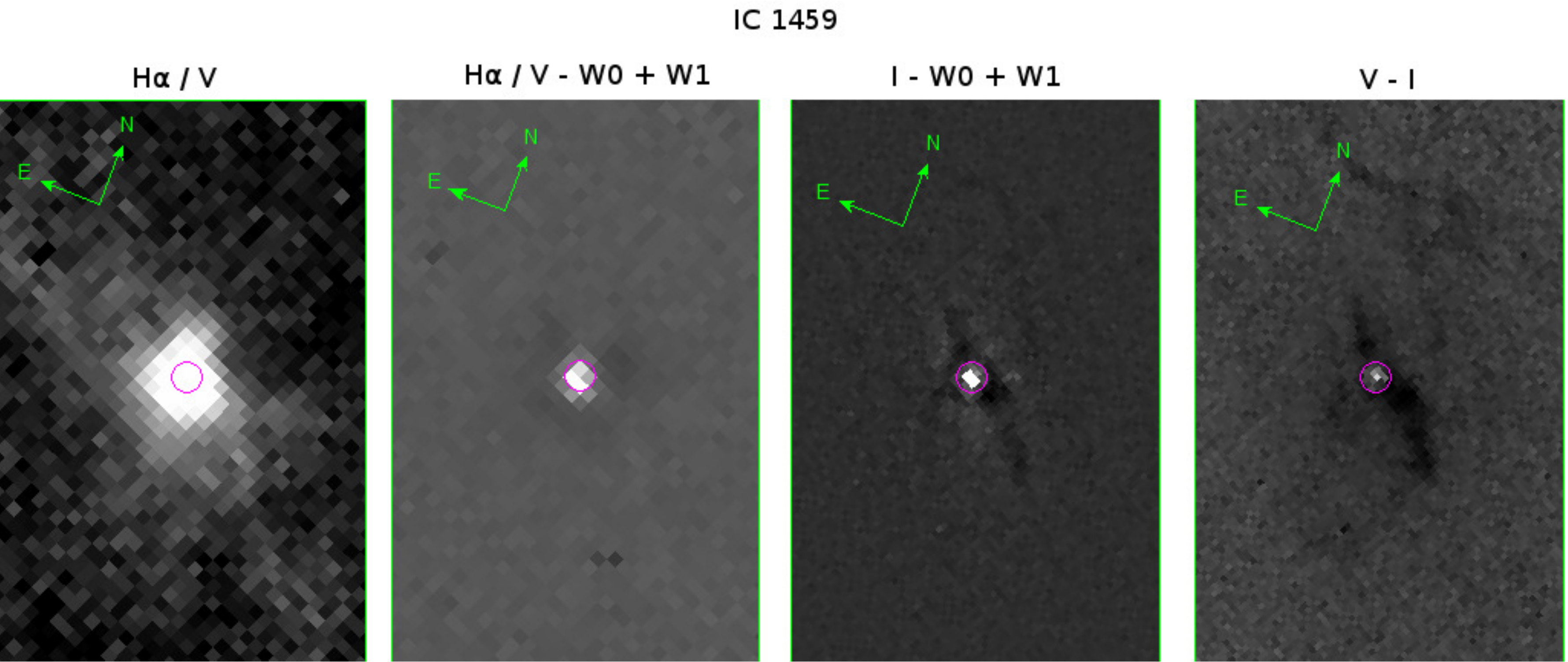}

\caption{HST data archive of NGC 1380, NGC 3136, ESO 208 G-21 and IC 1459. In all galaxies, the images have the same FOV and spatial orientation as the gas cubes. For some images, we show the sum of the wavelet components W0 and W1. This components are associated with the high spatial frequencies of the images. The magenta circles mark the nuclear region of the bulges of the galaxies. In V - I colour images, the dark regions are associated with regions with large extinctions. \label{HST_science}
}
\end{center}
\end{figure*}

Observations made with the HST may provide additional information to the results obtained with the gas cubes, especially because of the superb spatial resolution of the HST images. ESO 208 G-21, NGC 1380, IC 1459 and NGC 3136 have public images\footnote{NGC 1380: GO 10240, PI - Aaron Barth; IC 1459: GO 6537, PI - Tim de Zeeuw, GO 5454, PI - Marijn Franx; NGC 3136: GO 6822, PI - Paul Goudfrooij; ESO 208 G-21: GTO/ACS 9293, PI - Holland Ford} available in the HST data archive. Images taken with the V and I filters and with a narrow band filter on the H$\alpha$+[N II]$\lambda \lambda$6548,6583 emission lines were taken from the data archive (see Table \ref{tab_HST_images} for more information). Both H$\alpha$/I and H$\alpha$/V ratios highlight the ionized gas emission of the galaxies. We are mostly interested in analysing compact regions of line emission as well as the central core of the bulges. In order to emphasize this, we performed a wavelet decomposition of the HST images. We used the ``a trous'' algorithm for the wavelet decomposition \citep{2002aida.book.....S}. In Fig. \ref{HST_science}, we show only the first two wavelet components, W0 and W1, which are related to the high spatial frequencies of the images. The colour V - I maps the intrinsic extinction of the galaxies and, eventually, points out the existence of a non-stellar continuum, as seen in IC 1459. In Fig. \ref{HST_science}, we show HST images of the four galaxies mentioned above, with the same FOV and the same spatial orientation as the gas cubes. The comparison between the HST images and the data cubes analysed throughout this paper and also in Papers I and II took as reference the nuclear region of the galactic bulges. 

 \begin{table}
 \scriptsize
 \begin{center}
  \caption{Summary of the HST images shown in Fig. \ref{HST_science} \label{tab_HST_images}
}

 \begin{tabular}{@{}lcc}
  \hline
  Galaxy name & Instrument & Filters \\
  \hline
  ESO 208 G-21 &ACS/WFC1& F658N(H$\alpha$+[N II]), F814W(\textit{I}-band) \\
  &&\\
  IC 1459 &WFPC2/PC& F555W(\textit{V}-band), FR680P15(H$\alpha$+[N II]),\\
  &&F814W(\textit{I}-band) \\
  NGC 1380 & ACS/HRC& F555W(\textit{V}-band), F658N(H$\alpha$+[N II]),\\
  && F814W(\textit{I}-band) \\
  NGC 3136 &WFPC2/PC&F547M(\textit{V}-band), FR680P15(H$\alpha$+[N II]),\\
  && F814W(\textit{I}-band)\\
  \hline
 \end{tabular}
 \end{center}

\end{table}

In IC 1459, the ionized gas emission image reveals an extended structure, probably associated with the gaseous disc, as shown in the H$\alpha$/V image (Fig. \ref{HST_science}). The high spatial frequency wavelet component of the ionized gas emission shows a compact object in the central region of the galaxy, in the same position that a blue compact object is seen in the V - I colour image. This is probably related with the non-stellar continuum of the AGN of IC 1459. This is in agreement with the detection of the featureless continuum of the AGN in the eigenspectrum that was associated with the AGN in paper I. A spiral structure, seen in the V - I colour image, probably formed by dust, is present along the southwest-northeast direction. Many of these HST results were already discussed by \citet{1997ApJ...481..710C} and \citet{2002ApJ...578..787C}. 

In NGC 1380, the ionized gas emission image shows three compact objects along the same FOV of the gas cubes. Two objects are separated by 0.2 arcsec along the southwest-northeast direction and are not spatially resolved in the gas cubes. However, the EW([N II]) map, shown in Fig. \ref{mapa_ew_gal_1}, reveals two objects separated by 0.7 arcsec. The separation between the two closest objects and the third object in the HST image is $\sim$ 0.7 arcsec. In the region of all three objects, [N II]/H$\alpha$ $>$ 1.2. These three objects could probably be classified as LINERs. A further analysis of the nature of these three objects will be discussed in a future paper. The H II regions, proposed in the sections above, are not seen in the HST images. With the V - I colour image, we detected a dust disc which extends along the entire north-south direction of the FOV, to the west of the nucleus, which may be related to the external disc of NGC 1380, since this galaxy is classified as S0 (see paper I). A smaller dust structure, with the same orientation of the external disc, is located in the nuclear region. This is in agreement with the high E(B-V) values found for the nucleus (paper II).

In ESO 208 G-21, an intense ionized gas emission is seen in the central region of the galactic bulge. We also detected an extinction along the east-west direction in the FOV, slightly to the south of the nucleus. 

In NGC 3136, the ionized gas emission is complex. We confirmed that a compact emission arises from the centre of the bulge. A weaker and slightly extended emission is seen northeast of the centre. Both objects are in the same position as the compact objects detected with PCA Tomography (paper I). We also noticed that the extended structures detected southeast of the nucleus in both the EW(H$\alpha$) and the EW([N II]) maps are related to two compact objects in the HST H$\alpha$/I image. The structure located southwest of the nucleus is at the border of the FOV of the gas cube, but it is extended in the H$\alpha$/I HST image (see also Fig. \ref{fig:HST_2MASS_N3136_v2} in Appendix \ref{ngc3136_comments}). Moreover, the entire FOV related to this HST image seems to be filled by a diffuse gas emission. The V - I colour image shows an extended structure with a high extinction located southeast of the nucleus which is in the same position as the structure in redshift of the ionization bicone. This result indicates that this structure in redshift is the far side of the bicone, since it is probably located behind a dust screen that causes the extinction detected in the V - I colour image.

\section{Discussion and Conclusions} \label{sec:conclusions}

In this paper, we analysed the circumnuclear emission of eight from a sample of 10 galaxies. The other two galaxies, NGC 1399 and NGC 1404, did not have circumnuclear emission detected with PCA Tomography (paper I). After performing stellar spectral synthesis and subtracting the stellar components, an attempt to search for some type of ionized gas emission was successful only for the nuclear region of both objects (paper II). As mentioned in Section \ref{FOV_properties}, we used PCA tomography (paper I) to detect circumnuclear emission of ionized gas as kinematic bipolar structures, where one pole is redshifted and the other is blueshifted relative to the centre of the galaxies. The tomograms associated with these emissions resemble the maps corresponding to the blue and red wings of the H$\alpha$ emission line, shown in Fig. \ref{RGB_Ha}. The kinematic bipolar structure is also seen in the radial velocity maps of the galaxies. When compared to the nucleus, the circumnuclear region is less dense, but the lower limits for the ionized gas mass are always higher than the ionized gas mass in the unresolved nuclear region. In paper I, we proposed that six galaxies of the sample have gaseous discs. The results presented in this paper revealed that only ESO 208 G-21, NGC 1380 and NGC 7097 seem to have pure gaseous discs, i.e., their motions are apparently Keplerian. Furthermore, in NGC 1380 and ESO 208 G-21 we detected a corotation between the gaseous disc and the stellar component, as opposed to NGC 7097, whose gas structure and stellar component are counterrotating. However, this only means that, while the gaseous discs of ESO 208 G-21 and NGC 1380 may have an internal origin, the gaseous disc of NGC 7097 probably has an external origin, since the stellar radial velocity of this galaxy is slower than the gas radial velocity \citep{1986ApJ...305..136C,ricci2013}. Moreover, the maps of the red and the blue wings of the H$\alpha$ emission line of these three objects have narrow and well-defined isophotes, in agreement with the geometry of a gaseous disc. On the other hand, it is very likely that IC 1459 and NGC 4546 also have gaseous discs, but their radial velocities are affected by non-Keplerian motions. This is probably associated with outflows within the central region of these objects. For IC 1459, the non-Keplerian effects along the P.A. of the main gas structure were also detected by \citet{2002ApJ...578..787C}. In NGC 4546, \citet{2006MNRAS.366.1151S} observed twists along the radial velocity map and also a spiral structure in the ionized gas distribution, similar to an integral sign. It is worth mentioning that both \citet{2006MNRAS.366.1151S} and \citet{1987ApJ...318..531G} suggested a counterrotation between the stellar and the gas components for this galaxy, while we proposed in paper I that both structures are not in the same plane. However, the twists were detected in a region beyond the spatial coverage of the data cubes analysed in this paper. The same spiral structure was detected in IC 5181, but in a spatial scale covered by the FOV of the GMOS data cube. \citet{2006MNRAS.366.1151S} suggested that this spiral structure derives from non-axisymetric potentials in the central region of some ETGs. Nevertheless, we do not discard the hypothesis of an outflow in the most internal regions ($<$ 0.7 arcsec) of IC 5181. Even for NGC 4546, the scenario of an outflow for the observed twists may be true. In the case of NGC 3136, the kinematic results together with the V - I colour map from the HST observations suggest the presence of a bicone, where the structure in redshift is the far side of the bicone, since an extinction was detected in the V - I colour image of the same region, which may indicate that this side of the bicone is behind a dust screen. In NGC 2663, we propose two hypotheses: a bicone or the presence of two AGNs. 

Emission line ratios taken from the spectra of the circumnuclear regions suggest LINER-like emission. Although the nuclear regions have higher [N II]/H$\alpha$ ratio on the maps of seven galaxies, the kinematic bipolar structures also have typical line ratios of LINERs along their entire spatial extension detected in the gas cubes. Only in NGC 1380 we did detect line ratios typical of H II regions in two positions of the gaseous disc. 

LINER-like spectra may be produced by several ionization sources (AGNs - \citealt{1983ApJ...264..105F,1983ApJ...269L..37H}; shocks - \citealt{1980A&A....87..152H}; pAGB stars - \citealt{1994A&A...292...13B,2008MNRAS.391L..29S,2011MNRAS.413.1687C,2012ApJ...747...61Y}). BLRs, which are a typical signature of an AGN, were detected in six galaxies of the sample. However, the results presented in this paper suggest that a single AGN is not enough to photoionize the gaseous discs. Following the work done by \citet{2010MNRAS.402.2187S} and \citet{2013A&A...558A..43S}, we used a simple model to predict the H$\alpha$ emission line flux distribution in an infinitisemaly thin and optically thin gas structure, with constant density and filling factor, which is ionized by an AGN. Since the simplistic assumptions of this model do not hold, caution should be exercised when comparing the predictions from it with our measurements along the gaseous disc: The gas is not optically thin to Lyman continuum radiation, the H$\alpha$ flux distribution is therefore not expected to decrease as $R^{-2}$. The same situation applies when the gas density and filling factor decreases with radius. On the other hand, the flux distribution will be flatter than $R^{-2}$ under the assumption of a geometrically thick disc, since one has to take into account the integrated light over the line-of-sight. However, \citet{2013A&A...558A..43S} have shown that profiles that result from the projection of an axisymmetric disc with a radially declining gas density still do not explain the extended emission from their sample galaxies. In an isotropically emitting-spheric symmetric volume, a luminosity density distribution decreasing as a power law with an exponent -$\alpha$ yields a projected power-law intensity profile with an exponent -$\alpha$+1 (see, e.g., \citealt{2013A&A...555L...1P}). Therefore, a $R^{-2}$ dilution of the radiation from a central point source (the AGN) will yield an intensity profile of the form $R^{-\alpha}$ with $\alpha=1.0$. We estimate $\alpha$ = 1.1 to 1.7 along the gaseous discs of our sample galaxies. In this scenario, the AGN may be playing an important role in photoionizing the gaseous discs, however one would have to take into account the negative radial gradient of the gas density and also dust inside the gaseous disc. It seems very likely that an alternative ionization source has to be taken into account in addition to the AGNs.

Photoionization models also show that the AGNs do not emit enough photons to explain the emission from the gaseous discs, since the line ratios of these structures are described by an ionization parameter of log U $\sim$ -3.4. Table \ref{tab:ionized_radius} shows that the sample galaxies have ionized radii beyond the limit that an AGN with such a value for U is able to produce. However, if we assume that the AGNs of the sample galaxies are able to photoionize regions with radii which are twice the values of the ionized radii calculated for log U = -3.4, then the ionization parameter in these farthest radii should be log U = -4.0. The same model we used in Section \ref{agn_flux_distribution} but assuming log U = -4.0 produces a [N II]/H$\alpha$ ratio of $\sim$ 0.74. Since the low-velocity emission of the galaxies has [N II]/H$\alpha$ $<$ 1.0, it seems reasonable that the AGN may be photoionizing these regions. However, it is worth mentioning that these photoionization models are also subject to errors. For instance, for the NLR radius calculation, we used the number of ionizing photons, which is proportional to the H$\alpha$ luminosity. However, the uncertainty associated with the absolute flux calibration of the data cubes is quite high, certainly $\geq$ 30\%. Also the gas density and filling factor are very uncertain for the circumnuclear regions. Thus, caution is needed when analysing these results.

\begin{figure*}
\begin{center}
\includegraphics[scale=0.75]{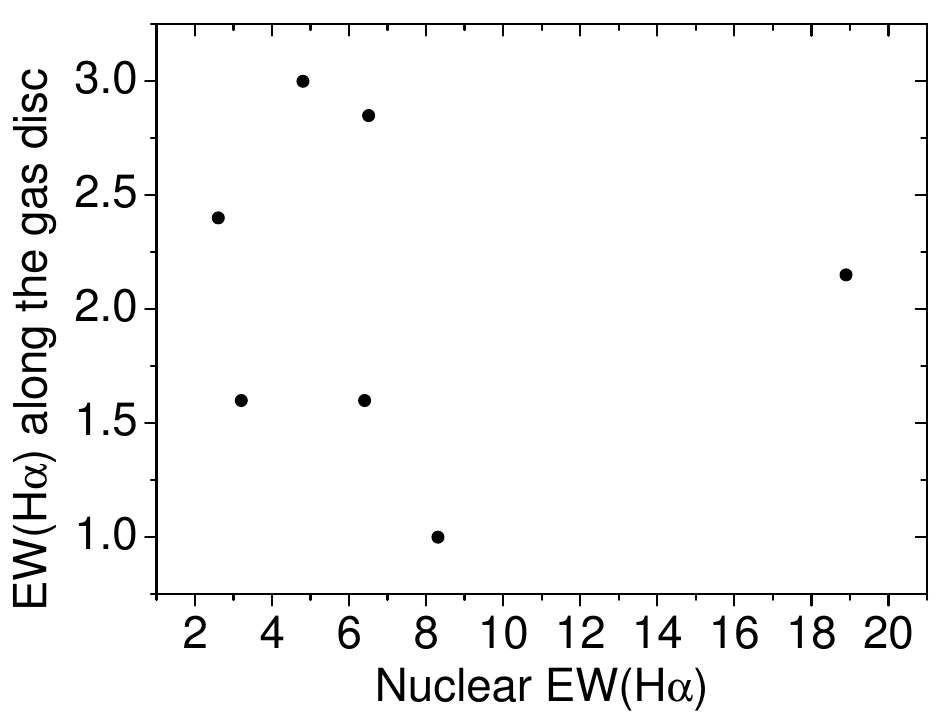}
\hspace{0.5cm}
\includegraphics[scale=0.75]{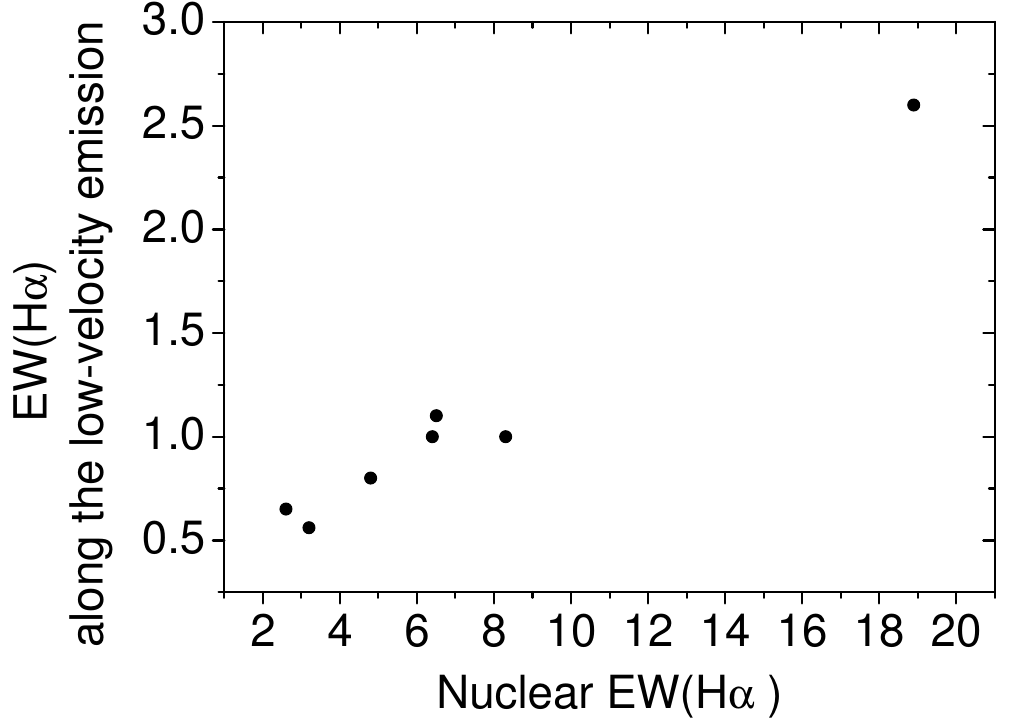}
\caption{Left: EW(H$\alpha$) along the gaseous discs versus nuclear EW(H$\alpha$). Right: EW(H$\alpha$) along the low-velocity emission (that is, extended emission perpendicular to the kinematic bipolar structure) versus nuclear EW(H$\alpha$). A correlation is seen only along the low-velocity emission. This may be an indication that AGN photoionization is important only along this region, while other ionization sources may be present along the gaseous discs. \label{EW_nuc_EW_ext}
 }
\end{center}
\end{figure*}

Another result that supports the scenario where photoionization by a central AGN does not hold for the gaseous disc but seems to be important along the low-velocity emission is presented in Fig. \ref{EW_nuc_EW_ext}. On the left of this figure, we show the relation between the typical EW(H$\alpha$) for the gaseous disc and the EW(H$\alpha$) of the nuclear region, while on the right we show the relation between the typical EW(H$\alpha$) for the low-velocity emission and the EW(H$\alpha$) of the nuclear region. We notice a correlation between the nuclear EW(H$\alpha$) and the EW(H$\alpha$) of the low-velocity emission, while no correlation is seen between the nuclear EW(H$\alpha$) and the EW(H$\alpha$) in the gaseous disc. Given these results, we propose a scenario where the low-velocity emission is related to ionization cones which arise in LINER-like AGNs, but the ionizing photons emitted by the AGN are collimated by some agent that is aligned with the gaseous disc. A scheme of this model was presented by \citet{2011ApJ...734L..10R} for NGC 7097 (see their Fig. 5). 

One hypothesis for the collimation agent is that a nuclear torus, within the context of the unified model \citep{1993ARA&A..31..473A}, exists in these galaxies and it is aligned with the gaseous discs. If this is true then, assuming that the inner radius of the torus is given by $R_{dust}=0.4L_{45}^{0.5}$ pc \citep{2008ApJ...685..160N}, where $L_{45}$ is the bolometric luminosity in units of 10$^{45}$ erg s$^{-1}$, we found that the $R_{dust}$ is smaller than 1 pc for all galaxies of the sample with a BLR (the bolometric luminosities of the sample galaxies are shown in the erratum of paper II). However, low-luminosity AGNs may lack a nuclear torus \citep{2006ApJ...648L.101E,2008ARA&A..46..475H,2009ApJ...701L..91E}. According to the model proposed by \citet{2009ApJ...701L..91E}, the BLR and the torus are formed by the same mechanism, i.e. winds emerging from the accretion disc. The boundary between the existence and the lack of a BLR/Torus structure occurs when an accretion disc around a SMBH has a very low radiative efficiency, which is the case for LINER-like AGNs \citep{2008ARA&A..46..475H}. For a BLR/Torus to exist, $L_{45} > 6.1\times10^{-17}(R_{Edd})^{-2}$, where $R_{Edd}$ is the Eddington ratio. All sample galaxies are above this limit by a factor of $\sim$ 100, except for NGC 1399 and NGC 1404. Thus, a nuclear torus is probably present in the eight galaxies analysed in this work. If the torus is aligned with the observed gaseous discs, then the nuclear ionizing radiation would not reach this structure, but it would be able to photoionize the regions that are perpendicular to the gaseous disc, i.e. the low-velocity emission. On the other hand, the assumption that the nuclear torus is always aligned with the gaseous disc may not hold. However, if the orientation of the nuclear torus is randomly distributed when compared to the gaseous disc orientation, then we should expect that the gas distribution along the the low-velocity emission is not ionized for some galaxies. However, we can draw no firm conclusions on this issue because our sample is statistically incomplete and small. Alternative hypotheses for the collimation agent in the sample galaxies are that the inner radius of the gaseous discs is optically thick or some structure somehow related to the gaseous discs is acting as a shield against the ionizing photons from the AGN. 


Below, we propose some hypotheses for the ionizing sources:

\begin{enumerate}

	\item \textit{AGN}: In paper II, we confirmed that an AGN is present in, at least, seven galaxies of the sample, since in six of them we detected a BLR and NGC 1399 is a radio galaxy \citep{2008MNRAS.383..923S}. In NGC 1380, NGC 1404 and NGC 3136, an AGN is likely to exist (papers I and II).

	\item \textit{Hot interstellar gas}: ETGs have a hot interstellar medium (ISM), with a temperature of $\sim$ 10$^7$ K. The ISM emits basically only X-ray photons and may be a source of photoionization for the circumnuclear region of the galaxies. We calculated the H$\alpha$ luminosity that would be emitted by a gas photoionized by the hot ISM in the galaxies IC 1459 and NGC 1380 using X-ray data analysed by \citet{2013ApJ...766...61S}. We found that the ratio between the calculated and the observed H$\alpha$ emission line luminosities is 0.02 for IC 1459 and 0.10 for NGC 1380. \citet{1989ApJ...346..653K}, using a similar procedure, found similar results for three other ETGs with LINER-like nuclei. Therefore, the combined contribution of a hot ISM and an AGN does not seem to be enough to explain the H$\alpha$ emission from the gaseous discs. 
	
	\item \textit{Shocks}: in NGC 3136, the presence of several structures whose spectra have line ratios typical of LINERs with a bicone increases the probability of shocks in the central region of this galaxy. Supernovae remnants (SNR) or superwinds may produce a mechanical source of energy which may cause shocks along this galaxy. 
	
	\item \textit{H II region contamination}: In NGC 1380, we detected H II regions at a projected distance beyond $\sim$ 1 arcsec from the nucleus. Although starbursts occupy a specific region in BPT diagrams, we should not discount the hypothesis that the light from the AGN is contaminating the H II region emission, which would characterize the observed spectrum as a transition object.
	
	\item \textit{Hot low-mass evolved stars (HOLMES)}: \citet{1994A&A...292...13B} proposed that part of the ionizing photons observed in LINERs could be provided by pAGB stars. Several works in the literature \citep{2008ARA&A..46..475H,2008MNRAS.391L..29S,2010ApJS..187..135E,2010MNRAS.402.2187S,2011MNRAS.413.1687C} have supported this idea. The values found for the EW(H$\alpha$) of the gaseous discs are smaller than 3\AA\ for all sample galaxies (see Fig. \ref{EW_nuc_EW_ext}). According to the WHAN diagram \citep{2011MNRAS.413.1687C}, this is an indication of photoionization caused by HOLMES. However, EW measurements may be affected by dilution effects caused by the stellar component in triaxial ETGs \citep{2013A&A...555L...1P}. As a result, the observed EW would mimic HOLMES photoionization even in cases where photoionization caused by an AGN dominates the emission, which may be the case of ESO 208 G-21 (nuclear EW(H$\alpha$) $<$ 3\AA, see Fig. \ref{mapa_ew_gal_1}). Nevertheless, one of the consequences of the HOLMES scenario is that the EW(H$\alpha$) should be constant along the extended structure that is photoionized by an extended source \citep{2010MNRAS.402.2187S}. In the 1D profile of EW(H$\alpha$) along the gaseous discs, shown in Figs. \ref{mapa_ew_gal_1} and \ref{mapa_ew_gal_2}, we see this trend in ESO 208 G-21, IC 1459, IC 5181, NGC 4546 and NGC 1380. In NGC 7097, the EW(H$\alpha$) values decrease until a projected distance of $\sim$ 1.5 arcsec and then rise again. Thus, it is very likely that these old hot stellar populations are contributing significantly along the kinematic bipolar structures. However, it is not clear why they contribute only along the kinematic bipolar structures and not in the perpendicular direction, since the stellar discs are not always coupled to the gas motions and the bulge of the galaxies is almost isotropic. In order to verify the importance of HOLMES, we used {\sc CLOUDY} to estimate the ionizing photon density emitted by HOLMES. Assuming a blackbody with T=150000K, $n_e$ = 200 cm$^{-3}$, a filling factor of 10$^{-3}$, a chemical composition of 2.5 times the solar metallicity and a log $U$ $\sim$ -3.3, we found [N II]/H$\alpha$ $\sim$ 0.91, [S II]/H$\alpha$ $\sim$ 0.80, [O I]/H$\alpha$ $\sim$ 0.08 and [O III]/H$\beta$ $\sim$ 3.36. Only the [O III]/H$\beta$ ratio does not match the observations. This is probably related to the fact that we assumed a blackbody radiation instead of using a spectrum of a real pAGB population. 
\end{enumerate}

It is worth mentioning that several works in the literature have discussed the importance of HOLMES in very extended regions, i.e., in scales of kpc \citep{2010MNRAS.402.2187S,2011MNRAS.413.1687C,2012ApJ...747...61Y,2013A&A...558A..43S}. Given the superb spatial resolution of the GMOS data cubes, we are able to propose that it is very likely that pAGB stars are also very important in spatial scales of $\sim$ 100 pc. Furthermore, it is important to emphasize that the greatest contributions of HOLMES is along the disc-like structure. This discovery was only made possible by the use of data cubes. 

Below, we summarize the main findings of this work:

\begin{itemize}
  \item We confirm the presence of kinematic bipolar structures in the circumnuclear region of the eight galaxies of the sample where PCA Tomography revealed this feature (paper I). When comparing the radial velocity maps to the tomograms shown in paper I, the direction of the gas kinematics as well as the P.A. of the structures are in agreement within the errors.
	\item In three galaxies, we detected pure gaseous discs in their circumnuclear regions. In other two objects, gaseous discs are probably present, but are probably affected by non-Keplerian motions (e.g. outflows). 
	\item In IC 5181, we detected a spiral structure of gas in the very central region of the galaxy. This may be the result of an outflow combined with a gaseous disc or even that the gas kinematics is the consequence of non-axisymetric potentials.
	\item In six galaxies, we detected low-velocity emission, perpendicular to the gaseous discs. 
	\item We detected an ionization bicone along the circumnuclear region of NGC 3136. We also detected five compact objects in this galaxy. They all have LINER-like emission. Although we did not ascertain the nature of the photoionization source of these objects, it is possible that they are associated with shocks (SNR or superwinds). 
	\item We propose that the circumnuclear emission along the disc-like structures are not photoionized by the central AGN. The best candidate for the photoionization source is hot, old stellar population (e.g. pAGB stars). 
	\item Along the low-velocity emission, the emission seems to be consistent with photoionization by AGNs.
	\item We propose a scheme for LINER-like circumnuclear regions where a low-velocity ionization cone, perpendicular to the disc, is formed by the collimation of the ionizing photons by a collimating agent aligned with the gaseous disc. 
\end{itemize}

\section*{Acknowledgements}

Some of the data presented in this paper were obtained from the Mikulski Archive for Space Telescopes (MAST). STScI is operated by the Association of Universities for Research in Astronomy, Inc., under NASA contract NAS5-26555. Support for MAST for non-\textit{HST} data is provided by the NASA Office of Space Science via grant NNX09AF08G and by other grants and contracts. This publication makes use of data products from the Two Micron All Sky Survey, which is a joint project of the University of Massachusetts and the Infrared Processing and Analysis Center/California Institute of Technology, funded by the National Aeronautics and Space Administration and the National Science Foundation.  This paper is based on observations obtained at the Gemini Observatory, which is operated by the Association of Universities for Research in Astronomy, Inc., under a cooperative agreement with the NSF on behalf of the Gemini partnership: the National Science Foundation (United States), the National Research Council (Canada), CONICYT (Chile), the Australian Research Council (Australia), Minist\'{e}rio da Ci\^{e}ncia, Tecnologia e Inova\c{c}\~{a}o (Brazil) and Ministerio de Ciencia, Tecnolog\'{i}a e Innovaci\'{o}n Productiva (Argentina). IRAF is distributed by the National Optical Astronomy Observatory, which is operated by the Association of Universities for Research in Astronomy (AURA) under cooperative agreement with the National Science Foundation. This research has made use of the NASA/IPAC Extragalactic Database (NED), which is operated by the Jet Propulsion Laboratory, California Institute of Technology, under contract with the National Aeronautics and Space Administration.

T.V.R, J.E.S and R.B.M. also acknowledge FAPESP for the financial support under grants 2008/06988-0 (T.V.R.), 2012/21350-7 (T.V.R.), 2011/51680-6 (J.E.S.) and 2012/02262-8 (R.B.M.). We also thank the anonymous referee for valuable suggestions that improved the quality of this paper. We thank Robert Proctor for carefully revising the manuscript.

\bibliographystyle{mn2e}
\bibliography{bibliografia}

\begin{thebibliography}{}

\bibitem[\protect\citeauthoryear{{Antonucci}}{{Antonucci}}{1993}]{1993ARA&A..3%
1..473A}
{Antonucci} R.,  1993, \araa, 31, 473

\bibitem[\protect\citeauthoryear{{Bacon}, {Copin}, {Monnet}, {Miller},
  {Allington-Smith}, {Bureau}, {Carollo}, {Davies}, {Emsellem}, {Kuntschner},
  {Peletier}, {Verolme} \& {de Zeeuw}}{{Bacon}
  et~al.}{2001}]{2001MNRAS.326...23B}
{Bacon} R.,  {Copin} Y.,  {Monnet} G.,  {Miller} B.~W.,  {Allington-Smith}
  J.~R.,  {Bureau} M.,  {Carollo} C.~M.,  {Davies} R.~L.,  {Emsellem} E.,
  {Kuntschner} H.,  {Peletier} R.~F.,  {Verolme} E.~K.,    {de Zeeuw} P.~T.,
  2001, \mnras, 326, 23

\bibitem[\protect\citeauthoryear{{Baldwin}, {Phillips} \&
  {Terlevich}}{{Baldwin} et~al.}{1981}]{1981PASP...93....5B}
{Baldwin} J.~A.,  {Phillips} M.~M.,    {Terlevich} R.,  1981, \pasp, 93, 5

\bibitem[\protect\citeauthoryear{{Beifiori}, {Maraston}, {Thomas} \&
  {Johansson}}{{Beifiori} et~al.}{2011}]{2011A&A...531A.109B}
{Beifiori} A.,  {Maraston} C.,  {Thomas} D.,    {Johansson} J.,  2011, \aap,
  531, A109

\bibitem[\protect\citeauthoryear{{Binette}, {Magris}, {Stasi{\'n}ska} \&
  {Bruzual}}{{Binette} et~al.}{1994}]{1994A&A...292...13B}
{Binette} L.,  {Magris} C.~G.,  {Stasi{\'n}ska} G.,    {Bruzual} A.~G.,  1994,
  \aap, 292, 13

\bibitem[\protect\citeauthoryear{{Caldwell}, {Kirshner} \&
  {Richstone}}{{Caldwell} et~al.}{1986}]{1986ApJ...305..136C}
{Caldwell} N.,  {Kirshner} R.~P.,    {Richstone} D.~O.,  1986, \apj, 305, 136

\bibitem[\protect\citeauthoryear{{Cappellari}, {Verolme}, {van der Marel},
  {Kleijn}, {Illingworth}, {Franx}, {Carollo} \& {de Zeeuw}}{{Cappellari}
  et~al.}{2002}]{2002ApJ...578..787C}
{Cappellari} M.,  {Verolme} E.~K.,  {van der Marel} R.~P.,  {Kleijn} G.~A.~V.,
  {Illingworth} G.~D.,  {Franx} M.,  {Carollo} C.~M.,    {de Zeeuw} P.~T.,
  2002, \apj, 578, 787

\bibitem[\protect\citeauthoryear{{Cardelli}, {Clayton} \& {Mathis}}{{Cardelli}
  et~al.}{1989}]{1989ApJ...345..245C}
{Cardelli} J.~A.,  {Clayton} G.~C.,    {Mathis} J.~S.,  1989, \apj, 345, 245

\bibitem[\protect\citeauthoryear{{Carollo}, {Franx}, {Illingworth} \&
  {Forbes}}{{Carollo} et~al.}{1997}]{1997ApJ...481..710C}
{Carollo} C.~M.,  {Franx} M.,  {Illingworth} G.~D.,    {Forbes} D.~A.,  1997,
  \apj, 481, 710

\bibitem[\protect\citeauthoryear{{Cid Fernandes}, {Mateus}, {Sodr{\'e}},
  {Stasi{\'n}ska} \& {Gomes}}{{Cid Fernandes}
  et~al.}{2005}]{2005MNRAS.358..363C}
{Cid Fernandes} R.,  {Mateus} A.,  {Sodr{\'e}} L.,  {Stasi{\'n}ska} G.,
  {Gomes} J.~M.,  2005, \mnras, 358, 363

\bibitem[\protect\citeauthoryear{{Cid Fernandes}, {Stasi{\'n}ska}, {Mateus} \&
  {Vale Asari}}{{Cid Fernandes} et~al.}{2011}]{2011MNRAS.413.1687C}
{Cid Fernandes} R.,  {Stasi{\'n}ska} G.,  {Mateus} A.,    {Vale Asari} N.,
  2011, \mnras, 413, 1687

\bibitem[\protect\citeauthoryear{{Coelho}, {Bruzual}, {Charlot}, {Weiss},
  {Barbuy} \& {Ferguson}}{{Coelho} et~al.}{2007}]{2007MNRAS.382..498C}
{Coelho} P.,  {Bruzual} G.,  {Charlot} S.,  {Weiss} A.,  {Barbuy} B.,
  {Ferguson} J.~W.,  2007, \mnras, 382, 498

\bibitem[\protect\citeauthoryear{{De Robertis}, {Dufour} \& {Hunt}}{{De
  Robertis} et~al.}{1987}]{1987JRASC..81..195D}
{De Robertis} M.~M.,  {Dufour} R.~J.,    {Hunt} R.~W.,  1987, \jrasc, 81, 195

\bibitem[\protect\citeauthoryear{{Dopita} \& {Sutherland}}{{Dopita} \&
  {Sutherland}}{2003}]{2003adu..book.....D}
{Dopita} M.~A.,  {Sutherland} R.~S.,  2003, {Astrophysics of the diffuse
  universe}

\bibitem[\protect\citeauthoryear{{Elitzur} \& {Ho}}{{Elitzur} \&
  {Ho}}{2009}]{2009ApJ...701L..91E}
{Elitzur} M.,  {Ho} L.~C.,  2009, \apjl, 701, L91

\bibitem[\protect\citeauthoryear{{Elitzur} \& {Shlosman}}{{Elitzur} \&
  {Shlosman}}{2006}]{2006ApJ...648L.101E}
{Elitzur} M.,  {Shlosman} I.,  2006, \apjl, 648, L101

\bibitem[\protect\citeauthoryear{{Eracleous}, {Hwang} \& {Flohic}}{{Eracleous}
  et~al.}{2010}]{2010ApJS..187..135E}
{Eracleous} M.,  {Hwang} J.~A.,    {Flohic} H.~M.~L.~G.,  2010, \apjs, 187, 135

\bibitem[\protect\citeauthoryear{{Falc{\'o}n-Barroso},
  {S{\'a}nchez-Bl{\'a}zquez}, {Vazdekis}, {Ricciardelli}, {Cardiel}, {Cenarro},
  {Gorgas} \& {Peletier}}{{Falc{\'o}n-Barroso}
  et~al.}{2011}]{2011A&A...532A..95F}
{Falc{\'o}n-Barroso} J.,  {S{\'a}nchez-Bl{\'a}zquez} P.,  {Vazdekis} A.,
  {Ricciardelli} E.,  {Cardiel} N.,  {Cenarro} A.~J.,  {Gorgas} J.,
  {Peletier} R.~F.,  2011, \aap, 532, A95

\bibitem[\protect\citeauthoryear{{Ferland}, {Korista}, {Verner}, {Ferguson},
  {Kingdon} \& {Verner}}{{Ferland} et~al.}{1998}]{1998PASP..110..761F}
{Ferland} G.~J.,  {Korista} K.~T.,  {Verner} D.~A.,  {Ferguson} J.~W.,
  {Kingdon} J.~B.,    {Verner} E.~M.,  1998, \pasp, 110, 761

\bibitem[\protect\citeauthoryear{{Ferland} \& {Netzer}}{{Ferland} \&
  {Netzer}}{1983}]{1983ApJ...264..105F}
{Ferland} G.~J.,  {Netzer} H.,  1983, \apj, 264, 105

\bibitem[\protect\citeauthoryear{{Filippenko} \& {Sargent}}{{Filippenko} \&
  {Sargent}}{1985}]{1985ApJS...57..503F}
{Filippenko} A.~V.,  {Sargent} W.~L.~W.,  1985, \apjs, 57, 503

\bibitem[\protect\citeauthoryear{{Filippenko} \& {Sargent}}{{Filippenko} \&
  {Sargent}}{1988}]{1988ApJ...324..134F}
{Filippenko} A.~V.,  {Sargent} W.~L.~W.,  1988, \apj, 324, 134

\bibitem[\protect\citeauthoryear{{Galletta}}{{Galletta}}{1987}]{1987ApJ...318.%
.531G}
{Galletta} G.,  1987, \apj, 318, 531

\bibitem[\protect\citeauthoryear{{Goudfrooij}, {Hansen}, {Jorgensen} \&
  {Norgaard-Nielsen}}{{Goudfrooij} et~al.}{1994}]{1994A&AS..105..341G}
{Goudfrooij} P.,  {Hansen} L.,  {Jorgensen} H.~E.,    {Norgaard-Nielsen} H.~U.,
   1994, \aaps, 105, 341

\bibitem[\protect\citeauthoryear{{Grier}, {Mathur}, {Ghosh} \&
  {Ferrarese}}{{Grier} et~al.}{2011}]{2011ApJ...731...60G}
{Grier} C.~J.,  {Mathur} S.,  {Ghosh} H.,    {Ferrarese} L.,  2011, \apj, 731,
  60

\bibitem[\protect\citeauthoryear{{Halpern} \& {Steiner}}{{Halpern} \&
  {Steiner}}{1983}]{1983ApJ...269L..37H}
{Halpern} J.~P.,  {Steiner} J.~E.,  1983, \apjl, 269, L37

\bibitem[\protect\citeauthoryear{{Heckman}}{{Heckman}}{1980}]{1980A&A....87..1%
52H}
{Heckman} T.~M.,  1980, \aap, 87, 152

\bibitem[\protect\citeauthoryear{{Heyer} \& {Schloerb}}{{Heyer} \&
  {Schloerb}}{1997}]{1997ApJ...475..173H}
{Heyer} M.~H.,  {Schloerb} F.~P.,  1997, \apj, 475, 173

\bibitem[\protect\citeauthoryear{{Ho}}{{Ho}}{2008}]{2008ARA&A..46..475H}
{Ho} L.~C.,  2008, \araa, 46, 475

\bibitem[\protect\citeauthoryear{{Ho}, {Filippenko} \& {Sargent}}{{Ho}
  et~al.}{1996}]{1996ApJ...462..183H}
{Ho} L.~C.,  {Filippenko} A.~V.,    {Sargent} W.~L.~W.,  1996, \apj, 462, 183

\bibitem[\protect\citeauthoryear{{Ho}, {Filippenko} \& {Sargent}}{{Ho}
  et~al.}{1997}]{1997ApJS..112..315H}
{Ho} L.~C.,  {Filippenko} A.~V.,    {Sargent} W.~L.~W.,  1997, \apjs, 112, 315

\bibitem[\protect\citeauthoryear{{Ho}, {Filippenko}, {Sargent} \& {Peng}}{{Ho}
  et~al.}{1997}]{1997ApJS..112..391H}
{Ho} L.~C.,  {Filippenko} A.~V.,  {Sargent} W.~L.~W.,    {Peng} C.~Y.,  1997,
  \apjs, 112, 391

\bibitem[\protect\citeauthoryear{{Kauffmann}, {Heckman}, {Tremonti},
  {Brinchmann}, {Charlot}, {White}, {Ridgway}, {Brinkmann}, {Fukugita}, {Hall},
  {Ivezi{\'c}}, {Richards} \& {Schneider}}{{Kauffmann}
  et~al.}{2003}]{2003MNRAS.346.1055K}
{Kauffmann} G.,  {Heckman} T.~M.,  {Tremonti} C.,  {Brinchmann} J.,  {Charlot}
  S.,  {White} S.~D.~M.,  {Ridgway} S.~E.,  {Brinkmann} J.,  {Fukugita} M.,
  {Hall} P.~B.,  {Ivezi{\'c}} {\v Z}.,  {Richards} G.~T.,    {Schneider} D.~P.,
   2003, \mnras, 346, 1055

\bibitem[\protect\citeauthoryear{{Kewley}, {Dopita}, {Sutherland}, {Heisler} \&
  {Trevena}}{{Kewley} et~al.}{2001}]{2001ApJ...556..121K}
{Kewley} L.~J.,  {Dopita} M.~A.,  {Sutherland} R.~S.,  {Heisler} C.~A.,
  {Trevena} J.,  2001, \apj, 556, 121

\bibitem[\protect\citeauthoryear{{Kewley}, {Groves}, {Kauffmann} \&
  {Heckman}}{{Kewley} et~al.}{2006}]{2006MNRAS.372..961K}
{Kewley} L.~J.,  {Groves} B.,  {Kauffmann} G.,    {Heckman} T.,  2006, \mnras,
  372, 961

\bibitem[\protect\citeauthoryear{{Kim}}{{Kim}}{1989}]{1989ApJ...346..653K}
{Kim} D.-W.,  1989, \apj, 346, 653

\bibitem[\protect\citeauthoryear{{Lucy}}{{Lucy}}{1974}]{1974AJ.....79..745L}
{Lucy} L.~B.,  1974, \aj, 79, 745

\bibitem[\protect\citeauthoryear{{Nenkova}, {Sirocky}, {Nikutta}, {Ivezi{\'c}}
  \& {Elitzur}}{{Nenkova} et~al.}{2008}]{2008ApJ...685..160N}
{Nenkova} M.,  {Sirocky} M.~M.,  {Nikutta} R.,  {Ivezi{\'c}} {\v Z}.,
  {Elitzur} M.,  2008, \apj, 685, 160

\bibitem[\protect\citeauthoryear{{O'Connell}, {Martin}, {Crane}, {Burstein},
  {Bohlin}, {Landsman}, {Freedman} \& {Rood}}{{O'Connell}
  et~al.}{2005}]{2005ApJ...635..305O}
{O'Connell} R.~W.,  {Martin} J.~R.,  {Crane} J.~D.,  {Burstein} D.,  {Bohlin}
  R.~C.,  {Landsman} W.~B.,  {Freedman} I.,    {Rood} R.~T.,  2005, \apj, 635,
  305

\bibitem[\protect\citeauthoryear{{Osterbrock} \& {Ferland}}{{Osterbrock} \&
  {Ferland}}{2006}]{2006agna.book.....O}
{Osterbrock} D.~E.,  {Ferland} G.~J.,  2006, {Astrophysics of gaseous nebulae
  and active galactic nuclei}.
2nd.~ed.~by D.E.~Osterbrock and G.J.~Ferland.~Sausalito, CA: University Science
  Books, 2006

\bibitem[\protect\citeauthoryear{{Papaderos}, {Gomes}, {V{\'{\i}}lchez} \& {et
  al.}}{{Papaderos} et~al.}{2013}]{2013A&A...555L...1P}
{Papaderos} P.,  {Gomes} J.~M.,  {V{\'{\i}}lchez} J.~M.,    {et al.} 2013,
  \aap, 555, L1

\bibitem[\protect\citeauthoryear{{Phillips}, {Jenkins}, {Dopita}, {Sadler} \&
  {Binette}}{{Phillips} et~al.}{1986}]{1986AJ.....91.1062P}
{Phillips} M.~M.,  {Jenkins} C.~R.,  {Dopita} M.~A.,  {Sadler} E.~M.,
  {Binette} L.,  1986, \aj, 91, 1062

\bibitem[\protect\citeauthoryear{{Pizzella}, {Morelli}, {Corsini}, {Dalla
  Bont{\`a}} \& {Cesetti}}{{Pizzella} et~al.}{2013}]{2013A&A...560A..14P}
{Pizzella} A.,  {Morelli} L.,  {Corsini} E.~M.,  {Dalla Bont{\`a}} E.,
  {Cesetti} M.,  2013, \aap, 560, A14

\bibitem[\protect\citeauthoryear{{Pogge}, {Maoz}, {Ho} \& {Eracleous}}{{Pogge}
  et~al.}{2000}]{2000ApJ...532..323P}
{Pogge} R.~W.,  {Maoz} D.,  {Ho} L.~C.,    {Eracleous} M.,  2000, \apj, 532,
  323

\bibitem[\protect\citeauthoryear{{Proxauf}, {{\"O}ttl} \&
  {Kimeswenger}}{{Proxauf} et~al.}{2014}]{2014A&A...561A..10P}
{Proxauf} B.,  {{\"O}ttl} S.,    {Kimeswenger} S.,  2014, \aap, 561, A10

\bibitem[\protect\citeauthoryear{{Ricci}}{{Ricci}}{2013}]{ricci2013}
{Ricci} T.~V.,  2013, PhD thesis, Universidade de S\~ao Paulo

\bibitem[\protect\citeauthoryear{{Ricci}, {Steiner} \& {Menezes}}{{Ricci}
  et~al.}{2011}]{2011ApJ...734L..10R}
{Ricci} T.~V.,  {Steiner} J.~E.,    {Menezes} R.~B.,  2011, \apjl, 734, L10+

\bibitem[\protect\citeauthoryear{{Ricci}, {Steiner} \& {Menezes}}{{Ricci}
  et~al.}{2014a}]{2014MNRAS.440.2442R}
{Ricci} T.~V.,  {Steiner} J.~E.,    {Menezes} R.~B.,  2014a, \mnras, 440, 2442

\bibitem[\protect\citeauthoryear{{Ricci}, {Steiner} \& {Menezes}}{{Ricci}
  et~al.}{2014b}]{2014MNRAS.440.2419R}
{Ricci} T.~V.,  {Steiner} J.~E.,    {Menezes} R.~B.,  2014b, \mnras, 440, 2419

\bibitem[\protect\citeauthoryear{{Richardson}}{{Richardson}}{1972}]{1972JOSA..%
.62...55R}
{Richardson} W.~H.,  1972, Journal of the Optical Society of America
  (1917-1983), 62, 55

\bibitem[\protect\citeauthoryear{{S{\'a}nchez}, {Kennicutt}, {Gil de Paz} \&
  {et al.}}{{S{\'a}nchez} et~al.}{2012}]{2012A&A...538A...8S}
{S{\'a}nchez} S.~F.,  {Kennicutt} R.~C.,  {Gil de Paz} A.,    {et al.} 2012,
  \aap, 538, A8

\bibitem[\protect\citeauthoryear{{Sarzi}, {Falc{\'o}n-Barroso}, {Davies},
  {Bacon}, {Bureau}, {Cappellari}, {de Zeeuw}, {Emsellem}, {Fathi},
  {Krajnovi{\'c}}, {Kuntschner}, {McDermid} \& {Peletier}}{{Sarzi}
  et~al.}{2006}]{2006MNRAS.366.1151S}
{Sarzi} M.,  {Falc{\'o}n-Barroso} J.,  {Davies} R.~L.,  {Bacon} R.,  {Bureau}
  M.,  {Cappellari} M.,  {de Zeeuw} P.~T.,  {Emsellem} E.,  {Fathi} K.,
  {Krajnovi{\'c}} D.,  {Kuntschner} H.,  {McDermid} R.~M.,    {Peletier} R.~F.,
   2006, \mnras, 366, 1151

\bibitem[\protect\citeauthoryear{{Sarzi}, {Shields}, {Schawinski} \& {et
  al.}}{{Sarzi} et~al.}{2010}]{2010MNRAS.402.2187S}
{Sarzi} M.,  {Shields} J.~C.,  {Schawinski} K.,    {et al.} 2010, \mnras, 402,
  2187

\bibitem[\protect\citeauthoryear{{Shaw} \& {Dufour}}{{Shaw} \&
  {Dufour}}{1995}]{1995PASP..107..896S}
{Shaw} R.~A.,  {Dufour} R.~J.,  1995, \pasp, 107, 896

\bibitem[\protect\citeauthoryear{{Shields}, {Rix}, {Sarzi}, {Barth},
  {Filippenko}, {Ho}, {McIntosh}, {Rudnick} \& {Sargent}}{{Shields}
  et~al.}{2007}]{2007ApJ...654..125S}
{Shields} J.~C.,  {Rix} H.-W.,  {Sarzi} M.,  {Barth} A.~J.,  {Filippenko}
  A.~V.,  {Ho} L.~C.,  {McIntosh} D.~H.,  {Rudnick} G.,    {Sargent} W.~L.~W.,
  2007, \apj, 654, 125

\bibitem[\protect\citeauthoryear{{Shurkin}, {Dunn}, {Gentile}, {Taylor} \&
  {Allen}}{{Shurkin} et~al.}{2008}]{2008MNRAS.383..923S}
{Shurkin} K.,  {Dunn} R.~J.~H.,  {Gentile} G.,  {Taylor} G.~B.,    {Allen}
  S.~W.,  2008, \mnras, 383, 923

\bibitem[\protect\citeauthoryear{{Singh}, {van de Ven}, {Jahnke} \& {et
  al.}}{{Singh} et~al.}{2013}]{2013A&A...558A..43S}
{Singh} R.,  {van de Ven} G.,  {Jahnke} K.,    {et al.} 2013, \aap, 558, A43

\bibitem[\protect\citeauthoryear{{Skrutskie}, {Cutri}, {Stiening} \& {et
  al.}}{{Skrutskie} et~al.}{2006}]{2006AJ....131.1163S}
{Skrutskie} M.~F.,  {Cutri} R.~M.,  {Stiening}   {et al.} 2006, \aj, 131, 1163

\bibitem[\protect\citeauthoryear{{Starck} \& {Murtagh}}{{Starck} \&
  {Murtagh}}{2002}]{2002aida.book.....S}
{Starck} J.-L.,  {Murtagh} F.,  2002, {Astronomical image and data analysis}

\bibitem[\protect\citeauthoryear{{Stasi{\'n}ska}, {Vale Asari}, {Cid
  Fernandes}, {Gomes}, {Schlickmann}, {Mateus}, {Schoenell}, {Sodr{\'e}} Jr. \&
  {Seagal Collaboration}}{{Stasi{\'n}ska} et~al.}{2008}]{2008MNRAS.391L..29S}
{Stasi{\'n}ska} G.,  {Vale Asari} N.,  {Cid Fernandes} R.,  {Gomes} J.~M.,
  {Schlickmann} M.,  {Mateus} A.,  {Schoenell} W.,  {Sodr{\'e}} Jr. L.,
  {Seagal Collaboration} 2008, \mnras, 391, L29

\bibitem[\protect\citeauthoryear{{Steiner}, {Menezes}, {Ricci} \&
  {Oliveira}}{{Steiner} et~al.}{2009}]{2009MNRAS.395...64S}
{Steiner} J.~E.,  {Menezes} R.~B.,  {Ricci} T.~V.,    {Oliveira} A.~S.,  2009,
  \mnras, 395, 64

\bibitem[\protect\citeauthoryear{{Su} \& {Irwin}}{{Su} \&
  {Irwin}}{2013}]{2013ApJ...766...61S}
{Su} Y.,  {Irwin} J.~A.,  2013, \apj, 766, 61

\bibitem[\protect\citeauthoryear{{Vazdekis}, {S{\'a}nchez-Bl{\'a}zquez},
  {Falc{\'o}n-Barroso}, {Cenarro}, {Beasley}, {Cardiel}, {Gorgas} \&
  {Peletier}}{{Vazdekis} et~al.}{2010}]{2010MNRAS.404.1639V}
{Vazdekis} A.,  {S{\'a}nchez-Bl{\'a}zquez} P.,  {Falc{\'o}n-Barroso} J.,
  {Cenarro} A.~J.,  {Beasley} M.~A.,  {Cardiel} N.,  {Gorgas} J.,    {Peletier}
  R.~F.,  2010, \mnras, 404, 1639

\bibitem[\protect\citeauthoryear{{Veilleux} \& {Osterbrock}}{{Veilleux} \&
  {Osterbrock}}{1987}]{1987ApJS...63..295V}
{Veilleux} S.,  {Osterbrock} D.~E.,  1987, \apjs, 63, 295

\bibitem[\protect\citeauthoryear{{Walcher}, {Coelho}, {Gallazzi} \&
  {Charlot}}{{Walcher} et~al.}{2009}]{2009MNRAS.398L..44W}
{Walcher} C.~J.,  {Coelho} P.,  {Gallazzi} A.,    {Charlot} S.,  2009, \mnras,
  398, L44

\bibitem[\protect\citeauthoryear{{Walsh}, {Barth}, {Ho}, {Filippenko}, {Rix},
  {Shields}, {Sarzi} \& {Sargent}}{{Walsh} et~al.}{2008}]{2008AJ....136.1677W}
{Walsh} J.~L.,  {Barth} A.~J.,  {Ho} L.~C.,  {Filippenko} A.~V.,  {Rix} H.-W.,
  {Shields} J.~C.,  {Sarzi} M.,    {Sargent} W.~L.~W.,  2008, \aj, 136, 1677

\bibitem[\protect\citeauthoryear{{Yan} \& {Blanton}}{{Yan} \&
  {Blanton}}{2012}]{2012ApJ...747...61Y}
{Yan} R.,  {Blanton} M.~R.,  2012, \apj, 747, 61

\end{thebibliography}

\appendix

\section{Comments on individual objects} \label{comments_on_objects}

Here, we summarize the main results from this work and also from papers I and II for each galaxy of the sample. Only for NGC 3136 a more detailed analysis is presented below. It is worth mentioning that the scenario proposed for the emission of ionized gas from the circumnuclear region discussed in Section \ref{sec:conclusions} is common for all sample galaxies, except for NGC 1399, NGC 1404 and NGC 2663. 

\subsection{ESO 208 G-21} \label{208g21_comments}

Both nuclear and circumnuclear spectra have emission lines typical of LINERs. An AGN is responsible for the nuclear emission, since a broad component is seen in the H$\alpha$ line (paper II). The circumnuclear gas component is probably a pure gaseous disc, as suggested by the radial velocity map (Section \ref{kinematics_extended_emission}) and also by the fact that this gaseous structure is corotating with the stellar component (paper I). We also detected a reddening of the stellar component in the nuclear region caused by the interstellar medium (ISM) of this object (paper I). A peculiarity of ESO 208 G-21 is the fact that the density map reveals two structures that are not related to the AGN. However, a deep study of the nature of these two structures is beyond the scope of this paper.

\subsection{IC 1459} \label{ic1459_comments}

IC 1459 has a well-known LINER-like AGN (see references cited in paper I), thus this object was used to validate the results obtained with PCA Tomography in paper I. We confirmed the presence of the AGN, since a broad component in the H$\alpha$ line is present in the nuclear spectrum (paper II). Furthermore, we detected a featureless continuum from the AGN with PCA Tomography (paper I). The gaseous component in the circumnuclear region also has a LINER-like spectrum. If it is a gaseous disc, then the kinematics are affected by non-Keplerian motions (e.g. outflows, see Section \ref{kinematics_extended_emission}) and it is counterrotating with respect to the stellar component (paper I). \citet{2002ApJ...578..787C}, with long-slit data, reached the same conclusions for the kinematics of the circumnuclear region.

\subsection{IC 5181} \label{ic5181_comments}

This object also has LINER-like spectra for both nuclear (paper II) and circumnuclear regions. A broad component is seen in the H$\alpha$ line (paper II). The gaseous component in the circumnuclear region has a spiral form, similar to an integral sign. This result may be related to an outflow in the nuclear region or to the fact that the gas is moving through a non-axisymmetric potential, which causes a gas concentration in certain regions (see Section \ref{discs_or_cones}). It is worth mentioning that the kinematics associated with this gaseous structure are perpendicular to the kinematics related to the stellar component (paper I). The same finding was also reported by \citet{2013A&A...560A..14P}. A reddening of the stellar component in the nuclear region, caused by the ISM of IC 5181, was also detected (paper I).

\subsection{NGC 1380} \label{ngc1380_comments}

Both the nuclear and circumnuclear regions have LINER-like spectra. Although we did not detect a broad component in any permitted lines from the nuclear spectrum, the HST image shown in Fig. \ref{HST_science} reveals three compact sources in the central region of this galaxy; one of the objects is located at the centre of the bulge and the others are separated by 0.2 and 0.7 arcsec from this central source. The EW([N II]) map in Fig. \ref{mapa_ew_gal_1} shows an extended structure that may be related to these compact objects, which indicates that they probably have LINER-like spectra. A deep analysis of these three structures will be performed in a future work, although we believe that the central object is probably related to an AGN. We also detected two H II regions near the centre of NGC 1380, seen in the EW(H$\alpha$) map (Fig. \ref{mapa_ew_gal_1}: see the discussion in Section \ref{EW_extended_emission}). A pure gaseous disc is present in NGC 1380 and it corotates with the stellar component (paper I). The reddening of the stellar component in the nucleus, caused by the ISM of this object, was also detected (paper I).

\subsection{NGC 1399} \label{ngc1399_comments}

This galaxy has a weak emission of ionized gas in the nuclear region (paper II). \citet{2005ApJ...635..305O} detected evidence of an AGN in a far-UV spectrum taken with HST. \citet{2008MNRAS.383..923S} showed that jets originated in an AGN created a cavity in the hot gas from the central region of NGC 1399. Thus, the emission lines detected by us in paper II are probably related to this AGN. In paper II, we showed that most of the energy released by the AGN is mechanical. No circumnuclear emission of ionized gas was detected. Also, we did not detect any kinematics related to the stellar component. 

\subsection{NGC 1404} \label{ngc1404_comments}

This is the only object where an AGN was not previously mentioned in the literature, although \citet{2011ApJ...731...60G} detected a point-like x-ray source with $L_{0.3-8keV} \sim 10^{40}$ erg s$^{-1}$. They mentioned that this measurement is probable associated with a low-luminosity AGN. We detected the H$\alpha$ and [N II] emission lines in the nuclear spectrum. The bolometric luminosity estimated for this AGN, using the H$\alpha$ line, matches the X-ray luminosity (paper II). We propose that NGC 1404 has a LINER-2 AGN. No circumnuclear emission of ionized gas is seen in this object. We detected a rotation of the stellar component with PCA Tomography (paper I). The reddening caused by the ISM was detected in a V - I colour image from the HST and also in eigenvector 2 of PCA Tomography applied to the 4325-6800 \AA\ spectral range.

\subsection{NGC 2663} \label{ngc2663_comments}

The gas emission detected in NGC 2663 is restricted to the nuclear region. It has a LINER-like spectrum with a broad component in the H$\alpha$ line. Although the emission is unresolved, a kinematic signature of the gaseous component is detected, but it is probably related to an outflow rather than a disc. No additional ionization source is needed to explain the gas emission of NGC 2663. When it comes to the stellar structure, no kinematic signature and also no reddening caused by an ISM were detected.

\subsection{NGC 3136} \label{ngc3136_comments}

\begin{figure*}
	\centering
		\includegraphics[scale=0.5]{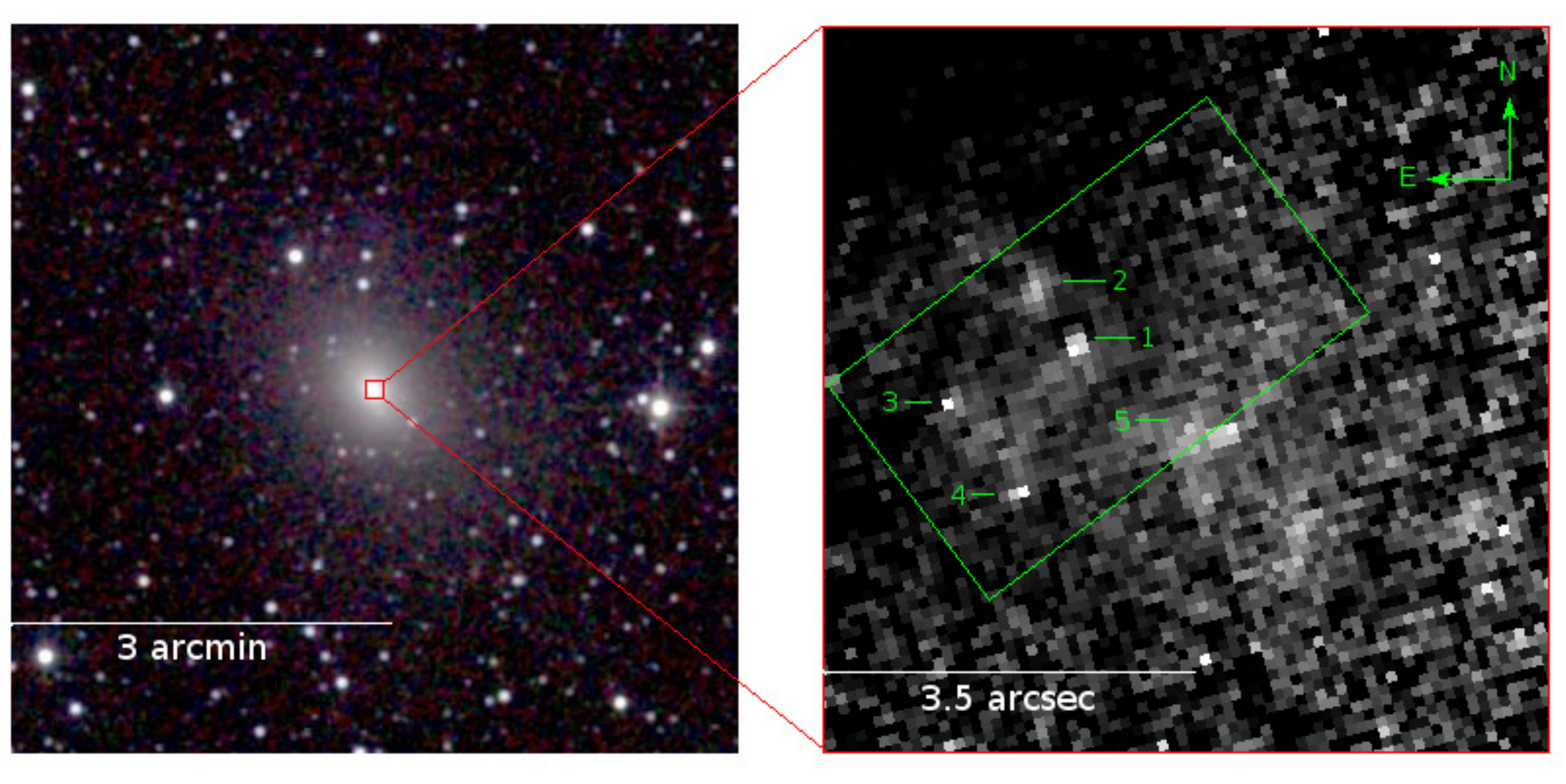}
		\caption{Left: Combined JHK infrared images of NGC 3136 obtained with 2MASS \citep{2006AJ....131.1163S}. Right: HST image of the (H$\alpha$+[N II])/I ratio. This is the same image shown in Fig. \ref{HST_science}, but with a FOV of 7 x 7 arcsec. The FOV of the data cube is delineated in green. The numbers correspond to the objects detected in the central region of NGC 3136. Object 1 is the central AGN proposed in papers I and II.  
	\label{fig:HST_2MASS_N3136_v2}}
\end{figure*}

Along this paper and also in paper I, we have proposed that NGC 3136 may contain other photoionization sources, in addition to the central AGN. PCA Tomography applied to the spectral region between 6250 and 6800 \AA\ of the data cube detected two compact objects (hereafter objects 1 and 2, see Fig. \ref{fig:HST_2MASS_N3136_v2}). One of them is redshifted and the other is blueshifted relative to the centre of NGC 3136. Moreover, both EW(H$\alpha$) and EW([N II]) maps revealed three extended structures, wherein the position of one of these structures corresponds to the objects 1 and 2. The structure southeastward from the nucleus is related to objects 3 and 4, while the structure southwestward from the nucleus was associated with object 5. In this section, we will try to ascertain the nature of these objects

We extracted spectra of all five objects assuming a Gaussian PSF with a FWHM equal to the seeing of the observation (paper I) and with an amplitude given by the flux in the spaxel where the respective object is centred. The coordinates relative to the central region of the FOV of each object is shown in Table \ref{tab_L_N3136}. In objects 2 and 4, the spectral regions corresponding to both [O I]$\lambda\lambda$6300, 6363 and H$\beta$ emission lines are dominated by noise. In object 3, we did not detect the [O I]$\lambda\lambda$6300, 6363 doublet. Where H$\beta$ is detected, the H$\alpha$ emission line luminosity shown in Table \ref{tab_L_N3136} was corrected for reddening effects as described in Section \ref{FOV_properties}. The emission lines fluxes are shown in Table \ref{medidas_fluxo_n3136}.

\begin{table}
 \scriptsize
 \begin{center}
  \caption{H$\alpha$ luminosities in erg s$^{-1}$ and the gas density in cm$^{-3}$ of the five objects detected in the central region of NGC 3136. The objects marked with a * did not have their H$\alpha$ luminosities corrected for reddening effects.  \label{tab_L_N3136}
}
 \begin{tabular}{@{}lccc}
  \hline
  Object & Spatial coordinates & log $L$(H$\alpha$)& $n_e$  \\
    \hline
  1 (nuc)& (0.0,-0.2) & 38.83$\pm$0.19 & 267 \\
  2 & (-1.0,0.0)&38.41$\pm$0.08* & 130\\
  3 & (-0.3,-1.4)&38.60$\pm$0.24 & $<$ 100\\
  4 & (0.2,-1.1)&38.34$\pm$0.08* & 120\\
  5 & (1.3,0.7)&37.92$\pm$0.34 & 140\\
  \hline
 \end{tabular}
 \end{center}

\end{table}

\begin{table*}
 \scriptsize
 \begin{center}
  \caption{Flux of the H$\alpha$ emission line in 10$^{-15}$ erg s$^{-1}$ cm$^{-2}$ and line rations for the five objects detected in NGC 3136. \label{medidas_fluxo_n3136}
}

 \begin{tabular}{@{}lccccccc}
  \hline
  Object & $f$(H$\alpha$)$_n$ & (H$\alpha$/H$\beta$)$_n$ & E(B-V) & [N II]/H$\alpha$ & [S II]/H$\alpha$ & [O I]/H$\alpha$ & [O III]/H$\beta$\\
  \hline

  1 (nuc) & 8.1$\pm$0.3 & 3.39$\pm$0.57&0.08$\pm$0.16&1.82$\pm$0.09&0.97$\pm$0.05&0.19$\pm$0.03&1.66$\pm$0.30 \\
  2  & 3.7$\pm$0.2&&&1.70$\pm$0.15&1.32$\pm$0.12&& \\
  3  & 2.7$\pm$0.2&4.27$\pm$0.92&0.31$\pm$0.21&1.57$\pm$0.17&1.16$\pm$0.12&&2.12$\pm$0.51\\
  4  & 3.2$\pm$0.2&&&1.49$\pm$0.15&1.03$\pm$0.09&& \\
  5 & 1.2$\pm$0.2&3.21$\pm$1.01&0.03$\pm$0.31&1.40$\pm$0.32&0.91$\pm$0.17&0.25$\pm$0.06&1.56$\pm$0.47 \\
  
  \hline
 \end{tabular}
 \end{center}
\end{table*}

\begin{figure*}
\begin{center}
\includegraphics[width=70mm,height=60mm]{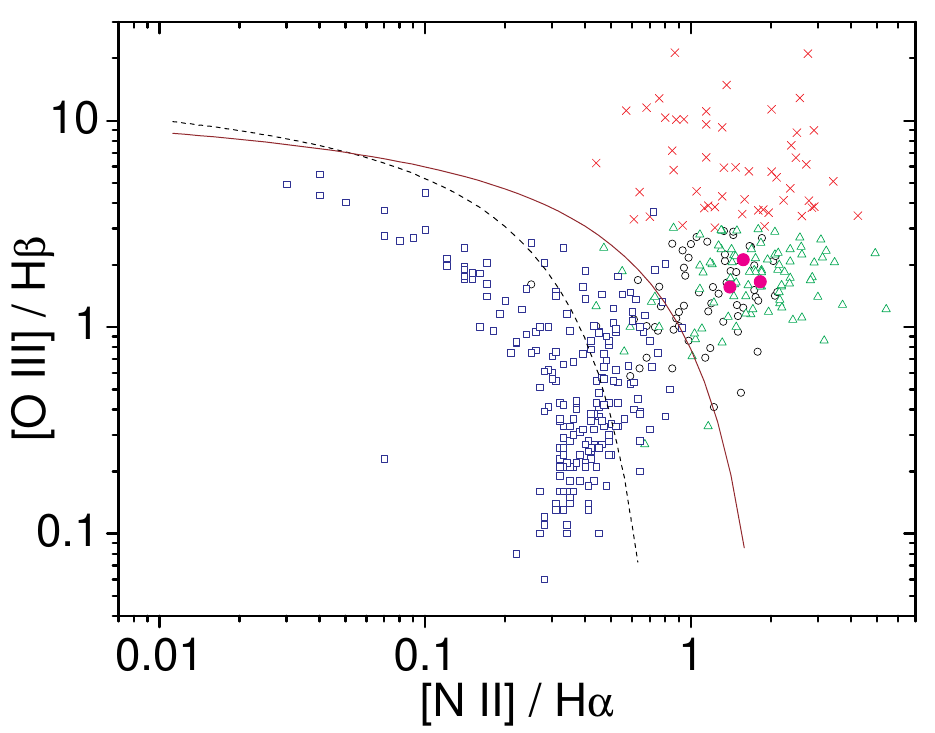}
\includegraphics[width=70mm,height=60mm]{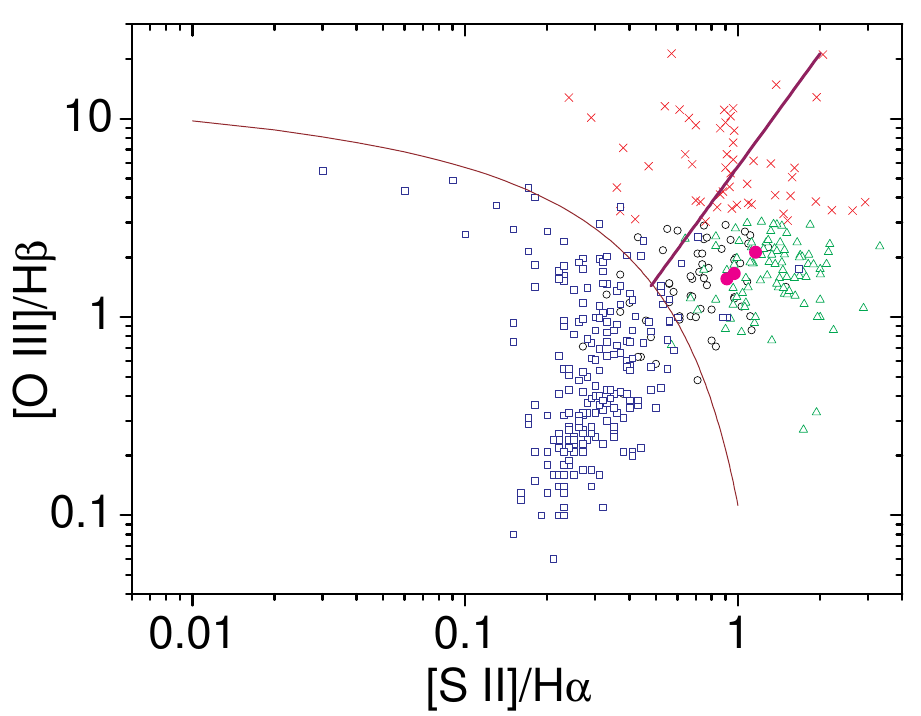}
\caption{BPT Diagrams with the results of the five objects found in the central region of NGC 3136. The line ratios are presented in Table \ref{medidas_fluxo_n3136}. According to the classification proposed by \citet{1997ApJS..112..315H}, the red crosses are Seyferts, green triangles are LINERs, hollow black circles are TOs and blue squares are H II regions. The thin brown line is the maximum starburst line proposed by \citet{2001ApJ...556..121K}, the dashed black line is the empirical division between H II regions and AGNs \citep{2003MNRAS.346.1055K} and the thick purple line is the LINER-Seyfert division suggested by \citet{2006MNRAS.372..961K}. All five objects are in the LINER region of the diagram.  \label{BPTn3136}
 }
\end{center}
\end{figure*}

Objects 1, 3 and 5 were plotted in BPT diagrams [N II]/H$\alpha$ versus [O III]/H$\beta$ and [S II]/H$\alpha$ versus [O III]/H$\beta$, which are shown in Fig. \ref{BPTn3136}. They may be classified as LINERs. Objects 2 and 4 did not have H$\beta$ detected, but their [S II]/H$\alpha$ and [N II]/H$\alpha$ are typical of LINERs. None of the objects have a broad H$\alpha$ component. However, the kinematic bipolar structure detected for this galaxy is probably associated with a bicone emerging from an AGN. Indeed, the V - I colour image taken with HST (see Fig. \ref{HST_science}) shows extinction in the region of the bicone that is redshifted relative to the centre of NGC 3136. This may indicate that the structure in redshift is the far side of the bicone, behind a dust screen. Object 1 is in the centre of NGC 3136 and is probably associated with the AGN of the galaxy. According to the PCA Tomography results (paper I), the H$\alpha$ + [N II] lines of object 2 are redshifted when compared to the same lines of object 1. One possible explanation is that both objects are LINER-like AGNs and they are rotating around their centre of mass. Another possibility is that objects 2, 3, 4 and 5 are associated with shocks. The presence of the bicone acting as the source of mechanical energy and the fact that shocks may produce LINER-like spectra \citep{1980A&A....87..152H,2003adu..book.....D,2008ARA&A..46..475H} make this hypothesis quite likely. H II regions may also play a role in the central region of NGC 3136. However, their [N II]/H$\alpha$ and [S II]/H$\alpha$ ratios are too high to be produced by young stellar populations, so one of the sources mentioned above (AGN or shocks) would have to coexist with this starburst regions in order to explain the high emission from low ionization elements.

\subsection{NGC 4546} \label{ngc4546_comments}

We detected emission of gas in both the nuclear and circumnuclear regions. The nuclear spectrum is in the border between LINERs and Seyferts and it also has a broad component in the H$\alpha$ line. The circumnuclear gas emission has a LINER-like spectrum. The structure of this gaseous component is probably related to a gaseous disc, but non-Keplerian motions are seen in the radial velocity map (Section \ref{kinematics_extended_emission}). We also detected a rotation of the stellar component which is perpendicular to the motion of the gaseous structure. \citet{2006MNRAS.366.1151S} found that both structures are counterrotating. However, they also reported that the gaseous component has a spiral form. Although this twist is located in a region beyond the FOV of our observations, our results are in accordance with those obtained by \citet{2006MNRAS.366.1151S}.

\subsection{NGC 7097} \label{ngc7097_comments} 

This galaxy also has LINER-like emission in both the nuclear and circumnuclear regions. In the nuclear spectrum, a broad component in the H$\alpha$ line is present. For the circumnuclear region, the gaseous structure is probably a pure gaseous disc, although it is counterrotating with respect to the stellar component. A reddening of the nuclear stars caused by the ISM was also detected. It is worth mentioning that the scenario for the circumnuclear gas emission proposed in this work (Section \ref{sec:conclusions}) was previously suggested for this object in \citet{2011ApJ...734L..10R}.

\end{document}